\documentclass[a4paper,11pt]{report}

\usepackage{graphicx}
\usepackage{natbib}
\usepackage{latexsym}
\usepackage{fancyhdr}
\usepackage{lscape}
\usepackage{rotating}
\usepackage{longtable}
\usepackage{subfigure}

\fancypagestyle{plain}{\pagestyle{fancy}}

\newcommand{\dif}{\mathrm{d}}

\begin{document} 

\pagestyle{empty}

\pagestyle{empty}

\begin{center}

\vspace{7cm}

{\textbf{\Huge{The Cluster and Large Scale Environments of Quasars \newline at z $<$ 0.9}}}

\vspace{3cm}

\huge{Kathryn Amy Harris}
\end{center}

\vspace{3cm}

\begin{center}
\Large{A thesis submitted in partial fulfilment for the requirements of the degree of Doctor of Philosophy at the University of Central Lancashire}
\end{center}

\vspace{3cm}

\begin{center}
\Large{September 2011}
\end{center}

\pagestyle{empty}

\begin{center}
\textbf{\large{Declaration}}
\end{center}

\vspace{2cm}
 I declare that while registered as a candidate for the research degree, I have not been a registered
candidate or enrolled student for another award of the University or other academic or professional
institution

 I declare that no material contained in the thesis has been used in any other submission for an
academic award and is solely my own work


\vspace{3cm}

Kathryn Amy Harris \newline
Doctor of Philosophy \newline
Jeremiah Horrocks Institute for Astrophysics and Supercomputing \newline
School of Computing, Engineering and Physical Sciences \newline
September 2011

\begin{center}
\textbf{\large{Abstract}}
\end{center}

In this thesis, I present an investigation into the environments of quasars with respect to galaxy clusters, and environment evolution with redshift and luminosity. The positions of quasars with respect to clusters have been studied using cluster and quasar catalogues available, covering the redshift range $0.2<z<1.2$. The 2D projected separations and the 3D separations have been found and the orientation of the quasar with respect to the major axis of the closest cluster calculated, introducing new information to previous work.

The positions of quasars with respect to clusters of galaxies will give an indication of the large scale environment of quasars and potentially clues as to which formation mechanisms are likely to dominate at various redshifts. For example, galaxy mergers are most likely to occur in galaxy group environments and will create luminous quasars. Galaxy harassment is more likely to occur on the outskirts of galaxy clusters and create lower luminosity AGN. Secular processes such as bar instability can also create AGN and are likely to be the cause of nuclear activity in isolated galaxies.  The aim of this work is to study the large scale environment over a large redshift range and study the evolution as well as any change in environment with quasar luminosity and redshift. Another aim of this work is to study the orientation of a quasar with respect to a galaxy cluster. If galaxy clusters lie orientated along filaments, the position of a quasar with respect to a cluster will give an indication as to where quasars lie with respect to the filament and therefore the large scale structure. 

There is a deficit of quasars lying close to cluster centres for $0.4<z<0.8$, indicating a preference for less dense environments, in agreement with previous work.
Studying the separations as a function of cluster richness, there was a change in quasars lying closer to poorer clusters for $z<0.2$ (\citealt{Lietzen2009}) to lying closer to richer clusters for $0.2<z<0.4$, though more clusters at low redshifts will be needed to confirm this. There is no obvious relation between the orientation angle between a quasar and the major axis of the closest galaxy cluster and 2D projected separations. Using faint ($M_r > -23.0$ mag) and bright ($M_r < -23.0$ mag) quasars, there is no difference between the two magnitude samples for the 2D separations or the cluster richness, in contrast to \citet{Strand2008} who found brighter quasars lying in denser environments than dimmer quasars. 
These is no change with redshift (over $0<z<1.2$) in the positions of the quasars with respect to the cluster or the cluster richness as a function of absolute quasar magnitude. There is also no preferred orientation between the quasar and the cluster major axis for bright or faint quasars. 

Spectra of a selection of 680 star forming galaxies, red galaxies, and AGN were taken by Luis Campusano and Ilona S\"ochting and 515 redshifts calculated. Though few of these galaxies turned out to be cluster members as was originally intended, it was possible to use these galaxies to study the environments of quasars with respect to star-forming galaxies and galaxy clusters. The objects were classified (33 classed as AGN), and star formation rates calculated and compared. Three AGN and 10 star forming galaxies lie at the same redshift ($z=0.29$) as three galaxy clusters. The three galaxy clusters have the same orientation angle and may be part of a filament along with the star forming galaxies and AGN. Further study will investigate the relation between AGN positions and filaments of structure.

A sample of quasar spectra taken by Lutz Haberzettl using Hectospec on the MMT were taken to increase the number of quasars used in this study. However, when studying the spectra, a number of high redshift quasars showed evidence of ultra-strong UV Fe\textsc{ii} emission in their spectra. The redshifts of these quasars were too high to be included in the main body of the study. However, a significantly large number of ultra-strong UV Fe\textsc{ii} emitting quasars have been found in the direction of three LQGs in the redshift range $1.1<z<1.6$, including the Clowes-Campusano Large Quasar Group (CCLQG). Ly$\alpha$ fluorescence can increase the UV Fe\textsc{ii} emission. However, Ly$\alpha$ emission from other quasars was found to be negligible compared to emission from the quasar's central source. Though there has been no previous indication that the LQG environment is unique, the high level of iron emission may indicate a difference in environment. Plans for future work based on these results are outlined.

\pagenumbering{roman}
\pagestyle{plain}
\clearpage{\cleardoublepage}	
\newpage
\addcontentsline{toc}{chapter}{Contents}
\tableofcontents
\listoftables
\listoffigures
\pagestyle{empty}

\begin{center}
\textbf{\large{Acknowledgements}}
\end{center}

I would like to thank the following:

my supervisors, Roger Clowes, Ilona S\"ochting, Gerry Williger, Anne Sansom, and Steve Howell, for their help, encouragement and patience,

the other members of the Large Quasar Group, especially Lutz Haberzettl, Luis Campusano, and Matthew Graham for their input and support,

Danielle Brewsher, for her amazing help and support (thanks for the coffee breaks and being a friend),

the Science and Technology Facilities Council and the Jeremiah Horrocks Institute for financial support,

my friends at the Dance Studio who kept me sane, especially Tricia and Katherine (thanks for the giggles!), 

and finally, and most of all, my family for their never ending encouragement, support, and utter faith in me! Without your support, this would never have happened. I am very lucky to have you on my side. Thank you.

\pagestyle{fancy}
\clearpage{\pagestyle{empty}\cleardoublepage}
\setcounter{page}{0}
\setcounter{chapter}{0}
\pagenumbering{arabic}


\chapter{Active Galactic Nuclei}
 
Galaxies occur in a variety of shapes and sizes. Most galaxies contain a super-massive black hole at their centre (\citealt{Richstone1998}). A super massive black hole refers to a black hole with mass, M$_{BH}>10^6 $M$_{\odot}$ (\citealt{Jogee2006}). For most galaxies, this black hole is quiescent, so no material is accreting onto the black hole. However, in some galaxies, material accretes onto the central black hole causing the galaxy to become active (\citealt{Lynden-Bell1969, Rees1985, Osterbrock1993}). This accretion releases large amounts of energy in a small compact area around the black hole, making these galaxies some of the brightest objects in the Universe. These galaxies are called Active Galactic Nuclei (AGN). 

The mass of the accreted material is converted into energy; the rate at which the energy is emitted gives the rate the energy is supplied via accretion to the nucleus. In a typical AGN, the nucleus is brighter than all the stars by a factor of 100 \citep{Peterson1997}. 
The luminosity of the AGN is determined by the rate at which energy is emitted by the nucleus, and is given by Equation \ref{AccretionLumin}

\begin{equation}
\label{AccretionLumin}
L = \eta \dot{M} c^2
\end{equation}

where $\eta$ is the efficiency factor (which depends on the nature of the accretion disk; \citealt{Jogee2006}), $\dot{M}$ is the rate of mass accretion, and $c$ is the speed of light. The mass accretion rate is given by Equation \ref{AccretionMass}.

\begin{equation}
\label{AccretionMass}
\dot{M} = \frac{L}{\eta c^2} \approx 1.8 \times 10^{-3} \left(\frac{L*}{\eta}\right) \mbox{M}_{\odot} \mbox{yr}^{-1} 
\end{equation}

where $L*$ is the characteristic luminosity of a field galaxy, $\sim10^{44}$ ergs s$^{-1}$. 
Using an efficiency factor $\eta$ =0.1 and $L=10^{46}$ ergs s$^{-1}$, the mass accretion rate is $\dot{M}$ = 2 M$_{\odot}$ yr$^{-1}$.  The Eddington rate (the mass accretion rate needed to sustain the Eddington luminosity) is given by Equation \ref{AccretionEdd}.

\begin{equation}
\label{AccretionEdd}
L_E = \eta \dot{M}_{E} c^2 = 1.51 \times 10^{38} \frac{M}{M_{\odot}} \mbox{erg s}^{-1}
\end{equation}

where $L_E$ is the Eddington luminosity and $M_{\odot}$ is a solar mass. The Eddington Luminosity is the luminosity at which the gravitational force matches the radiation pressure force. It follows from Equation \ref{AccretionEdd} that the high luminosities seen in AGN must be created by a minimum central mass (\citealt{Sparke2000}).  
This represents the maximum accretion rate possible for mass $M$ (using a simple spherical accretion model), though this rate can be exceeded if the accretion occurs in a disk. For a bright quasar, the black hole must consume roughly 1\% of the stellar mass of a bright elliptical galaxy or 10\% of a bright spiral during their lifetime \citep{Lake1999}. The Eddington ratio is defined as $\lambda = L_{bol}/L_{E}$ where $L_{bol}$ is the bolometric accretion luminosity of the system.

\subsection{AGN Signatures}
AGN show strong emission over a wide wavelength range, including radio, $\gamma$-ray and X-ray, where most galaxies barely radiate (\citealt{Sparke2000}). One of the most prominent features in AGN spectra is the emission lines, which are stronger than those seen in stars and other galaxies. Sometimes these emission lines are broad, emitted from gas travelling at high speeds ($\sim$10,000 kms$^{-1}$), which is faster than the speed of stars orbiting within the galaxy. 

AGN can be distinguished from inactive galaxies by their position on a Baldwin, Phillips and Terlevich (BPT) plot \citep{Baldwin1981}. This plot uses the ratios of lines ([O\textsc{iii}]$\lambda$5007/H$\beta$ and [N\textsc{ii}]$\lambda$6583/H$\alpha$) to classify objects by distinguishing between black-body and power-law ionising spectra.  

\subsection{Classes of AGN}
There are different classes of AGN, mainly defined by their flux output as well as the emission lines seen and other data. Table \ref{tab:AGNproperties} shows some of the properties associated with the different classes of AGN. Point-like refers to whether the host galaxy can be resolved, and variable indicates whether the output from the central black hole is variable. 

\begin{sidewaystable}[!ht] 
\caption{\small{Classes and properties of AGN. }}
\centering
\begin{tabular}{ c | c c c c c c }
               & Point like  & Broad emission lines & Narrow emission lines  & Radio  & Variable & Typical $L_{bol}$  \\
               &            & (FWHM$\sim10^4$km s$^{-1}$) & (FWHM$\sim400$ km s$^{-1}$) &        & (ergs s$^{-1}$)  \\\hline 
Quasars        & Yes        & Yes             & Yes              & Yes    & Yes      & $10^{46}-10^{47}$ \\  10-100                   \\
Seyfert Type 1 & Yes        & Yes             & Yes              & Weak   & Some     & $10^{42}-10^{44}$ \\ 
Seyfert Type 2 & No         & No              & Yes              & Weak   & No       & $10^{42}-10^{44}$ \\ 
LINERs         & No         & No              & Yes              & No     & No       & $10^{41}-10^{42}$ \\ 
BL Lacs        & Yes        & No              & No               & Yes    & Yes      & $10^{44}-10^{46}$ \\ 
OVV            & Yes        & Yes             & Yes              & Yes    & Yes      & $10^{44}-10^{46}$      
\end{tabular}
\label{tab:AGNproperties}
\end{sidewaystable}

\subsubsection{Seyfert Galaxies}
Seyferts galaxies show strong nuclear emission and prominent emission lines with an absolute magnitude in the V-band of $M_V > -22.5$  or $L < L^{10}L_{\odot}$ \citep{Sparke2000}. This magnitude boundary is simply a convention that has arisen and has no special meaning.
The host galaxy containing the black hole at its centre can be spatially resolved due to the central source having a low enough luminosity to allow the host to be viewed. There are two types of Seyferts. Type 1 Seyfert galaxies have both narrow and broad lines within their spectra while Type 2 contain only narrow lines.  
Often the terms AGN, Seyferts and quasars are used interchangeably \citep{Osterbrock1986}.

\subsubsection{Quasars}
Quasars are regarded as the brighter version of Seyfert galaxies, with an absolute magnitude in the V-band of $M_V < -22.5$ or $L > 10^{11}L_{\odot}$ \citep{Sparke2000}. Quasars are the most luminous objects known and have been found up to redshift of $z\sim7$ (\citealt{Mortlock2011}). The quasar host galaxy cannot be spatially resolved due to the brightness of the central source. Some quasars (~5-10\%) are radio strong sources, with the majority being radio-weak.

\subsubsection{LINERs}
Low Ionisation Nuclear Emission Line Region Galaxies (LINERs) \citep{Heckman1980} are similar to Seyfert Type 2s and show AGN signatures \citep{GonzalezMartin2009}, but have strong low-ionisation lines (such as [O\textsc{i}]$\lambda$6300 and [N\textsc{ii}]$\lambda\lambda$6548,5483). These objects are very common and dominate the population of active galaxies in the present universe and may be detected in nearly half of all spiral galaxies \citep{Ho1994}. These are distinguishable from H\textsc{ii} regions by their larger values of [N\textsc{ii}]$\lambda$6583/H$\alpha$ and lower values of [O\textsc{iii}]$\lambda$5007/H$\beta$. This puts them in a distinct area on the BPT plot. LINERS may be different to other AGN due to complex absorbing structures along the line of sight \citep{GonzalezMartin2011}.

\subsubsection{Radio Loud and Quiet}
AGN can also be split into radio loud and radio quiet objects. Radio loud quasars have powerful jets of material coming out from the central black hole, and are only found in elliptical galaxies. Radio quiet AGN do not have jets, have less radio emission, and are found in a variety of spiral galaxies.
Radio loud galaxies can be split into broad line radio galaxies (BLRG) and narrow line radio galaxies (NLRG) which are analogous to Type 1 and Type 2 Seyferts respectively.

\subsubsection{BAL Quasars}
A sub-category of quasars is Broad Line Absorption quasars (BAL) which show broad absorption lines within the optical spectra, and are found in roughly $\sim$10\% of quasars. The line widths show evidence of high Doppler broadening in the ranges of 0.01-0.1$\times c$, the speed of light \citep{Robson1996}, which are indicative of massive outflows of material from the quasar centre \citep{Hopkins2008}. There is also a category of low-ionisation BAL (LoBAL), which make up only 1.5-2.1\% of the entire quasar population (\citealt{Dai2010}). These quasars show absorption from low-ionisation lines such as Mg\textsc{ii} and Fe\textsc{ii}.

\subsubsection{BL Lacs and OVVs}
There are other AGN types, which can be grouped together due to strong similarities in their radio-loud flat spectra, the variability in the optical output, and are strongly beamed (\citealt{Fan1997}), called BL Lac and OVV (\citealt{Angel1980}). 

BL Lacs (originally thought to be variable stars) are high-luminosity Type 1 radio loud galaxies. It is believed these objects lie with their jets close to our line of sight as they show superluminal motion, evidence for synchrotron radiation within a cone, and are beamed towards the observer’s line of sight. Emission and absorption lines are very weak or absent in BL Lac spectra, leaving the spectrum featureless. They also have strong and rapid variable radiation (on the time-scale of hours and longer).

Optically violent variable (OVV) quasars are similar to BL Lacs. OVV quasars are also radio loud and are very rare, but they tend to show prominent broad emission lines on the spectra and are high luminosity sources \citep[][and references therein]{Basu2001}. The flux output is highly variable (in the orders of magnitudes) and varies erratically, with time-scales ranging from days to years. 

\subsubsection{ULIRGs}
Luminous or Ultra-Luminous Infra-red galaxies (LIRGs/ULIRGs) are believed to be the dust-enshrouded phase of a quasar \citep{Sanders1988a, Sanders1988b}, emitting most of their energy in the infra-red with luminosities of $L_{IR} > 10^{12}L_{\odot}$ \citep{Meng2010}. These galaxies show evidence of strong interactions (most likely the advanced stages of a major merger) \citep{Rich2011, Krolik1999}, and are roughly as numerous as AGN of comparable luminosity \citep{Sanders1996}. More than 95\% of ULIRGs show evidence of morphological disruptions such as tidal tails, double nuclei, bridges and overlapping disks (\citealt{Veilleux2001}). It is believed ULIRGs may be the first stages of a quasar \citep{Meng2010} and may evolve into elliptical galaxies. Once the dust surrounding the AGN has been consumed or swept away, the optical AGN is revealed.

\clearpage
\section{Structure}
\begin{figure}[!h]
\centering
\includegraphics[scale=0.55]{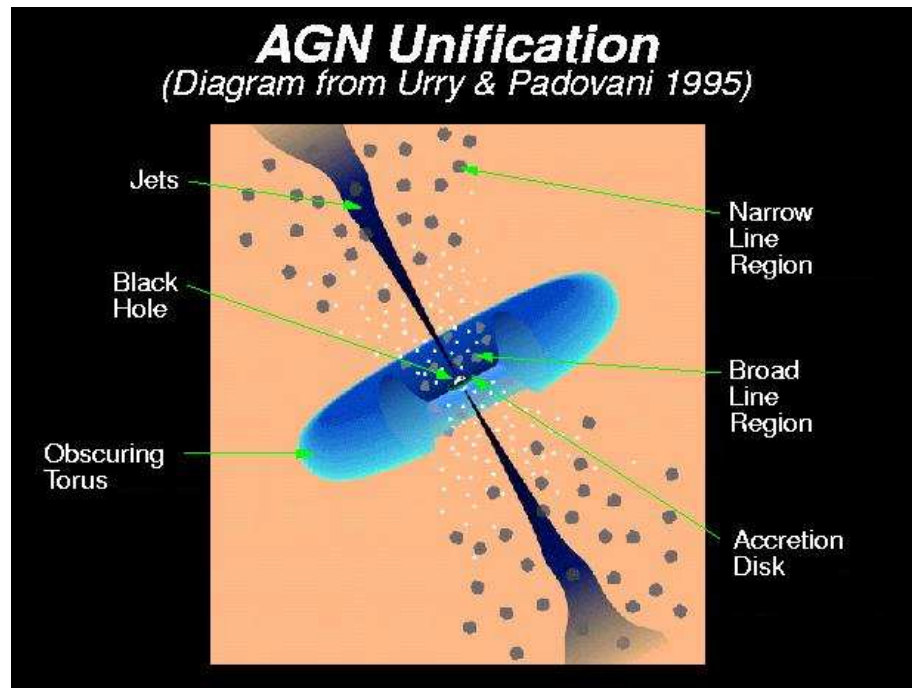}
\caption[Structure of an AGN]{\small{The structure of an AGN is shown in this figure, highlighting different regions. Diagram from \citet{Urry1995}; annotated by M. Voit.)}}
\label{fig:unification}
\end{figure}

Figure \ref{fig:unification} shows the structure of an AGN (\citealt{Urry1995}; annotated by M. Voit). 

The centre of the AGN consists of a super massive black hole, which is very hot and luminous and photoionises the surrounding area \citep{Osterbrock1986}. The size of the accretion disk (for a 10$^8M_{\odot}$ black hole) is roughly the size of our solar system. Surrounding this central source is the broad line region (BLR) associated with broad emission lines. Beyond this lies the narrow line region (NLR).

AGN appear to be axially rather than spherically symmetric. There is likely to be an optically thick torus of dust around the quasar obscuring the unresolved radiation. This permits the radiation to only escape along the torus axes.

The Broad Line Region (BLR) lies beyond the central black hole and accretion disk. Lines emitted from this region typically have a full-width-half-maximum (FWHM) of $\sim$10,000 kms$^{-1}$, although can be up to 15,000 kms$^{-1}$ \citep{Robson1996}. The BLR has a typical radius of $\sim$0.07-1.0 pc \citep{Osterbrock2006} and is comprised of solar-like abundances. The exact dynamics and kinematics of the BLR are still not clear due to the inability to spatially resolve this region. The density is estimated to be 10$^{9}$ to 10$^{10}$cm$^{-3}$.

The BLR is comprised of a number of distinct optically thick clouds, the energy source for which is photoionisation by the continuum radiation from the central source \citep{Peterson1997}.  Most of the emission from the BLR arises from these clouds, although they occupy only a small fraction of the volume of the BLR and are assumed to be arranged in spherical shells around the central source. There are estimated to be around $5 \times 10^4$ clouds in the BLR, with radii of 400$R_{\odot}$.

The Narrow Line Region (NLR) lies outside the BLR at 10-100 pc and is the largest spatial scale where ionising radiation from the central source dominates. The NLR also lies outside the dust torus. This region is several orders of magnitude more massive than the BLR (although the emission is often comparable). The FWHM of lines can lie between 200 $< \Delta z <$ 900 kms$^{-1}$, though most lie within 350-400 kms$^{-1}$.

Like the BLR, the NLR is also clumpy in nature, containing clouds of gas which move at a slower speed which produces narrower spectral lines than seen in the BLR. 
This region has electron densities between 10$^{2}$ cm$^{-3}$ to 10$^5$ cm$^{-3}$, and temperatures 10,000 to 25,000 K, with an average of 16,000 K \citep{Koski1978}. 

The torus is a thick band of obscuring material around the central source but inside the NLR so the BLR is hidden \citep{Konigl1994, Elitzur2006}. This allows the AGN radiation to only escape via the torus axes, defined by ionisation cones. The dust in the torus is likely to be in the form of high-density clumpy clouds \citep[e.g.][]{Krolik1988, Nenkova2002, Deo2011}, containing 10$^9$ M$_{\odot}$ of dust and molecular gas, and most of this material will be very hot ($\sim$1000K). The torus is a few hundred pc across, with the central torus hole being a few pc. This allows the central engine and the BLR to be obscured unless viewed face-on. 
This torus is essential for the unification models of the varying AGN types, which uses the theory that the AGN are all similar, but simply viewed from different angles \citep{Antonucci1993}.

\section{Radio sources}
Radio emission is created by synchrotron radiation, when relativistic electrons interact with a magnetic source and lose their energy via radiation. In AGN, this is created by outflowing plasma which produces bow shocks when it collides with the ambient NLR gas. One way of estimating the strength of the radio sources is using the radio optical flux at 6 cm (5 GHz) and 4400\AA\ (680 THz), $R_{r-o}$. A radio quiet quasar (RQQ) has $0.1<R_{r-o}<1.0$, whereas a radio loud quasar (RLQ) has $R_{r-o}>$10 \citep{Kellermann1989}. 

Radio loud quasars are only a small proportion of the AGN population except at the high end of the luminosity distribution.  It is possible quasars with radio axes close to the plane of the sky are not detected as quasars but as radio galaxies. It is also thought radio quiet quasars may be the remnants of radio loud quasars \citep{Marecki2011}.

It has been proposed that the two radio types have different black hole spins, with the radio loud quasars having high spin black holes and radio quiet AGN having lower spins \citep{Sikora2007, Wu2011}. RLQ and RQQ reside in different galaxy host morphologies with radio loud AGN lying in early type red galaxies \citep{Ledlow1996}, and RQQ lying in disk galaxies \citep{Lawrence1999}.

\section{X-ray sources}
The most common characteristic of AGN is that they are all X-ray bright sources, a fact which is used to find radio-quiet AGN in surveys. The X-ray emission comes from the central core region and extends to $<$1 pc \citep{Elvis1978}.  X-ray surveys are also very useful in finding quasars and AGN which are optically obscured by dust, as the X-ray regime is not as affected by dust. This wavelength is more sensitive to less luminous AGN compared to using optical selection.

\section{Host galaxy}
Most Seyferts are hosted by spiral galaxies, and tend to be (though are not always) early-type spirals. Generally, radio quiet galaxies and Seyferts are found in disk galaxies, while radio loud and broad line radio galaxies (BLRGs) are found in elliptical galaxies \citep{Lawrence1999}. \citet{Georgakakis2009} state that disk-dominated host galaxies contribute 30 $\pm$ 9 \% of the total AGN space density at $z\sim$1. 

Irregular morphological features in the host galaxies are often linked to tidal interactions. It is more difficult to assess the morphology of quasar hosts due to the brightness of the central source overwhelming the starlight from the host galaxy. However, not all quasars are point-like sources and in low redshift quasars, about 50\% of hosts show evidence of morphological peculiarities such as tidal features \citep{Peterson1997}. The host galaxy luminosity correlates with the quasar luminosity \citep{Lawrence1999} with brighter AGN  found in more luminous galaxies.

The colours of the host galaxies are generally consist with their morphological type. Though the colour distribution has been seen to be dependent on the influence of large scale structure (on the scales of $\sim$10 Mpc) \citep{Silverman2008}. Silverman et al. suggest that AGN prefer bluer hosts at $z>0.8$ than AGN at $z<0.6$. It has also been suggested that the AGN has an impact on the host galaxy by halting the star formation due to AGN feedback \citep[e.g.][]{Power2011, Blecha2011}.

\section{Unification theory}
Most of the work in unification theory focuses on the morphology of the AGN and the angle at which the AGN is viewed. AGN will appear different when viewed from different angles, because of the dust torus preventing emission being seen from certain areas. Table \ref{tab:unification} shows the types of AGN seen when viewed from different angles.
\clearpage
\begin{table}[!h]
\caption[Unification models]{\small{AGN types with respect to the orientation of how they are viewed \citep{Peterson1997}. }}
\centering
\begin{tabular}{ l | c c}
Radio &\multicolumn{2}{c}{Orientation}\\
\cline{2-3}
Properties    & Face-on    & Edge-on        \\ \hline
Radio Quiet   & Seyfert 1  & Seyfert 2      \\
              & QSO        & FIR galaxy?    \\
Radio Loud    & BL Lac     & FR I           \\
              & BLRG       & NLRG           \\
              & Quasar/OVV & FR II          
\label{tab:unification}
\end{tabular}
\end{table}

\begin{figure}[!h]
\centering
\includegraphics[scale=0.55]{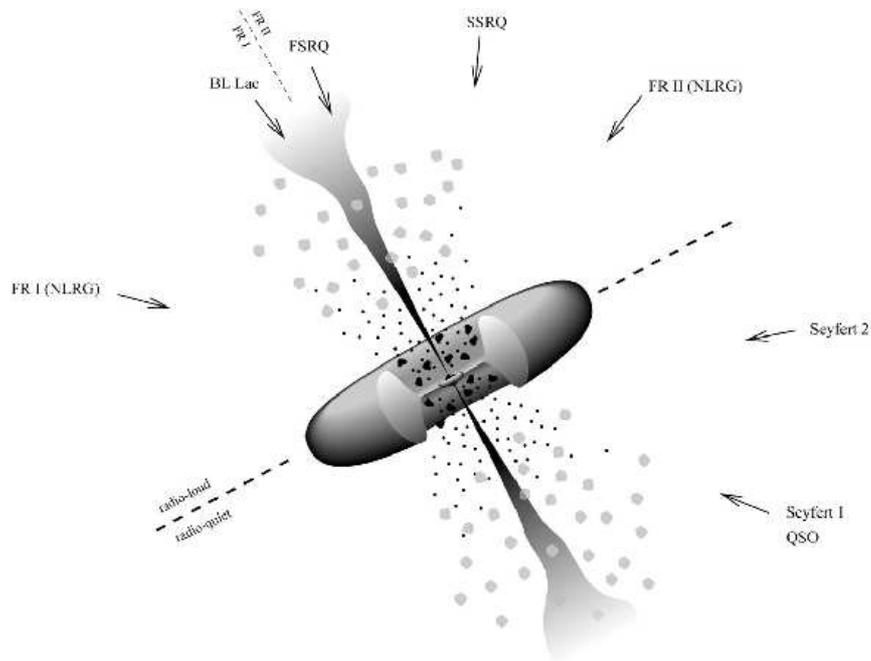}
\caption[Unification model]{\small{The AGN types found by viewing the AGN at different angles \citep{Torres2004}. }}
\label{fig:unification2}
\end{figure}

Figure \ref{fig:unification2} shows an example of how different types of AGN can be found by viewing the AGN from different angles. FRI galaxies are weak radio sources with a bright centre and decreasing surface brightness. FR2 are more luminous radio galaxies, are much more powerful (occurring on scales of kpc) and have steep radio emission found in the inner regions \citep{Hughes1991}. FR stands for Fanaroff and Riley who first classified these radio galaxies (\citealt{FR1974}). FSRQ and SSRQ stand for flat-spectrum and steep-spectrum radio quasars, respectively.

The main difference between Seyferts and quasars is the luminosity of the central source. The Seyfert Type 1 and Type 2 are thought to be the same objects, viewed from different angles \citep{Antonucci1993}. At least some Seyfert Type 2s are definitely Seyfert Type 1 with an obscured BLR. Spectra from the narrow line region are indistinguishable between Seyfert Type 1 and 2. The torus must block out 3/4 of the sky seen by the central source (\citealt{Peterson1997}), as estimated from the number of Type 1 and Type 2 Seyferts. Evidence for a torus has been found in other wavelengths. \citet{Corral2011} looked at absorption in X-ray spectra and found larger amounts of intrinsic material for Type 2 than Type 1, which, if this is a line-of-sight effect, suggests the presence of a dust torus. Also all classes of Seyfert have been found to show the same nuclear continuum \citep{Ricci2011}.

However, it is likely that not all Type 2s are Type I viewed from a different angle, and the unification model breaks down. All quasars (high luminosity AGN) have Type 1 spectra. If all Type 2 were Type 1 viewed from a different angle, we would still expect some quasars with Type 2 spectra. (The reason we do not could be because high luminosity sources either do not have an obscuring torus or the torus is thin.) Quasars with Type 2 spectra could exist but so far have been classed as far-infra-red galaxies (FIR), which have a quasar-like luminosity and narrow line spectra. Also the continua from Type 2 Seyferts are not generally polarised which suggests the absence of a scattering medium, which was suggested should be seen. In the X-ray, the fraction of exceptions to the unified model was found to be 5\% \citep{Corral2011}.

\section{Fuelling mechanisms}
The main problem in fuelling a quasar is moving the material from further out in the galaxy into the central parsec and removing the angular momentum of the material \citep{Peterson1997}. 

The specific angular momentum of fuel ($L=v\times r$) at the last stable orbit of the black hole of mass, $M_8$ (10$^8$M$_{\odot}$), is several times 10$^{24}$M$_8$ cm$^{2}$s$^{-1}$. However, material in the galaxy's disk, with an orbit of 10 kpc, has angular momentum of several times 10$^{29}$ M$_8$ cm$^{2}$s$^{-1}$ \citep{Jogee2006}. Therefore, the material needs to be moved into the centre of the galaxy and its angular velocity must be reduced for it to be able to join the accretion disk which has a small radius. 

There are various suggestions for fuelling an AGN, which may produce different luminosities and be dominant at different cosmic times and in different environments. For example, major mergers offer the most plausible mechanism for the triggering of brightest quasars, and dominate AGN evolution at early times ($z=2-3$). At later times, the main fuelling mechanisms are more likely to be secular processes (such as bar instabilities) and minor mergers \citep{vanBreukelen2009, Cisternas2011, Ryan2010}. 

Different mechanisms may also be dominant in different environments. 

\subsection{Mergers}

There are two types are mergers: major and minor. 

A major merger is often described as the main method for fuelling AGN. This refers to the merger of two disk galaxies with a mass ratio of 3:1 or less. These mergers can induce a large scale inflow of gas ($\sim$a few percent of the galaxy's gas) into the inner kpc and cause starbursts and AGN activity \citep{Kauffmann2000}. This is believed to be the main (if not only) mechanism for the very brightest AGN.  Major mergers remain the most commonly accepted method for triggering high-luminosity AGN, though there is little evidence they are also responsible for low-luminosity AGN. 

Galaxies in clusters have a high chance of merging and have frequent interactions. However, the galaxies in the centre of a cluster are gas-poor. The velocity dispersion in the centre of a cluster is also too high for major mergers to occur (\citealt{Binney2008, Martini2007}). In less dense regions and in isolated galaxies, the galaxies are more gas-rich but the number density is lower, making interactions and mergers less likely. An intermediate environment might be in groups where there are neighbour galaxies, where there is still enough cold gas available and the galaxy velocity dispersion is low enough to enable mergers to take place \citep{Arnold2009}. Mergers are likely to be rare in cluster environments. 

During the early stages of a merger, tidal interactions cause an increase in star formation and accretion onto the central black hole, though the effect is weak. During the final stages of merging, large inflows of gas will trigger strong starbursts, which can be seen in ULIRGS and sub-millimetre galaxies. The inflows also feed the black hole, but the central black hole is obscured in the optical due to dust. Finally the gas (and dust) is consumed by the black hole and starbursts or blown out of the system by AGN feedback. This causes the quasar to become visible in the optical leaving a red sequence host and bright quasar \citep{Hopkins2008}. 

Merger rates increase with redshift, which has been suggested to explain some of the increase in quasar activity and the activity peak at z$\sim$2-3 \citep{Carlberg1990} but not all \citep{Kauffmann2000}. The decrease of activity towards lower redshifts is also likely to be affected by a decrease in the fuel available to the black holes. It is believed that the accretion efficiency changes with redshift so black holes accrete at slower rates at later times \citep{Kauffmann2000}.

As the shape of the merging galaxies is distorted by the merger, if mergers are a dominant fuelling mechanism, it is expected that the host of the AGN would show evidence of these distortions. Some authors find evidence for tidal interactions and mergers \citep[e.g.][]{Bahcall1997, Hutchings2003, Bennert2008} while others suggest the hosts of AGN are indistinguishable from those of isolated elliptical galaxies which are not interacting \citep[e.g.][]{Dunlop2003, Cisternas2011}. Most AGN hosts ($>$85\%) show no evidence of strong distortions and there is no significant difference in the number of galaxies with distortion features between active and inactive galaxies \citep{Dunlop2003, Cisternas2011}. This suggests active galaxies are involved in mergers at the same rate as inactive galaxies. In the redshift range $0.3<z<1$, \citet{Cisternas2011} found over 50\% of the AGN hosts were disk dominated suggesting the AGN was formed by a triggering mechanism which would not destroy the disk as a major merger would.

A minor merger consists of a galaxy and a satellite or dwarf galaxy with a mass ratio of 10:1 or greater, and may result in less luminous AGN than those produced in major mergers. These are likely to be more common than major mergers. In fact, more galaxies are likely to have accrued a large percentage of their mass through minor mergers of discrete subunits \citep[e.g][]{Ostriker1975}, compared to 20\% at most which have been through a major merger \citep[e.g.][]{Hernquist1995}. Minor mergers can ``drive structural evolution in disks without completely destroying them'' \citep[][and references therein]{Hernquist1995}. The disk may be warped or heated and this may be the origin for the ``thick'' disk \citep[e.g.][]{Walker1996}. Minor mergers can also drive material into the centre of the host galaxies, fuelling an AGN.

\subsection{Interactions and Galaxy Harassment}
Galaxy harassment caused by close interactions of galaxies can create dynamical instabilities in the galaxy and rapidly channel gas into the centre of a low luminosity host. During the first encounter, a bar instability is formed, stronger than that induced by the cluster's tidal field alone. Within a few billion years, 90\% of the gas in a galaxy can be driven into the central 500 pc \citep{Lake1999}.

The strongest encounters do not necessarily occur in the centre of the cluster (\citealt{Lake1998}). The impact of the galaxy harassment depends on the square of the masses of the largest galaxies encountered. If galaxies are tidally limited, the more massive galaxies will lie on the edges of the cluster. Also, the velocity of the galaxy decreases in the outer regions of a cluster, which makes the encounter stronger (\citealt{Lake1998}). \citet{Alonso2007} determine that, in an interaction, the luminosity of the paired galaxy may be important in determining the AGN activity.

The infall of field galaxies peaks between redshifts of 0.3 and 0.5 \citep{Kauffmann1995} so if harassment is the cause of nuclear activity in quasars in sub-L* galaxies, the frequency of AGN in clusters should also peak in this redshift range in clusters, which is shown in the Butcher-Oemler effect (\citealt{Lake1998}). The Butcher-Oemler effect \citep{Butcher1978} suggests that the cluster core of rich clusters at intermediate redshifts ($z\sim0.3-0.4$) will contain more blue galaxies than lower redshift clusters.

\subsection{Hot gas}

Another approach is to consider that AGN could be formed during the host galaxy formation and the main source of fuel is the interstellar medium formed as the galaxy collapses \citep{Nulsen2000}. The first galaxies collapse, which forms a hot gas and then the first quasars form shortly after. During the collapse, the radiative cooling is quicker than the shock heating so the gas is cooled quickly. In the collapse of large systems, some gas can form a hot atmosphere after the collapse. As the cooling time is less than the time needed for the collapse, the hot gas will start to cool and forms a cooling flow \citep{Fabian1994}, from which the black hole accretes hot gas. The black hole growth is determined by the temperature of the gas and the Mach number of the cooling flow.

This hot gas is depleted as time goes on and the accretion rate will drop to where the efficiency of accretion plummets causing the quasar to shut off. The depletion of hot gas does not, however, explain the lack of luminous AGN at the current epoch as there are nearby ellipticals which have a supply of accretable hot gas but have low accretion luminosities. This could be due to the accretion flow becoming advection-dominated and therefore, having a low efficiency rate \citep{Reynolds1996, DiMatteo1997}.

This model, however, fails to account for the number of luminous AGN at $z\sim$2 and earlier. This model over-predicts the number of quasars with respect to the optical luminosity function but is consistent with models from the X-ray background (\citealt{Nulsen2000, Somerville2008}). 

\subsection{Bars}

Stellar bars can be seen in abundance in spiral galaxies (possibly out to z$\sim$1). They vary in strength, exert a gravitational torque, and alter the mass and angular momentum distribution of material in the galaxies. 30\% of spiral galaxies have strong bars (in the optical), a figure which increases to 50\%, if weak bars are included. Bars represent a strong non-axisymmetric distortion of the galaxy mass distribution \citep{Binney2008}. They contain prominent dust lines on the leading edge of the bar so are more prominent in near IR images. 

\citet{Mulchaey1997} found no excess of bars in Seyfert galaxies, while \citet{Jenkins2011} found almost 80\% of Seyfert Type 2s are barred spirals. Not all barred spirals show evidence of AGN but due to the lifetime of AGN activity, this would not be expected. Also not all AGN in spirals have bars.

In strong bars, the net gas-flow rate is typically $<$1 kms$^{-1}$, which though small, is enough to transport most of the gas in a galaxy into the centre within a galaxy's lifetime \citep{Binney2008}. Once the gas has been transported to the centre, it gathers in circular orbits and creates nuclear rings, which have typical radii of a few hundred pc. These rings are possible reservoirs for the accretion disks, though another mechanism is then needed to move the gas onto the accretion disk region.

\subsection{Choosing between fuelling mechanisms}

Studying the properties of the host galaxies and environment can determine the likelihood of each fuelling mechanism occurring.
A major merger will create a luminous AGN in an elliptical galaxy. There may be evidence of tidal features such as shells and tails in the host \citep{Bennert2008} (though not always as these features may decay, \citealt{Schawinski2010}). The luminous AGN are likely to lie in areas with a low velocity dispersion and an intermediate density (\citealt{Arnold2009}). 
Major mergers are likely to be the cause of bright AGN and be dominant at higher redshifts.

Minor mergers and galaxy harassment cause instabilities in the host galaxy and the size of the interaction depends on the mass of the largest galaxy. Harassment is also likely to create ellipticals \citep{Lake1998} (though this will depend on the strength of the harassment) while in minor mergers, the disk can survive \citep{Hernquist1995}. Secular processes such as bar instability are likely to be more dominant in the local universe and create lower luminosity AGN. 

Different mechanisms may be dominant at different times and in different environments.

\section{Quasar formation}

To create an observed luminosity of 10$^{12}$L$_{\odot}$, the quasars must have an accretion rate of 2M$_{\odot}$ yr$^{-1}$. (This assumes the standard efficiency rate of $\epsilon \sim$ 0.1.) 

The highest redshift quasar found is $z=7.085$, which has a luminosity of $6.3\times 10^{13}$L$_{\odot}$ \citep{Mortlock2011}. The spectrum for this quasar is similar to lower redshift quasars of the same luminosity. This quasar is estimated to have a black hole of mass $2\times 10^9 $M$_{\odot}$, which will place strong limitations on black hole formation and accretion mechanisms, as formation mechanisms must account for a $2\times 10^9$ M$_{\odot}$ black hole only 0.77 Gyr after the Big Bang. 
The quasar formation mechanism for small black holes ($M\sim10^5 $M$_{\odot}$) may be different to that for more massive AGN \citep{Haehnelt1998}, though it is currently not possible to detect black holes with M$<10^6 $M$_{\odot}$.

In the early universe ($z>6$), the galaxy systems were rich with cold gas, had rotation-dominated dynamics, and contain a small ``seed'' central black hole. They were clumpier and more turbulent than present day blue galaxies. The size of the dark matter halo in which optical quasars are found (M$_{halo}\sim3 \times 10^{12} $M$_{\odot}$) remains constant with redshift. At least some black holes formed early on \citep{Silk1997}. \citet{Shen2009} modelled major mergers and predicted most of the black holes with M$>10^{8.5}$M$_{\odot}$ will be in place by $z=1$ and 50\% in place by $z=2$. (For lower mass black holes, the processes are likely to be secular and assembled more recently.)

\section{Large Scale Structures}

Large Scale Structure (LSS) is the product of the mass distribution of the early Universe, observed today as filaments and clumpy structures connected by galaxy clusters \citep{York2000, Colless2001} and in place at high redshifts \citep{Bond2010}. Structures have been found at a range of redshifts (e.g. $z=0.55$, \citet{Tanaka2009}; $z=0.73$, \citet{Guzzo2007}; $z=0.985$, \citet{LeFevre1994}, to name a few) and the evolution with redshift has been studied \citep{Choi2010}.

Clusters lie along filaments or mostly commonly lie on the nodes of structures with prominent filamentary structures around them \citep{Bond1996, Springel2005}. The filaments provide pathways in which to accrete matter onto the galaxy clusters \citep[e.g.][]{Tanaka2007}. 

Superclusters (for example, the Sloan Great Wall, the Shapley Supercluster and the Sculptor Supercluster) are comprised of a number of clusters or groups in a network of filaments on the scale of 10-100 Mpc \citep{Kocevski2009}. These were the sites for early star formation and formed earlier than smaller structures. In rich superclusters, the core of the structure will contain more early type red galaxies and richer groups than the outskirts of the supercluster, and contain a larger fraction of X-ray clusters. These differ from poor superclusters by the presence of a high density core. Galaxies in rich clusters have lower star formation rates than galaxies in poor clusters \citep{Porter2005, Porter2007, Einasto2008}. The environment of a supercluster affects properties of the galaxy groups and clusters located within it. 

The largest known structures in the Universe are Large Quasar Groups (LQG) which can cover 50-200$h^{-1}$ Mpc and contain between 4 and 25 quasars \citep[e.g.][]{Crampton1987, Clowes1991, Clowes1994}. These clusters of quasars exist at high redshifts, presumably trace the mass distribution, and are potentially the precursors of the large structures seen at the present epoch, such as superclusters \citep{Komberg1994}. There are $\sim$40 published examples of LQGs. 

\section{Environments}

At radii between 25 kpc and 1 Mpc from the galaxy centre, quasars are found in higher density regions than L* galaxies, with the overdensity being greatest closest to the quasar \citep{Serber2006}. Observational studies have found a small-scale excess at scales below $\sim$100 kpc $h^{-1}$ \citep{Hennawi2006, Myers2007}, and are supported by simulations \citep{Degraf2011}.

On scales of between 1 and 10 Mpc, AGN and quasars have been suggested to lie in environments similar to that of L* galaxies \cite[e.g.][]{Smith1995, Croom1999}. On scales of 10 Mpc and greater, quasars are more strongly clustered than galaxies but less than rich clusters \citep[][and references within]{Serber2006}. In nearby quasars, underdensities of bright galaxies in the environments around quasars were found at a few Mpc (\citealt{Lietzen2009}). 
\citet{Hutchings1993} and \citet{Tanaka2001} found an excess of faint red galaxies around a quasar at $z\sim$1.1, extending for $\sim20h_{50}^{-1}$ Mpc. 

There are different conclusions as to whether AGN and quasars lie in dense regions and are therefore, affected by their environment. For example, \citet{Coldwell2006} suggest the galaxy number density around AGN and quasars is similar to that around typical galaxies so there is no relation between the AGN activity and its environment. \citet{Miller2003} also find no difference in the local density of AGN and field galaxies. However, other authors \citep[e.g.][]{Serber2006} have found an increase in the local density around quasars greater than that around typical L* galaxies. 
This discrepancy could be explained by the fact that luminous AGN do avoid high density areas but low-luminosity AGN do not (\citealt{Kauffmann2004,Kocevski2009, Lietzen2009}). AGN are preferentially located 1-2 Mpc from the centres of the clusters \citep{Johnson2003, Sochting2004}. This excess may increase with redshift. 

Dim AGN in the redshift range $0.3<z<0.8$ have the same clustering properties as typical local galaxies \citep{Shirasaki2009}. Dim AGN in the range $0.8<z<1.5$ show evidence of lying in denser environments than typical galaxies, as do bright AGN in the redshift range $1.5<z<1.8$, which suggests a redshift evolution in the density preferred by both bright and dim AGN (\citealt{Strand2008}). Assuming AGN are the result of major mergers, the assembly of large systems will occur more frequently in denser areas so the bright AGN should be seen in denser environments. However, the mass assembly of large systems stops at an earlier time than small systems and small scale assembly continues so bright AGN can be produced via low-mass assembly at a later epoch and lie in sparser regions (\citealt{Shirasaki2009}).

At low redshifts, many quasars are on the edges of rich clusters \citep{Oemler1972, Green1984, Yee1987, Sochting2002}, though some lie in the centres of clusters \citep{Schneider1992, Yee1990}. The AGN fraction may also be higher in clusters with low velocity dispersions as mergers become more likely (\citealt{Gavazzi2011}). 

The general consensus is that galaxies in denser environment are less likely to host an AGN (\citealt{Kauffmann2004,vonderLinden2010,Gavazzi2011}). 

\section{Current Standing and Motivation}

Currently, the roles of mergers, harassment and secular process are still in debate. However, it is believed that different mechanisms dominate at different times.

Some authors have found AGN in overdense regions (e.g., \citealt{Serber2006,Georgakakis2007}), while others found no difference between the environments of AGN and fields galaxies (e.g., \citealt{Miller2003,Waskett2005,Martini2007}), or that AGN avoid overdensities (e.g., \citealt{Popesso2006}).
This discrepancy could be explained by the fact that luminous AGN do avoid high density areas but low-luminosity AGN do not (\citealt{Kauffmann2004}). This result also depends on the wavelength used to observe the AGN as different types of AGN may reside in different environments (\citealt{Lietzen2011}). For example, radio AGN are strongly clustered and reside in high density regions, while AGN detected in the IR are weakly clustered (\citealt{Hickox2009}).

However, a general consensus is developing that AGN prefer intermediate density regions, such as galaxy groups (e.g. \citealt{Waskett2005, Gilmour2007, Silverman2009}). In this environment, galaxy mergers are likely to occur. Mergers are more frequent in groups than clusters, due to the lower velocity dispersion and high density (\citealt{Popesso2006,Lin2010}).
Mergers are thought to create high luminosity quasars, as a merger can quickly drive large amounts of material into the centre of the galaxy. Mergers are also likely to dominate high mass systems, M$_{gal}>10^{11}$ M$_{\odot}$ (\citealt{Hopkins2008}). However, \citet{Cisternas2011} found over 50\% of the AGN hosts were disk dominated in the redshift range $0.3<z<1$. This suggests major mergers can not be a dominant mechanism as a major merger would destroy the disk.

Galaxy harassment can create lower luminosity AGN, as harassment will drive less gas into the galaxy centre and onto the black hole. Galaxy harassment is also likely to occur where the relative velocity of the encounters is decreased, but also potentially in higher density environments (\citealt{Silverman2009}). 
Harassment can also occur in the centre of a cluster where the cluster's tidal field will have a strong effect on the galaxy.

There is also still much debate as to whether there is any evolution with redshift (\citealt{Fanidakis2010})
The merger rate is higher at higher redshifts ($z>2$), as at lower redshifts, the gas supply in the galaxies has been depleted. However, \citet{Williams2011} found few additional mergers occurring at $z=1-2$ than at lower redshifts. Galaxy harassment has been proposed for lower redshifts to account for the number of lower-luminosity AGN at low redshifts (\citealt{Silverman2009}). While secular processes are most likely to be dominant in the present universe and in small galaxies (\citealt{Hopkins2008}). 

\citet{Strand2008} found that brighter quasars lay in denser environments than dimmer quasars on small scales, and \citet{Hasinger2005} found a peak in the X-ray AGN luminosity function at $z\sim0.7$. 
Bright AGN show a stronger evolution with redshift, with a space density peak at $z\sim2$ as opposed to fainter AGN, which show less evolution with redshift and have a peak in space density at lower redshifts, $z<1$ (\citealt{Hasinger2005,Fanidakis2010}). However, others (e.g., \citealt{Adelberger2005}) found no evidence of luminosity dependence in the clustering properties of AGN and galaxies.  

However, the impact of environment on AGN and quasars and their evolution with redshift and luminosity are still controversial subjects. The aim of this work is to study the large scale environment over a large redshift range and study any potential evolution as well as any change in environment with luminosity.

\section{Outline of the Thesis} 

Chapter 2 describes the data samples and surveys used in this thesis, as well as the methods created to analyse the data. 

Chapter 3 studies the proximity of quasars with respect to galaxy clusters and any evolution of the distance between a quasar and the closest cluster with redshift. This chapter also contains a study of the distance between a quasar and the closest cluster with respect to other cluster parameters such as the cluster richness, and the orientation of the quasar with respect to the cluster major axis.

In Chapter 4, the evolution of the position of the quasar as a function of the quasar luminosity is studied. Again, the orientation of the quasars with respect to the cluster major axes is studied, along with the influence of cluster richness on the quasar luminosity. The luminosities of quasars lying within a cluster have been discussed.

Chapter 5 describes the properties of a set of spectra from star-forming galaxies, red galaxies and AGN. This chapter uses spectra selected by \citet{Haberzettl2009} and observed by Luis Campusano and Ilona S\"ochting. The data reduction is described, the objects have been classified and star formation rates have been calculated and discussed. 

Using the AGN and star-forming galaxies classified in Chapter 5, the environments of AGN have been studied with respect to the star forming galaxies in Chapter 6.

Chapter 7 contains the study of a set of quasars with ultra-strong UV Fe\textsc{ii} emission within Large Quasar Groups. The spectra for these quasars was taken by Lutz Haberzettl on the Hectospec instrument. 

Chapter 8 contains the summary and an outline of future work.

\chapter{Data Samples}

Studying the large Megaparsec scale environments of Active Galactic Nuclei (AGN) and quasars and their positions with respect to the Large Scale Structure (LSS) and galaxy clusters may determine which AGN formation mechanisms are most likely. It will also determine whether these mechanisms change over time and whether the mechanism changes with AGN luminosity. To do this, large samples of clusters and quasars are required to study the relation between objects. 

This chapter will look at the data samples used, how they are selected, and any selection biases in the quasar and galaxy cluster samples used. The distances between AGN and galaxy clusters, and the methods and reasoning involved are also discussed in this chapter. The control field used to test the significance of the results is also presented.

The table containing the data used in Chapters 2-4 is described in Appendix 2 and can be found in the disk attached. This catalogue contains all of the input parameters used from the various catalogues, and all of the parameters derived from the methods described in this Chapter. 

\section{Large AGN and Galaxy Surveys}

For this work, two independent cluster data samples have been used. The clusters were identified in data taken from the Cosmic Evolution Survey (COSMOS survey; \citealt{Scoville2007}) and Stripe 82 in the Sloan Digital Sky Survey (SDSS) (\citealt{York2000}). 

COSMOS is a Hubble Treasury Project to survey a continuous field of 2 deg$^{2}$ at the celestial equator, centred on $RA(J2000)$=10:00:28.6 and $DEC(J2000)$=+02:12:21.0, which covers a comoving area of $50\times50$ Mpc at $z=0.5$ (\citealt{Scoville2007}). The aim of this project is to study the LSS, and map the morphology of galaxies as a function of the local epoch and environment, over a range of redshifts ($0.2<z<3$) (\citealt{Mobasher2004}), and to study the formation and environments of galaxies, dark matter, and quasars. The initial survey was undertaken by the Hubble Space Telescope, with additional data coming from observations on Subaru, the Very Large Array (VLA) radio telescopes, and the XMM X-ray telescope.
Photometric redshifts (mostly from ground based imaging) were obtained using the Subaru telescope, the Canada-France-Hawaii Telescope (CFHT), the United Kingdom Infra-Red Telescope (UKIRT), and the National Optical Astronomy Observatory (NOAO), giving the redshifts for $\sim2\times10^5$ galaxies at $z<1.2$, and an accuracy of $\Delta z/(1+z)\leq0.02$ (\citealt{Scoville2007}). Spectroscopic redshifts are given by the zCOSMOS survey using the VLT and the Magellan telescopes (\citealt{Lilly2007}), which provide $\sim$37,500 galaxy redshifts and several thousand quasar redshifts, with a precision of $\Delta z\sim$0.0003 in redshift for $z<1.2$ for quasars. This large number of spectroscopic redshifts for galaxies and quasars makes this sample perfect for analysing quasar and galaxy cluster positions.

Stripe 82 is an equatorial area of 270 deg$^2$ covered in depth by the SDSS to a limit of 2 magnitudes fainter than the rest of the SDSS field (\citealt{York2000}). The database for Stripe 82 is comprised of 303 runs, between $-50<RA<59$, and $-1.25<DEC<1.25$ with the whole area covered approximately 80 times (\citealt{York2000}). In total, this gives a deeper visual survey than for the rest of the SDSS data. This area is also covered by the Faint Images of the Radio Sky at Twenty-centimetres (FIRST) on the Very Large Array (VLA), the Atacama Cosmology Telescope, and the UKIRT Infra-red Deep Sky Survey (UKIDSS), giving radio, microwave and near-infra-red data respectively. Though this field covers a large area, it is long and thin, limited in $DEC$.

\section{Cluster Samples}\label{sect:clustsamples}

The cluster catalogue from the COSMOS field contains 1497 galaxy clusters within the redshift range $0.2<z<1.356$ and 1370 galaxy clusters within the redshifts $0.2<z<1.2$. Though the COSMOS survey does extend beyond $z\sim1.2$, galaxies become increasingly faint at higher redshifts, limiting this catalogue, so clusters with $z>1.2$ have been excluded from this study. The clusters contain between 8 and 235 members (one cluster has 649 members) and the centre of the cluster is defined as the mean position of all the cluster members (see \citealt{Sochting2011} for more details).

The Stripe 82 Cluster Catalogue (\citealt{Geach2011}) contains 4098 clusters with redshift range $0.038<z<0.832$, with the majority at $z<0.5$, 49 clusters with $z>0.6$, and between 5 and 173 members per cluster. 
The centre of the cluster is also defined as the mean position of all the cluster members.

The cluster detection methods are briefly described in Section \ref{sect:clustermethods}.

The COSMOS field has been used because of its depth. This data goes to high redshifts (z$\sim$1.2), allowing a large redshift range to be used. 
The Stripe 82 field covers a large area (with a large number of spectroscopic observations), which allows the lower redshift area to be probed though it is not as deep as the COSMOS field. The Stripe 82 field also contains a large number of clusters, increasing the size of the data sample and improving statistics.

Table \ref{tab:clusterdata} shows the properties of the cluster data for the COSMOS and Stripe 82 fields, comparing redshifts, area and errors. 

\begin{table}[!ht] 
\caption[Properties of COSMOS and Stripe 82 clusters]{\small{Properties of the cluster data sets used from COSMOS and Stripe 82 in SDSS. }}
\centering
\begin{tabular}{ c | c c c c c}
Survey     &   Area       & Number of          & Redshift   & median       & Redshift   \\                  
           &              & clusters           & range      & redshift     & errors   \\       \hline
COSMOS     &  2 deg$^2$   & 1370               & 0.201 - 1.2       & 0.72	     & 0.2-5.8\% \\
Stripe 82  &  270 deg$^2$ & 4098               & 0.038 - 0.832     & 0.32	     & 5-9\%    
\end{tabular}
\label{tab:clusterdata}
\end{table}

\subsection{Cluster Selection Method}\label{sect:clustermethods}
The clusters were identified in the COSMOS (\citealt{Sochting2011}) and Stripe 82 (\citealt{Geach2011}) samples using slightly different cluster detection methods. 

The cluster selection method for the COSMOS field consists of two parts. The first part uses relatively narrow slices in the photometric redshifts. This will find the seeds of possible clusters in overlapping cuts. 

In the first part of the method, narrow slices in redshift of thickness $\dif z=0.02\times(1.0+z)$ are used to select the initial galaxies. The slices overlap and are moved in steps of 0.02*(1.0+$z$)/2.0. (The figure 2.0 in the denominator is suggested by \citealt{Ilbert2009} to match the uncertainty in the photometric redshifts.) The limit of $i^+_{AB} < 3.0\times\mbox{zphot}+21.5$ is applied to take into account the affects of cosmological dimming of galaxies which leads to galaxies becoming more sparse and to avoid over-saturation by dwarf galaxies at lower redshifts. 

\begin{figure}[!ht]
\centering
\includegraphics[scale=0.6,angle=-90]{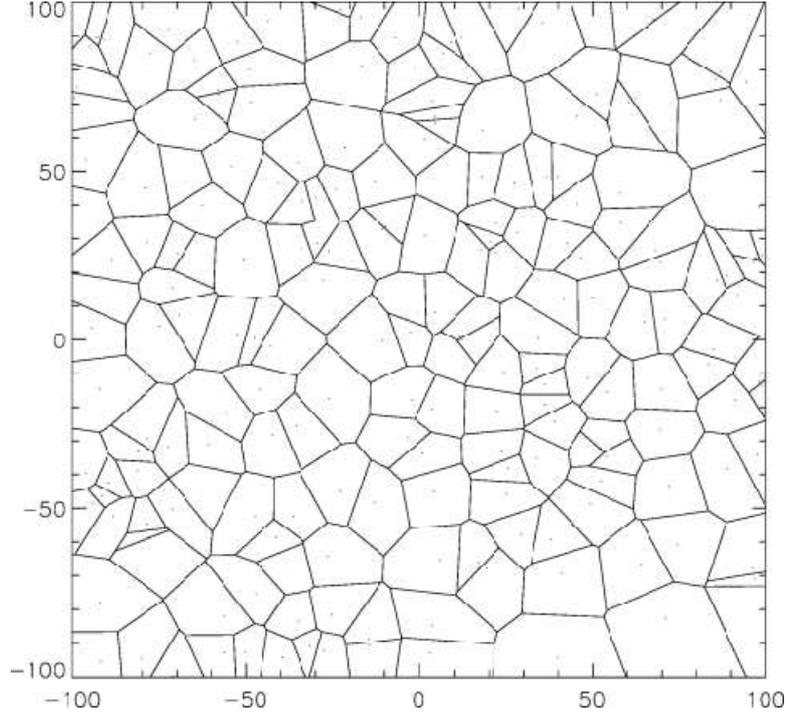}
\caption[Example of Voronoi Tessellations]{\small{An example of using Voronoi Tessellations to find overdensities due to galaxy clusters. Taken from \citealp{Sochting2006}.}}
\label{fig:Voronoi}
\end{figure}

The second part of the method determines the exact number of members within the clusters and the properties of the detected cluster.
The cluster seeds are detected separately in each of the redshift slices using Voronoi Tessellations. Voronoi Tessellations are used to map the density of regions and make no assumptions about the shape of the overdensity. Each cell contains one galaxy. The cell boundary lies equidistant between adjacent galaxies. (See Figure \ref{fig:Voronoi} for an example of Voronoi Tessellations; \citealt{Sochting2006}.)  For the COSMOS cluster catalogue, only cells with densities greater than twice the mean density level were selected. As each cell is added, the density of the cluster is assessed and the edge of the cluster is determined as when the average density for the cluster falls below twice the average field density. Only clusters with at least eight adjacent galaxies above the density threshold were selected as clusters. To avoid projection contaminations, cluster members were limited to $\pm 0.02*(1.0+z_{cl})$, where $z_{cl}$ is the median cluster redshift. 

ee members were combined. This allows the seeds of the same clusters to be combined but avoids overmerging.

For the Stripe 82 cluster catalogue (\citealt{Geach2011}), galaxies were selected using three colour cuts to select potential cluster members; ($g-r$), ($r-i$) and ($i-z$). The width and gradient of the slices are fixed by fitting colour magnitude in each filter for the richest cluster in Stripe 82. Once galaxies have been selected, Voronoi tessellations are applied to find the over-dense regions where a cluster lies. For the Stripe 82 catalogue, a minimum number of 5 adjacent cells was used to class a cluster and the edge of the cluster is defined as when the cluster density falls below 10$\times$ the average field density.

For more details, see \citet{Geach2011} and \citet{Sochting2011}.

\subsection{Selection Criteria and Biases}\label{sect:cluster_selection}

Due to the small area of sky covered by the COSMOS field, the physical area covered by the field is small at low redshift compared to the relative size of rich clusters, creating a bias against clusters at $z<0.2$. This should not be the case in the Stripe 82 field as the area covered is larger, allowing more low redshift clusters to be found. However the Stripe 82 field is long and thin, covering a large RA range but is limited in DEC, which will lead to some limitations on the cluster size at low redshifts.

The Stripe 82 cluster catalogue detects the majority of clusters up to $z\sim0.5$, with a few of higher redshift clusters (311 with $z>0.5$, and only 8 with $z>0.7$). The median redshift of this catalogue is $z=0.32$ compared to the deeper data of COSMOS where the median redshift is $z=0.72$. 
Because of these redshift limits, the Stripe 82 clusters will mainly be used for low and intermediate redshifts, while the COSMOS clusters can be mainly used for intermediate and high redshift comparisons.

\begin{figure}[!ht]
\centering
\includegraphics[scale=0.8]{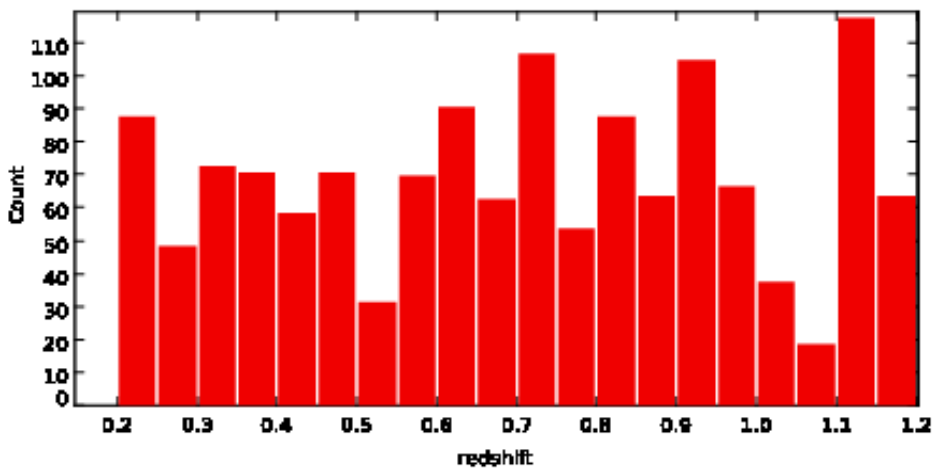}
\caption[Cluster redshift distribution in COSMOS]{\small{The distribution of redshifts for the COSMOS cluster catalogue in the redshift range $0<z<1.2$.}}
\label{fig:COSMOSzdistrib}
\end{figure}

\begin{figure}[!ht]
\centering
\includegraphics[scale=0.8]{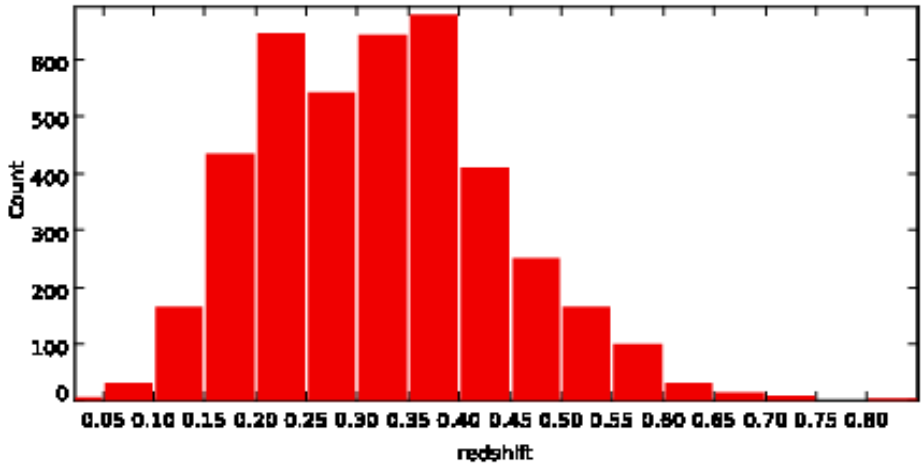}
\caption[Cluster redshift distribution in Stripe 82]{\small{The distribution of redshifts for the Stripe 82 cluster catalogue.}}
\label{fig:Stripe82zdistrib}
\end{figure}

The redshift distributions of the galaxy clusters in the COSMOS and Stripe 82 cluster catalogues can be seen in Figures \ref{fig:COSMOSzdistrib} and \ref{fig:Stripe82zdistrib}. Figure \ref{fig:COSMOSzdistrib} shows a fairly even distribution of clusters over the redshift range $0<z<1.2$. Peaks in the number of clusters at some redshifts may be due to the underlying LSS. Figure \ref{fig:Stripe82zdistrib} shows a peak in the distribution of redshifts in the Stripe 82 cluster catalogue with the number of clusters decreasing for $z>0.4$. This decrease is a selection effect due to the magnitude detection limits of the Stripe 82 area, which will limit the number of galaxies found at higher redshifts. \citet{Geach2011} claim that SDSS data allows for detection of galaxy clusters up to $z\sim0.5$. In the Stripe 82 field, the detection of clusters with $z<0.15$ is likely to be affected by the limited field size.

The redshift errors on the Stripe 82 clusters are estimated to be $\sim$5-9\% (at the median redshift) based on spectroscopic confirmation of 1549 galaxies within the clusters, and cluster redshifts are accurate to $z\sim0.5$ (\citealt{Geach2011}). There are some clusters with higher redshifts than this, and the errors on these are likely to be greater, so the maximum error value of 9\% has been used throughout.

The redshift errors on the COSMOS clusters have been estimated individually for each cluster, based on the standard deviation of the distribution of cluster member photometric redshifts (\citealt{Sochting2011}). These have been found to be between 0.2\% and 5.8\% of the cluster redshift (with some dependency on the redshift), with a mean value of 1.36\%.

\begin{figure}[!ht]
\centering
\includegraphics[scale=0.5,angle=-90]{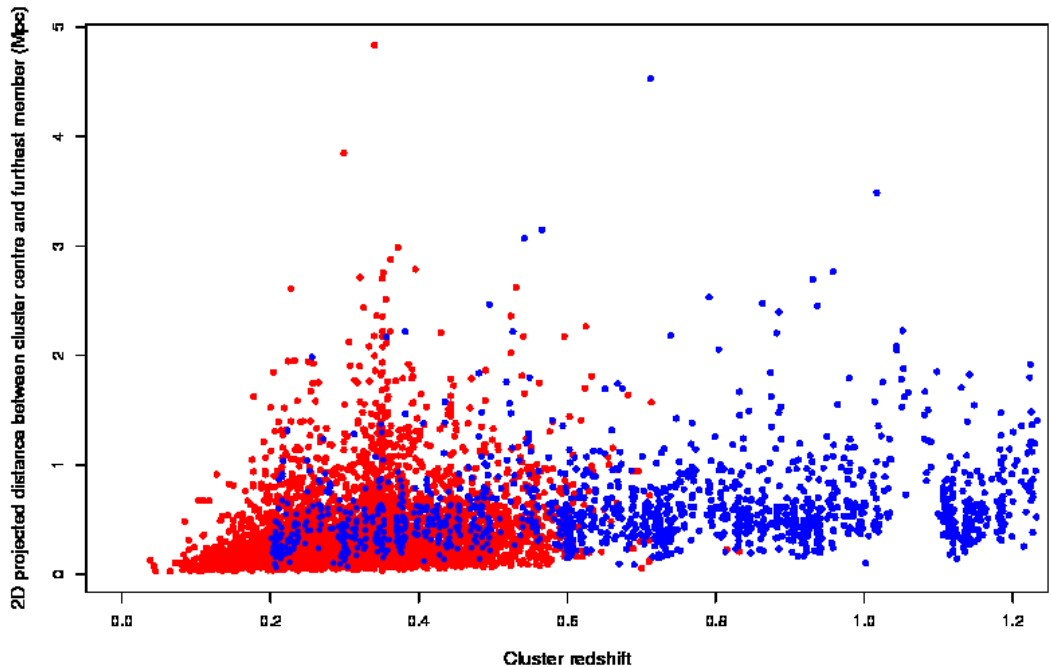}
\caption[Furthest cluster member]{\small{The distance from the centre of the galaxy cluster to the furthest galaxy. The red points show the Stripe 82 galaxy clusters, and the blue points are the COSMOS galaxy clusters.}}
\label{fig:clustersize}
\end{figure}

There are differences between the cluster catalogues in some definitions. For example, \citet{Geach2011} restrict the cluster boundary when the density reaches 10$\times$ the average field density, while \citet{Sochting2011} use a value of $2\times$ the average field density as the cluster limit. To test for any differences in cluster size, the distance from the cluster centre to the furthest cluster member has been calculated (Figure \ref{fig:clustersize}). For this, the distance to each cluster member from the cluster centre is found and galaxy with the largest distance is classed as the furthest cluster member. This distance is used to indicate the size of the cluster.

For Stripe 82, the minimum cluster size increases from 0.27 Mpc at $z=0.07$ to $\sim$0.3Mpc at $z=0.72$. The mean cluster size is 0.32 Mpc with a standard deviation of 0.26 Mpc. For the COSMOS clusters, the minimum cluster size increases from 0.96 Mpc at $z=0.21$ to 1.17 Mpc at $z=1.19$. The mean cluster size is 1.23 Mpc, with a standard deviation of 0.70 Mpc, which is larger than the values for Stripe 82. Therefore, the COSMOS cluster catalogue contains larger clusters than the Stripe 82 cluster catalogue due to the selection criteria. The difference in the definition of the cluster boundary is the cause of the difference in the maximum number of cluster members (178 for Stripe 82 and 235 for COSMOS, with COSMOS also having a cluster with 649 members).

Figure \ref{fig:clustersize} shows the distance from the centre of the galaxy cluster to the furthest galaxy. The red points and the blue points mark the Stripe 82 galaxy clusters and the COSMOS galaxy clusters, respectively. An increase in the minimum cluster size with redshift can be seen for both the Stripe 82 and COSMOS samples. The volume of the field increases with redshift, allowing larger clusters to be found at larger redshifts. 

There are peaks with the distribution for the Stripe 82 clusters in Figure \ref{fig:clustersize}. These suggest that at some redshift (such as $z\sim$0.35), there is an excess of clusters with a large distance between the cluster centre and the distance to the furthest cluster member. There is also a gap in the number of smaller clusters in the COSMOS data at $z\sim$1.05. The reasons for both of these effects are unknown.

This figure also shows the difference in cluster sizes selected by the two catalogues, indicating that the cluster size is affected by the selection criteria. Table \ref{tab:clustersizecomp} shows the size of the smallest cluster (using the distance to the furthest cluster member as an indication of cluster size) for a range of redshifts for both the COSMOS and Stripe 82 cluster samples. This shows the differences in the distance to the furthest cluster member and the increase in cluster size with redshift for both cluster catalogues for the smallest cluster found.

\begin{table}[!ht] 
\caption[Comparison of furthest cluster member distances]{\small{Comparison of distance to furthest cluster member for the smallest cluster for the COSMOS and Stripe 82 cluster samples. }}
\centering
\begin{tabular}{ c | c c c c }
Cluster Sample  & $z\sim0$ & $z\sim0.2$ & $z\sim0.6$ & $z\sim1.2$  \\ \hline
COSMOS          &  -       & 0.19Mpc   & 0.21Mpc    & 0.25        \\
Stripe 82       & 0.03Mpc  & 0.05Mpc   & 0.15Mpc    & -           \\
\end{tabular}
\label{tab:clustersizecomp}
\end{table}

Given that the Stripe 82 catalogue restricts the cluster size at 10$\times$ the average field density, it is possible that this method will only find the cluster cores and will miss possible cluster members further out. As the COSMOS data is deeper than the Stripe 82 field, this data will also find fainter members than found in Stripe 82. Differences in the selection process will not have an effect when comparing distances to cluster centres, as the mean RA and DEC of the cluster are unlikely to be affected by this. However, this will have an impact when looking at the distance to the closest galaxy and the distance compared to the size of the cluster. The richness is estimated by the number of galaxies within the cluster which have magnitudes between the magnitude of the Brightest Cluster Galaxy (BCG) and the BCG magnitude + 3. The richness estimate will also be affected by this density selection effect. When analysing the separation between a quasar and the closest cluster galaxy, and any separation and the richness, the COSMOS and Stripe 82 data will be studied separately.

\section{Quasar samples}

The Stripe 82 quasars were taken from the SDSS DR7 quasar catalogue (DR7QSO) (\citealt{Schneider2010}). The DR7QSO catalogue selects quasars with at least one broad emission line (therefore only Type I sources) and includes both quasars and lower luminosity sources such as Seyfert galaxies (\citealt{Richards2006}). When using the word quasar in this catalogue they often mean AGN in general.
This catalogue contains 105,783 objects with absolute magnitudes of $-30.28>M_i>-22.0$ ( and apparent magnitudes of $14.86<i<22.36$) and redshifts of $0.065<z<5.461$. The DR7QSO catalogue quotes redshift errors of $\Delta z\sim0.004$. In the Stripe 82 area, there are 1891 quasars with redshifts in the range $0.08<z<0.9$. The DR7QSO catalogue also covers the COSMOS field, containing 15 quasars within the redshift range $0.34<z<1.18$.

For the COSMOS field, the Large Quasar Astrometric Catalogue (LQAC) (\citealt{Souchay2009}) was used which contains 113,666 quasars, compiled from 12 quasar catalogues (four radio selected and eight optically selected, with SDSS DR6 being the largest). All of the quasars used within the COSMOS field were from either the SDSS Data Release 6 (\citealt{AdelmannMcCarthy2008}) or \citet{Veron2006}. For the LQAC, the redshifts errors are $\Delta z=0.01$ for $z<1$, and do not go above $\Delta z= 0.03$ for redshifts above 1.0 (\citealt{Souchay2009}). The redshift errors are larger for the LQAC catalogue as this redshift error is for the entire catalogue, which includes other databases of quasars with larger errors than the errors for SDSS. There is data from the $u, b, v, g, r, i, z, J, K$ photometric bands and from radio fluxes at 1.4GHz, 2.3GHz, 5.0GHz, 8.4GHz, and 24GHz. The absolute magnitudes in both the $i$ and $b$ bands can be found in the catalogue and the faintest magnitude being $i$mag = 20.31.

Overall, this has resulted in 47 quasars, 32 from the LQAC and 15 from DR7QSO, over a redshift range of $0.132<z<1.188$ in the area covered by the COSMOS field.  

In the COSMOS field, there are two X-ray point source catalogues, from \citet{Cappelluti2009} and \citet{Lusso2010}, taken as part of the COSMOS survey. The X-ray sources were taken from XMM-Newton data in the 0.5-2 keV, 2-10 keV and 5-10 keV energy bands, plus some Chandra observations of the central 0.9 deg$^2$ of the COSMOS field. The spectroscopic redshifts are from optical data from the Inamori Magellan Areal Camera and Spectrograph (IMACS) on the Magellan telescope, the Multi Mirror Telescope (MMT) observations, zCOSMOS, or are available in SDSS \citep{Brusa2009}. The spectroscopic redshifts have an accuracy of $\Delta z=0.004$ (\citealt{Lilly2007}). The limiting X-ray fluxes are $\sim1.7\times10^{-15}\mbox{erg cm}^{-2} s^{-1}$, $\sim9.3\times10^{-15}\mbox{erg cm}^{-2} s^{-1}$, and $\sim1.3\times10^{-14}\mbox{erg cm}^{-2} s^{-1}$ in the $0.5-2$ keV, $2-10$ keV and $5-10$ keV bands respectively. 65\% of these objects have properties typical of Type 1 quasars, and 15\% have Type 2 properties. 1887 independent point sources were detected in at least one band, 1032 of which have spectroscopically confirmed quasar redshifts, giving a photometric accuracy of $\sigma_{\Delta z/(1+z_s)}=0.014$ at $i^*_{AB}<22.5$ (\citealt{Salvato2009}) in the redshift range $0.103<z<4.251$. 98\% of the X-ray sources in this catalogue have optical counterparts \citep{Brusa2010} so some are likely to appear in the other optical catalogues. These X-ray quasars will be used to supplement the quasars in the COSMOS field, with the X-ray quasars being used only if there are no optical counterparts in the LQAC and DR7QSO catalogues. No distinction will be made between the optical and X-ray sources.

\subsection{Photometric Quasar Accuracy}\label{sect:qsophotz}

The COSMOS and Stripe 82 fields are covered by photometric and spectroscopic redshift surveys. Using quasars with both photometric and spectroscopic redshifts would increase the quasar sample size. However, before the quasars with photometric redshifts can be used, the accuracy of the redshifts must be tested. Within each field, comparisons were made between objects which have both photometric and spectroscopic redshifts.

\begin{figure}[!h]
\centering
\includegraphics[scale=0.5,angle=-90]{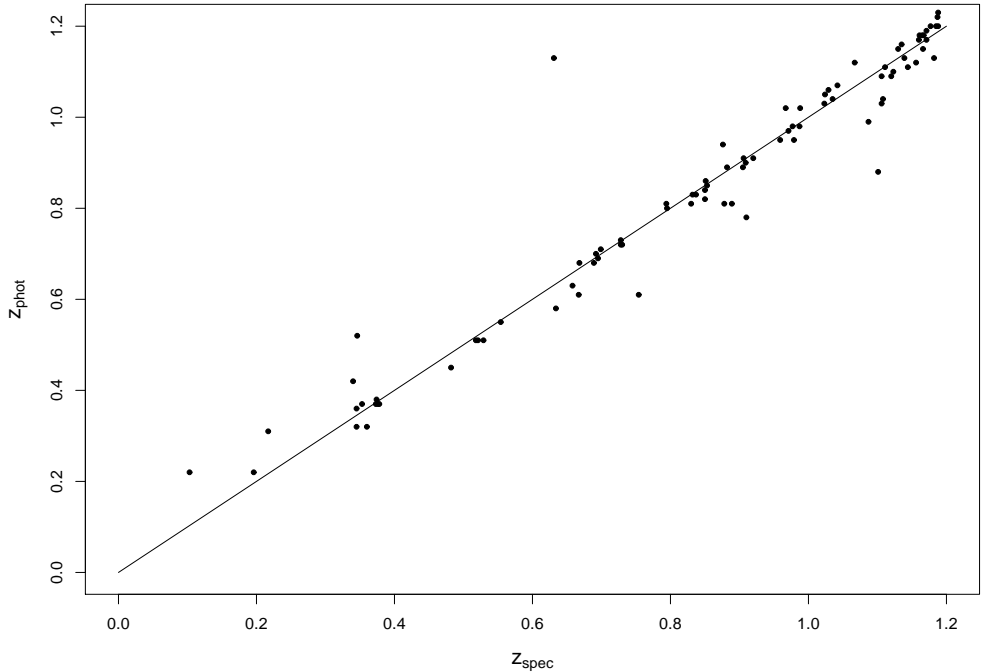}
\caption[Photometric and spectroscopic redshift from X-ray quasars]{\small{Photometric ($z_{phot}$) and spectroscopic ($z_{spec}$) redshifts for X-ray quasars within the COSMOS field. The line is the $z_{spec}=z_{phot}$ line.}}
\label{fig:xrayqsocomp} 
\end{figure}

Within the redshift range of 0.$103<z<1.2$, there are 99 X-ray quasars with spectroscopic and photometric redshifts within the COSMOS field area.  Figure \ref{fig:xrayqsocomp} shows photometric and spectroscopic redshifts for the X-ray quasars within the COSMOS field. The line is the $z_{spec}=z_{phot}$ line. Figure \ref{fig:xrayqsocomp} shows some scatter but a general good agreement between the two redshifts.

To assess the accuracy of the photometric redshifts for the X-ray quasars, Equation \ref{eq:deltaz} is used to calculate the discrepancies on the photometric redshifts.

\begin{equation}
\Delta z = \frac{z_{phot}-z_{spec}}{1+z_{spec}}
\label{eq:deltaz}
\end{equation}

The errors can also be found using Equation \ref{eq:deltaz2}, which is used by \citealp{Salvato2009} when comparing quasar redshifts. This uses the normalised median absolute deviation (NMAD; \cite{Hoaglin1983}). The NMAD is a more robust measure of variability than that used in Equation \ref{eq:deltaz} as it uses the sample median so is less influenced by outliers. 
\begin{equation}
\Delta z = 1.48 \times \mbox{median}\left((z_{phot}-z_{spec}) \over 1+z_{spec}\right)
\label{eq:deltaz2}
\end{equation}

As we wish to include the effect of any outliers on the samples, Equation \ref{eq:deltaz} will be used. 

\begin{figure}[!h]
\centering
\includegraphics[scale=0.5,angle=-90]{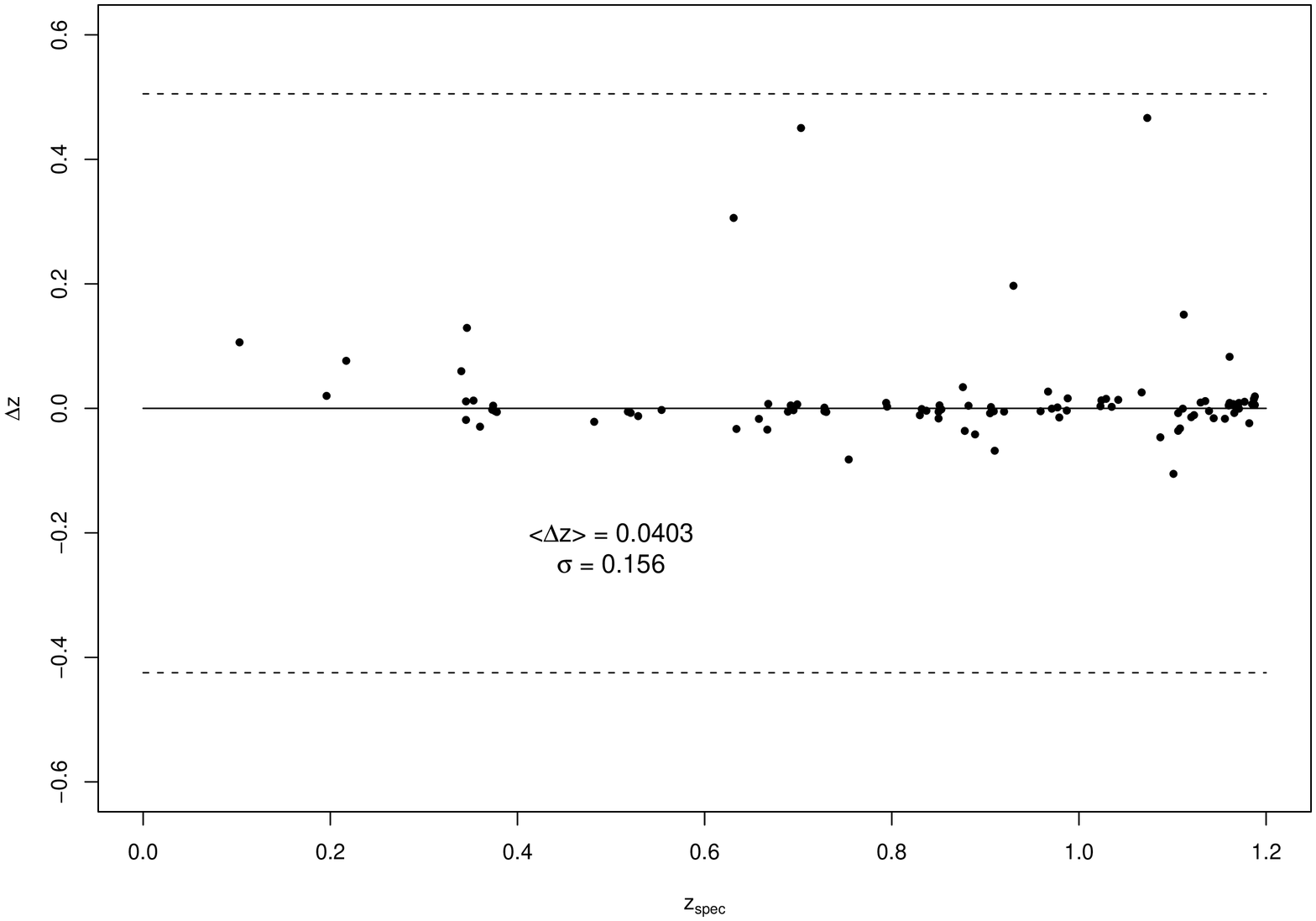}
\caption[Error on photometric X-ray quasars]{\small{The errors between the photometric and spectroscopic redshifts for X-ray quasars in the COSMOS field verses the spectroscopic redshifts. The dashed lines show the $3\sigma$ limits and the solid line shows the line $\Delta z=0$.}}
\label{fig:xrayqsoerr}
\end{figure}

Figure \ref{fig:xrayqsoerr} shows the redshift error against the spectroscopic redshift, using Equation \ref{eq:deltaz}. The dashed lines show the $3\sigma$ spread and the solid line marks $\Delta z=0$. The values for the $\Delta z$ and the standard deviation, $\sigma_{\Delta z}$ are based on the redshift range shown. This value is larger than that quoted in \citet{Salvato2009} because of the difference in the definition of $\Delta z $ used (Equation \ref{eq:deltaz2}). Figure \ref{fig:xrayqsoerr} show the spread in differences between photometric and spectroscopic redshifts. 

\begin{figure}[!h]
\centering
\includegraphics[scale=0.5,angle=-90]{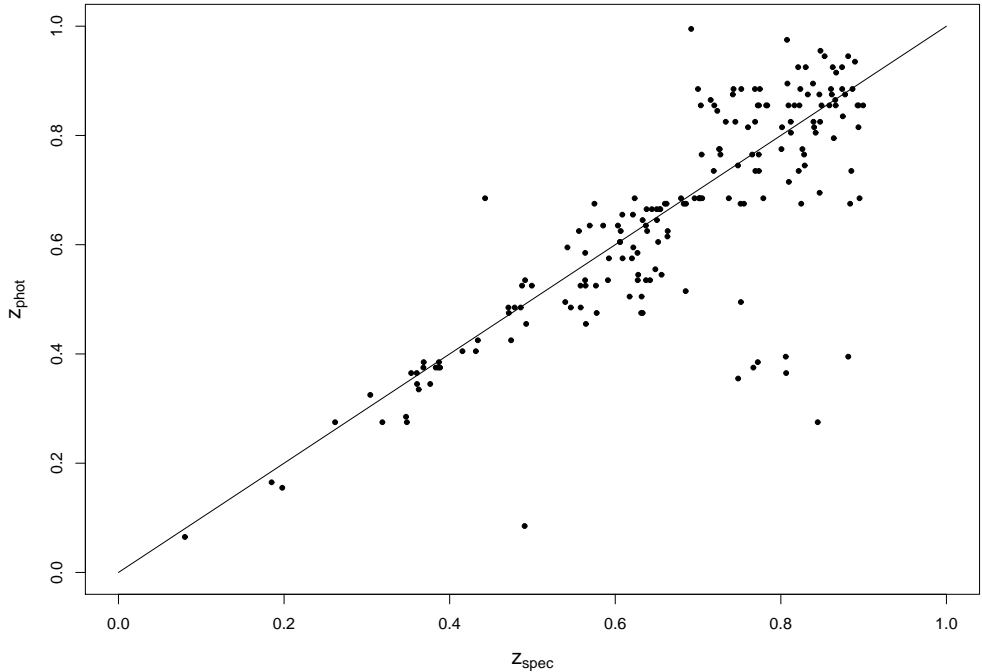}
\caption[Photometric and spectroscopic redshift from SDSS quasars]{\small{Photometric and spectroscopic redshifts for SDSS quasars within the Stripe 82 field. The line is the $z_{spec}=z_{phot}$ line.}}
\label{fig:stripe82qsocomp} 
\end{figure}

From SDSS, a catalogue of quasars with photometric redshifts has been created by \citet{Richards2009}. Comparing this photometric catalogue to quasars with spectroscopic redshifts from the DR7QSO catalogue gives an estimate on the errors on the photometric redshifts. Figure \ref{fig:stripe82qsocomp} shows the comparison of the photometric and spectroscopic redshifts. The solid line shows the $z_{spec}=z_{phot}$ relationship. The discrepancies between the two redshift estimates increase at $z>0.6$. 

\begin{figure}[!h]
\centering
\includegraphics[scale=0.5,angle=-90]{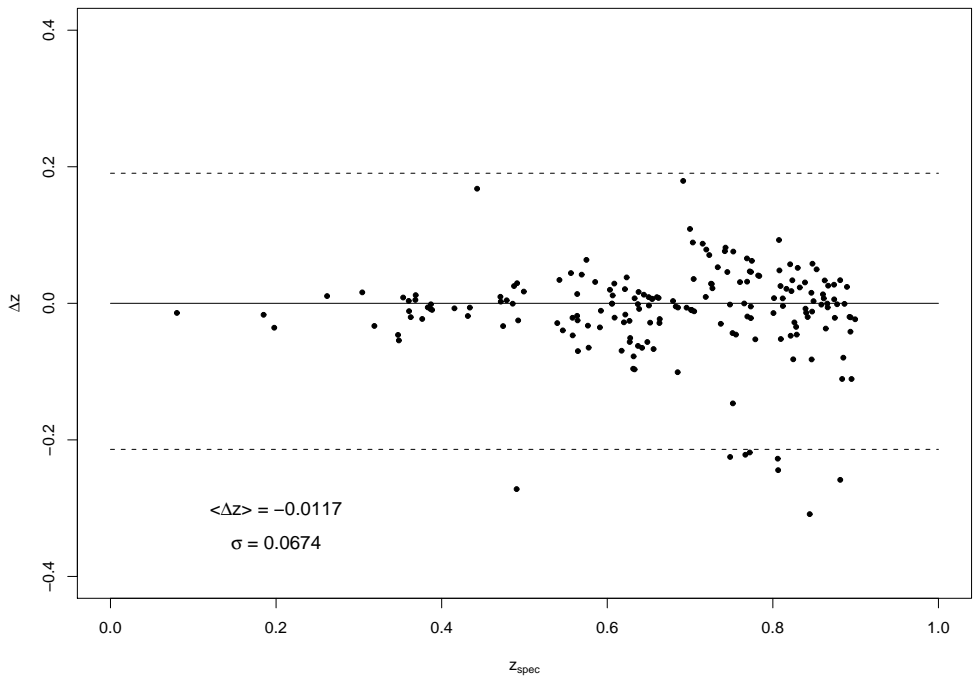}
\caption[Error on photometric SDSS quasars]{\small{The errors between the photometric and spectroscopic redshifts for SDSS quasars in the Stripe 82 field against the spectroscopic redshift. The dashed lines show the $3\sigma$ limits and the solid line shows the line $\Delta z=0$.}}
\label{fig:stripe82qsoerr}
\end{figure}

Figure \ref{fig:stripe82qsoerr} shows the calculated errors using Equation \ref{eq:deltaz}. The dashed lines show $3\sigma$ and the solid line shows the line $\Delta z=0$. These errors are similar to those for the X-ray quasars, though the standard deviation is less. There is still a difference in the photometric and spectroscopic redshift estimates. The distribution of points appears non-Gaussian so any points around $\Delta z = 0$ have been used to calculate $\sigma$.

Accurate redshifts are needed to accurately match quasars and clusters. There were sufficient quasars in the COSMOS and Stripe 82 fields to exclude quasars with only photometric redshifts because of the larger errors on the photometric redshifts. Table \ref{tab:qsodata} shows the data properties of quasar catalogues used.

\begin{table}[!ht]
\caption[Quasar catalogues]{QSO data properties. The number of quasars is the total number within the catalogue.}
\centering
\begin{tabular}{ c | c c c c c}
Survey                   & Magnitude limit    & Area          & Number     & Redshift errors \\ \hline
SDSS DR7QSO     & $i = 21.1$	      & 9380 deg$^2$  & 105,783    & 0.004           \\
LQAC                  & $i\sim 23$ & 9380 deg$^2$  & 113,666    & 0.01            \\
X-ray Catalogue             & $i^+ < 22.5$    & 2.13 deg$^2$  & 326        & 0.004            \\
\end{tabular}  
\label{tab:qsodata}
\end{table}\

\subsection{Selection Criteria and Biases}\label{sect:qsobiases}

Quasar classification within SDSS is based solely on the presence of broad emission lines (\citealt{Schneider2010}). Therefore, the SDSS DR7QSO catalogue selects the quasars which have at least one emission line, thereby eliminating any Type 2 quasars.

The overall DR7QSO catalogue contains objects with luminosities greater than $M_i=-22.0$. This means that, at $z\approx0.4$, an object with $M_i=-22$ will be rejected due to the catalogue selection criteria. These limits are not applied to the Stripe 82 area as this field is designed to be deeper than this and can be considered as a separate sample. However, this selection criteria will affect the COSMOS field, limiting the lower luminosity objects from the SDSS database at high redshifts.

\citet{Schneider2010} stress that the DR7QSO catalogue is not a statistical sample because the quasar selection process does not produce a uniform and homogeneous sample. The algorithm used in the selection process varies over time for the different SDSS data releases, so there is no uniform set of selection criteria (\citealt{Schneider2002}). Some of the quasars included have also been found during other surveys (i.e., not targeted by the spectroscopic SDSS quasar selection algorithm), so are not subject to the same criteria.

The LQAC contains objects from the SDSS DR6QSO release so some quasars may be found in both catalogues. Even though DR7 will include some of the quasars from DR6, some may have been removed from the new catalogue due to modifications in the selection criteria. The objects in DR6 are still valid quasars. 

The quasars in X-ray catalogues have been found in the main SDSS database to obtain the magnitudes so the quasar magnitudes can be compared.

\section{Sample windowing}

\begin{figure}[!h]
\centering
\includegraphics[scale=0.7,angle=0]{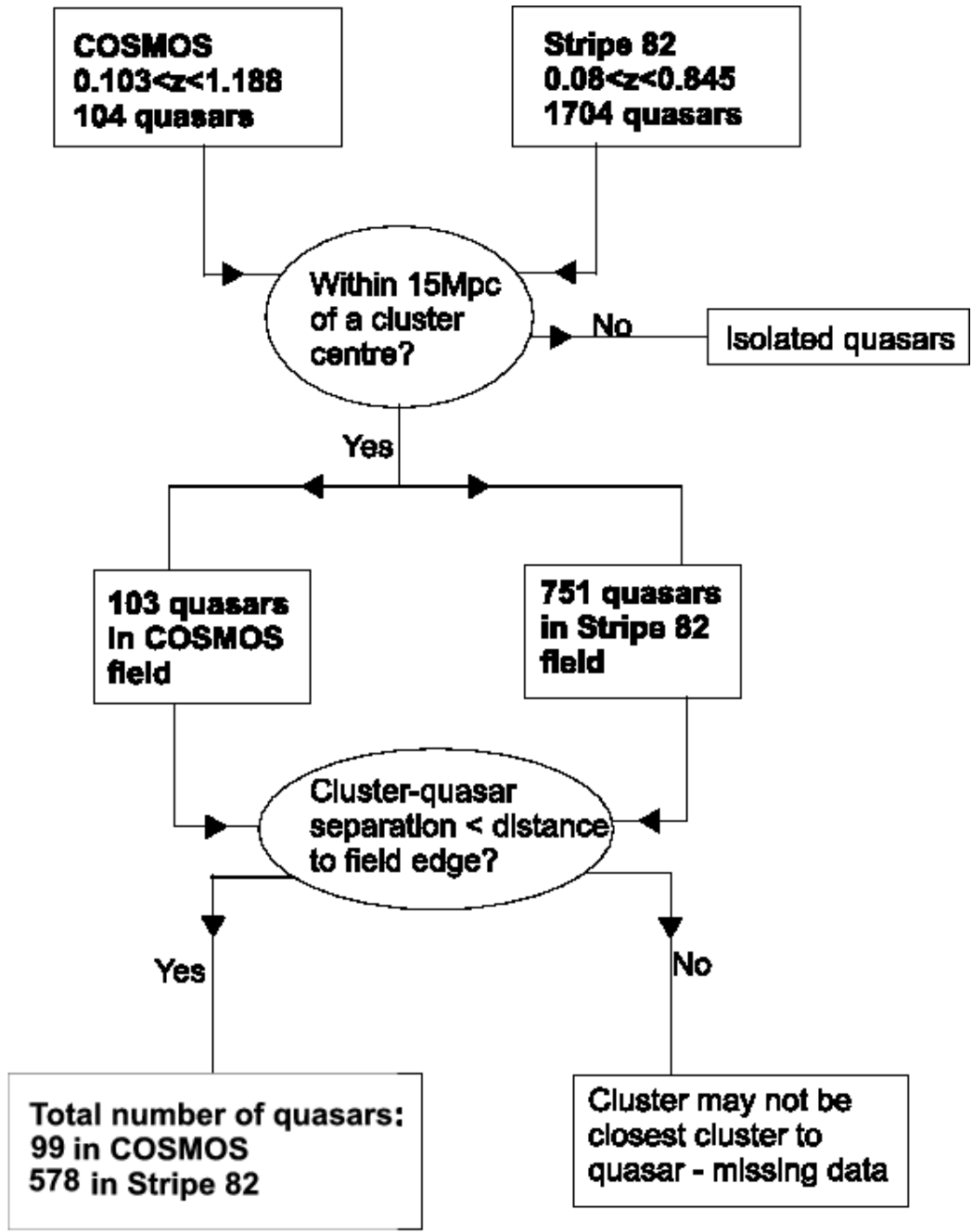}
\caption[Quasar-cluster pair selection]{\small{Flow chart showing the process used to select and eliminate quasars in final sample.}}
\label{fig:flowchartqso}
\end{figure}

Figure \ref{fig:flowchartqso} shows a flow chart of the process used to select the quasars from the catalogues. 

Initially, there were 104 quasars in the COSMOS field. These were taken from the SDSS DR7QSO catalogue, LQAC, and from the X-ray catalogues. Some quasars appeared in two or all of these catalogues. If this was the case, quasars from the DR7QSO catalogue were given priority (as it is the most up-to-date), then the LQAC quasar data, with the X-ray quasar data only being used if the quasar was not found in the other two catalogues. All of these quasars are within the redshift range $0.103<z<1.188$. 
In the Stripe 82 field, there were 1704 quasars within the redshift range $0.08<z<0.845$. This smaller redshift range is used as there is no cluster data above this redshift for this field. 

When the quasars were matched to their nearest cluster, there were 103 in the COSMOS field and 751 in the Stripe 82 field within a 2D projected distance of 15 Mpc  of the closest cluster, calculated at the epoch of the quasar. A maximum separation of 15 Mpc was chosen as this is roughly $\sim3\times$ the size of the largest cluster as shown in Figure \ref{fig:clustersize}. Therefore quasars lying at separation $>$ 15 Mpc will be beyond the sphere of influence of the largest cluster. To ensure that the cluster selected is the closest cluster to the quasar, the separation between the quasar-cluster pair must be less than the distance to the nearest field edge. The cluster-quasar pairs were discarded if the quasar lay closer to the field edge than the closest cluster, as it would not be possible to determine if there could be another cluster closer to the quasar outside the field.

This selection process reduced the numbers to 99 quasars in the COSMOS field and 578 in the Stripe 82 field.

\section{Cluster-Quasar Separations}\label{sect:separations}

The separation between the galaxy cluster and quasars are found using the redshift of the quasar and the RA and DEC positions (see Section \ref{sect:2dsep}).

Initially all clusters lying within a specific redshift range of the quasar were selected. This redshift range was given by the error on the cluster redshifts. For the COSMOS galaxy clusters, the error is given individually for the COSMOS clusters and is estimated at 9\% of the cluster redshift for the Stripe 82 clusters (see section \ref{sect:clustsamples} for more details). This error is to equivalent of 18 Mpc at $z=0.1$ and 92 Mpc at $z=0.7$. The distances from the quasars to each of these clusters was calculated, and the clusters lying within a 2D projected distance of 15 Mpc of each quasar were selected. The distances were found as the proper distance at the epoch corresponding to the quasar redshift. The distance at the redshift of the quasar was used as the quasar redshift was the most accurate. This distance would also give the most logical distance to compare as this gives the physical distance at the epoch of the quasar.

To give an indication of where the quasar is lying with respect to the cluster, the distances to the closest galaxy and the Bright Cluster Galaxy (BCG) were found to asses any relationship between the quasar and BCG, and the quasar and closest galaxy.

A ``separation ratio" (sepRatio) is defined as the separation using the quasar redshift at the redshift of the quasar, divided by the average radius of the cluster. This will give an indication as to whether the quasar lies within the cluster (SepRatio $<$ 1) or outside the cluster (sepRatio $>$ 1). With this method, there will also be some line of sight problems. Given the errors on the cluster redshift, quasars which appear to lie within the cluster boundaries may lie in front or behind the cluster. For quasars seen as lying outside the clusters, this effect will not exist. 

The 3D separations have been calculated which will give some indication as to whether the quasar is likely to lie within the cluster or if it is a line of sight effect. The errors on the 3D separations have been calculated. The projected 2D separations do not take into account the cluster redshift, except to ensure the cluster and quasar are within a set redshift range. This means the cluster to which a quasar is attached to as the closest cluster may be different for the 2D and the 3D separations. The information about each quasar is included in the final database (see Section \ref{databaseinfo} for details on the final database).

\subsection{Calculating the 2D Separations}\label{sect:2dsep}

The 2D projected separation between the cluster centre and the quasar at the redshift of the quasar is given by Equation \ref{eq:sep}. q and g denote the quasar and galaxy cluster respectively. 

\begin{equation}
\mbox{projected separation}  = R_{tq} \times r_q \times \psi
\label{eq:sep}
\end{equation}

where the scale factor $R_{tq}$ = $1 \over(1+z_q)$,  $z_q$ is the redshift of the quasar and $\psi$ is the angular separation.

The comoving distance, $r_q$, was found using Equation \ref{eq:comovingdist} (\citealt{Peacock1999}) and assuming that $\Omega_M + \Omega_\Lambda = 1$ and $\Omega_R = 0$, where $\Omega$ is the density parameter.

\begin{equation}
r_q = \frac{c}{H_0}\int_0^z (\Omega_M (1+z)^3 + 1 - \Omega_M )^{-0.5} \dif z
\label{eq:comovingdist}
\end{equation}

The angular separation, $\psi$, between the cluster and the quasar is given using spherical trigonometry (Equation \ref{eq:angsep}) (\citealt{Smart1977}).

\begin{equation}
\cos\psi = \cos\theta_g\cos\theta_q + \sin\theta_g\sin\theta_q\cos(\phi_g - \phi_q)   
\label{eq:angsep}
\end{equation}

where $\theta$ = (${\pi}\over{2}$ - $DEC$) and $\phi$ = $RA$ ($DEC$ and $RA$ are both in radians). 

The projected separations between the cluster centres (the mean of the RA and Dec of all cluster members) and the quasars have been calculated at the redshift of the quasar. 

The 2D projected separations between the quasar and the BCG, and the quasars and closest cluster galaxy have also been calculated using this method.

\subsection{Calculating the 3D Separation}\label{3Dsep}
The 2D separations are only projected separations and do not give information about where, in 3D space, the quasar lies in relation to the galaxy cluster. The 3D separation takes into account the differences in redshift of the quasar and cluster and calculates the actual physical separation in Mpc. Finding the 3D separations is limited to only the distance between the quasar and the cluster centre. Without individual redshifts for each galaxy within the cluster, it is not possible to determine the physical 3D distance to the closest cluster member or the edge of the cluster. For these values, only the 2D projected distance can be estimated. 

The 3-dimensional separations are found using the positions and redshifts of both the cluster and the quasar. The distances to each cluster and quasar were calculated, and the separation between the quasar and cluster centre calculated using simple trigonometry. The separations are much larger than the 2D projected distances given the differences in redshift of the cluster and the quasar, and the errors involved are also larger due to the errors associated with the cluster redshifts. The separations were calculated at the redshift of the quasar.

The 3D distance was calculated using Equation \ref{eq:3Dsep} (\citealt{Kibble2004}).
\begin{equation}
\mbox{sep\_3D} = \sqrt{ (x1 - x2)^{2} + (y1 - y2)^{2} + (z1 - z2)^{2} }  
\label{eq:3Dsep}
\end{equation}

where  $x1$,\ $y1$\ and $z1$\ are given by Equations \ref{eq:prop1_1} - \ref{eq:prop1_2}, which use the quasar redshift and position. 
\begin{equation}
x1 = R_{t0} \times r_q \times \sin\theta_q \times \cos\phi_q
\label{eq:prop1_1}
\end{equation}
\begin{equation}
y1 = R_{t0} \times r_q \times \sin\theta_q \times \sin\phi_q 
\end{equation}
\begin{equation}
z1 = R_{t0} \times r_g \times \cos\theta_q 
\label{eq:prop1_2}            
\end{equation}

$x2$,\ $y2$\ and $z2$\ are calculated using Equations \ref{eq:prop2_1} - \ref{eq:prop2_2} and use the redshift and position of the cluster centre. 

\begin{equation}
x2 = R_{t0} \times r_g \times \sin\theta_g \times \cos\phi_g 
\label{eq:prop2_1} 
\end{equation}
\begin{equation}
y2 = R_{t0} \times r_g \times \sin\theta_g \times \sin\phi_g 
\end{equation}
\begin{equation}
z2 = R_{t0} \times r_g \times \cos\theta_g  
\label{eq:prop2_2}           
\end{equation}

$r_g$ and $r_q$ are the comoving distances for the galaxy cluster and the quasar respectively, given by Equation \ref{eq:comovingdist}, and $R_{t0}$ is the scale factor at the present epoch and is equal to 1.0.

\subsection{Separation Errors}\label{seperrors}

Given that the 2D separations rely on only the redshift of the quasar and the positions of both the quasar and cluster, the errors are relatively small. The errors on the quasar redshift can be found in Table \ref{tab:qsodata}.

The errors on the 3D separations are larger due to the increased error on the redshifts of the clusters which is also dependent on the cluster redshift. For the errors on the galaxy clusters, see section \ref{sect:cluster_selection}.  The 3D separations are useful as they give information about the actual physical separations between the quasar and the galaxy cluster centre. 

Errors on the 3D separations were calculated using the cluster redshift errors. To do this, the physical distance represented by the cluster redshift error is estimated. For this, the distance to the cluster has been calculated. Then the distance to the cluster is calculated if the cluster were to be lying at $z\pm\Delta z$, using Equation \ref{eq:3Dsep}. The difference in the two distances estimates the physical distance which corresponds to the error on the cluster redshift. This gives a good estimate on the error on the 3D separations.

For the COSMOS sample, the mean error on the 3D separations due to the error on the cluster redshift is $<\Delta \mbox{sep3D}>=29.7$ Mpc with a standard deviation of 8.9 Mpc. For Stripe 82 clusters, the mean error on the 3D separations is $<\Delta \mbox{sep3D}>=139.17$ Mpc with a standard deviation of 27.8 Mpc. This is expected due to the larger errors on the Stripe 82 cluster redshifts.

\section{Cluster Shape}\label{sect:clustMorph}

The length of the major and minor axes of the galaxy clusters has been estimated using the inertia tensor. This method will also provide information about the orientation of the cluster in the sky. This information is used when studying the separation ratio, which uses the mean of the major and minor axes as an estimate of the cluster radius. The orientation of the cluster is used to look at large scale structure and to study the positions of the quasars with respect to the orientation of galaxy clusters.

\subsection{Inertia Tensor}
The inertia tensor was used to calculate the moment of inertia for each cluster member. We have no information about the masses of the individual galaxies so have assumed a point mass for each galaxy and have also assumed an elliptical 2D distribution. Given that most clusters are elliptical, for the inertial tensor, this is a good assumption. A 2D distribution has been used as the redshift information for each individual galaxy in a the cluster is not available, giving no information about the 3D mass distribution. With only a few galaxies in some clusters, calculating the density of the cluster would be unreliable so a uniform distribution has been used. Using the mean of the moment of inertia for each cluster member, the overall inertial tensor for the cluster is found. The centre of the galaxy cluster is taken as the mean position of all the cluster members. The tensor is given by:

   \[ I = \left| \begin{array}{cc}
     I_{xx} & I_{xy}  \\
     I_{yx} & I_{yy}  \end{array} \right|\] 

where 
\begin{eqnarray}
I_{xx} & = & \Sigma ((y- \overline y)^2 ) \nonumber \\
I_{yy} & = & \Sigma ((x- \overline x)^2 ) \nonumber \\
I_{xy} = I_{yx} & = & - \Sigma (x - \overline x) \times ( y - \overline y ) \nonumber
\end{eqnarray}

The array is then diagonalised and, from the resultant array, the major axis is given by $i_a = \sqrt{4 I'_{yy}\over \mbox{n}}$ and the minor axis is given by $i_b = \sqrt{4 I'_{xx}\over \mbox{n}}$, where n is the number of galaxies within the cluster, and $I'_{xx}$ and $I'_{yy}$ are from the diagonalised array.

The major and minor axes found using the inertia method are only the axes of the cluster as seen projected on the sky. Due to the lack of redshift information on each cluster member, the shape of the cluster along the line of sight can not be estimated. 

\begin{figure}[!h]
\centering
\includegraphics[scale=0.7,angle=0]{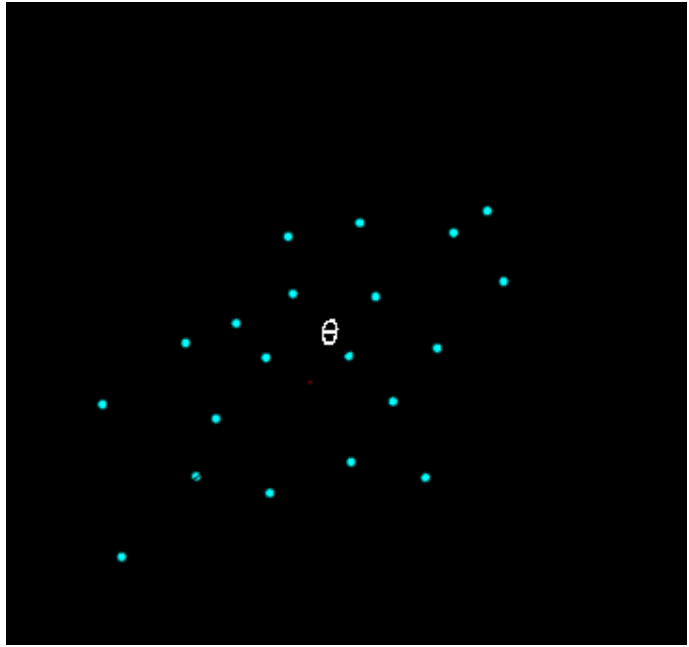}
\caption[Angle from inertia tensor]{\small{The angle and major and minor axes calculated using the inertial tensor method.}}
\label{fig:inerttensor}
\end{figure}

Figure \ref{fig:inerttensor} shows the major and minor axes, and the angle for an example cluster. The vertical line is set at the RA of the centre of the cluster. 
The orientation angle between the major axis of the cluster and the vertical line through the RA of the cluster centre is taken as $\pm$90 deg with 0 degrees lying at the RA of the cluster centre. This angle will also be a projected angle between the cluster's major axis and the vertical line of constant RA (i.e., this is not a solid angle estimate).

\subsection{Inertia Tensor Errors}\label{sect:angerr}

To calculate the errors on the angle of orientation of the cluster, the inertia tensor is calculated for each cluster with a single member removed and repeated for each cluster member. The spread in the angles of the orientation of the galaxy cluster shows the effect a single galaxy has on that cluster. Clusters with a larger standard deviation are more affected by single members. This method does work for most of the clusters. However, for clusters with a small number of members (e.g., $<10$ members), this method is less reliable as each cluster member has a larger affect on the overall angle value than in a cluster with a large number of members.

\begin{figure}[!ht]
\centering
\includegraphics[scale=0.5,angle=-90]{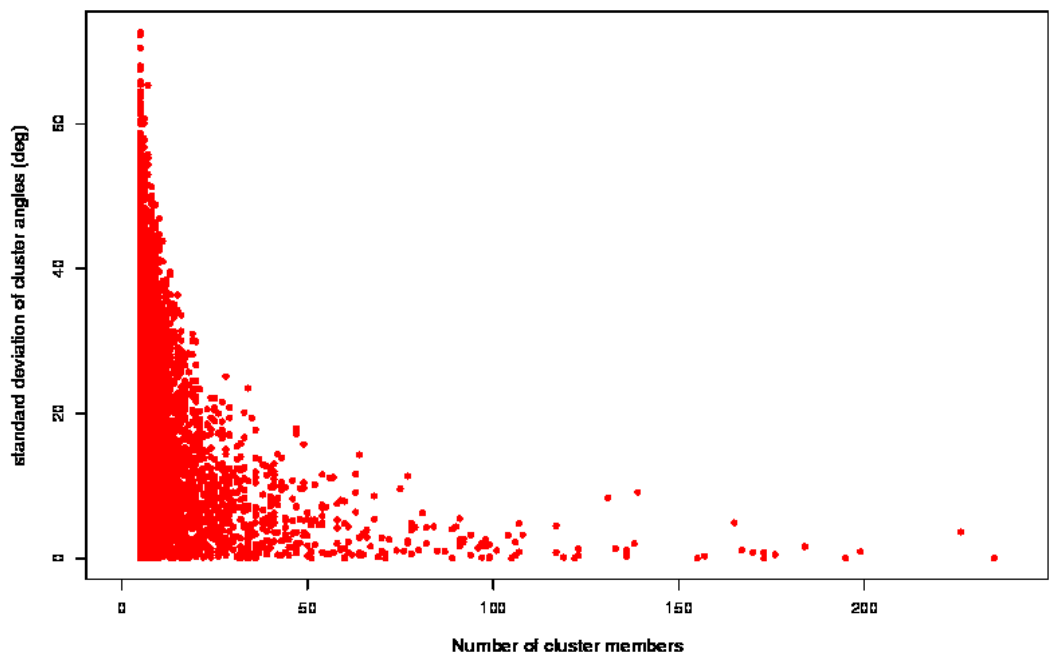}
\caption[Error on angle from inertia tensor]{\small{The standard deviation of the angle of the orientation of the galaxy cluster found when a single member is removed from a cluster.}}
\label{fig:inerterr}
\end{figure}

Figure \ref{fig:inerterr} shows the standard deviation in the errors on the angles of the clusters against the number of cluster members. The spread in the estimated angle of orientation decreases as the number of member galaxies increases as a single member has less impact on the estimated orientation of the cluster. On average, the errors were found to be within 20$^{\circ}$ ($2\sigma$) of the quoted angle of orientation from the inertia tensor.

\clearpage

This method relies on the cluster being elliptical. The errors on the angle of the cluster will be larger for more circular clusters as their major axes are less well defined. To test for this effect, the ratio of the minor to major axes has been calculated, to assess the ellipticity of the clusters. The average ratio between the minor and major axis is 0.46$\pm$0.19, indicating on average, the clusters are elliptical. In Stripe 82, 4\% of the clusters have a ratio between the minor and major axis of 0.80. This percentage is slightly higher for the COSMOS field at 6\%. However, this does indicate that the errors on the orientation of the cluster for most clusters will not be affected by ellipticity of the cluster.  

\begin{figure}[!ht]
\centering
\includegraphics[scale=0.5,angle=-90]{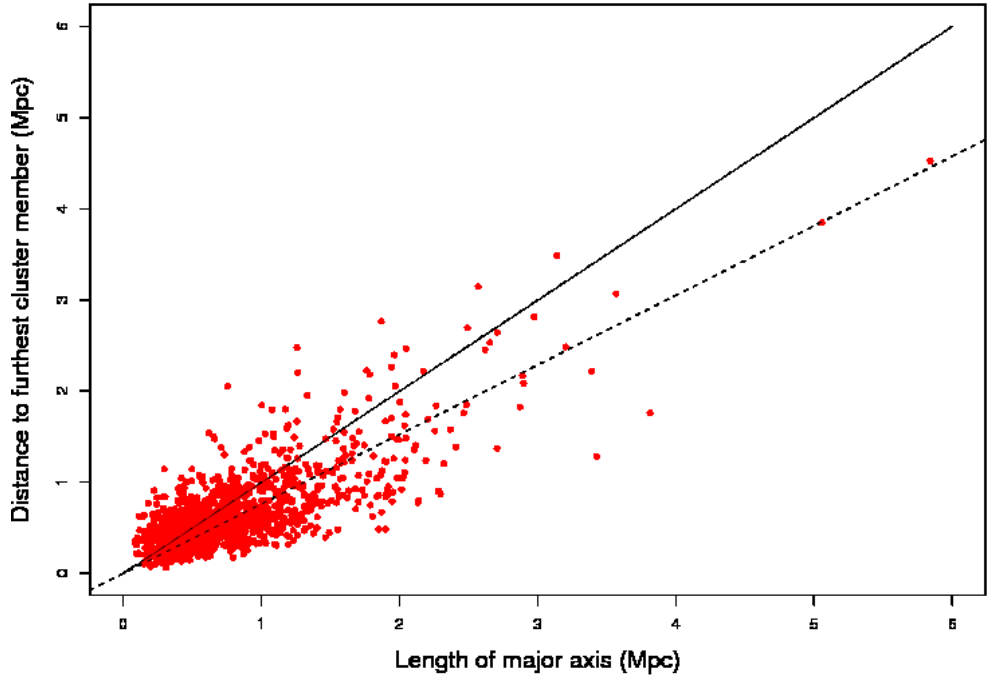}
\caption[Comparison of major axis and furthest cluster member for COSMOS]{\small{The distance to the furthest cluster member plotted against the major axis estimate for the COSMOS cluster sample. The solid line is the distance to the furthest member = length of the major axis. The dashed line is the line of best fit between the major axis values and the distance to the furthest cluster member and is found to be the distance to the furthest cluster member = 0.76 $\times$ the length of the major axis.}}
\label{fig:COSMOSsizecomp}
\end{figure}

\begin{figure}[!h]
\centering
\includegraphics[scale=0.5,angle=-90]{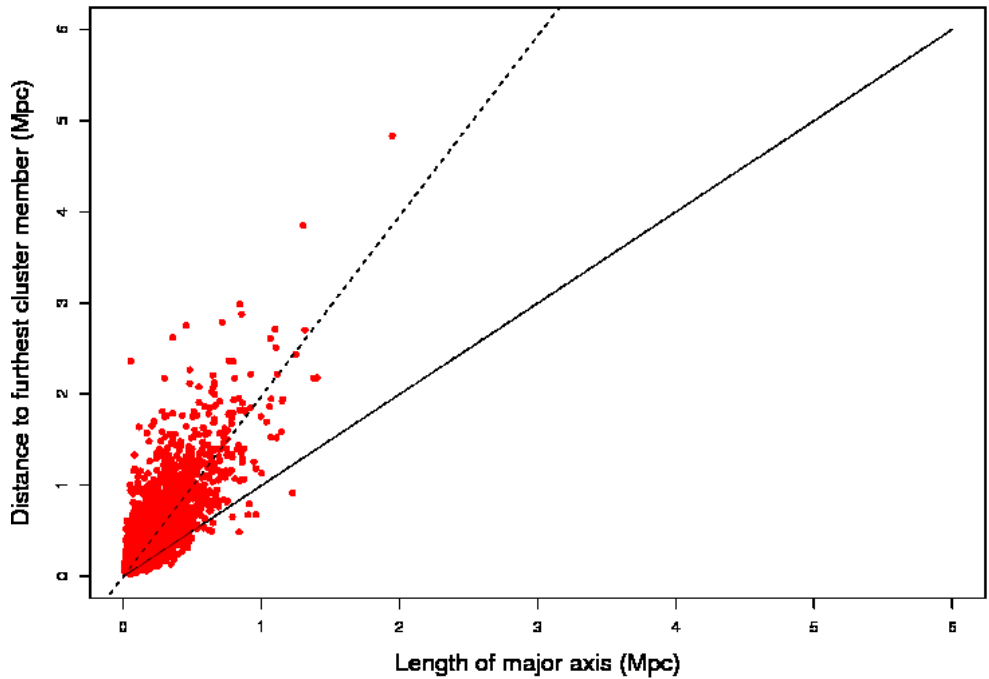}
\caption[Comparison of major axis and furthest cluster member for Stripe 82]{\small{The distance to the furthest cluster member plotted against the major axis estimate for the Stripe 82 cluster sample. The solid line is the distance to the furthest member = length of the major axis. The dashed line is the line of best fit between the major axis values and the distance to the furthest cluster member and is found to be the distance to the furthest cluster member = 1.98 $\times$ the length of the major axis.}}
\label{fig:strip82compsize}
\end{figure}

To check the accuracy of the length of the major axis of the clusters from the inertia tensor method, the projected 2D separations from the cluster centre to the furthest cluster member have been calculated and plotted against the estimated major axis length for the COSMOS sample (Figure \ref{fig:COSMOSsizecomp}) and for Stripe 82 (Figure \ref{fig:strip82compsize}). As the galaxy clusters are mostly elliptical, the distance to the furthest cluster member should correlate roughly to the length of the major axis estimated from the inertial tensor method. 

For COSMOS clusters, there is a large amount of scatter (Figure \ref{fig:COSMOSsizecomp}). The solid line is the $x=y$ line. The dotted line is the line of best fit, where the distance to the furthest galaxy member is 0.76$\times$ the length of the major axis. 

However, for the Stripe 82 sample (Figure \ref{fig:strip82compsize}), the distance to the furthest cluster member is calculated to be larger than major axis estimate. The solid line is the line of $x=y$ and the dotted line is the line of best fit, where the distance from the cluster centre to the furthest galaxy member is 1.98$\times$ the length of the major axis. This difference is likely due to the selection criteria used in selecting the cluster members, as discussed in Section \ref{sect:cluster_selection}. The clusters identified in Stripe 82 also have a lower minimum number of members, which will increase the error associated with the inertial tensor.

\clearpage
\section{Control Sample}\label{sect:simulations}
To check any results found are not due to random alignments, a sample of control quasars has been created. This sample was created by taking the existing samples and randomly selecting the RAs and DECs, without replacement. This created a random sample of quasars with the same distributions of RA and DEC as the observed sample. The separations between quasars from the random quasar sample and clusters from the COSMOS and Stripe 82 cluster samples were calculated using the methods described in Section \ref{sect:separations}. The cluster samples were not altered for this.

If observed results match the simulated results obtained using this random sample, any patterns seen in the observed sample should be due to random coincidence and not a physical process. However, if there are any large structures within the field, there is no way to avoid some of these structures potentially appearing in the random sample.

The absolute magnitude was calculated using (\citealt{Peacock1999}):

\begin{equation}
M_r = r - 2.5(\alpha-1)\log(1+z) - 5\log\left(\frac{L_D}{r_{pc}}\right)
\label{eq:absmag}
\end{equation}

where $r$ = the apparent $r$ magnitude, $r_{pc}=10$pc and $L_D$ = the luminosity distance which is given by:

\begin{equation}
L_D = \sqrt{\frac{L}{4\pi F}}
\label{eq:lumindist}
\end{equation}

where L is the luminosity, and F is the flux. 

 $\alpha$ is related to the power law of the Spectral Energy Distribution (SED) of the quasar, using the form $f_{\nu}$ $\propto$ $\nu^{- \alpha}$ where $\alpha=0.5^{+0.5}_{-0.05}$ (\citealt{VandenBerk2001,Zheng1997,Schneider2001}). To study the evolution of the environments of AGN and quasars, a large redshift range has been used. Because of this, at different redshifts different parts of the spectra will be seen in the $r$ band.

As the control sample uses the same selection methods for finding quasar-cluster pairs as the observed sample, the two samples will not nessarily have the same sample size. For example, a smaller control sample would indicate more quasars lie at distances $<15$ Mpc in the observed sample than in the control sample. When using the statistical tests in Chapters \ref{chap:3}, \ref{sect:chap4}, and \ref{sect:chap6}, the effect of the sample size and of differing sample sizes has been investigated, and numbers have been normalised where needed.  

All parameters (e.g., separation ratio, distances to BCG and closest galaxy, etc.) have been calculated for the simulated data. Results from these control catalogues will be compared to results obtained from the observations and will be used to assess the significance of any results. The control samples are included in the attached disk and further details on the database file can be found in Appendix 2. 

All the results are presented in Chapters \ref{chap:3} and \ref{sect:chap4}.

\section{Summary}
The quasar and cluster catalogues cover a redshift range of $0<z<1.2$, though in the range $0.8<z<1.2$, there are only redshifts in the COSMOS field. Quasar-cluster pairs have been selected where the quasar lies at the same redshift of the cluster (within the cluster redshift errors) and within a 2D projected distance of 15 Mpc. There are 99 quasar-cluster pairs in the COSMOS field and 578 in the Stripe 82 field. The 2D projected separations at the epoch of the quasar have been found between the quasar and the closest cluster centre, the closest cluster member, and the BCG of the cluster. The 3D distances between the quasar and closest cluster centre have also been found, which uses the redshifts of both the quasar and the cluster. The errors on the 3D separations have been found using the errors on the cluster redshift.

The orientation of the cluster is found using the inertia tensor. This method also gives an estimate of the length of the major and minor axes of the cluster. The orientation angle between the quasar and the cluster major axis have been calculated to study the orientation of the quasar with respect to the closest cluster. 

There are small differences in the definitions of the boundaries of a cluster for the 2 fields, which mean that the clusters selected have different properties such as cluster size. The fields have been studied together and separately to account for these differences.

\chapter{Quasar-Cluster Proximity as a Function of Redshift}\label{chap:3}

This chapter investigates the distances between quasars and clusters as a function of redshift. Different mechanisms can trigger quasar activity and dominate in different environments. The distance a quasar lies from a cluster gives an indication of that environment and studying the change in the environment over a range of redshift allows the evolution of the quasar triggering mechanisms to be studied. The distances to the nearest galaxy and the 3D distances have also been studied. 

The results are also compared to studies of the control field which uses a random sample of quasars (see Section \ref{sect:simulations} for more details).

\section{Quasar-Cluster Distances}

Different distances have been calculated (Sections \ref{sect:2dsep} and \ref{3Dsep}):
\begin{itemize}
\item projected 2D separation between the quasar and the closest cluster centre at the redshift of the quasar;
\item projected 2D separation between the quasar and the closest cluster member at the redshift of the quasar, and
\item 3D separation between the quasar and the closest cluster centre
\end{itemize}
Each of these values will be studied separately and compared to the same distance in the control field. This control field will be used to compare the observed data to a random sample, making it possible to determine whether any trends seen are real. 

To study the evolution of quasar-cluster separations with redshift, redshift slices have been taken with low redshift defined as $0<z<0.4$, intermediate redshift as $0.4<z<0.8$ and high redshift as $0.8<z<1.2$. 

The number of quasar-cluster matches within 15 Mpc, median redshift of the quasars and number of quasar-cluster matches to the closest cluster within the redshift ranges described above can be seen in Table \ref{tab:fin_datasample}.

\begin{table}[!h]
\caption[Numbers of quasar-cluster pairs]{\small{Information on the final number of quasar-cluster matches.}}
\centering
\begin{tabular}{c | c c c c c }
Survey           & QSO-cluster & median z & $0<z<0.4$ & $0.4<z<0.8$ & $0.8<z<1.2$ \\
                 & pairs       &          &           &             &              \\ \hline
Both fields      & 677         & 0.487    & 192       & 420         & 65     \\
COSMOS           & 99          & 0.905    & 11        & 24          & 64     \\                         
Stripe 82        & 578         & 0.467    & 181       & 396         & 1      
\label{tab:fin_datasample}
\end{tabular}
\end{table}

\subsection{Separation from the Closest Cluster Centre}\label{sect:centre_seps}

\begin{figure}[!ht]
\centering
\includegraphics[scale=0.5,angle=-90]{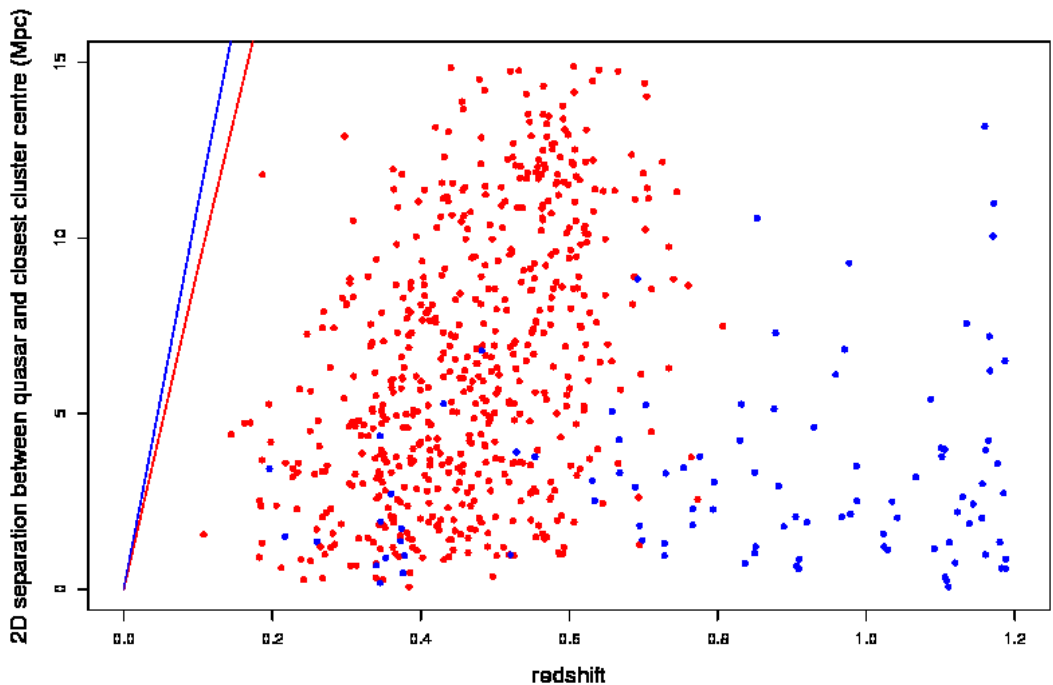}
\caption[2D separations to closest cluster centre]{\small{The quasar-closest cluster centre separations for the redshift range $0<z<1.2$. The red points are from Stripe 82 and the blue points are from COSMOS. The lines indicate the field edge for the COSMOS (blue) and Stripe 82 (red) fields. }}
\label{fig:sep_each}
\end{figure}

Figure \ref{fig:sep_each} shows the separations between a quasar and the closest cluster centre as a function of redshift for the COSMOS (blue points) and Stripe 82 (red points) fields. This clearly shows the area where each survey dominates. There are no quasars in the top left of the figure as would be expected due to the limited field sizes. However, this would not affect the top right of the figure, where at high redshifts, there is a deficit of quasars lying at large separations from a cluster centre. 

There appears to be a ``spike'' in the number of quasars at $z\sim1.15-1.2$. This indicates an increase in the density in this area, possibly the present of the LQG.  

The lines on Figure \ref{fig:sep_each} show the physical distance to the edge of the field for a quasar lying at the centre of that field, to give an indication of the approximate size of the field. The COSMOS field is square so at the centre of the field, the minimum distance to the edge of the field will be the same in the RA and DEC directions. The Stripe 82 field, however, is rectangular. The width of Stripe 82 is only 3$^{\circ}$ in $DEC$ and 118$^{\circ}$ in $RA$, so therefore limited most by the $DEC$ range. The distance from $DEC$=0 to the closest edge has been used.

\begin{figure}[!ht]
\centering
\includegraphics[scale=0.5,angle=-90]{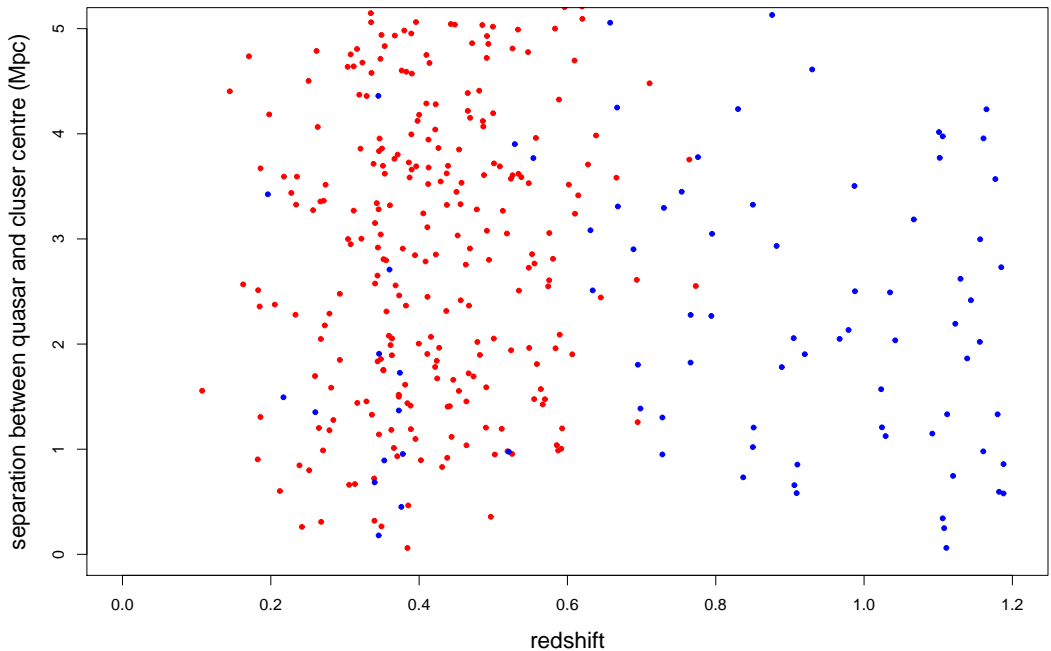}
\caption[Small 2D separations to closest cluster centre]{\small{The separations between quasar and the closest cluster centre for the redshift range $0<z<1.2$. The red points are from Stripe 82 and the blue points are from COSMOS.}}
\label{fig:all_sep_close}
\end{figure}

Figure \ref{fig:all_sep_close} shows the separations up to 5 Mpc for the COSMOS and Stripe 82 fields. The red points are from Stripe 82 and the blue points are from the COSMOS field. There appears to be a deficit of quasars lying close to a cluster centre ($<$1 Mpc) in the intermediate redshift range, $0.4<z<0.8$. In this redshift range, only 33\% of quasars lie at separations $<$5 Mpc and only 2.4\% lie closer than 1 Mpc. This is potentially significant. This deficit is most evident in the COSMOS sample as the COSMOS field covers the whole redshift range. It can also be seen to some extent in the Stripe 82 sample, though this sample becomes less complete for $z>0.6$. This deficit of quasars at close distances to galaxy clusters should not be caused by the field size as the deficit at larger separations will be. 

At low redshifts, $z<0.4$, the COSMOS field will be affected by the field edge so the results for separations at redshifts $z<0.4$ should not be considered complete for the COSMOS sample. However, at higher redshifts, the limited field size should not affect the results. For redshifts $z>0.8$, 75.8\% of the quasars lie at distances $<$5Mpc with 19.4\% lying within 1Mpc. For the Stripe 82 field, above $z=0.6$, the number of quasar-cluster pairs decreases. This is due to the reduction in the completeness of the Stripe 82 sample (\citealt{Geach2011}) and should not be considered as a physical effect.

Table \ref{tab:sepdata} shows the fractions of quasars lying $<1$ Mpc and $<5$ Mpc from the centre of a cluster for low ($0<z<0.4$), intermediate ($0.4<z<0.8$) and high ($0.8<z<1.2$) redshift ranges, for the combined COSMOS and Stripe 82 sample, and for the COSMOS and Stripe 82 fields separately. At low redshifts, all of the quasars in the COSMOS field lie $<5$ Mpc from a cluster centre. This is an effect of the limited field size which means the field is too small to get quasar-cluster pairs with larger separations. The Stripe 82 data shows a decrease in the fraction of quasars lying close to a cluster centre as the redshift increases. This can also be seen to a lesser extent in the COSMOS data. At higher redshifts, the fraction of quasars lying close to a cluster centre appears to increase again. This would mean quasars avoid the higher density regions at intermediate redshifts ($0.4<z<0.8$) so secular processes would be more dominate in this redshift range.

\begin{table}[!ht]
\caption[Separations for different redshifts]{\small{The percentage of quasars at distances from the closest cluster centre for the COSMOS and Stripe 82 fields, at low, intermediate and high redshifts.}}
\centering
\begin{tabular}{c | c c | c c | c c }
 &\multicolumn{2}{c|}{$0<z<0.4$}&\multicolumn{2}{c|}{$0.4<z<0.8$}&\multicolumn{2}{c}{$0.8<z<1.2$}\\
 &$<1$ Mpc&$<5$ Mpc &$<1$ Mpc&$<5$ Mpc &$<1$ Mpc&$<5$ Mpc \\
\hline
Both fields     & 10.4\% & 68.9\% & 2.4\%  & 33.0\% & 19.0\% & 74.6\%  \\
COSMOS field    & 38.5\% & 100\%  & 8.3\%  & 79.2\% & 19.4\% & 75.8\%  \\
Stripe 82 field &  8.3\% & 66.7\% & 2.0\%  & 30.2\% & -      & -       
\label{tab:sepdata}
\end{tabular}
\end{table}

\subsection{3D Separations}

\begin{figure}[!ht]
\centering
\includegraphics[scale=0.5,angle=-90]{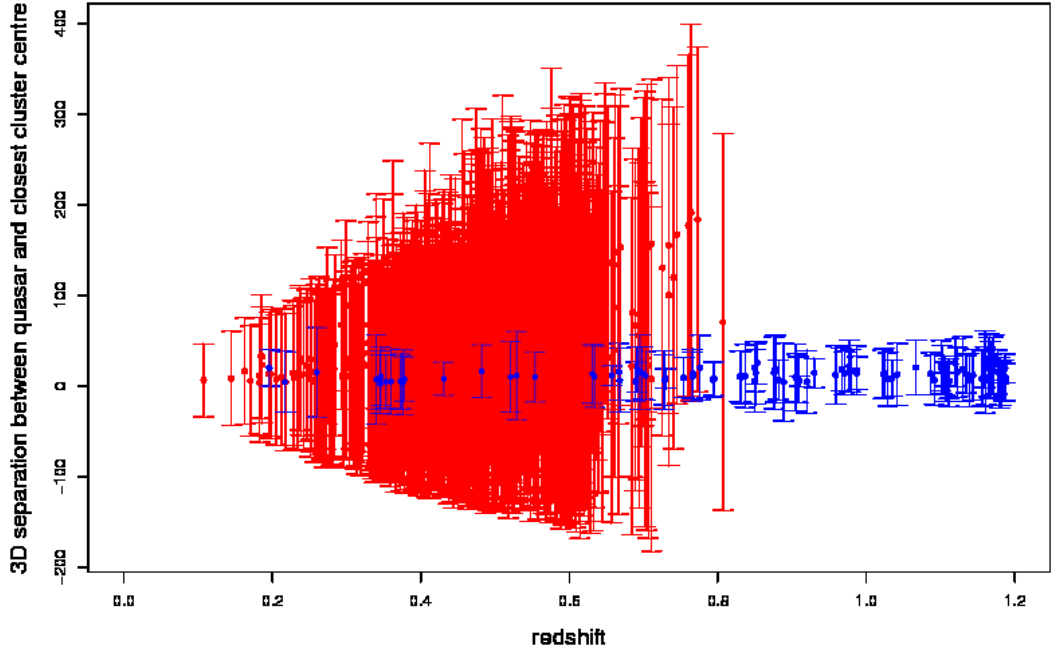}
\caption[3D separations to closest cluster centre]{\small{The closest 3D separations between quasar and the closest cluster centre for the redshift range $0<z<1.2$ and errors on the 3D separations. The red points shows quasars from the Stripe 82 data while the blue points mark COSMOS quasars.}}
\label{fig:3Dsep_sep}
\end{figure}

Figure \ref{fig:3Dsep_sep} shows 3D separations between quasars and the closest cluster centre as a function of redshift for the Stripe 82 field (red points) and COSMOS field (blue points) with error bars showing the error on the calculated 3D separation. As mentioned in Section \ref{seperrors}, the errors have been calculated for the 3D separations using the error on the cluster redshifts. Figure \ref{fig:3Dsep_sep} shows the errors on the 3D separatios in Mpc. The errors on the COSMOS cluster (1-5\%) are significantly lower than those for the Stripe 82 clusters (9\% of the cluster redshift). The errors increase with redshift, which is expected as the error on the cluster redshift is a fraction of the redshift.

As can be seen in Figure \ref{fig:3Dsep_sep}, the errors on the 3D separations are too large to allow conclusions to be drawn from the 3D separations. However, the 3D separations will be included in later Chapters for completeness. 

\subsection{Separation to Closest Cluster Galaxy}

The 2D projected separations between the quasar and the closest cluster member have been calculated. This distance is a projected 2D distance as the redshift information for individual cluster members is not available. 

\begin{figure}[!ht]
\centering
\includegraphics[scale=0.5,angle=-90]{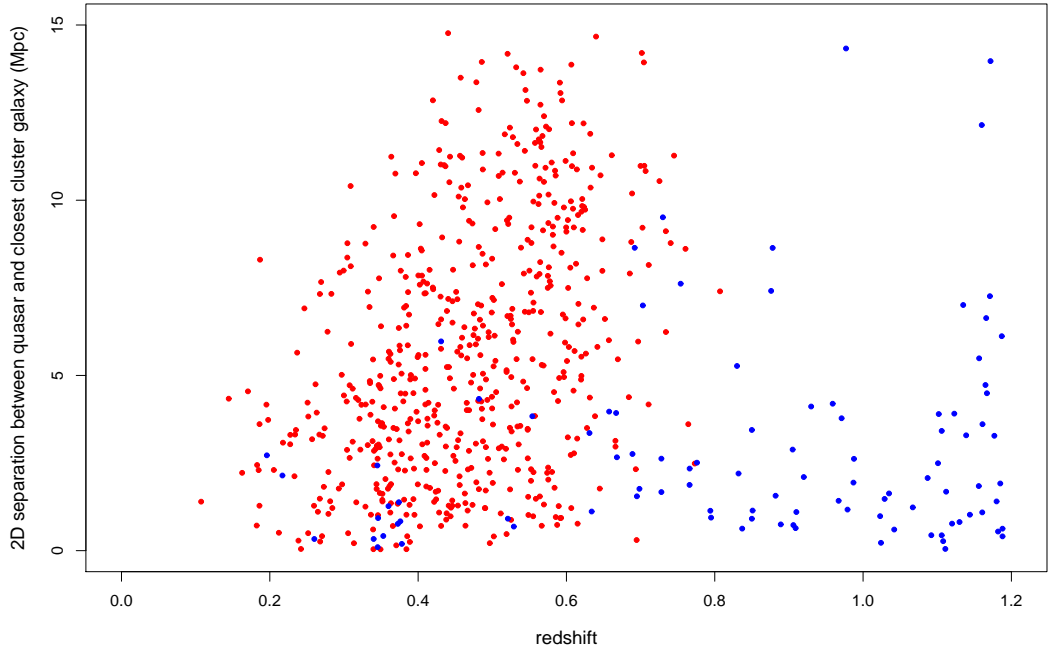}
\caption[2D separations to closest cluster galaxies]{\small{The closest separations between quasar and the closest cluster member for the redshift range $0<z<1.2$. The red points are quasars from Stripe 82 and the blue points are quasars from COSMOS.}}
\label{fig:sep_closestgal}
\end{figure}

Figure \ref{fig:sep_closestgal} shows the 2D projected separations between the quasar and the closest cluster member for the Stripe 82 field (red) and the COSMOS field (blue). In this Figure, the deficit of quasars lying close to clusters at intermediate redshifts is not so apparent. There appears to be no evolution with redshift of the distance a quasar lies from the closest galaxy. 
A larger data sample at higher redshifts is needed to further test any evolution with redshift. The significance of this will be tested in Section \ref{sect:stats}.

\subsection{Control Field}

Section \ref{sect:simulations} describes the method used to create a control sample of quasars. This involves randomly re-sampling the RAs and DECs of quasars to create a random sample with the same distribution of positions as the observed sample. The positions and redshifts of the galaxy clusters have been left the same. The field edge effects are the same for the control sample as for the observed sample, as the two datasets have been treated the same. 

\begin{figure}[!ht]
\centering
\includegraphics[scale=0.5,angle=-90]{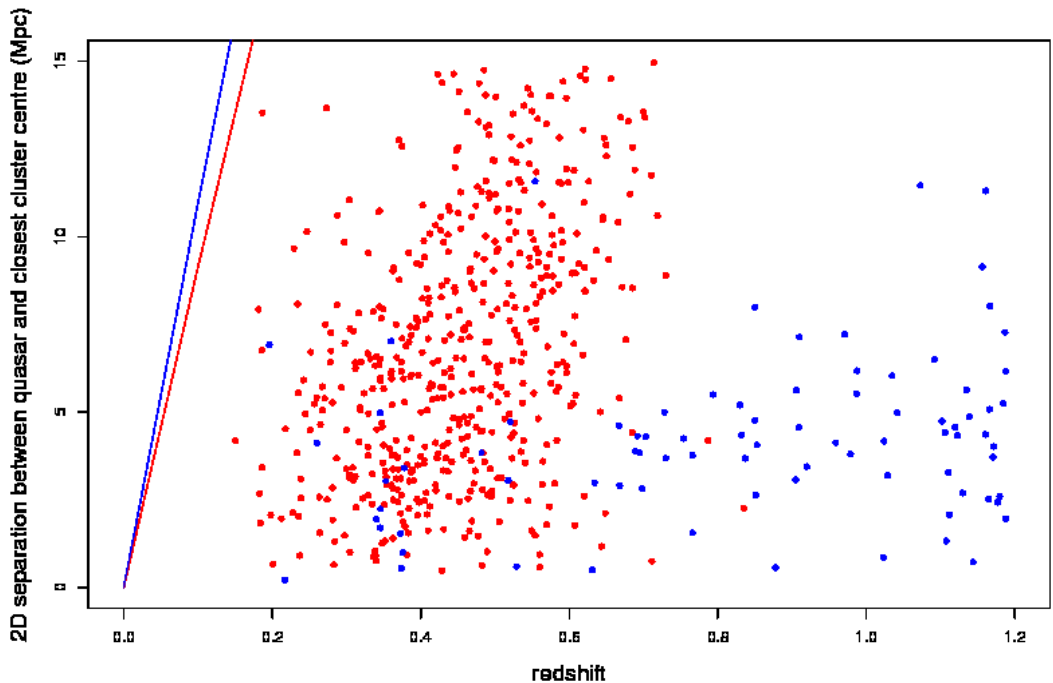}
\caption[2D separations for control fields]{\small{The separations between a control quasar and the closest cluster centre for the redshift range $0<z<1.2$. The red points are from Stripe 82 and the blue points are from COSMOS. The lines indicate the field edges for the COSMOS (blue) and Stripe 82 (red) fields. }}
\label{fig:sep_closestsim}
\end{figure}

Figure \ref{fig:sep_closestsim} shows the 2D projected separations between the samples of control quasars and the closest cluster centres. The red points show separations using the control quasars with randomly selected positions in the Stripe 82 field whereas the blue point mark the control quasars in the COSMOS field. The lines on Figure \ref{fig:sep_closestsim} show the physical distance to the edge of the field for a quasar lying at the centre of that field, to give an indication of the approximate size of the field. These lines are the same as those in Figure \ref{fig:sep_each}.

This Figure shows the distribution of the control quasar-cluster separations with redshift is similar to that of the COSMOS and Stripe 82 observations shown in Figure \ref{fig:sep_each}. For the COSMOS sample (blue points), there appear to be more quasars lying further away from the closest cluster centre in the observed sample than in the control sample. This suggests, for the COSMOS field, quasars prefer to lie in less dense regions.

For the Stripe 82 data, between $0.2<z<0.4$, there appear to be more quasars lying further away from the closest cluster centre than are seen in the observed data sample. 

In Figure \ref{fig:sep_closestsim}, there is no evidence of the deficit of quasars lying close to a cluster centre at $0.4<z<0.8$, which is seen in Figure \ref{fig:sep_each}. In the COSMOS control sample at high redshifts, there is a lack of quasars lying at large distances from the closest cluster centre. This lack is not seen in the observed samples. More control samples are needed to investigate this deficit.

The errors on the 3D separations will have the same distribution as for the observed data. It is not possible to comment on any trends from the calculated 3D separations as the errors are too large.

Statistical tests will test the validity of these results and observations based on visual inspection (see Section \ref{sect:stats}).

\section{Angular Separations}\label{angsep}

\begin{figure}[!ht]
\centering
\includegraphics[scale=0.7]{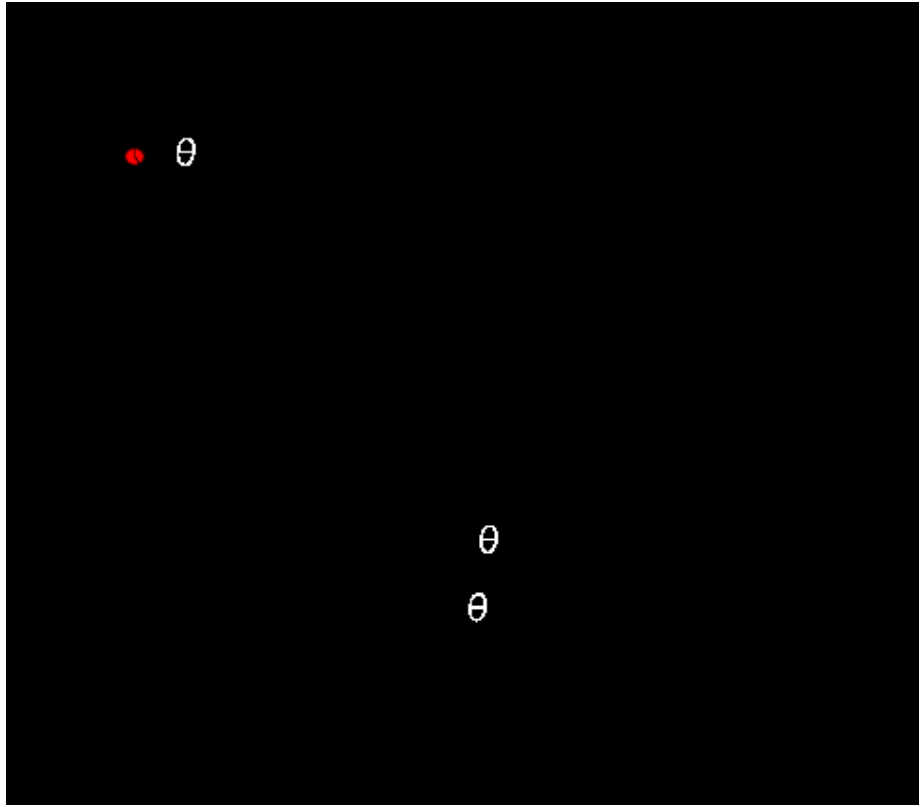}
\caption[Calculating orientation angle]{\small{The angles used to calculate the angle between a quasar and the major axis of the closest cluster.}}
\label{fig:ang_calc}
\end{figure}

We define the angular separation as the angle between the line connecting the cluster centre with the quasar position and the major axis of the cluster. First the angle between the line connecting the cluster centre with the quasar position and the Declination axis intersecting the quasar position is calculated ($\theta_q$ in Figure \ref{fig:ang_calc} which shows the geometrical arrangement). Second the angle between the major axis and the Declination axis intersecting the cluster centre, $\theta_c$, is found. These are then used in Equation \ref{eq:phi_cq} to derive the angular separation of the quasar with respect to the major axis of the galaxy cluster, $\theta_{cq}$.

\begin{equation}
\theta_{cq} = 180 - \theta_q + \theta_c
\label{eq:phi_cq}
\end{equation}

\begin{figure}[!ht]
\centering
\includegraphics[scale=0.65,]{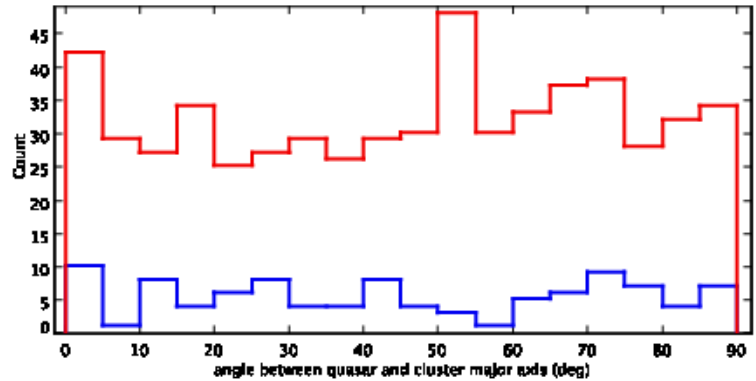}
\caption[Histograms of orientation angles]{\small{Histogram of angles between a quasar and the major axis of the closest galaxy cluster for COSMOS (blue) and Stripe 82 (red). }}
\label{fig:anghist}
\end{figure}

An angle of 0$^{\circ}$ indicates that the quasar lies in line with the major axis of the cluster. An angle of 90$^{\circ}$ indicates the quasar lies perpendicular to the major axis of the cluster (i.e., in line with the cluster minor axis). The distribution of the angles between a quasar and the major axis of the closest cluster can be seen in Figure \ref{fig:anghist} for COSMOS (blue) and Stripe 82 (red). Section \ref{sect:angerr} describes the method used to calculate the errors on the major axis of the cluster. The errors on the angle are 13$^{\circ}\pm 10^{\circ}$. This error is not large enough to have a significant impact on the overall distribution of the angles. 
  
\begin{figure}[!ht]
\centering
\includegraphics[scale=0.65]{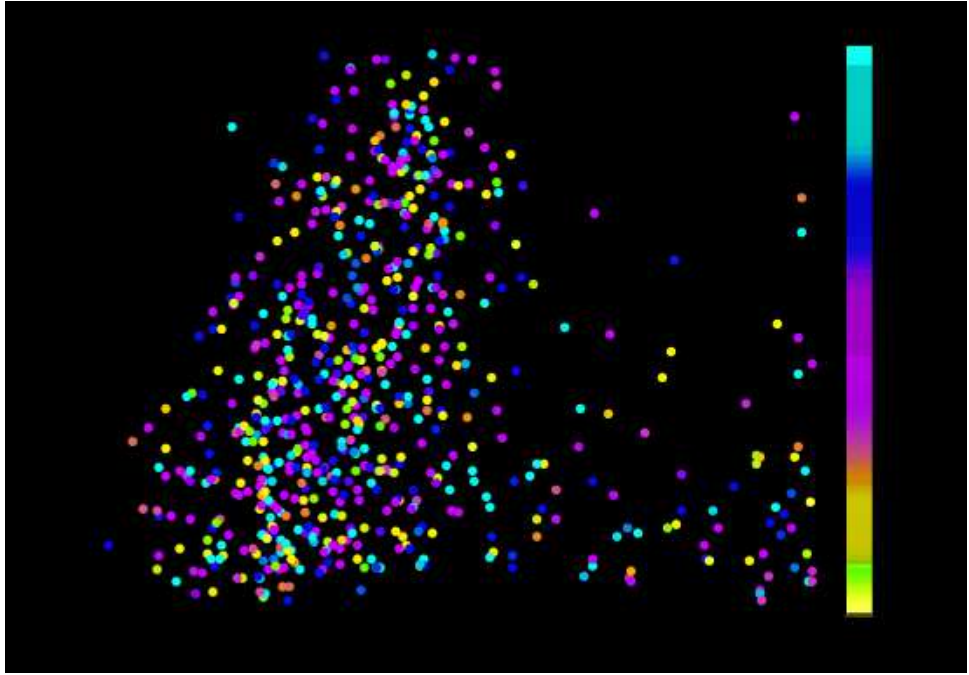}
\caption[Separations as a function of orientation angle for observed data]{\small{The separations between a quasar and the closest cluster centre for the redshift range $0<z<1.2$. The points are colour coded with respect to the angle between the quasar and the cluster major axis.}}
\label{fig:sepz}
\end{figure}

Figure \ref{fig:sepz} shows the separations between the quasar and the closest cluster centre over the redshift range $0<z<1.2$. The points are colour coded to show the angle between the cluster major axis and the quasar. The figure shows no pattern in the distributions of angles with either redshift or separation distance, which suggests quasars lie at no preferential angle to the cluster orientation. Again, the errors on the orientation angle are not large enough to have a significant impact on this result.

These observations are based on the visual inspection of the plots. A statistical analysis can be found in Section \ref{sect:stats}.
\clearpage

\subsection{Control Field}

The angles between the control quasars and the cluster major axis were found using the method described in Section \ref{angsep}.

\begin{figure}[!ht]
\centering
\includegraphics[scale=0.65]{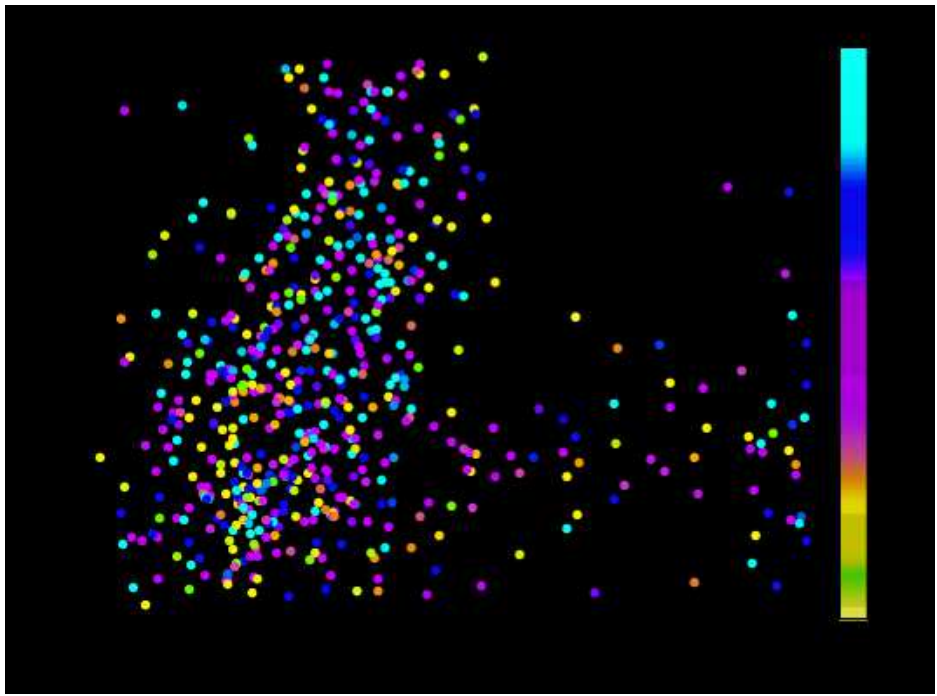}
\caption[Separations as a function of orientation angle for control data]{\small{The closest separations between a control quasar and the closest cluster centre for the redshift range $0<z<1.2$. The points are colour coded with respect to the angle between the quasar and the cluster major axis.}}
\label{fig:ang_sep_sim}
\end{figure}

Figure \ref{fig:ang_sep_sim} shows the distribution of separations between the control quasars and the closest cluster centre as a function of redshift. The distribution of angles between a control quasar and the major axis of the closest cluster is shown by the colours of the points. This will be compared to observed data (seen in Figure \ref{fig:ang_sep}) in Sections \ref{sect:compangles} and \ref{sect:stats}.

\subsection{Comparing Angles in Observed and Control Fields}\label{sect:compangles}
For each of the histograms in this section, the solid line shows data from the observed field and the dotted line shows the data from the control field.

\begin{figure}[!ht]
\centering
\includegraphics[scale=0.75]{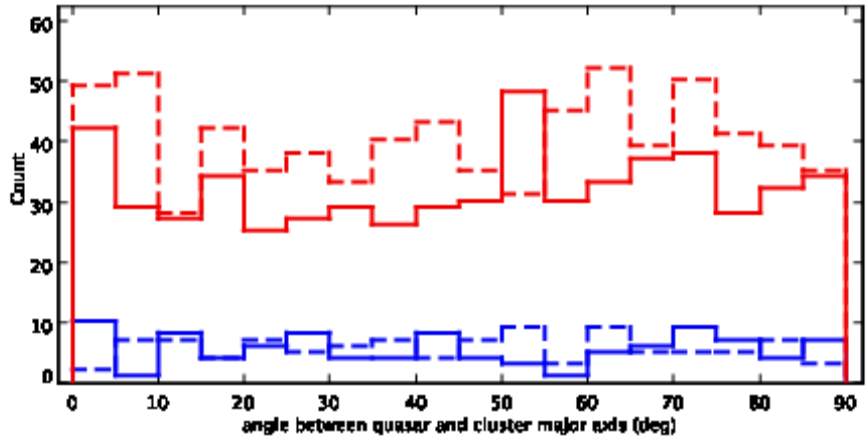}
\caption[Comparing orientation angles in observed and control data]{\small{The orientation angle of the quasar with respect to the cluster major axis for Stripe 82 (red) and COSMOS (blue) for the observed (solid) and control (dotted) fields.}}
\label{fig:ang_field}
\end{figure}

\begin{figure}[!ht]
\centering
\includegraphics[scale=0.75]{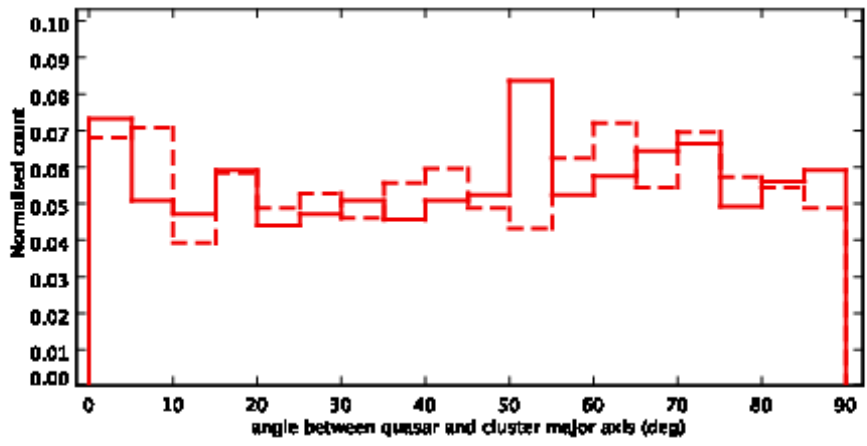}
\caption[Comparing Stripe 82 orientation angles]{\small{The orientation angle of the quasar with respect to the cluster major axis for Stripe 82 for the observed (solid) and control (dotted) fields.}}
\label{fig:ang_stripe82}
\end{figure}

Figure \ref{fig:ang_field} shows the distributions of the orientation angle of the quasar with respect to the cluster major axis for Stripe 82 (red) and COSMOS (blue), for the the observed (solid) and control (dotted) fields. There are no significant differences between the overall distributions of the angles for the observed and control fields for either the COSMOS or the Stripe 82 field. The sample size is different for the control and observed sample, which is more noticeable in the Strip 82 sample.
Figure \ref{fig:ang_stripe82} shows just the angle distribution for Stripe 82. where the count has been normalised. This shows that there is no difference in the distributions so the difference seen in Figure \ref{fig:ang_field} is due to different sample sizes.

\begin{figure}[!ht]
\centering
\includegraphics[scale=0.75]{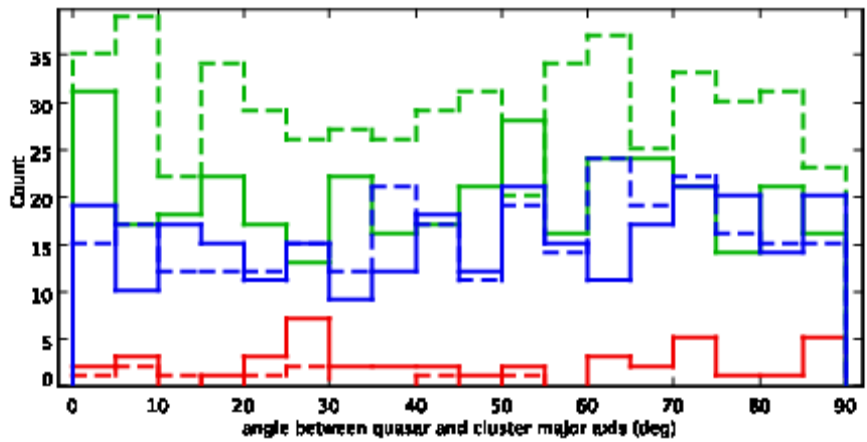}
\caption[Orientation angle as a function of 2D separations for control and observed data]{\small{The orientation angle of the quasar with respect to the cluster major axis for the separations $<1$ Mpc (red), 1-5 Mpc (blue) and $>$5 Mpc (green) for the observed (solid) and control (dotted) fields.}}
\label{fig:ang_sep}
\end{figure}

\clearpage
Figure \ref{fig:ang_sep} shows the distributions of the orientation angle of the quasar with respect to the cluster major axis for a range of 2D projected separations: $<1$ Mpc (red), 1-5 Mpc (blue) and $>$5 Mpc (green). At close separations ($<$1 Mpc), more quasars in the observed fields prefer to lie at angles $>60^{\circ}$ than in the control fields. However, the sample of quasars at close separations is small for both the observed and control samples. A larger number of objects would be needed to further assess this difference. At intermediate (1-5 Mpc) and larger ($>5$ Mpc) separations, there is no observable difference between the angles for the observed and control fields. 

\begin{figure}[!ht]
\centering
\includegraphics[scale=0.75]{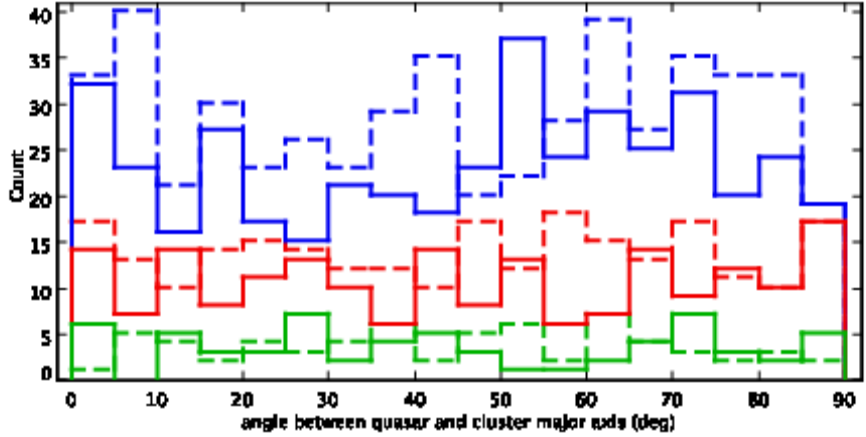}
\caption[Orientation angle as a function of 2D separations for control and observed data]{\small{The orientation angle of the quasar with respect to the cluster major axis as a function of the quasar redshift for the redshift ranges $0<z<0.4$ (red), $0.4<z<0.8$ (blue) and $0.8<z<1.2$ (green) for the observed (solid) and control (dotted) fields.}}
\label{fig:ang_z}
\end{figure}

Figure \ref{fig:ang_z} shows the distributions of the orientation angle of the quasar with respect to the cluster major axis for different redshift ranges: $0<z<0.4$ (red), $0.4<z<0.8$ (blue) and $0.8<z<1.2$ (green). There appears to be no difference between the distributions of angles from the observed and and control fields, for any of the redshift ranges.   

These observations are based on the visual inspection of the plots. A statistical analysis can be found in Section \ref{sect:stats}.

\section{Statistics}\label{sect:stats}

\subsection{Statistics Tests}
\subsubsection{One Dimensional Kolmogorov Smirnov Test}\label{1dks}

The Kolmogorov Smirnov (K-S) test is used to determine if two datasets are significantly different. This test is used when the data is continuous (i.e., unbinned) and makes no assumption about the underlying distribution of the data. The test works by creating a cumulative distribution function for each dataset and comparing the two distributions. The value D is the maximum value of the absolute difference between the two cumulative distribution functions. The p-value given is the significance of any non-zero value of D returned. The cumulative probability distribution used to calculate the p-value is given by (\citealt{Press1992}):
\begin{equation}
P_{KS}(z) = 1 - 2 \displaystyle\sum\limits_{j=1}^{\infty} (-1)^{j-1} \mbox{exp}(-2j^2z^2)
\label{eq:cpf_KS}
\end{equation}
where $P_{KS}(z)$ is the cumulative distribution function which finds the probability of the variable $z$, and $j$ is an integer between 1 and $\infty$.

$Q_{KS}(z) = 1-P_{KS}(z)$ is the cumulative probability function and is also used. The limiting values for both cumulative distribution functions are:
\begin{eqnarray}
P_{KS}(0) = 0 & & P_{KS}(\infty) = 1 \nonumber \\
Q_{KS}(0) = 1 & & Q_{KS}(\infty) = 0 \nonumber 
\end{eqnarray}

In terms of $Q_{KS}$, the p-value of an observed D is approximated by:
\begin{equation}
\mbox{Probability} (D > D_{obs}) \approx Q_{KS} \left( \left[ \sqrt{N_e} + 0.12 + \frac{0.11}{\sqrt{N_e}} \right] D \right)
\label{eq:Q_KS}
\end{equation}
where $N_e$ is the effective number of data points. This approximation is good as long as $N_e \geq 14$ (\citealt{Press1992}). 

The K-S test is most sensitive around the median of values and less sensitive towards the extreme ends of the distribution. 

The null hypothesis is that the two distributions are drawn from the same continuous distribution and a p-value of 1 indicates that the null hypothesis is likely to be the correct one. This test uses a two-sided alternative hypothesis. This tests whether the distributions are different, but not how. For example, it will not test whether distribution $x$ is shifted to the right or left of distribution $y$.

\subsubsection{Two Dimensional Kolmogorov Smirnov Test}\label{2dks}
To assess the evolution of the projected 2D separations between a quasar and the closest cluster centre with redshift, a two dimensional KS test is used.

\begin{figure}[!ht]
\centering
\includegraphics[scale=0.4]{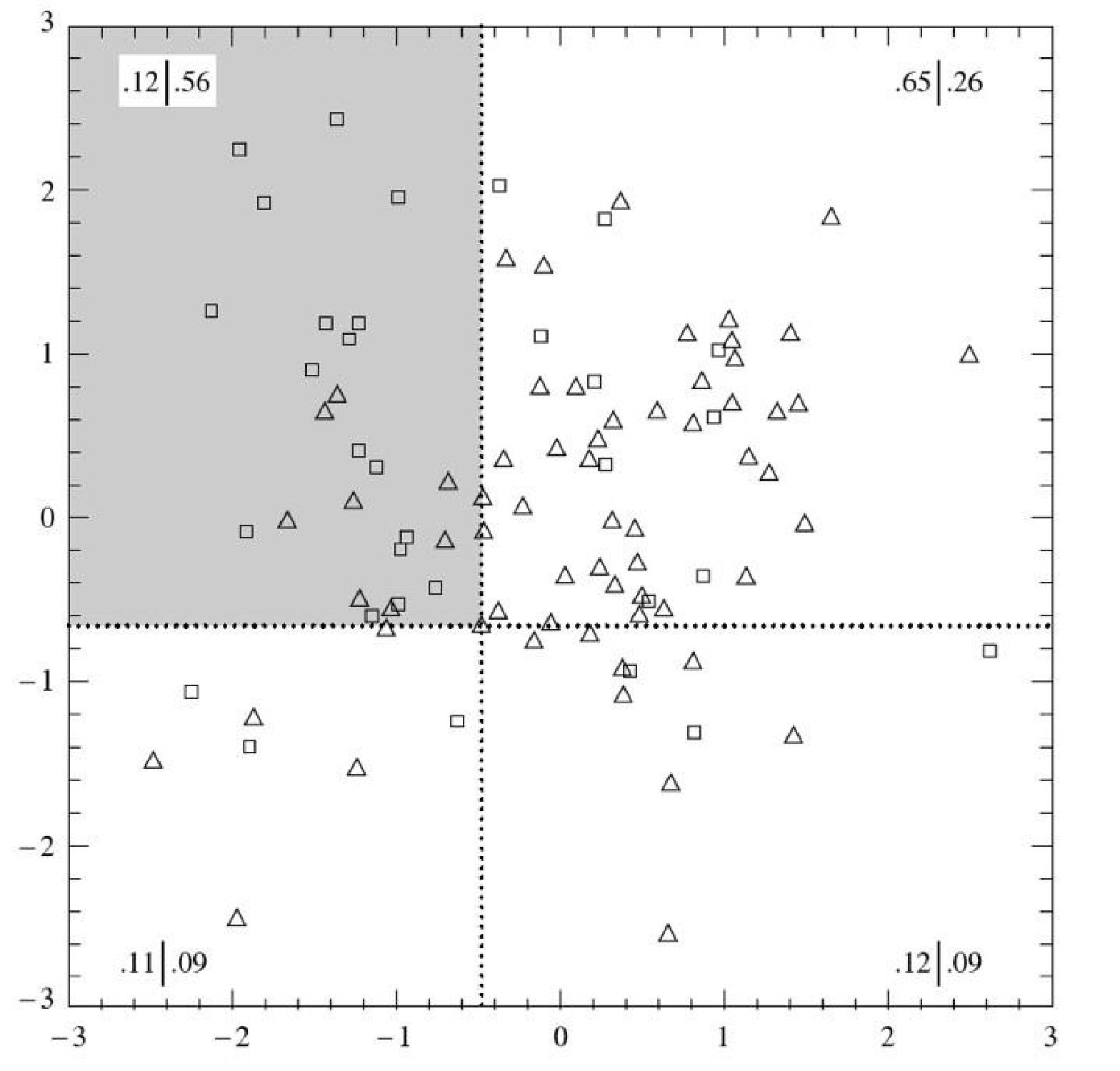}
\caption[Example of 2D KS test]{\small{An example of the 2 dimensional KS test (\citealt{Press1992}). }}
\label{fig:2dkstest}
\end{figure}

Each data point is given a pair of coordinates $(x,y)$, which, in this case, will be the redshift and the projected 2D separation. The $(x,y)$ plane is then split into four quadrants around a given point $(x_i,y_i)$ which is the point which maximises the difference between the number of points in each quadrant. Then the integrated probability in each of the four natural quadrants is found. An example of these quadrants can be found in Figure \ref{fig:2dkstest}. The squares and triangles show the two different samples and the dotted lines show the quadrants with the shaded quadrant being the quadrant where the maximum difference between the number of squares and triangles occurs.

The KS statistic D is now taken as the maximum difference between corresponding integrated probabilities (\citealt{Press1992}). The significance of the difference between quadrants is determined by Equation \ref{eq:2dKS}.
\begin{equation}
\mbox{Probability}(D > D_{obs}) \approx Q_{KS} \left( \frac{\sqrt{N}D} { 1+\sqrt{1-r^2} (0.25- \frac{0.75}{\sqrt{N}} )} \right)
\label{eq:2dKS}
\end{equation} 
where $Q_{KS}= 1-P_{KS}(z)$ is the cumulative probability function ($P_{KS}(z)$ is from Equation \ref{eq:cpf_KS}), and $N=$$ {N_1N_2} \over{N_1 + N_2}$ where $N_1$ and $N_2$ are the numbers in each sample.

The 2D KS test is slightly dependent on the distribution so the probability given is only an estimate. This test is also affected by sample size, so statistics for the COSMOS field will be less accurate than the statistics for the Stripe 82 field which has more quasar-cluster pairs due to the larger sky area covered.

From this test, a small probability indicates that the two samples are significantly different. This means that any trends seen in the observed data are likely to be real and not due to random chance (\citealt{Press1992}).  

In the one dimensional KS test, a large value for D indicates there is a large maximum difference between the cumulative probability distributions of the observed and control samples. (A small value of D means the cumulative probabilities are similar.) For a two dimensional KS test, a large value for D indicates the maximum differences between the quadrants. Therefore, a large value of D indicates a large difference between the control and observed distributions and any trends seen within the observed samples are significant.

\subsection{Separation Statistics} 
The distributions of characteristics of the quasar-cluster pairs can be seen in Table \ref{tab:stats}, which shows the D and p-values using the two dimensional KS test for different characteristics of quasar-cluster pairs. This table shows the statistics for all of the observed data and does not distinguish between the COSMOS and Stripe 82 fields. 

\begin{table}[!h]
\caption[2D KS test results for all quasar-cluster pairs]{\small{D and p-values from two dimensional KS test for different characteristics of quasar-cluster pairs.}}
\centering
\begin{tabular}{c | c c }
Distribution                                                   & D                    & p-value \\ \hline
2D quasar-cluster centre separation - redshift                 & 0.115                & 0.006   \\
2D quasar-cluster centre separation - quasar orientation angle & 0.086                & 0.074   \\
2D quasar cluster centre separation - cluster richness         & 0.067                & 0.274   \\
2D quasar-closest galaxy separation - redshift                 & 0.107                & 0.013   \\ 
2D quasar-BCG separation            - redshift                 & 0.120                & 0.004   \\ \hline
3D quasar-cluster centre separation - redshift                 & 0.129                & 0.013   \\
3D quasar cluster centre separation - cluster richness         & 0.067                & 0.276   \\
quasar orientation angle            - redshift                 & 0.084                & 0.089   \\
quasar orientation angle            - cluster richness         & 0.063                & 0.343     
\label{tab:stats}
\end{tabular}
\end{table}

\subsubsection{2D Projected Separations}
To study any changes in separations between a quasar and the closest cluster centre, the two dimensional KS test is used with one variable as the 2D projected separations and the other as the redshift, angle between quasar and cluster major axis, or the cluster richness.

Using a significance level of p=0.01, the 2D projected separations between the quasar and cluster centre as a function of redshift are shown to have significantly different distributions for the observed and control fields. The observed data is shown in Figure \ref{fig:sep_each} and the control data is shown in Figure \ref{fig:sep_closestsim}. In Figure \ref{fig:sep_each}, there is a possible deficit of quasars lying close to a cluster centre in the redshift range $0.4<z<0.8$, which is not as pronounced in the control data in Figure \ref{fig:sep_closestsim}. This, as well as the 2D KS test result, suggests that there is a potential evolution of the position of quasars with respect to galaxy clusters as a function of redshift.

\begin{figure}[!ht]
\centering
\includegraphics[scale=0.5,angle=-90]{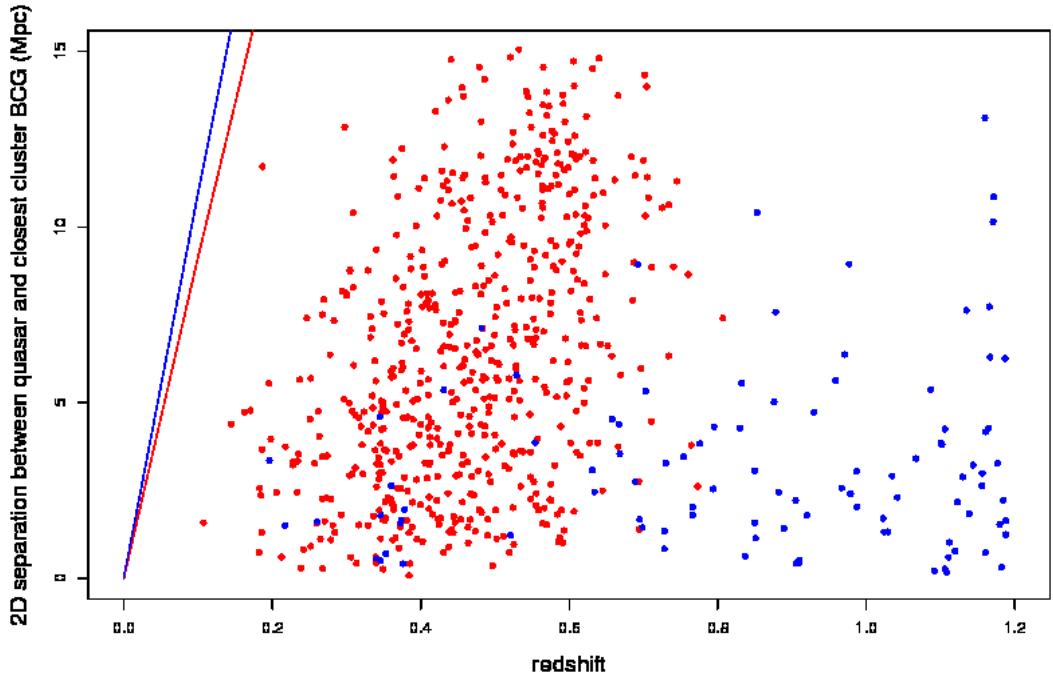}
\caption[2D separations to the closest cluster BCG for observed data]{\small{The 2D projected separation between an observed quasar and the closest cluster BCG for the COSMOS (blue) and Stripe 82 (red) fields. The lines indicate the field edges for the COSMOS (blue) and Stripe 82 (red) fields. }}
\label{fig:sepbcg_obs}
\end{figure}

\begin{figure}[!ht]
\centering
\includegraphics[scale=0.5,angle=-90]{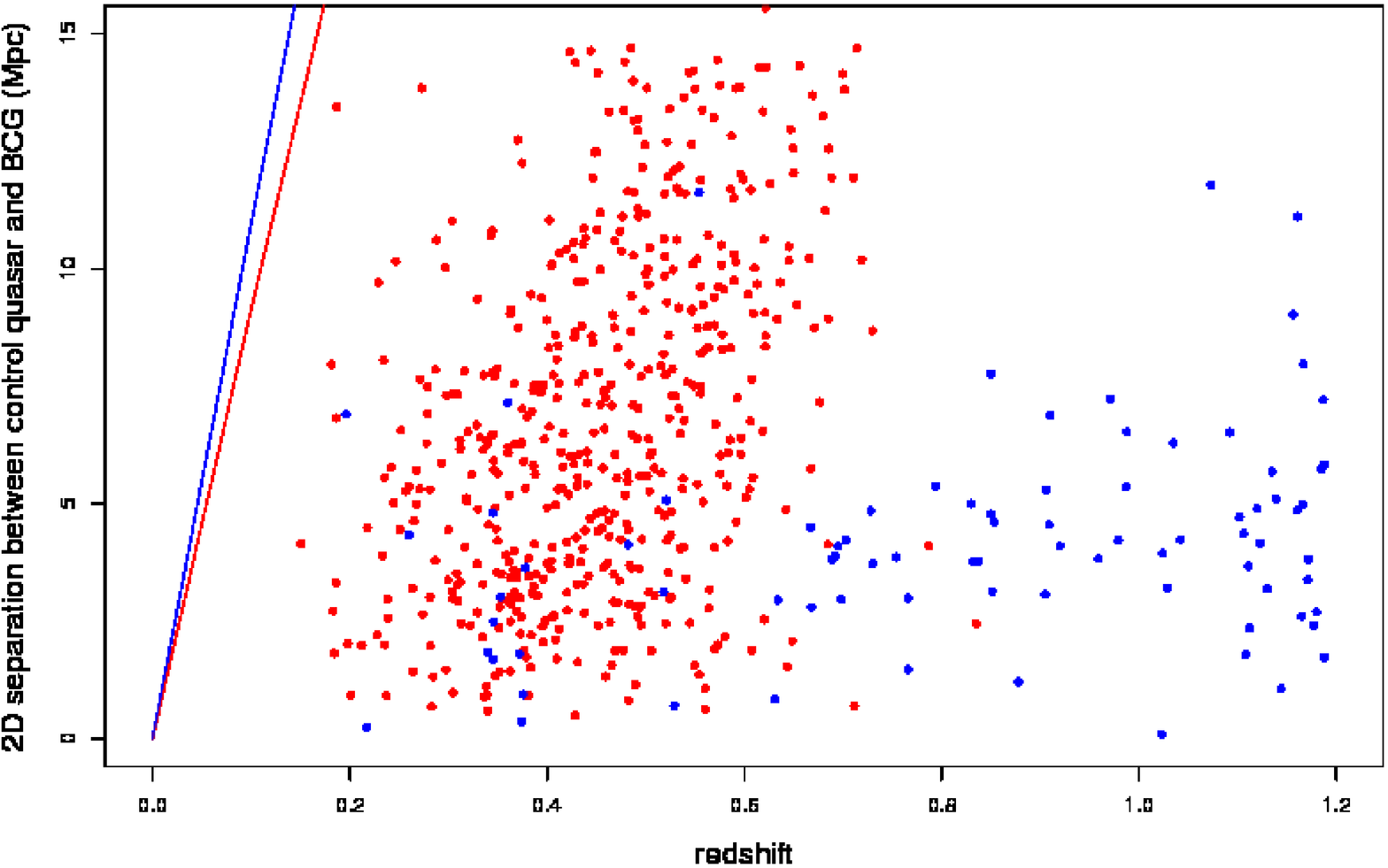}
\caption[2D separations to the closest cluster BCG for control data]{\small{The 2D projected separation between a control quasar and the closest cluster BCG for the COSMOS (blue) and Stripe 82 (red) fields. The lines indicate the field edges for the COSMOS (blue) and Stripe 82 (red) fields.}}
\label{fig:sepbcg_sim}
\end{figure}

The distributions for the observed and control data for the 2D projected separations between the quasar and the cluster BCG also significantly different, according to the 2D KS test. Figure \ref{fig:sepbcg_obs} shows the 2D projected separation between a quasar and the BCG of the closest galaxy cluster for the COSMOS (blue) and Stripe 82 (red) fields. Figure \ref{fig:sepbcg_sim} shows the 2D projected separation between a control quasar and the BCG of the closest galaxy cluster for the COSMOS (blue) and Stripe 82 (red) fields. For both Figures, the lines indicate the edges of the fields. In the observed data, the quasar prefer to lie closer to the BCG than in the control data sample.

\subsubsection{Cluster Richness}
\citet{Lietzen2009} found that groups of galaxies with a quasar lying at less than 2Mpc tended to be poorer on average. In contrast, in the COSMOS and Stripe 82 samples, the average cluster richness for clusters with a quasar lying at $<$2 Mpc is 11.6 compared to a richness of 8.9 for clusters lying at $>$2Mpc from a quasar. This suggests, for the richest clusters, quasars lie closer to the cluster centre. However, the quasar sample used by \citet{Lietzen2009} used only quasars in the redshift range $0.078<z<0.172$, which is below the redshift range for most of this work. 

\begin{figure}[!ht] 
\centering
\includegraphics[scale=0.50,angle=-90]{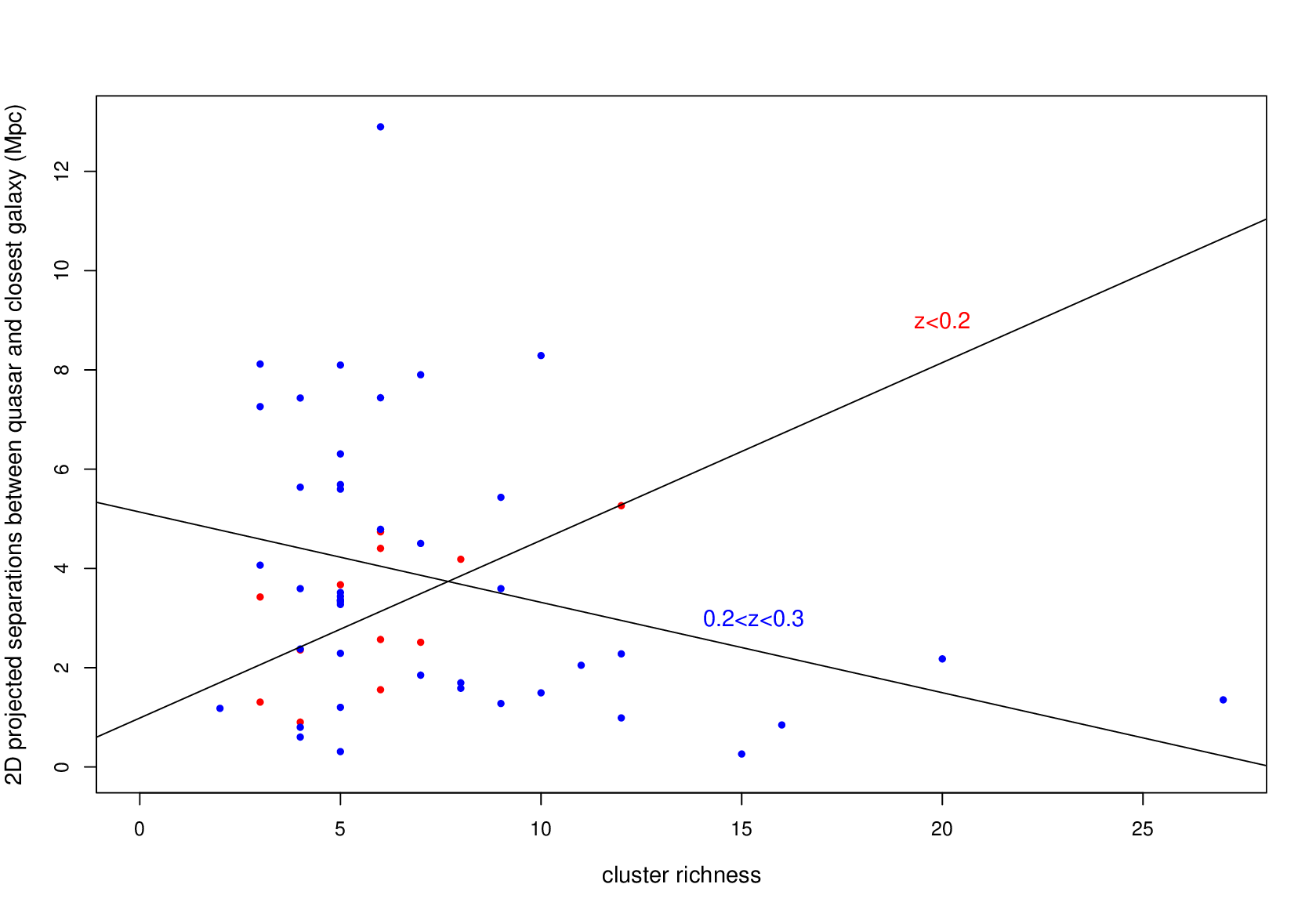}
\caption[2D separations as a function of cluster richness]{\small{2D projected separations between quasar and closest cluster centre against cluster richness as a function of cluster redshift for $z<0.2$ (red) and $0.2<z<0.3$ (blue). The lines show the best fit for separation = richness + c for $z<0.2$ and $0.2<z<0.3$.}}
\label{fig:sep_lowz_rich} 
\end{figure}

Figure \ref{fig:sep_lowz_rich} shows the 2D projected separations between the quasar and closest cluster centre as a function of the richness of the galaxy cluster for $z<0.2$ (red) and $0.2<z<0.3$ (blue). The lines shown are best fit lines for separation = richness + c for the different redshift ranges. This shows a change in the trend for quasars to lie closer to a poorer cluster at $z\sim$0.2.

\citet{Lietzen2009} also found that the richest clusters lay between 5 and 15$h^{-1}$ Mpc from a quasar. In contrast, in the COSMOS and Stripe 82 samples, 13 out of 14 of the richest clusters (richness $>$ 40) lie at 2D projected separations of $<5.5$ Mpc from a quasar; the 14th lies at 8.5Mpc from a quasar. Given the small redshift range covered by \citet{Lietzen2009}, this result is likely to be larger than the cluster richness's they measured.

\subsubsection{3D Separations}
The two dimensional KS test values for the distribution of the 3D separations between quasar-cluster pairs and the quasar redshift, angle between the quasar and cluster major axis, and cluster richness are not significant. However, given the large errors on the 3D separations, it is not possible to come to any conclusions about these distributions.

\subsection{Angle Statistics}\label{angstats}

Comparing the 2D projected separation between quasars and the closest cluster centre and the quasar orientation angle gives a p-value of 0.074 and D = 0.086. This indicates the distribution of the angles with respect to the separation between the quasar and cluster centre in the control sample and the observed sample are not significantly different. This suggests that there is no link between the orientation of the cluster major axis and the separation distance between the quasar and the cluster centre.

Comparing the redshift with the angle between the quasar and cluster major axis gives a p-value of 0.089 and D = 0.084. This indicates that the angles from the control sample and the angles from the observed samples as a function of redshift are not significantly different. 

Therefore, the orientation angle of the quasar with respect to the major axis of the closest cluster is not dependent on redshift or quasar-cluster separation.

\subsubsection{1D KS Test}
To test whether there is a pattern in the distribution of angles between the quasar and the major axis of the closest cluster, a 1D KS test can be used.
The null hypothesis for the K-S test is that the distribution of angles for the observed samples could be from the same distribution as the angles taken from the control field. 

The p-value from the K-S test was found as 0.696 with D = 0.039. Therefore the null hypothesis, that the samples have the same distribution, should be accepted.
This means that quasars do not lie in any preferred direction with respect to the cluster axis of the closest cluster. This is supported by results from the 2D KS test.

\subsection{Comparing Fields}

The COSMOS and Stripe 82 fields have also been considered separately; the statistics of which are shown in Tables \ref{tab:cosmos_stats} and \ref{tab:stripe82_stats} for COSMOS and Stripe 82, respectively.

\begin{table}[!h]
\caption[2D KS test for COSMOS quasar-cluster pairs]{\small{D and p-values from the two dimensional KS test for different characteristics of quasar-cluster pairs for the COSMOS field.}}
\centering
\begin{tabular}{c | c c }
Distribution                                                   & D                    & p-value    \\ \hline
2D quasar-cluster centre separation - redshift                 & 0.290                & 0.006 \\
2D quasar-cluster centre separation - quasar orientation angle & 0.319                & 0.002 \\
2D quasar cluster centre separation - cluster richness         & 0.327                & 0.001  \\
2D quasar-closest galaxy separation - redshift                 & 0.302                & 0.003  \\ 
2D quasar-BCG separation            - redshift                 & 0.230                & 0.049  \\ \hline
3D quasar-cluster centre separation - redshift                 & 0.180                & 0.205  \\
3D quasar cluster centre separation - cluster richness         & 0.206                & 0.105   \\
quasar orientation angle            - redshift                 & 0.170                & 0.264   \\
quasar orientation angle            - cluster richness         & 0.177                & 0.226                
\label{tab:cosmos_stats}
\end{tabular}
\end{table}

\begin{table}[!h]
\caption[2D KS test for Stripe 82 quasar-cluster pairs]{\small{D and p-values from dimensional KS test for distribution of different characteristics of quasar-cluster pairs for the Stripe 82 field.}}
\centering
\begin{tabular}{c | c c }
Distribution                                                   & D                    & p-value \\ \hline
2D quasar-cluster centre separation - redshift                 & 0.124                & 0.005   \\
2D quasar-cluster centre separation - quasar orientation angle & 0.087                & 0.111 \\
2D quasar cluster centre separation - cluster richness         & 0.089                & 0.103 \\
2D quasar-closest galaxy separation - redshift                 & 0.131                & 0.002 \\ 
2D quasar-BCG separation            - redshift                 & 0.300                & 0.004 \\ \hline
3D quasar cluster centre separation - cluster richness         & 0.076                & 0.199  \\ 
3D quasar cluster centre separation - cluster richness         & 0.064                & 0.415   \\
quasar orientation angle            - redshift                 & 0.086                & 0.119   \\
quasar orientation angle            - cluster richness         & 0.066                & 0.376                
\label{tab:stripe82_stats}
\end{tabular}
\end{table}

For the COSMOS field, all of the relations for the 2D projected separation between a quasar and the closest cluster centre are statistically significant, as well as the separation between a quasar and the closest cluster member as a function of redshift. This is using a p-value of 0.01.

For the Stripe 82 field, the observed distributions for the 2D projected separations for the cluster centre, closest cluster member and cluster BCG as functions of redshift are statistically different to the control sample distributions. The distributions for the 2D projected quasar-cluster centre separations as a function of cluster richness and orientation angle are not significantly different from the control samples. 
The observed distribution of the 2D separation between the quasars and the closest cluster centre and the quasar orientation angle with respect to the cluster major axis is significantly different from the control sample for the COSMOS field but not for the Stripe 82 field.

This could indicate a difference in the COSMOS and Stripe 82 fields. However, the COSMOS field has fewer quasar-cluster pairs, which will effect the statistics, so caution should be taken when comparing the p-values and distributions from the different fields.

Neither the COSMOS or Stripe 82 fields show any relation between the quasar orientation angle and the quasar redshift, supported by a small value for D (the maximum difference between the quadrants is small) and a large p-value (they are from significantly different distributions). This means that there is no relation between the quasar orientation with respect to the cluster major axis and the quasar redshift in either the COSMOS or Stripe 82 field.

\section{Two Point Correlation Function}\label{corrfunc}

When studying the clustering of galaxies and AGN, the two-point correlation function is often used. The two-point correlation function gives a measure of the strength of the clustering. However, this correlation function does not contain any directional information. It will be used here to compare the COSMOS and Stripe 82 data samples to previously published data.  

There are two ways to find the correlation function, angular and spatial. The angular correlation function is used when there is no redshift information (\citealt{Cress1996}). However, this would remove any information about redshift evolution. Therefore, for this section, the spatial correlation function in two dimensions has been used, with the redshift of the quasar used for the redshift of the quasar-cluster pair, and the 2D projected separations as the separation, $r$.

The two point correlation function is a simple way to characterise clustering. It is defined as a measure of the probability of finding a cluster in a volume $\delta$V at a separation of $r$ from a quasar,
\begin{equation}
\dif P = n \dif V
\label{eq:tpcf_1}
\end{equation} 

where $n$ is the number density of either quasars or clusters.

The correlation function is given by: 

\begin{equation}
\dif P = n^2(1+ \xi(r)) \delta V_1 \delta V_2
\label{eq:tpcf_2}
\end{equation}

where $\xi(r)$ is the estimator and $\delta V_1$ and $\delta V_2$ are volume elements. This is the probability $\dif P$ of finding, simultaneously, a quasar and a cluster at a separation $r$ from each other within 2 volume elements $\delta V_1$ and $\delta V_2$ in a sample with number density $n$ of clusters. As it is not possible to use the 3D distances due to the large uncertainties created by the redshift errors, the 2D projected distances will be used and the number of objects per area element used. If there is clustering at distance $r$, $\xi>0$ and if there is no clustering and quasars and cluster avoid each other, $\xi<0$. Larger positive values of $\xi$ indicate stronger levels of clustering. At small distances, the correlation function acts as a power law, so the estimator is :
\begin{equation}
\xi(r) = \left( \frac{r}{r_0} \right) ^{\gamma}
\label{eq:xi_1}
\end{equation}
where $r_0$ is the correlation length and $\gamma$ is the power law slope. For galaxy distributions, $\gamma$ = -1.77 and when separation is $r<$10h$^{-1}$Mpc, $r_0 \sim$5h$^{-1}$Mpc (\citealt{Coil2004}). 

There are various different ways to estimate $\xi(r)$, including the standard Davis-Peebles estimator (\citealt{Davis1983}), the Hamilton estimator (\citealt{Hamilton1993}) and the Landy-Szalay estimator (\citealt{Landy1993}). All of the estimators used the following variables:
\begin{itemize}
\item $QG(r)$: the number of quasar-cluster pairs in the observed sample within 2D projected separation $r$
\item $QR(r)$: the number of pairs, within 2D projected separation $r$, between an observed quasar and a cluster from a catalogue where the positions have been randomly selected
\item $RR(r)$: the number of pairs, within 2D projected separation $r$, between a quasar and a cluster, both from catalogues where the positions have been randomly selected
\item $N$: the number of quasar-cluster pairs in total from the observed data catalogues (within all 2D projected separations $r$)
\item $N_R$: the number of quasar-cluster pairs from the catalogues containing quasars and clusters with randomly selected positions (within all 2D projected separations $r$)
\end{itemize}

\subsection{Davis-Peebles Estimator}
The standard Davis-Peebles estimator (\citealt{Davis1983}) is given by:
\begin{equation}
\xi_{DP}(r) = \frac{2N_R}{N-1} \frac{\mbox{QG(r)}}{\mbox{QR(r)}} -1
\label{eq:xi_2}
\end{equation}

where $QG(r), QR(r)$ and $N$ are described in Section \ref{corrfunc}. 
However, this was found to have systematic biases due to finite sampling, though is still often used (\citealt{Landy1993,Mullis2004}). The estimators have been compared in Section \ref{sect:compests}. 

\subsection{Hamilton Estimator}\label{Hamestimator}

The Hamilton estimator (\citealt{Hamilton1993}) was suggested:
\begin{equation}
\xi_{Ham}(r) = \frac{4NN_R}{(N-1)(N_R-1)}\frac{QG(r) \times RR(r)}{[QR(r)]^2} - 1
\label{eq:xi_3}
\end{equation}
where $QG(r)$, $QR(r)$, $RR(r)$, $N$ and $N_r$ are described in Section \ref{corrfunc}.

\subsection{Landy-Szalay Estimator}\label{sect:LS}

The Landy-Szalay estimator (\citealt{Landy1993}) is given by:
\begin{equation}
\xi_{LS}(r) = \frac{N_R(N_R-1)}{N(N-1)}\frac{QG(r)}{RR(r)} - \frac{(N_R-1)}{N}\frac{QR(r)}{RR(r)} + 1
\label{eq:xi_4}
\end{equation}

where $QG(r)$, $QR(r)$, $RR(r)$, $N_r$ and $N$ are given in Section \ref{corrfunc} (\citealt{Mullis2004}).

\subsection{Comparing Estimators}\label{sect:compests}
The Hamilton estimator (\citealt{Hamilton1993}) and the Landy-Szalay estimator(\citealt{Landy1993}) show less variance than the Davis-Peebles (\citealt{Labatie2010}). The Davis-Peebles estimator is more sensitive to fluctuations in the number of galaxies used. The Landy-Szalay estimator has the lowest variance (though the Hamilton is close). The Hamilton is potentially more sensitive to density of random points (\citealt{Kerscher2000}). The Landy-Szalay shows the smallest deviations on large scales ($r\sim30h^{-1}$Mpc), therefore deals with the edge corrections the best (\citealt{Kerscher2000}).

Field edge corrections should not be a large source of error in this analysis. When selecting the quasar-cluster pairs, a quasar-cluster pair was only selected if the distance between the quasar and the closest cluster centre was less than the distance to the nearest field edge. This ensures the cluster selected as closest to a quasar is the actual closest cluster, though the field edge will still be an affect for low redshift due to the limited field size. This selection process was also repeat on the random re-sampled catalogues.

\subsection{Application}

To create the random points, the positions of either the quasar or the cluster or both were randomly selected without replacement so the overall distribution of positions would be the same as the distributions of positions in the observed data.

To calculate the number of pairs $QR(r)$ between a quasar and a random cluster, the positions of the quasars were fixed and the cluster RA and DEC were re-sampled at random. For the number of pairs $RR(r)$ between a random quasar and random cluster, the positions of both the quasars and the clusters were selected at random. To increase the sample size and decrease the noise, the sample containing random matches (i.e., $RR(r)$) was run three times. The estimators include a normalising coefficient to take the difference in sample size into account.

To select clusters within the redshift range of the quasar, the redshift error on the clusters was again used. For the Stripe 82 field, the redshift error was set as 9\% of the cluster redshift, and was found individually for the clusters in the COSMOS field (Section \ref{sect:cluster_selection}).  

Values for each of the estimators (Davis-Peebles, Hamilton, and Landy-Szalay) have been calculated for the whole sample to allow a comparison. The estimators have also been found for redshift slices, to allow any evolution with redshift to be studied. These slices are taken as low ($0<z<0.4$), intermediate ($0.4<z<0.8$), and high ($0.8<z<1.2$) redshifts.  
Table \ref{tab:estimators} shows the values for the Davis-Peebles, Hamilton and Landy-Szalay estimators using 2D projected separation bins of 1 Mpc. There are discrepancies between the values for the different estimators. However, the trend of decreasing values as the distance $r$ increases can be seen in each estimator.

\begin{table}[!h]
\caption[Two point correlation estimators]{\small{Comparing correlation function estimators for 2D projected separations $r$ in 1 Mpc bins over all redshifts for the Stripe 82 and COSMOS fields together.}}
\centering
\begin{tabular}{c|ccc}
 $r$      & $\xi_{DP}$                &  $\xi_{HAM}$ &  $\xi_{LS}$  \\ \hline     
  0-1   & 24.15$^{+3.88}_{-4.53}$     & 34.50$^{+5.48}_{-6.39}$    & 17.40$^{+2.75}_{-3.21}$   \\
  1-2   & 13.56$^{+1.64}_{-1.83}$     & 17.15$^{+2.04}_{-2.28}$    & 11.07$^{+1.31}_{-1.47}$   \\
  2-3   &  8.58$^{+1.21}_{-0.92}$     & 11.47$^{+1.57}_{-1.58}$    &  6.83$^{+0.93}_{-1.05}$   \\
  3-4   &  8.94$^{+1.09}_{-1.22}$     &  9.97$^{+1.20}_{-1.34}$    &  8.20$^{+1.14}_{-1.10}$   \\
  4-5   &  5.59$^{+0.91}_{-1.04}$     &  7.31$^{+1.14}_{-1.31}$    &  4.65$^{+0.72}_{-0.82}$   \\
  5-6   &  5.79$^{+0.91}_{-1.04}$     &  5.58$^{+0.88}_{-0.96}$    &  5.95$^{+0.94}_{-1.07}$   \\
  6-7   &  6.69$^{+1.07}_{-1.21}$     &  7.29$^{+1.14}_{-1.31}$    &  6.29$^{+0.98}_{-1.13}$   \\
  7-8   &  6.57$^{+1.08}_{-1.24}$     &  6.46$^{+1.07}_{-1.23}$    &  6.66$^{+1.10}_{-1.27}$   \\
  8-9   & 12.60$^{+2.05}_{-2.38}$     & 18.45$^{+2.93}_{-3.41}$    &  9.11$^{+1.43}_{-1.67}$   \\
  9-10  &  5.02$^{+1.25}_{-1.40}$     &  4.32$^{+1.03}_{-1.25}$    &  5.54$^{+1.31}_{-1.58}$   \\
  10-11 &  9.54$^{+1.83}_{-2.18}$     & 10.90$^{+2.07}_{-2.46}$    &  8.56$^{+1.62}_{-1.93}$   \\
  11-12 & 12.55$^{+2.12}_{-2.47}$     & 12.87$^{+2.17}_{-2.53}$    & 12.28$^{+2.07}_{-2.41}$   \\
  12-13 &  9.32$^{+1.95}_{-2.92}$     & 12.11$^{+2.48}_{-2.99}$    &  7.55$^{+1.53}_{-1.85}$   \\
  13-14 &  7.38$^{+2.24}_{-2.92}$     &  7.98$^{+2.40}_{-3.13}$    &  6.96$^{+2.68}_{-2.72}$   \\
  14-15 &  6.06$^{+1.89}_{-2.46}$     &  3.97$^{+1.33}_{-1.73}$    &  8.18$^{+2.68}_{-3.43}$   \\
   0-15 &  4.37$^{+0.21}_{-0.20}$     &  9.62$^{+0.41}_{-0.39}$    &  2.70$^{+0.61}_{-0.10}$   \\
COSMOS 0-15 & 4.87$^{+0.65}_{-0.59}$  & 11.03$^{+1.32}_{-1.34}$    &  2.89$^{+0.29}_{-0.32}$   \\
Stripe 82 0-15 & 4.29$^{+0.22}_{-0.23}$ & 9.41$^{+0.43}_{-0.45}$   &  2.67$^{+0.12}_{-0.11}$   
\end{tabular}
\label{tab:estimators}
\end{table}

\begin{figure}[!ht] 
\centering
\includegraphics[scale=0.50,angle=-90]{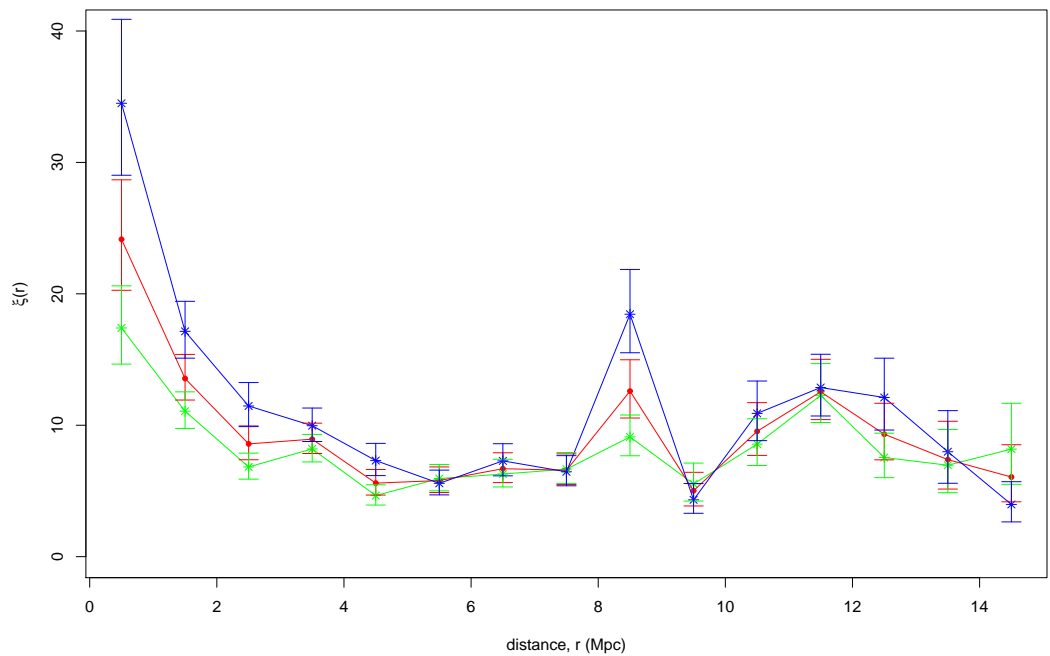}
\caption[Comparing two point correlation function estimators]{\small{The Davis-Peebles (red), Hamilton (blue), Landy-Szalay (green) estimators.}}
\label{fig:xi_comp} 
\end{figure}

Figure \ref{fig:xi_comp} shows the Davis-Peebles, Hamilton and Landy-Szalay estimator values calculated using the whole sample (i.e, both Stripe 82 and COSMOS fields) and redshift range $0<z<1.2$. Each of the estimators shows an increase in clustering close to the quasar with the clustering decreasing at increased separations. At separations $4<r<8$ Mpc, the estimators remain constant and at values just above zero, suggesting there is only weak clustering at separations $4<r<8$ Mpc. This result does not depend on the estimator used. However, there is an increase in $\xi(r)$ for all of the estimators for $r=8-9$ Mpc. This suggests an increase in clustering at this distance. There is also an increase in clustering between $10<r<13$ Mpc. This result may be affected by small numbers and also may the limited size of the fields.  

The Hamilton is considered more sensitive to the number of random points used and the Davis-Peebles estimator shows increased error and systematic biases. The Landy-Szalay estimator shows the least errors, the smallest deviations, and has the best compensation for edge corrections of the three estimators (\citealt{Kerscher2000}). Therefore the Landy-Szalay estimator will be used to compare the different fields and the clustering at different redshifts. 

The Landy-Szalay estimator as a function of 2D projected separation, $r$, can also be compared with results from \citet{Mullis2004} who found similar values for this estimator. \citet{Mullis2004} used a sample of 445 X-ray selected AGN with a redshift range of $z = 0.026 - 3.89$ and a median redshift of $z = 0.41$ (compared to the sample from COSMOS and Stripe 82 which has $z<1.2$ but a similar median of $z = 0.49$). They studied a range of separations between 5 and 60 $h^{-1}$Mpc (where H$_0$ is set as 100 $h^{-1} $km s$^{-1}$ Mpc$^{-1}$ to allow for comparisons to previous results). At $r\sim$5$h^{-1}$Mpc, using a Landy-Szalay estimator, they found $\xi_{LS} \simeq 1.6$ (Figure 6 in \citealt{Mullis2004}). Using the COSMOS and Stripe 82 samples, a higher value of $\xi_{LS} \simeq 5.95$ was found. 

The errors on the Landy-Szalay Estimator are usually estimated using a Poisson estimate (\citealt{Croom2004}), in Equation \ref{eq:est_err}.

\begin{equation}
\Delta \xi (r) = \frac{1 + \xi(r)}{\sqrt{QG(r)}}
\label{eq:est_err}
\end{equation}

However, for $QG(r)<10$, Equation \ref{eq:est_err} does not give the correct upper and lower confidence limits. Instead, the formulae for small-number statistics in \cite{Gehrels1986} are used, using Equations \ref{eq:upper} and \ref{eq:lower} to calculate the upper and lower limits for the number of observed quasar-cluster pairs at separation, $r$, where $\lambda_u$ and $\lambda_l$ are the upper and lower limits, $n$ is the number of events (in this case, $QG(r)$) and $S$ is the equivalent Gaussian number and has been set at 1$\sigma$ (so $S=1$).

\begin{equation}
\lambda_u \approx n + S\sqrt{n + 1} + \frac{S^2 + 2}{3} \approx N + 1 + \sqrt{n + 1} 
\label{eq:upper}
\end{equation} 
\begin{equation}
\lambda_l \approx n - S\sqrt{n} + \frac{S^2 -1}{3} \approx n - \sqrt{n}
\label{eq:lower}
\end{equation}

\begin{figure}[!ht]
\centering
\includegraphics[scale=0.50,angle=-90]{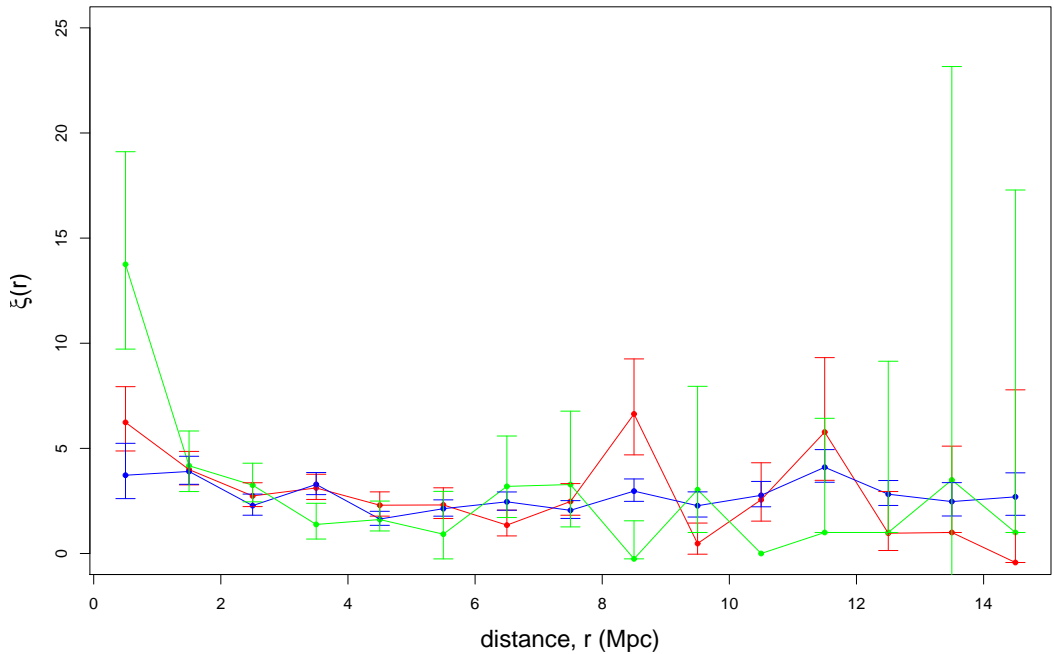}
\caption[Clustering estimator as a function of redshift]{\small{The Landy-Szalay estimator, $\xi_{LS}$, as a function of 2D projected separation $r$ for different redshift ranges; low ($0<z<0.4$, red), intermediate ($0.4<z<0.8$, blue) and high ($0.8<z<1.2$, green).}}
\label{fig:xi_zslices2} 
\end{figure}

The observed sample of quasar-cluster matches has also been split into redshift slices. Figure \ref{fig:xi_zslices2} shows the Landy-Szalay estimator for the redshift slices $0<z<0.4$ (red), $0.4<z<0.8$ (blue), and $0.8<z<1.2$ (green). At close separations, the intermediate redshift sample still shows a deficit of quasars. At separations $3<r<6$ Mpc, all 3 redshift samples show the same clustering. For separations $r>6$ Mpc, the clustering in the high redshift sample varies, which is due to the small numbers of quasar-cluster pairs in the sample and the limited field size of the COSMOS field, which limits the possible number of pairs with large separations. For separations $r>6$ Mpc, the clustering is comparable for the low and intermediate redshift slices. The decrease in the low redshift sample at larger separations is also due to limit field sizes at low redshifts. However, the increase in clustering at $r=8-9$ Mpc is still visible and appears in the lower redshift range, $0<z<0.4$.   

\begin{figure}[!ht]
\centering
\includegraphics[scale=0.50,angle=-90]{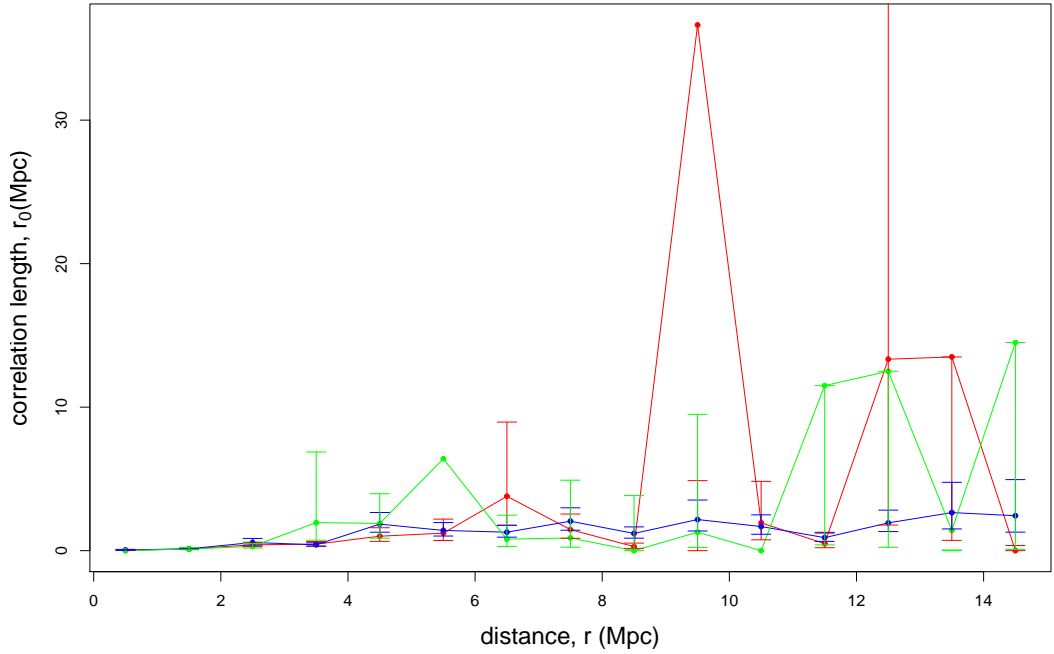}
\caption[Correlation length as a function of redshift]{\small{The correlation length $r_0$ as a function of 2D projected separation $r$ for both COSMOS and Stripe 82 together; low ($0<z<0.4$, red), intermediate ($0.4<z<0.8$, blue) and high ($0.8<z<1.2$, green). }}
\label{fig:r_0_all} 
\end{figure}

To compare to previous literature, the correlation length, $r_0$ has been calculated (Equation \ref{eq:xi_1}). A fixed value of $\gamma$ = 1.8 has been used (e.g., \citealt{LeFevre1996,Croom2001,Mullis2004,Coil2004}). Figure \ref{fig:r_0_all} shows the correlation length, $r_0$, as a function of 2D projected separation, $r$. The low redshift sample is shown in red, the intermediate in blue and high in green. The large oscillations in the correlation length in the high redshift sample is partly due to the small numbers of objects available in this redshift range. It is also caused by negative values of $\xi_{LS}$, which suggests that quasars and clusters are avoiding each other, therefore, the correlation length has no meaning.

The values for both $r_0$ and $\xi_{LS}$ become unreliable at large separations for the high and low redshift samples due to the limited number of pairs with large separations in these redshift samples. This is due to the limited field size of both samples at low redshift, and the limited size of the COSMOS field at high redshifts. 

As the cluster detection method differs slightly, the COSMOS and Stripe 82 fields have been compared separately as well. This will use allow any differences in the clustering in the fields to be studied. The Landy-Szalay estimator has been used as this is the most commonly used estimator in the literature as this estimator is considered the least biased to the number of points used (\citealt{Kerscher2000}).

\subsubsection{Stripe 82}

\begin{figure}[!h]
\centering
\includegraphics[scale=0.50,angle=-90]{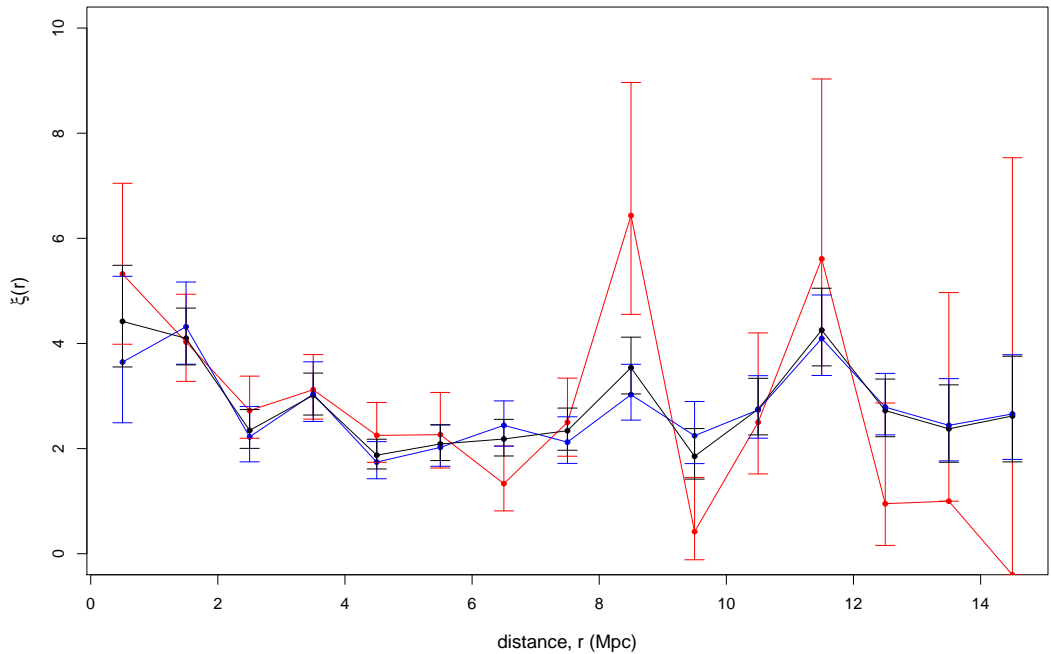}
\caption[Clustering estimator for Stripe 82 as a function of redshift]{\small{The Landy-Szalay estimator as a function of 2D projected separation $r$ for different redshift ranges for the Stripe 82 field. The red points and line show $0<z<0.4$, the blue points and line show the range $0.4<z<0.8$, and all redshifts ($0<z<0.8$, black).}}
\label{fig:xi_zslicesStripe82} 
\end{figure}

Figure \ref{fig:xi_zslicesStripe82} shows the Landy-Szalay estimator as a function of 2D projected separation, $r$, for two different redshift ranges, low ($0<z<0.4$) and intermediate ($0.4<z<0.8$) redshifts for the Stripe 82 field. The figure shows that the clustering close to the quasar is greater at lower redshifts than at intermediate redshifts. At $r=8-9$ Mpc, there is a large increase at in $\xi(r)$ for the lowest redshift range, $0<z<0.4$. This may be an affect of the field size. There is no evidence of an increased number of quasars with 2D projected separations between the quasar and cluster centre seen in Section \ref{sect:centre_seps} and Figure \ref{fig:sep_each} in particular. This suggests this result from the correlation function may be due to binning the data and problems encountered when there are only a small number of objects per bin.

\subsubsection{COSMOS}

\begin{figure}[!ht]
\centering 
\includegraphics[scale=0.50,angle=-90]{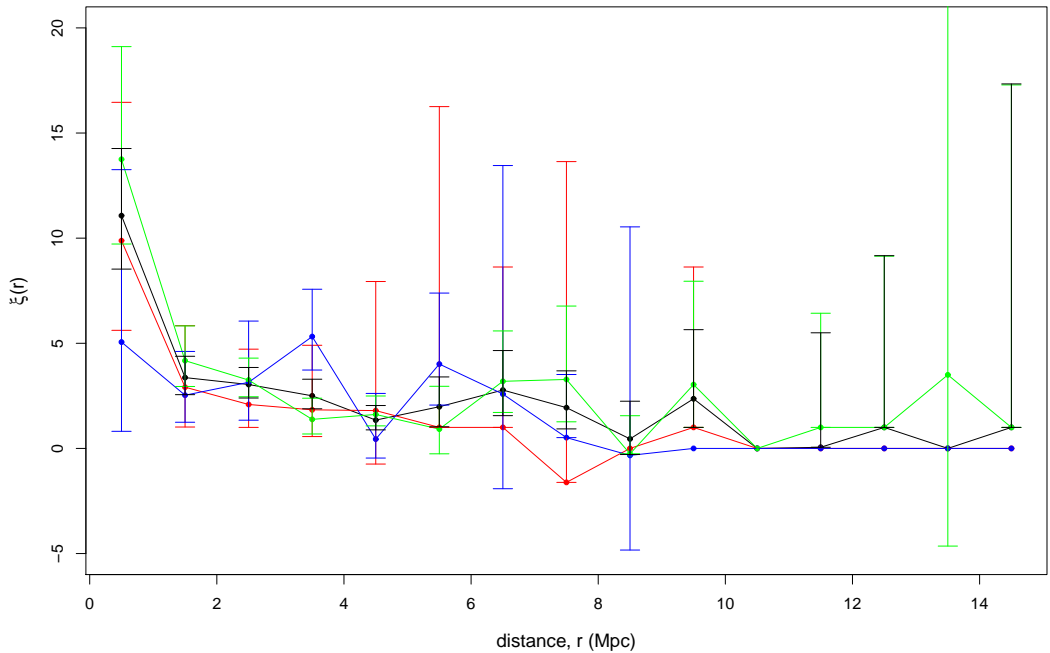}
\caption[Clustering estimator for COSMOS as a function of redshift]{\small{The Landy-Szalay estimator as a function of 2D projected separation $r$ for different redshift ranges for the COSMOS field; low ($0<z<0.4$, red), intermediate  ($0.4<z<0.8$, blue), high ($0.8<z<1.2$, green) and all redshifts ($0<z<1.2$, black).}}
\label{fig:xi_zslicesCOSMOS} 
\end{figure}

Figure \ref{fig:xi_zslicesCOSMOS} shows the Landy-Szalay estimator as a function of 2D projected separation, $r$, for three different redshift ranges, low redshifts ($0<z<0.4$) in red, intermediate redshifts ($0.4<z<0.8$) in blue and high redshifts ($0.8<z<1.2$) in green for the COSMOS field. The clustering at close separations is stronger for at lower redshift than at intermediate or high redshifts. There is no peak in the clustering at $r=8-9$ Mpc to correspond to the increase seen in the Stripe 82 field. However, there are less quasar-cluster pairs in the redshift range $0<z<0.4$ in the COSMOS data. The COSMOS field is also smaller, so naturally restricts the number of quasar-cluster pairs with large separations at low redshifts.

\section{Summary}

Using the 2D KS test on both fields together, the distribution for the observed data for the 2D projected separation between the quasar and the closest cluster centre as a function of redshift proved to be significantly different from the control data sample. In Figure \ref{fig:all_sep_close}, there appears to be a deficit of quasars lying close to cluster centres for $0.4<z<0.8$, which is not seen as prominently in the control sample (Figure \ref{fig:sep_closestsim}). When the fields are taken separately, this difference in the distributions is still seen for each field. However, this redshift is at the edge of the redshift limits for the Stripe 82 cluster sample and may not be complete. The clusters in the COSMOS sample do cover this redshift range, however the area covered by this sample is smaller. To investigate the significance of this result, a larger sample of quasars and clusters in the redshift range $0.4<z<0.8$ is needed. 

The observed distribution of the 2D projected separation between the quasar and the BCG in closest cluster (Figure \ref{fig:sepbcg_obs}) is also significantly different from the distribution in the control sample (Figure \ref{fig:sepbcg_sim}). The observed quasars prefer to lie further away from the BCG than in the control sample. When the fields are taken separately, this significant difference is still found in the Stripe 82 but not in the COSMOS field. For the 2D separation between the quasar and the BCG of the closest cluster, the difference between observed and control samples is significant for the Stripe 82 field, but not for the COSMOS field. This difference in the fields may be due to the redshift ranges of the fields. 

The observed distribution of the 2D separation between the quasars and the closest cluster centre and the quasar orientation angle with respect to the cluster major axis is significantly different from the control sample for the COSMOS field but not for the Stripe 82 field. 

Figure \ref{fig:sep_lowz_rich} shows a change in the trend for quasars to lie closer to a poorer cluster at $z\sim$0.2. However, richer clusters are found at higher redshifts due to the limited field size, so this change in the trend may be due to selection effects. Larger fields would be needed, so richer cluster at lower redshift could be found, to test this result. If there is a change in the positions of quasars with respect to the cluster richness at $z\sim0.2$, this would not be shown in the COSMOS field, as the cluster redshift range for the COSMOS field is $0.201<z<1.2$. 

Using the separation ratio (which is the ratio of the 2D projected separation between the quasar and the closest cluster centre and the mean radius of the cluster), the quasar is estimated to lie inside a cluster for 34 out of 677 quasar-cluster pairs (i.e., 5\%). This is likely to be an overestimate, as the separation ratio relies on the 2D projected distance between the quasar and the cluster centre. Given the large errors on the 3D separations, it is not possible to determine whether the quasar does lie in the cluster or whether the separation ratio is due to a projection effect. 
However, this result still supports the idea that quasars avoid the highest density areas (e.g. \citealt{Sochting2002,Lietzen2009,Strand2008}).

The Landy-Szalay estimator shows that at small separations between the quasar and the closest cluster centre, the clustering is greater in the COSMOS field than in the Stripe 82 field. There is also an increase in the clustering at $r=8-9$ Mpc in the Stripe 82 field at low redshifts ($0<z<0.4$), which is not shown in the COSMOS field. This may be a result of the smaller field size for COSMOS.

\chapter{Quasar-Cluster Proximity as a Function of Luminosity}\label{sect:chap4}

This Chapter investigates the positions and angles between quasars and clusters as a function of quasar magnitude/luminosity. The luminosity of a quasar is dependent on the accretion rate of material onto the central black hole and this accretion rate is likely to be affected by the quasar triggering mechanism. More violent triggering mechanisms will likely produce higher accretion rates, and therefore brighter luminosities.

The results from COSMOS and Stripe 82 are also compared to studies of the control field which uses a random sample of quasars (see Section \ref{sect:simulations} for more details).

\section{Absolute Magnitudes}
\subsection{Quasar SED}
\begin{figure}[!h]
\centering
\includegraphics[scale=0.4,angle=-90]{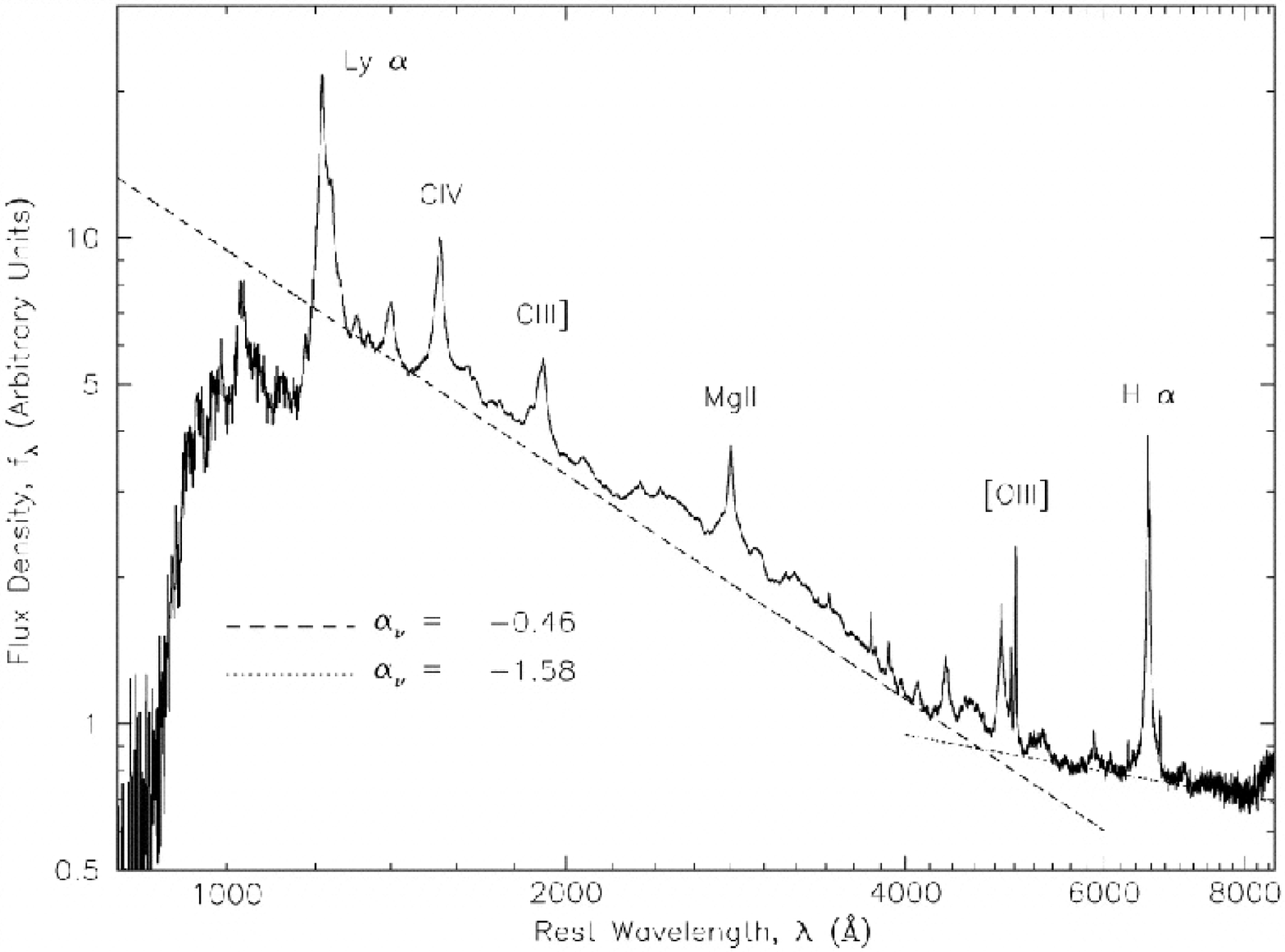}
\caption[Example of quasar SED]{\small{The SED of a quasar showing the power laws (\citealt{VandenBerk2001}). }}
\label{fig:quasarSED}
\end{figure}

The Spectral Energy Distribution (SED) of the quasar can be fit using a power law of the form $f_{\nu}$ $\propto$ $\nu^{ \alpha}$ where $\alpha=-0.5$ (\citealt{VandenBerk2001}). Figure \ref{fig:quasarSED} (\citealt{VandenBerk2001}) shows the power law fit to the quasar SED. As can be seen in Figure \ref{fig:quasarSED}, the power law changes at $\sim$5000\AA\ from $\alpha\approx-0.5$ to $\alpha\approx-2.5$. The change in the slope is likely caused by a combination of contamination from the host galaxy and a real change within the quasar continuum (\citealt{VandenBerk2001}).

To calculate the absolute magnitude, Equation 2.13 has been used (see Section \ref{sect:simulations} for further details). This equation uses $\alpha$ to correct for the slope of the quasar SED .

\subsection{Filter Selection}

\begin{figure}[!h]
\centering
\includegraphics[scale=0.5,angle=-90]{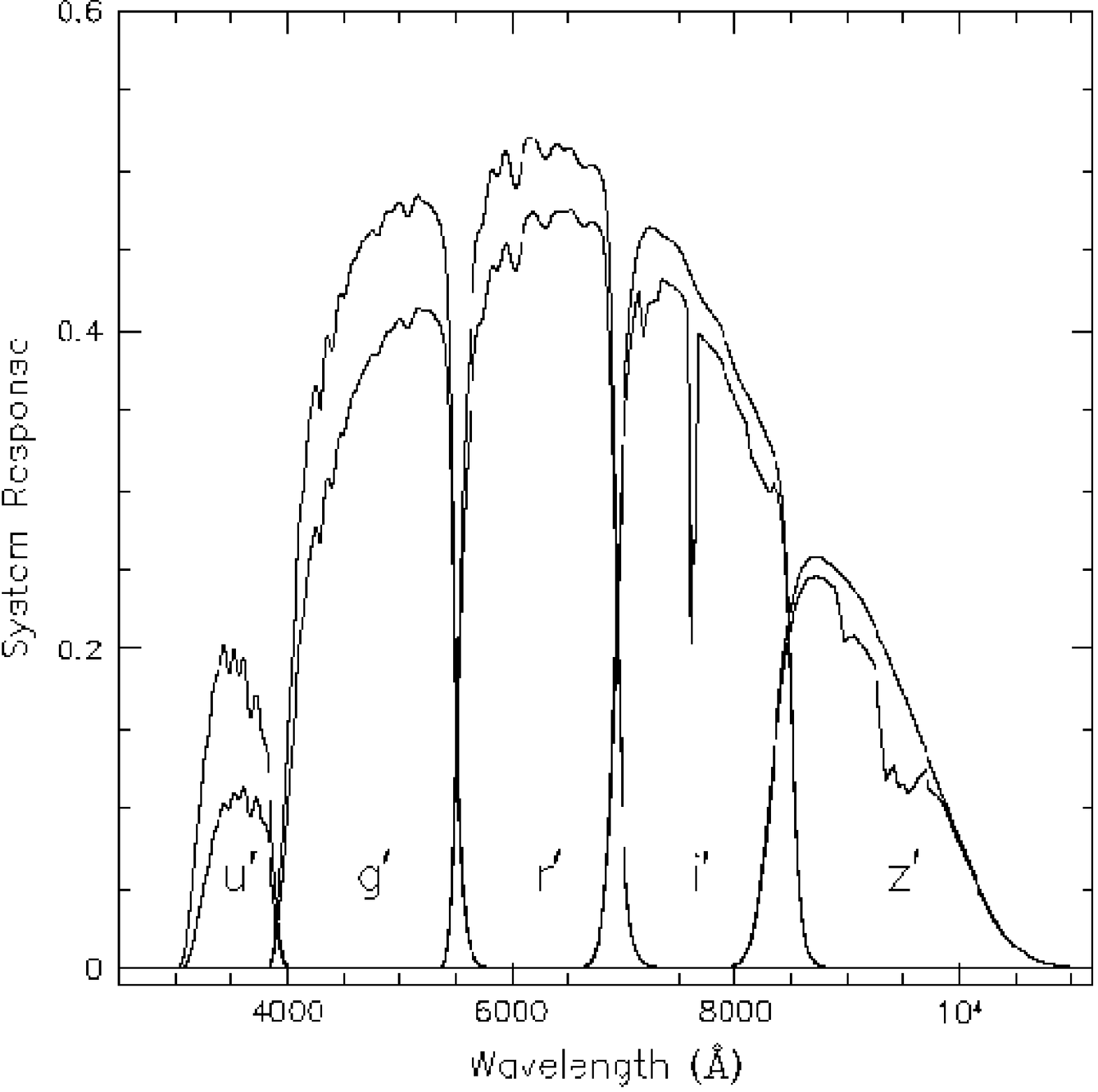}
\caption[SDSS filter response curves]{\small{Response curves of the SDSS filters (\citealt{Fukugita1996}). }}
\label{fig:response curves}
\end{figure}

The SDSS survey uses a system of 5 filters; $u', g', r', i'$ and $z'$. These are not the standard colour system but were chosen to produce 2 more orders of magnitude of photometry than previously existed. Figure \ref{fig:response curves} shows the response curves for the SDSS filters (\citealt{Fukugita1996}). 

Table \ref{tab:Delimits} shows the limiting magnitudes, central wavelengths, and widths of the SDSS filters. The limiting magnitude is less for the $z'$ filter so fainter galaxies and quasars can not be seen using this filter.

\begin{table}[!h]
\caption[SDSS filter details]{\small{SDSS filters information.}}
\centering
\begin{tabular}{ c | c c c }
Filter   & Limiting Apparent Magnitude (mag) & Central $\lambda$ (\AA) & FWHM (\AA)   \\ \hline
$u'$       & 22.3                     & 3540                    & 570        \\
$g'$       & 23.3                     & 4770                    & 1370       \\
$r'$       & 23.1                     & 6230                    & 1370       \\
$i'$       & 22.3                     & 7630                    & 1530       \\
$z'$       & 20.8                     & 9130                    & 950        \\
\end{tabular}
\label{tab:Delimits}
\end{table}

Ideally, a section of the spectrum containing no emission lies would be used to calculate the magnitude as the magnitude should be calculated from the continuum. However, selecting an area with no emission lines is difficult and in this case, a large redshift range makes finding a range with no emission lines for all redshifts not possible. Therefore, the wavelength range chosen has as few emission lines as possible.

\begin{figure}[!ht]
\centering
\subfigure[Filter $g$]{
\includegraphics[scale=0.35,angle=-90]{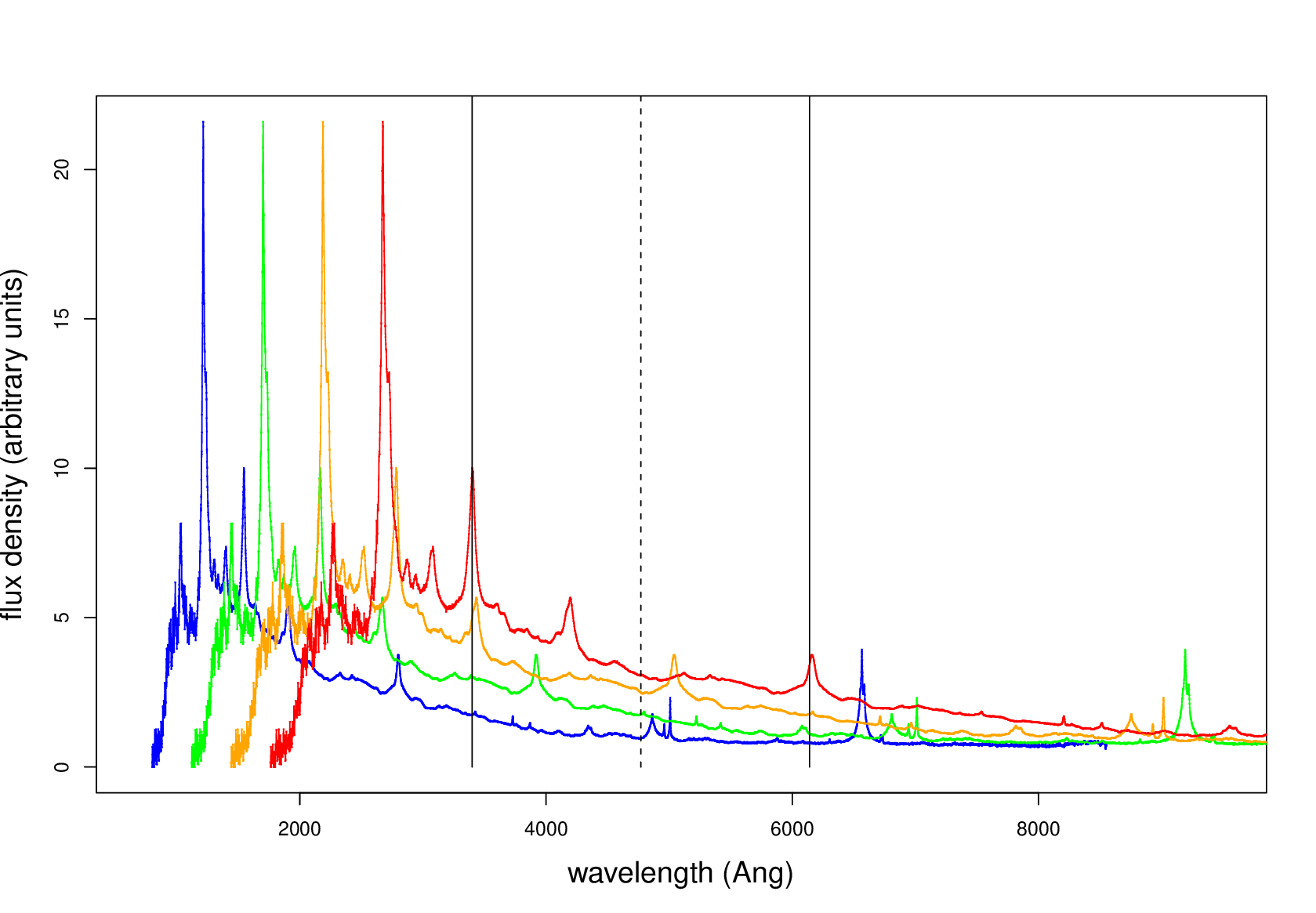}
\label{fig:filterg}
}
\subfigure[Filter $r$]{
\includegraphics[scale=0.35,angle=-90]{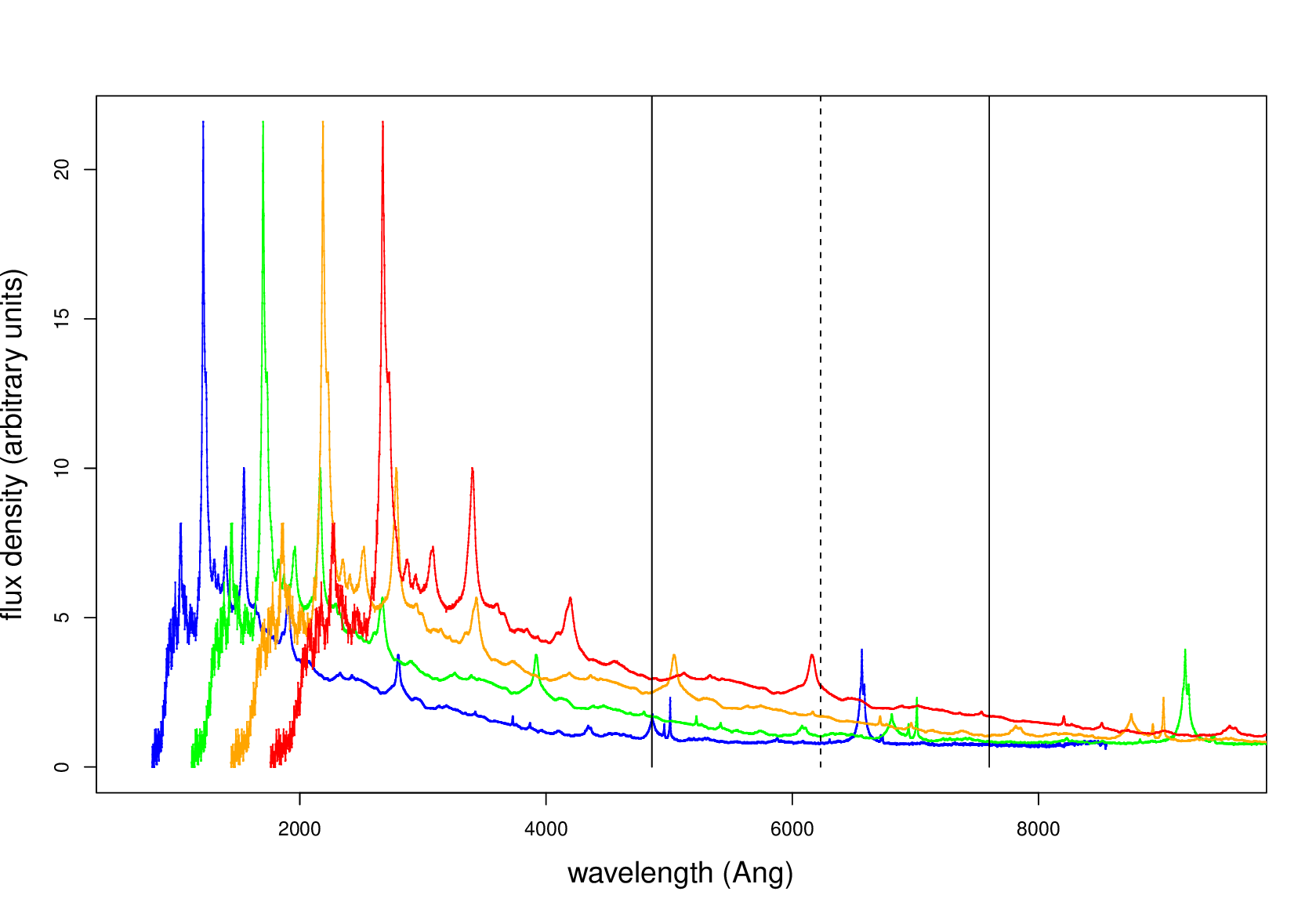}
\label{fig:filterr}
}
\subfigure[Filter $i$]{
\includegraphics[scale=0.35,angle=-90]{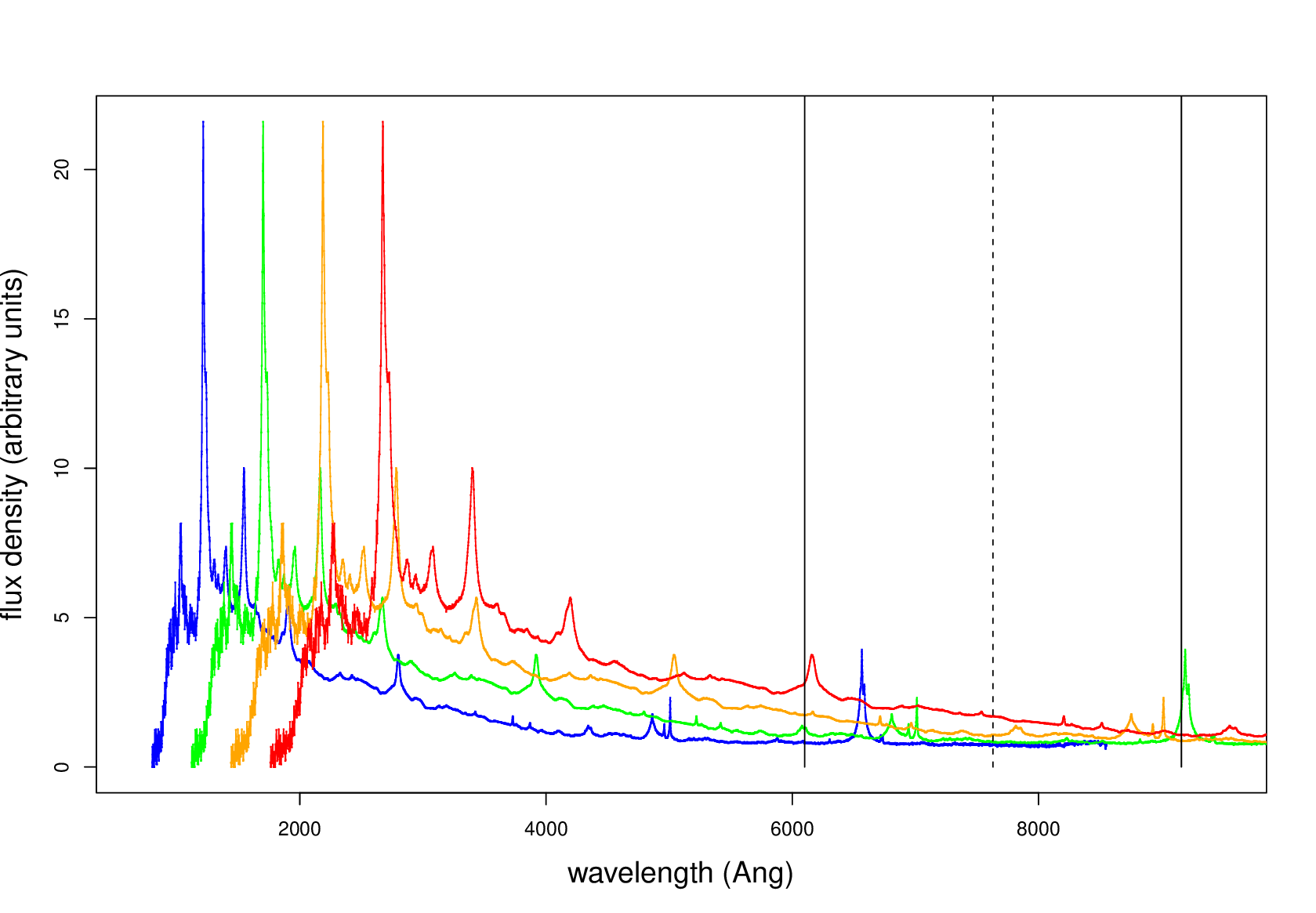}
\label{fig:filteri}
}
\caption[Quasar spectrum at varying redshifts]{\small{SDSS composite quasar spectrum (\citealt{VandenBerk2001}) for redshifts, $z$ = 0.0 (blue), 0.4 (green), 0.8 (orange) and 1.2 (red) for \subref{fig:filterg} filter $g'$, \subref{fig:filterr} filter $r'$ and \subref{fig:filteri} filter $i'$. The dotted lines show the central wavelentgh of the filter and the solid lines show $\pm$ the FWHM.}}
\label{fig:sdssfilter_limits}
\end{figure}

Figure \ref{fig:sdssfilter_limits} shows the SDSS composite spectrum, created by \citet{VandenBerk2001}, for a range of redshifts, $z$=0.0 (blue), 0.4 (green), 0.8 (orange) and 1.2 (red). Figure \ref{fig:sdssfilter_limits} also shows the position of the central wavelength (dotted line) $\pm$ the FWHM (solid lines) of (a) filter $g'$, (b) filter $r'$ and (c) filter $i'$ in the observed frame for quasars at different redshifts. The $u'$ and $z'$ filters have not been used as they both have a poor response and the $u'$-band is also affected by atmospheric emission. 

\clearpage

\clearpage

Figure \ref{fig:filteri} shows that for low redshift quasars, the wavelength range of the $i'$-band filter falls outside the observed wavelength range and estimates the magnitude from the redder end of the spectra which is partly contaminated by emission from the host galaxy. So, although the $i'$-band filter is the widest filter, it is not the best to use with this data. Figure \ref{fig:filterg} shows, for higher redshift quasars, the emission lines from C\textsc{iv}$\lambda 1546$\AA\ and C[\textsc{iii}]$\lambda 1906$\AA\ fall into the filter wavelength range. 

Though some emission lines such as O[\textsc{iii}] and H$\beta$ fall into the wavelength range of the $r'$-band filter (Figure \ref{fig:filterr}, these  are not as strong as some other emission lines. The $r'$-band filter was chosen to calculate the absolute magnitudes as it covers an area of the spectra with minimal emission lines over a wide redshift range. The $r'$-band filter also has a lower limiting magnitude than the $i'$-band filter so fainter quasars can be seen.

The absolute magnitude was calculated using the $r'$-band apparent magnitude and Equation \ref{eq:absmag}. The method for calculating the absolute magnitude is described in Section \ref{sect:simulations}.

\subsection{Selection Effects - Magnitude Limits}

\begin{figure}[!ht]
\centering
\includegraphics[scale=0.5,angle=-90]{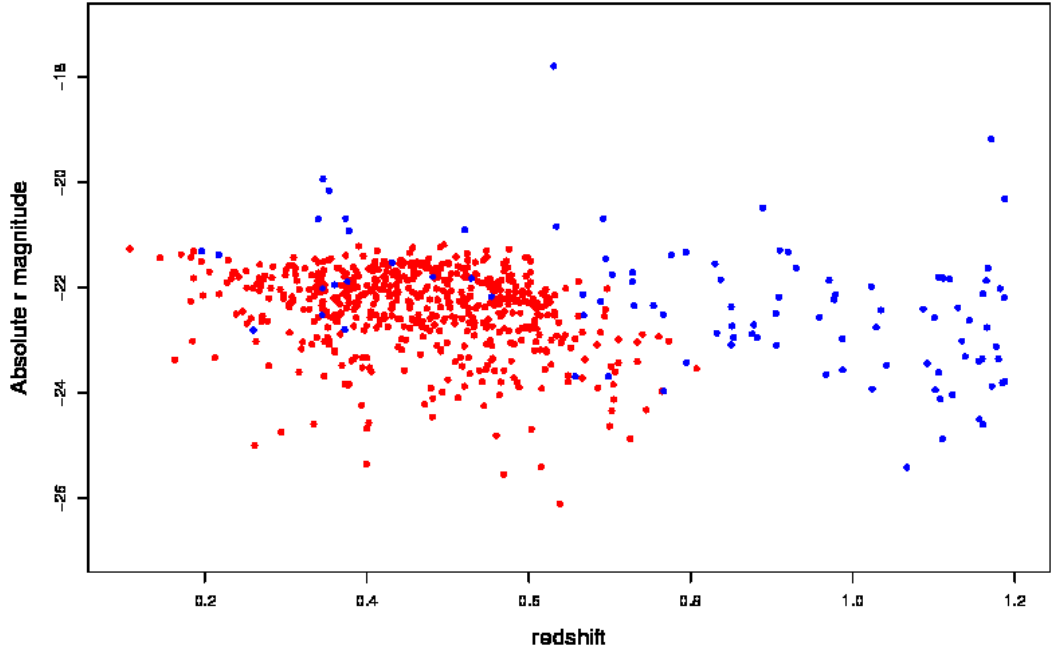}
\caption[Absolute magnitude as a function of redshift]{\small{The absolute $r'$ magnitude of quasars as a function of quasar redshift for the COSMOS (blue) and Stripe 82 (red) fields. }}
\label{fig:magz}
\end{figure}

Figure \ref{fig:magz} shows the distribution of the quasar magnitudes with redshift for COSMOS (blue) and Stripe 82 (red) quasars, which are part of a quasar-cluster pair. Though there are quasars in the Stripe 82 field with $z>0.8$, the quasars used have been limited to $z<0.8$ to match the cluster redshifts. In the Stripe 82 cluster catalogue, the completeness decreases for $z>0.6$. At $z\sim0.6$ in the Stripe 82 field, the number of faint quasars decreases. The reason for this is due to selection effects in the DR7QSO catalogues and is discussed in Section \ref{sect:qsobiases}.
The absolute magnitude also decreases with redshift for the COSMOS quasars, though the decrease is more gradual across the redshift range. The quasars in the COSMOS field are from 3 different sources, which do not have the same imposed magnitude limits which effect the DR7QSO quasar selection. 

When studying the Stripe 82 field, this limiting of the quasar magnitude at $z>0.6$ should be considered. For this reason, two redshift ranges have also been considered: $z<0.6$ and $z>0.6$. 

\subsection{Within a Cluster Environment}

The separation ratio (the ratio of the 2D projected separation between a quasar and the closest cluster centre and the mean radius of the cluster) is used to give an estimate as to whether a quasar lies within a cluster or outside. If a quasar-cluster pair have a separation ratio of $>$1, the quasar lies outside the cluster. If the separation ratio is $<$1, in 2D projection, the quasar lies within the cluster. As the 3D separations have large errors, it is not possible to say whether the quasar does lie within the cluster. However, for these purposes, it will be assumed that if a quasar-cluster pair has a separation ratio of $<$1, the quasar does lie within the cluster (though in 3D this may not be the case).

\begin{figure}[!ht]
\centering
\includegraphics[scale=0.5,angle=-90]{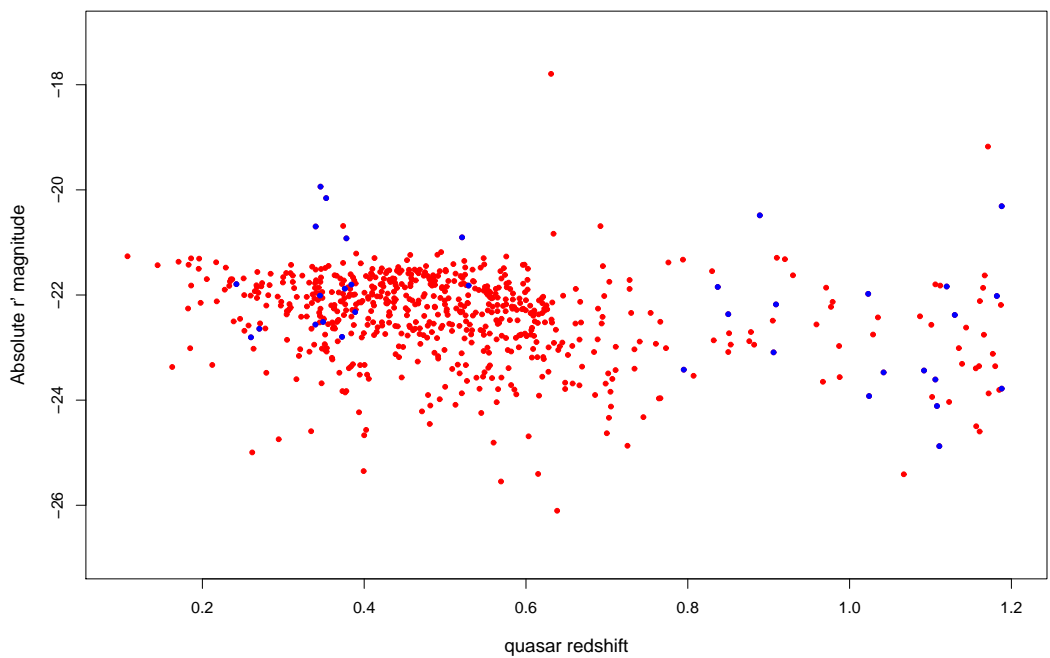}
\caption[Absolute magnitude of quasars lying inside and outside of clusters]{\small{The absolute $r$ magnitude of the quasar as a function of redshift for quasars lying inside a cluster (blue) and outside a cluster (red).}}
\label{fig:mag_inclust}
\end{figure}

Figure \ref{fig:mag_inclust} shows absolute $r'$ magnitude of a quasar as a function of the quasar redshift for the quasar lying inside (blue) and outside a cluster (red). Quasars lying within a cluster at lower redshifts appear slightly dimmer than those not lying within a cluster. At higher redshifts, the quasars lying within a cluster are brighter than at lower redshifts. As brighter quasars are more likely to be selected at higher redshifts than dimmer quasars, this may be a selection effect.

Table \ref{tab:inclust_mags} shows the average absolute $r'$ magnitude of quasars lying inside and outside clusters, for the redshift ranges $0.1<z<0.6$ and $0.8<z<1.2$. The increase in brightness with redshift is greater for quasars lying within clusters. However, this increase is still within the error limits of both averages. There is also a lack of quasars lying within clusters for the redshift range $0.6<z<0.8$, suggesting quasars prefer to lie in less dense regions at this redshift range. 

In Figure \ref{fig:mag_inclust}, there appear to be more quasars lying ``within'' clusters at higher redshifts. This may be due to larger clusters being found due to the selection criteria in the COSMOS field, which covers the high redshift range. Larger clusters will increase the chance of a line-of-sight projection occurring between a quasar and a cluster.

\begin{table}[!h]
\caption[Average magnitudes of quasars inside and outside clusters]{\small{Average absolute $r'$ magnitude for quasars in the redshift ranges $0.1<z<0.6$ and $0.8<z<1.2$. The errors show 1 standard deviation.}}
\centering
\begin{tabular}{ c | c c }
                  & $0.1<z<0.6$       & $0.8<z<1.2$         \\ \hline
Inside a cluster  & -21.72 $\pm$ 0.90 & -22.73 $\pm$ 1.19   \\
Outside a cluster & -22.39 $\pm$ 1.44 & -22.72 $\pm$ 1.04
\end{tabular}
\label{tab:inclust_mags}
\end{table}

The p-values using the t-test to compare the means of the absolute magnitudes at different redshift ranges can be seen in Table \ref{tab:inclust_stats}. The redshift ranges $0.1<z<0.6$ and $0.8<z<1.2$ have been used as there are no quasars lying inside to 2D boundary of a cluster for $0.6<z<0.8$. If the redshift range is changed to $0.6<z<1.2$, the p-value does not change. Though the p-value for the lower redshift range is lower than for the higher redshift range, neither p-value is significant. 

\begin{table}[!h]
\caption[T-test results for quasars inside and outside clusters]{\small{The p-value results using the t-test for quasars lying inside and outside clusters for the redshift ranges $z<0.6$ and $z>0.8$.}}
\centering
\begin{tabular}{ c | c }
redshift range    & p-value        \\ \hline
$0<z<0.6$         & 0.055   \\
$0.8<z<1.2$       & 0.459   
\end{tabular}
\label{tab:inclust_stats}
\end{table}

\section{Quasar Magnitude Relations}

To study differences in environments as a function of the quasar brightness, the quasars from the COSMOS and Stripe 82 fields have been split into two magnitude samples: bright ($M_r < -23$ mag) and faint ($M_r > -23$ mag) quasars. Technically, the faint quasars should be called AGN. However, the term quasar will be used to be consistent throughout and classed as either bright or faint. 
Table \ref{tab:mean_mags} shows the number of objects in each sample and the range covered in absolute $r'$ magnitude, as well as the mean 2D separations and mean richness of the closest cluster. The separations are found at the epoch of the quasar. There are no major differences between means for the bright and faint quasar samples. 
\clearpage
\begin{table}[!h]
\caption[Details of bright and faint magnitude quasars samples]{\small{Values for magnitude samples for bright ($M_r < -23$) and faint ($M_r > -23$) quasars. The errors show 1 standard deviation. }}
\centering
\begin{tabular}{ c | c c }
                                         & Bright quasars & Faint quasars   \\ 
                                         & $M_r < -23$    & $M_r > -23$      \\ \hline
Number in sample                         & 147            & 530                    \\
Absolute $r'$ magnitude range            & $-23<M_r<-26.17$ & $-17.79<M_r<-23$ \\
Mean 2D quasar-cluster centre            & 6.49 $\pm$ 3.86 & 5.76 $\pm$ 3.75      \\
separation (Mpc)                         &                 &                      \\
Mean 2D quasar-cluster member            & 5.72 $\pm$ 3.67 & 5.10 $\pm$ 3.54      \\
separation (Mpc)                         &                 &                      \\
Mean 2D quasar-cluster BCG               & 6.50 $\pm$ 3.85 & 5.78 $\pm$ 3.75      \\
separation (Mpc)                         &                 &                      \\
Cluster richness                         & 9.39 $\pm$ 12.10 & 9.42 $\pm$ 11.95
\end{tabular}
\label{tab:mean_mags}
\end{table}

The quasars from control fields have also been split into the same magnitude ranges. 
Table \ref{tab:mean_mags_sim} shows the same variables as seen in Table \ref{tab:mean_mags} for quasars from the control sample. The separations are also found at the epoch of the quasar. There appears to be no difference between the values for the faint and the bright control samples either. 

\begin{table}[!h]
\caption[Details for bright and faint control quasars samples]{\small{Values for magnitude samples for bright ($M_r < -23$) and faint ($M_r > -23$) control quasars. The errors show one standard deviation.}}
\centering
\begin{tabular}{ c | c c }
                                         & Bright quasars & Faint quasars   \\ 
                                         & $M_r < -23$    & $M_r > -23$      \\ \hline 
Number in sample                         & 120            & 483                    \\
Absolute $r'$ magnitude range            & $-23<M_r<-25.83$ & $-17.79<M_r<-23$  \\
Mean 2D quasar-cluster centre            & 6.51 $\pm$ 3.65 & 6.37 $\pm$ 3.57      \\
separation (Mpc)                         &                 &                      \\
Mean 2D quasar-cluster member            & 6.11 $\pm$ 3.64 & 5.94 $\pm$ 3.47      \\
separation (Mpc)                         &                 &                      \\
Mean 2D quasar-cluster BCG               & 6.53 $\pm$ 3.66 & 6.37 $\pm$ 3.57      \\
separation (Mpc)                         &                 &                      \\
Cluster richness                         & 9.28 $\pm$ 8.16 & 9.63 $\pm$ 14.54
\end{tabular}
\label{tab:mean_mags_sim}
\end{table} 

The Student's t-test has been used to study the difference in the mean values. The t-test is used to compare two small data sets, when the two samples are collected independently of one another, for example using random re-sampling. An un-paired t-test was used as the samples have different sizes, and was two-sided, so the alternative hypothesis was that the mean values were different. The null hypothesis was the difference between the means of two samples is zero. 

Table \ref{tab:t_test} shows the p-values when the mean values for bright and faint for the 2D separations and the cluster richness are compared. Using a significance level of 0.01, the difference in the means for the bright and faint quasars is not significant for the separations or the cluster richness. 

\begin{table}[!h]
\caption[T-test p-values comparing bright quasars and faint quasars]{\small{p-values using the Student's t-test to compare the mean for bright and faint quasars.}}
\centering
\begin{tabular}{ c | c }
                                         & p-value   \\ \hline 
Mean 2D quasar-cluster centre separation & 0.043     \\
Mean 2D quasar-cluster member separation & 0.069     \\
Mean 2D quasar-cluster BCG separation    & 0.048     \\
Cluster richness                         & 0.974
\end{tabular}
\label{tab:t_test}
\end{table}

Table \ref{tab:t_test_sim} shows the p-values when the means from the bright and faint observed quasar samples are compared with the means from the bright and faint quasars in the control samples. For the observed bright quasars, the means for the separations and the cluster richness for the observed quasars are not significantly different from the means from the control quasars. For the observed faint quasars, the means for the 2D projected separation between the quasar and the closest cluster centre, and the quasars and the closest cluster member are significantly different from the means in the control sample. 

\begin{table}[!h]
\caption[T-test p-values comparing bright and faint quasars to control quasars]{\small{p-values using the Student's t-test to compare the mean for observed bright and faint quasars and the mean from control bright and faint quasars.}}
\centering
\begin{tabular}{ c | c c }
                                           & Bright p-value & Faint p-value\\ \hline 
Mean 2D quasar-cluster centre separation   & 0.972          & 0.008   \\
Mean 2D quasar-cluster member separation   & 0.386          & 0.0002      \\
Mean 2D quasar-cluster BCG separation      & 0.940          & 0.011        \\
Cluster richness                           & 0.928          & 0.808   
\end{tabular}
\label{tab:t_test_sim}
\end{table}
 
\begin{figure}[!h]
\centering
\includegraphics[scale=0.8]{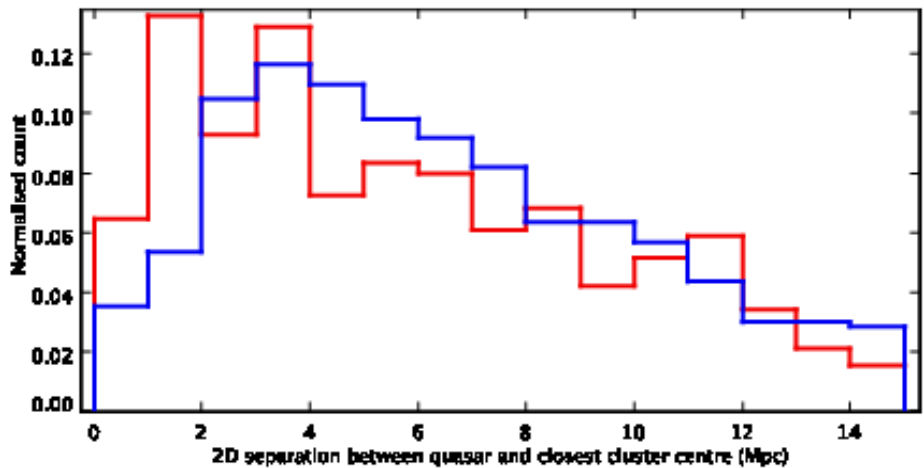}
\caption[2D separations of faint observed and control quasar-cluster pairs]{\small{2D projected separation between the quasars and the closest cluster centre for faint observed (red) and control (blue) quasars.}}
\label{fig:hist_faint}
\end{figure}

Figure \ref{fig:hist_faint} shows the distributions of 2D projected separation between the quasar and the closest cluster centre for faint observed (red) and control (blue) quasars. This Figure shows more observed faint quasars lying closer to the nearest galaxy cluster than in the control sample. This suggests that some faint quasars do lie in a preferred position with respect to the closest cluster and the closest cluster galaxy.

\begin{figure}[!h]
\centering
\includegraphics[scale=0.8]{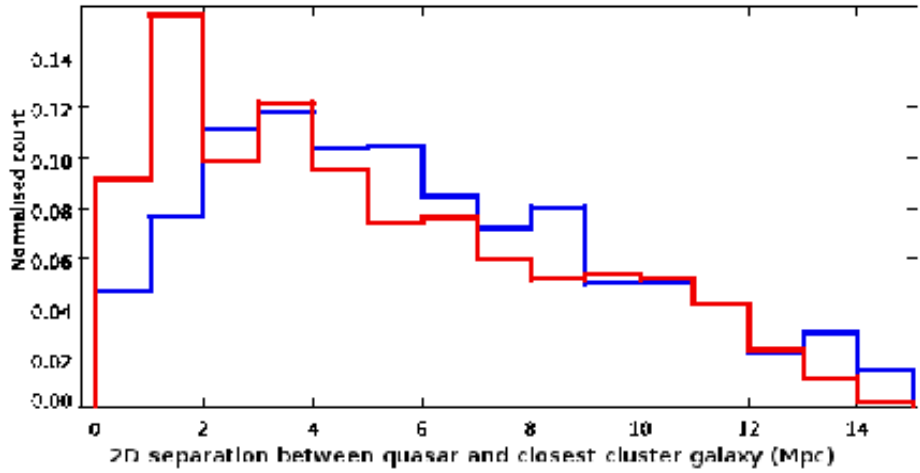}
\caption[2D separations of faint observed and control quasar-cluster galaxy pairs]{\small{2D projected separation between the quasars and the closest cluster galaxy for faint observed (red) and control (blue) quasars.}}
\label{fig:hist_galfaint}
\end{figure}

Figure \ref{fig:hist_galfaint} shows the distribution for the 2D separations between a faint quasar and the closest cluster member for observed (red) and control (blue) quasars. This is a shift in the observed quasars with respect to the control sample of quasars lying close to closest cluster member. Figure \ref{fig:hist_galfaint} supports the result in Table \ref{tab:t_test_sim}, which gives a small p-value (0.0002) suggesting that the observed distribution is significantly different from the control distribution.

\begin{figure}[!h]
\centering
\includegraphics[scale=0.8]{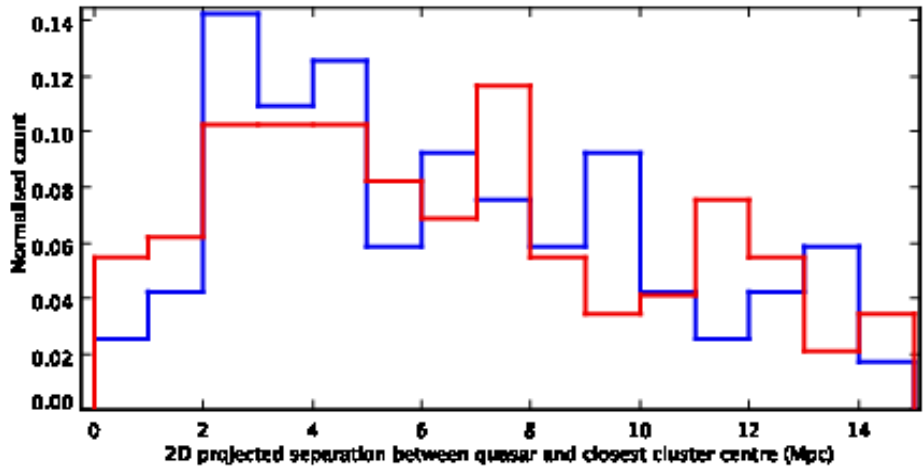}
\caption[2D separations of bright observed and control quasar-cluster pairs]{\small{2D projected separation between the quasars and the closest cluster centre for bright observed (red) and control (blue) quasars.}}
\label{fig:hist_bright}
\end{figure}

Figure \ref{fig:hist_bright} shows the distributions of 2D projected separation between a bright quasar and the closest cluster centre for bright observed (red) and control (blue) quasars. This is a slight increase in the number of observed bright quasars lying near the closest cluster, though this increase is smaller than for faint quasars.

Though the positions of the bright and faint quasars are indistinguishable, there is a difference for the faint quasars with respect to the control sample. There is no difference for the bright quasars between the observed and control sample. A larger sample of bright quasars would be needed to investigate this.  
 
The bright and faint quasar samples have different sample sizes, with the faint sample containing $\sim4\times$ the number of quasar-clusters pairs than the bright sample. To test the effect of the sample size on the t-test, a random sample (selected without replacement) was taken of 147 faint observed quasars and 120 faint control quasars, to match the sample sizes used in the bright quasar sample. The p-value from the t-test for the differences in the mean for the 2D projected separation between the quasar and the cluster closest centre become 0.265, which is no longer significant. The sample size also had an effect on the distributions for the 2D projected separations between the quasar and the closest cluster member, increasing the p-value to 0.034. This indicates the sample size has a large effect on the result from the t-test. However, Figures \ref{fig:hist_faint}, \ref{fig:hist_galfaint} and \ref{fig:hist_bright} do show some potential differences between the observed and control samples for both the faint and to a lesser extent brighter quasars. Larger sample sizes for both faint and bright quasars would be needed to fully assess the significance.

\section{Evolution with Redshift}

To study any evolution with redshift, the magnitude samples have also been split into high ($z>0.6$) and low ($z<0.6$) redshift ranges. Table \ref{tab:mean_mags_bright} shows the values for the two redshift ranges for bright quasars ($M_r < -23$). There appears to be no difference between the positions of bright quasars at high and low redshifts with respect to 2D separations or cluster richness.

\begin{table}[!h]
\caption[p-values comparing bright observed to control quasars in different redshift ranges]{\small{Values for magnitude samples for bright ($M_r < -23$) quasars for low ($z<0.6$) and high ($z>0.6$) redshifts. The errors show one standard deviation.}}
\centering
\begin{tabular}{ c | c c }
                                         & $z<0.6$     & $z>0.6$      \\ \hline 
Number in sample                         & 82          & 65                    \\
Mean 2D quasar-cluster centre            & 6.60 $\pm$ 3.45 & 6.35 $\pm$ 4.32      \\
separation (Mpc)                         &                 &                      \\
Mean 2D quasar-cluster member            & 5.87 $\pm$ 3.20 & 5.54 $\pm$ 4.18      \\
separation (Mpc)                         &                 &                      \\
Mean 2D quasar-cluster BCG               & 6.61 $\pm$ 3.45 & 6.35 $\pm$ 4.29      \\
separation (Mpc)                         &                 &                      \\
Cluster richness                         & 7.63 $\pm$ 5.25 & 11.60 $\pm$ 16.96
\end{tabular}
\label{tab:mean_mags_bright}
\end{table}

Table \ref{tab:mean_mags_faint} shows the values for the redshift ranges for faint quasars ($M_r > -23$). There appears to be no difference between the positions of high and low faint quasars.

\begin{table}[!h]
\caption[p-values comparing faint observed to control quasars in different redshift ranges]{\small{Values for magnitude samples for faint ($M_r > -23$) quasars for low ($z<0.6$) and high ($z>0.6$) redshifts. The errors show one standard deviation.}}
\centering
\begin{tabular}{ c | c c }
                                         & $z<0.6$     & $z>0.6$      \\ \hline 
Number in sample                         & 433         & 97                    \\
Mean 2D quasar-cluster centre            & 5.74 $\pm$ 3.71 & 5.86 $\pm$ 3.93      \\
separation (Mpc)                         &                 &                      \\
Mean 2D quasar-cluster member            & 5.11 $\pm$ 3.43 & 5.54 $\pm$ 4.18      \\
separation (Mpc)                         &                 &                      \\
Mean 2D quasar-cluster BCG               & 5.77 $\pm$ 3.72 & 5.04 $\pm$ 3.62      \\
separation (Mpc)                         &                 &                      \\
Cluster richness                         & 8.39 $\pm$ 7.70 & 14.05 $\pm$ 22.12
\end{tabular}
\label{tab:mean_mags_faint}
\end{table}

These tables suggest the position of a quasar, with respect to galaxy clusters, as a function of quasar absolute magnitude for bright and faint quasars is not dependent on redshift. This is supported by the p-values seen in Table \ref{tab:t_test_z}, which shows the results from the t-test comparing faint and bright quasars at high ($z>0.6$) and low ($z<0.6$) redshifts. There are no significant p-values, suggesting that there is no change with redshift over the redshift range $0<z<1.2$.

\begin{table}[!h]
\caption[Comparing bright to faint observed quasars at different redshifts]{\small{p-values using the Student's t-test to compare the mean for bright and faint quasars with $z<0.6$ and the bright and faint quasars with $z>0.6$.}}
\centering
\begin{tabular}{ c | c c }
                                           & Bright p-value & Faint p-value\\ \hline 
Mean 2D quasar-cluster centre separation   & 0.712          & 0.791   \\
Mean 2D quasar-cluster member separation   & 0.601          & 0.861    \\
Mean 2D quasar-cluster BCG separation      & 0.696          & 0.819      \\
Cluster richness                           & 0.075          & 0.015  
\end{tabular}
\label{tab:t_test_z}
\end{table}

\section{Angular Separations}

\begin{figure}[!ht]
\centering
\includegraphics[scale=0.65]{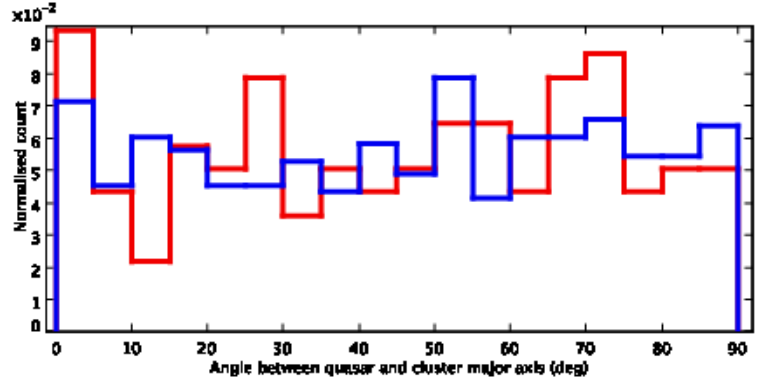}
\caption[Orientation angle for bright and faint quasars]{\small{The orientation angle between a quasar and the major axis of the closest cluster for bright (red) and faint (blue) quasars. }}
\label{fig:mag_ang}
\end{figure}

The mean orientation angle between the quasar and the major axis of the closest cluster is 45$^{\circ}$ for both the bright and faint quasar samples, for high and low redshifts, suggesting a uniform distribution.
Figure \ref{fig:mag_ang} shows the distribution of the orientation angle between a quasar and the major axis of the closest cluster for bright ($M_r < -23$, red) and faint ($M_r > -23$, blue) quasars.
 
There are differences in the distributions of the orientation angle for the two magnitude samples. For example, for angle $>60^{\circ}$, the distribution for the faint quasars is even. However, for the bright quasars, there is a spike at $65-75^{\circ}$. Overall, the distribution for the faint quasars appears to be more uniform. The significance of these differences will be tested in Section \ref{sect:magks}.

\begin{figure}[!ht]
\centering
\includegraphics[scale=0.65]{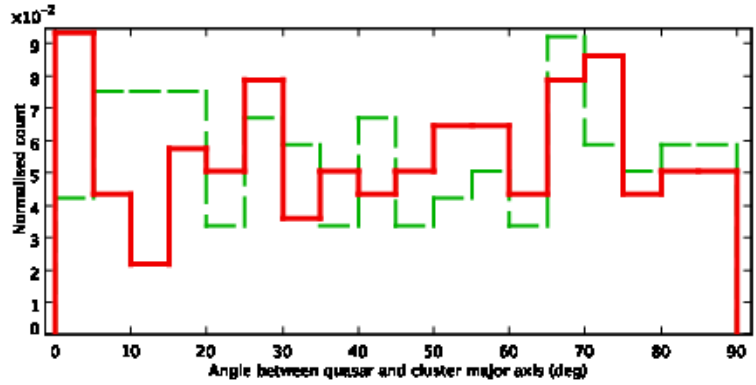}
\caption[Orientation angle for bright observed and control quasars]{\small{The orientation angle between a quasar with $M_r < -23$ and the major axis of the closest cluster for observed (solid red) and control (dashed green) quasars. }}
\label{fig:mag_ang_bright}
\end{figure}

\begin{figure}[!ht]
\centering
\includegraphics[scale=0.65]{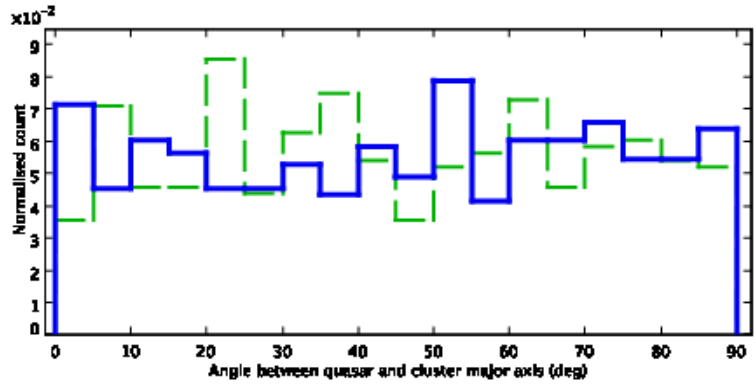}
\caption[Orientation angle for faint observed and control quasars]{\small{The orientation angle between a quasar with $M_r > -23$ and the major axis of the closest cluster for observed (solid blue) and control (dashed green) quasars. }}
\label{fig:mag_ang_faint}
\end{figure}

Figures \ref{fig:mag_ang_bright} and \ref{fig:mag_ang_faint} show the distribution of the orientation angle between a quasar and the major axis of the closest cluster for bright and faint quasars, respectively. The solid lines show the distribution for observed quasars, while the green dashed lines show the distribution for quasars from the control sample.

There are differences in the distributions of the orientation angle for the observed and control samples. For the bright quasars, the largest difference between the observed and control field is for $<20^{\circ}$. This shows more control quasars lying at this orientation than observed quasars. For the faint quasars, this difference (with more control quasars lying at the orientation than observed quasars) is for angles between $20^{\circ}$ and $40^{\circ}$. However, the errors on the orientation angles are $\sim13\pm10^{\circ}$ (Section \ref{angsep}), making these differences less significant. 

The significance of these differences will also be tested in Section \ref{sect:magks}.

\section{2D KS Test Results}\label{sect:magks} 

The two dimensional KS test (Section \ref{2dks}) has been used to determine whether the distributions of the following variables are the same with respect to the control data:
\begin{itemize}
\item Absolute magnitude of the quasar and the 2D projected separation between the quasar and the closest cluster centre,
\item Absolute magnitude of the quasar and the 2D projected separation between the quasar and the closest cluster galaxy,
\item Absolute magnitude of the quasar and the 2D projected separation between the quasar and the BCG,
\item Absolute magnitude of the quasar and the cluster richness, and
\item Absolute magnitude of the quasar and the orientation angle between the quasar and the cluster major axis.
\end{itemize}

The D and p-values of these distributions are shown in Table \ref{tab:stats_mag} for the whole magnitude range. 

\begin{table}[!h]
\caption[2D KS test for absolute magnitude for all quasar-cluster pairs]{\small{D and p-values from two dimensional KS test for different characteristics of quasar-cluster pairs with respect to the quasar absolute magnitude.}}
\centering
\begin{tabular}{c | c c }
Distribution                                         & D                    & p-value   \\ \hline
Quasar $M_r$ - 2D quasar-cluster centre separation   & 0.112                & 0.008    \\
Quasar $M_r$ - 2D quasar-cluster galaxy separation   & 0.122                & 0.003 \\
Quasar $M_r$ - BCG separation                        & 0.115                & 0.006 \\ 
Quasar $M_r$ - cluster richness                      & 0.068                & 0.260 \\
Quasar $M_r$ - quasar orientation angle              & 0.074                & 0.177 
\label{tab:stats_mag}
\end{tabular}
\end{table}

The two dimensional KS test shows that the distributions using the 2D projected separations for the observed sample are significantly different from the same distributions in the control sample. However, the observed distributions for the richness and orientation angle as a function of the absolute magnitude are not significantly different to the distributions in the control samples.

\subsection{Comparing Magnitudes}

The 2D KS test in Section \ref{sect:magks} indicates that the 2D projected separations as a function of quasar absolute $r'$ magnitude are significantly different in the observed and control samples. To assess the difference between bright and faint quasar environments, the 2D KS test has been performed on the  bright ($M_r  < -23.0$ mag) and faint ($M_r > -23.0$ mag) magnitude samples, used in the previous sections. 
The D and p-values for the bright quasars are shown in Table \ref{tab:stats_magsbright} and for the faint quasar in Table \ref{tab:stats_magfaint}. These observed distributions are compared to the distributions from the control samples. 

\begin{table}[!h]
\caption[2D KS test for absolute magnitude for bright quasar-cluster pairs]{\small{D and p-values from two dimensional KS test for different characteristics of bright quasar-cluster pairs with respect to the bright quasars with absolute magnitude $M_r < -23.0$.}}
\centering
\begin{tabular}{c | c c }
Distribution                                         & D                    & p-value              \\ \hline
Quasar $M_r$ - 2D quasar-cluster centre separation   & 0.115                & 0.115   \\
Quasar $M_r$ - 2D quasar-cluster galaxy separation   & 0.143                & 0.266   \\
Quasar $M_r$ - BCG separation                        & 0.119                & 0.488 \\        
Quasar $M_r$ - cluster richness                      & 0.115                & 0.532 \\
Quasar $M_r$ - quasar orientation angle              & 0.138                & 0.307
\label{tab:stats_magsbright}
\end{tabular}
\end{table}

None of the distributions for the bright quasars have significant p-values, indicating that the distributions for the observed samples are not different from those for the control samples.

\begin{table}[!h]
\caption[2D KS test for absolute magnitude for faint quasar-cluster pairs]{\small{D and p-values from two dimensional KS test for different characteristics of faint quasar-cluster pairs with respect to the faint quasars with absolute magnitude $M_r > -23.0$.}}
\centering
\begin{tabular}{c | c c }
Distribution                                         & D                    & p-value              \\ \hline
Quasar $M_r$ - 2D quasar-cluster centre separation   & 0.126                & 0.007   \\
Quasar $M_r$ - 2D quasar-cluster galaxy separation   & 0.142                & 0.002   \\
Quasar $M_r$ - BCG separation                        & 0.130                & 0.005   \\        
Quasar $M_r$ - cluster richness                      & 0.078                & 0.235   \\
Quasar $M_r$ - quasar orientation angle              & 0.008                & 0.174
\label{tab:stats_magfaint}
\end{tabular}
\end{table}
\clearpage
The distribution of 2D projected separations for observed faint quasars are significantly different from the same distributions for faint quasars from the control samples. This supports the result seen in Table \ref{tab:t_test_sim}, where the mean of the distribution of 2D separations between the observed quasars and the closest cluster centre is significantly different from the mean of the same distribution for the control sample.

\begin{figure}[!ht]
\centering
\includegraphics[scale=0.45,angle=-90]{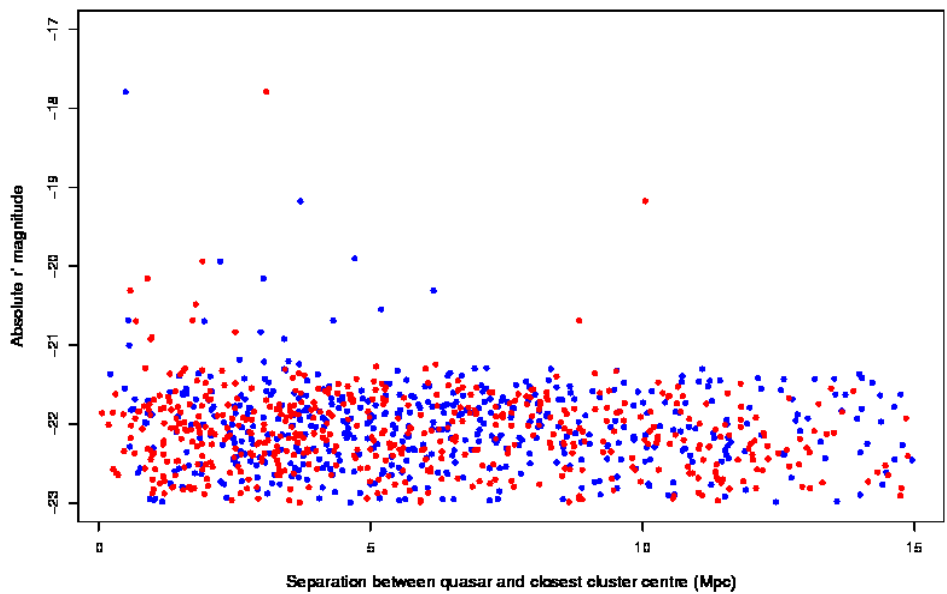}
\caption[2D separation as a function of magnitude for faint observed and control quasars]{\small{2D projected separation between the quasar and the closet cluster centre as a function of the quasar absolute $r'$ magnitude for faint observed (red) and control (blue) quasars. }}
\label{fig:sep_abs_faint}
\end{figure}

Figure \ref{fig:sep_abs_faint} shows the distribution of 2D projected separations between a quasar and closest cluster centre as a function of the absolute $r'$ magnitude for the faint quasars. For quasars with $M_r>-21$, there are more observed quasars lying closer to the cluster centre than control quasars.

\section{Summary}

Using the separation ratio, at low redshifts ($z<0.6$), the average quasar absolute magnitude is fainter (-21.72 $\pm$ 0.90) for quasars lying within a cluster than outside the cluster (-22.39 $\pm$ 1.44), though the difference is within errors. At higher redshifts ($0.8<z<1.2$), the magnitudes of quasars inside and outside clusters are more comparable. However, the separations used  are 2D projected separations. Using the t-test, this difference is not significant. In 3D, the quasars and clusters may not lie at the same redshifts.
In the redshift range $0.6<z<0.8$, there are no quasar lying within clusters, suggesting at this redshift range, quasars prefer to lie in less dense environments. 

Using the t-test and comparing the means for faint ($M_r > -23.0$ mag) and bright ($M_r < -23.0$ mag) quasars, there is no difference between the two magnitude samples for the 2D separations or the cluster richness. 

For the bright quasar sample, there is no difference between the observed and control samples from either the t-test or the 2D KS test. This suggests that quasars with $M_r < -23$ mag do not lie in any preferred position with respect to clusters or with respect to the cluster richness. This may be due to the limited number of quasars with $M_r < -23$ mag in the COSMOS and Stripe 82 quasar samples. 

However, there is a difference between the distributions for the 2D separations for faint quasars in the observed and control samples, shown in the results from both the t-test and the 2D KS test. This suggests that faint quasars lie in preferred positions with respect to galaxy clusters. The 2D KS test shows that there is a difference between the distribution of the 2D projected distance between the quasar and the closest cluster centre as a function of the quasar absolute $r'$ magnitude for faint observed and control quasars. Figure \ref{fig:sep_abs_faint} shows that observed quasars with $M_r>-21$mag prefer to lie closer to the closest cluster centre than quasars from the control sample.

These is no change with redshift (over the range $0<z<1.2$) for the positions of the quasars with respect to the cluster or the cluster richness as a function of absolute quasar magnitude. There is also no preferred orientation between the quasar and the cluster major axis for either bright or faint quasars.

\chapter{Star Forming Galaxies}

Spectra have been taken of a group of star forming galaxies and candidate $z\sim0.8$ early type galaxies in the direction of the Clowes-Campusano Large Quasar Group (CCLQG) using the Inamori-Magellan Areal Camera and Spectrograph (IMACS) instrument on the Magellan Baade Telescope. This data was taken by Luis Campusano and Ilona S\"ochting and was initally intended to be used to provide spectroscopic redshift confirmation for cluster members in this area to enable the clusters to be used in the studies in Chapters 3 and 4, and to study the environments of quasars with respect to star-forming galaxies. Unfortunately few galaxies were cluster members. However, it was possible to study the environments of some of quasars with respect to star-forming galaxies and galaxy clusters. The results can be found in Chapter 6.

This chapter will explain the data reduction process as well as discuss the redshift determination, classification, and extinction correction of the observed objects. The Star Formation Rate (SFR) has been calculated using different methods and compared. Also details of the determination of the redshifts of galaxy clusters known in the area are being presented.

The IMACS catalogue is described in Appendix 2 and available on the attached disk. This catalogue contains all of the input parameters used, and all of the parameters derived from the methods described in this Chapter. 

\section{Star Forming Galaxies}

Star forming galaxies are an important stage in the evolution of galaxies. Environmental effects may show the key to triggering star formation, through events such as galaxy mergers or galaxy harassment. Star formation rates can vary, with some galaxies having rates as high as hundreds of solar masses per year.  These are often called starburst galaxies. Though the definition of a starburst galaxy is not fixed, generally a galaxy with SFR $>10$ M$_{\odot}$yr$^{-1}$ can be considered a starburst galaxy (\citealt{Kennicutt2005}). The star formation can occur across the whole galaxy or in areas around the galaxy nucleus. 

Lyman Break Galaxies (LBGs) are a subset of high redshift, actively star-forming galaxies with relatively low dust obscuration compared to typical galaxies (\citealt{Coppin2007}) and are identified by their colours in the Far-UV around the 912\AA\ Lyman continuum discontinuity (\citealt{Giavalisco2002}). Overdensities of Lyman Break galaxies have been found in the region of the CCLQG (\citealt{Haberzettl2009}) suggesting they trace the Large Scale Structure (LSS). 

\section{Selecting Star Forming Galaxies}
The star forming galaxies were selected by \citet{Haberzettl2009} and observed by Luis Campusano and Ilona S\"ochting.
Two slightly overlapping fields of 1.2 deg were imaged in the direction of the CCLQG. NUV and FUV observations were taken using the Galaxy Evolution Explorer (GALEX)\footnote{http://www.galex.caltech.edu/researcher/techdoc-ch1.html} as part of the Guest Investigator program cycle 1 proposal 35 and consisted of two $\sim$20,000s exposures in the FUV and two $\sim$35,000s exposures in the NUV. Point sources were extracted and their positions cross-correlated with the SDSS DR5 catalogue. The parameter \textit{PhotoTypesvalues} within the SDSS DR5 catalogue was used to discriminate between stars and galaxies, and objects marked \textit{GALAXY} were selected. 

A selection of objects were selected from GALEX data in the Far- and Near-UV to identify star-forming galaxies and Lyman Break galaxies in particular. Additional complementary data for these galaxies was found in the SDSS database. This resulted in magnitude values from 7 pass-bands, $u, g, r, i, z$ and the NUV and FUV, for each galaxy.

To identify LBGs with $z>0.5$, the FUV dropout techniques by \cite{Burgarella2006} was used, which select objects with $m_{NUV} < 23.5$ mag and $m_{FUV} - m_{NUV} > 2$ mag, where $m_{NUV}$ and $m_{FUV}$ are the NUV and FUV magnitudes. For full details about the selection criteria, see \citet{Haberzettl2009}.

\section{Spectroscopic Observations}

On February 25$^{\mbox{th}}$ and 26$^{\mbox{th}}$ 2007, multi-object spectroscopy was taken by Luis Campusano and Ilona S\"ochting on the IMACS instrument on the 6.5m Baade Magellan telescope, at Las Campanas Observatory in Chile. The aims of these observations were to:
\begin{itemize}
\item obtain redshifts for Lyman Break galaxies (LBGs),
\item confirm the suspected large scale structure at $z\sim0.8$ (\citealt{Haines2004}),
\item obtain additional quasars for studies of the large scale environment of quasars.
\end{itemize}

At $z\sim0.8$, there is evidence of a potential large scale structure (\citealt{Haines2004}), however, spectroscopic redshifts of clusters within this structure are needed to confirm how the large scale structure relates to galaxies and their properties.

The galaxy clusters have been found by Ilona S\"ochting in the R and z deep images from the MegaCam instrument at the Canadian-France Hawaiian Telescope (CFHT). The technique used Voronoi Tessellations for density sampling and colour discrimination (slices in the $R-z$ versus $z$ in colour space) that takes advantage of the existence of the cluster red sequence (\citealt{Gladders2000}). The redshift of the cluster can be deduced from the colour of the red sequence if the spectroscopic redshifts are available for some of the clusters across the full redshift range. Calibration of this relation for clusters in this field is addressed in this work (Section \ref{sect:LSS}).

The redshifts from the IMACS observations were used in \citet{Haberzettl2009} to investigate potential sheets of structures using the over-densities of LBGs. The redshifts allow the accuracy of the photometric redshifts to be determined and to be restricted to objects brighter than $m_{NUV} = 23.5$ mag due to an increase in photometric redshift error at fainter magnitudes.

\subsection{The IMACS Instrument}
The spectra were taken on the short f/2 camera. The camera has a 27.4 arcsec diameter field of view with a central scale of 0.2 arcsec per pixel.
For the spectral observations, the 150 grism (which has 150 lines per mm) was used, which has a wavelength range of 3200-10,000\AA, central wavelength of 7200\AA\, and a blaze angle of 10.8$^{\circ}$. The dispersion of this grism is 2.63 \AA\ per pixel, which gives a spectral resolution of R = 26.3\AA.  

Three separation multislit masks were used. Two have 7 exposures of 1800 sec each, giving a total exposure time of 12600 secs. Due to observational time constraints, the data for the last mask only has 5 exposures of 1800 secs, giving a 9000 sec exposure time in total. Each mask contains $\sim$200 slits.

Table \ref{tab:obslog} shows the details about the observations, including the exposures times, dates and positions of the centres of the fields.

\begin{table}[!ht]
\caption[IMACS observation log]{\small{Observing log for IMACS observations.}}
\centering
\begin{tabular}{c | c c c c c }
Mask    &  Date    & Exposures (secs) & Slits          & RA          & DEC         \\ \hline
GALEX 1 & 25.02.07 & 3 $\times$ 1800  & 240            & 10:46:59.4  & +05:23:30.1 \\
GALEX 1 & 26.02.07 & 4 $\times$ 1800  & 240            & 10:46:59.4  & +05:23:30.1 \\
GALEX 3 & 25.02.07 & 7 $\times$ 1800  & 218            & 10:47:53.4  & +05:39.12.0 \\
GALEX 4 & 26.02.07 & 5 $\times$ 1800  & 206            & 10 47:53.4  & +05:22:47.9 
\label{tab:obslog}
\end{tabular}
\end{table}

\subsection{Data reduction}

Most of the data reduction was completed using Cosmos\footnote{The Cosmos software was written for the data reduction of IMACS data. This is not be confused with the Cosmos Evolution Survey, COSMOS.} software, designed for the IMACS instrument. The last steps of the data reduction process were completed using IRAF\footnote{IRAF is a data reduction package, written and supported by the NOAO, which is operated by AURA, Inc.} software.

The positions of the slits were located using \textit{align-mask}\ on the image of the slits. The program locates the edges of the slits and uses the peak of the flux in the slit to locate the slit centre. This defines the slit position, and creates an estimate of the aperture size and position across the CCD.  \textit{align-mask}\ is also performed on the images of the comparison arcs and uses the previous slit position and the aperture size to predict where the arc lines should be. The predicted lines are compared to their observed positions and a list of offsets created. 

\begin{figure}[!h]
\centering
\includegraphics[scale=0.25]{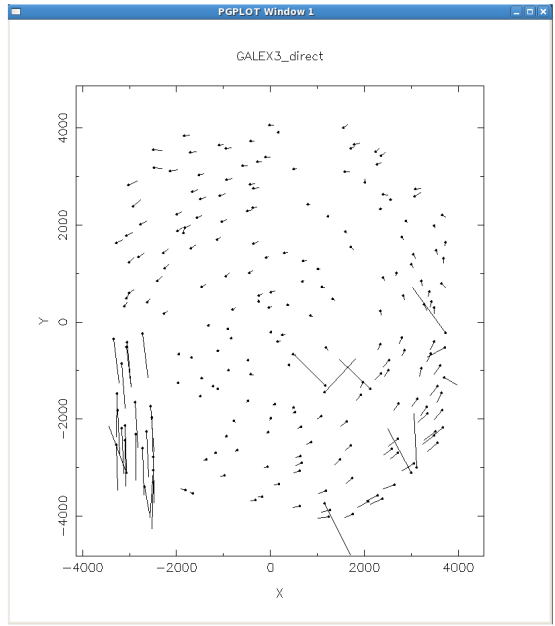}
\caption[Output from \textit{align-mask}]{\small{The output from \textit{align-mask}. The lines indicate the offsets between the predicted and observed positions.}}
\label{fig:align-mask}
\end{figure}

Figure \ref{fig:align-mask} shows the errors on the offsets of the positions of the lines (i.e. the difference between the predicted positions from the image of the slit mask and the actual image of the arc spectral lines). The distortions within the instrument can be seen in the areas of small systematic errors. The large error lines are due to problems such as bad pixels, which create bright lines which are mistaken for spectral lines and will be removed using a bad pixel mask later in the data reduction. This process is repeated, each time refining the slit position, until the errors have been reduced as much as possible.

he first rough slit position definition. As these values were wrong, it produced the wrong positions for the slits.

The next step is to use \textit{map-spectra} to give an approximate wavelength calibration. This program uses a standard configuration file for the instrument which estimates the wavelength calibration using the known positions of the slits. This is only an initial estimate and must be refined using \textit{adjust-map}, which uses the comparison arc spectral lines to map the wavelength calibration on the CCDs. The wavelengths of the arc spectral lines are known. A list of the wavelengths of spectral lines from the comparison arc is included within the software and contains a list of bright isolated lines. If dim closely clustered lines are used, the wrong line can be selected and the errors on the final positioning increased. The more lines used, the more accurate the wavelength calibration will be. For each slit, a map of the wavelength per pixel across the width of the aperture is created.

The wavelength calibration should be checked to ensure the correct wavelength positions of the arc spectral lines are accurately predicted. Running \textit{spectral-map} will create a list of the predicted positions of the spectral lines. The positions of these lines can be plotted over the image of the comparison arc. If the positions do not match, the program is matching the wrong wavelength on the line list to a spectral line. Further adaptation of the line list to include only a selection of the brightest lines may correct this problem. If this does not work, it may be necessary to rerun \textit{adjust-map}. These parameters in \textit{adjust-map} allow various elements such as the width of the slit, errors, the tilt of the slit, and the height and width of the lines to be adjusted.

All of the above procedures create files which are used to locate and define the aperture sizes, and to create the wavelength calibration for the spectroscopic images. 

At this point, it is necessary to create a bad pixel mask. The initial file used is from the Cosmos data reduction package and contains the positions of any bad pixels or bad columns in the CCDs. The program \textit{badorders} adds to this file the positions of the zero'th order contamination. When used in other programs such as \textit{Sflats} and \textit{subsky}, this mask will remove the affected areas by replacing the area with a blank section. This also removes any other data in this area. The bad pixel mask can be edited to ensure only the zero'th order contamination is removed and all possible information is retained. 

\begin{figure}[!h]
\centering
\includegraphics[scale=0.5]{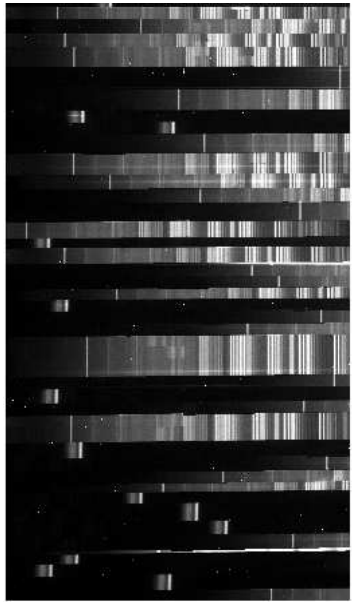}
\caption[Science frame]{\small{The flat fielded and bias subtracted science frame showing the first order spectra with sky lines and the zero'th order contamination.}}
\label{fig:flat_field}
\end{figure}

The science images were bias subtracted and flat fielded. The bias frames were combined using an IRAF routine, \textit{zerocombine} and the flat fields combined in Cosmos using the function \textit{Sflats}. The bad pixel mask is included to mask out bad areas of the CCDs. The bias subtraction and flat fielding of the science (and arc) images is done in the program \textit{biasflat} which requires the map file created in \textit{adjust-map} to show the positions of the slits (the product of which can be seen in Figure \ref{fig:flat_field}). It is important to bias subtract the arc images as this will reduce the noise level and prevent any of the weaker lines being missed. 

The sky lines are removed using \textit{subsky}, which uses the procedure by \citet{Kelson2003}. It is essential that the comparison arcs are correctly matched to the science frames and the wavelength calibration is accurate. If not, the positions of the sky lines are incorrectly predicted and removed. 

At this point, the spectra are still in a single image per mask, each image containing over 200 spectra each. To separate these, the program \textit{extract} is used to cut the slits and put the spectra into a multi-extension fits file which can be viewed in the program \textit{gaia}, in IRAF or using the Cosmos program \textit{viewspectra}. This is completed for each of the separate exposures for each mask. Using the program \textit{sumspec}, the exposures are summed to give a single data cube for each mask, scaled with exposure time. Cosmic ray removal is completed by removing any data points with value greater than 4$\sigma$ above the mean value summed across the spectra. The value was obtained through trial and error as the level must be selected such that the emission lines are left but the cosmic rays, which have a much larger flux, are removed. 

\subsection{Redshift Measurements}

Using the IRAF software, the individual spectra were removed from the multi-extension files using \textit{scopy} and the 1D spectra were extracted using \textit{apall}. The \textit{apall} routine allows the user to interactively select the area of the 2D spectra to extract, and multiple spectra can be removed from single slits if two objects were observed in the same slit. From this, the 1D spectra are extracted and run through \textit{rvidlines}, which identifies spectral emission and absorption lines from a given line list. 

The redshift is estimated by an average of the redshifts from each emission line labelled in the spectra. The error on the redshift is given by the range of these individual redshifts. In most cases, more than four lines were used to make a positive redshift estimate. In some cases, the zero'th order contamination caused spectral lines to be removed, though it was possible to determine which lines were missing and the redshift could still be found.

For the red objects in the masks, the redshift estimate was mainly taken from the 4000\AA\ break which in most cases was very clear. The  4000\AA\ break is the onset of absorption features at wavelengths shorter than 4000\AA\ and produces a noticeable drop in the continuum level. It is also a good indication that star formation is occurring (\citealt{Gorgas1999}). 

A subjective quality value was also given to the spectra. This value ranges from 1 (indicating good clear spectra, low noise, and little doubt on the redshift value given) to 4 (indicating noisy spectra with few clear spectral lines, and lower confidence in the redshift estimate due to the lack of clear lines or fewer lines). 264 of the redshifts have a quality value of 1, 158 have a quality of 2, with the 72 having quality 3 and only four spectra with a quality of 4. 

\begin{figure}[!h]
\centering
\includegraphics[scale=0.7]{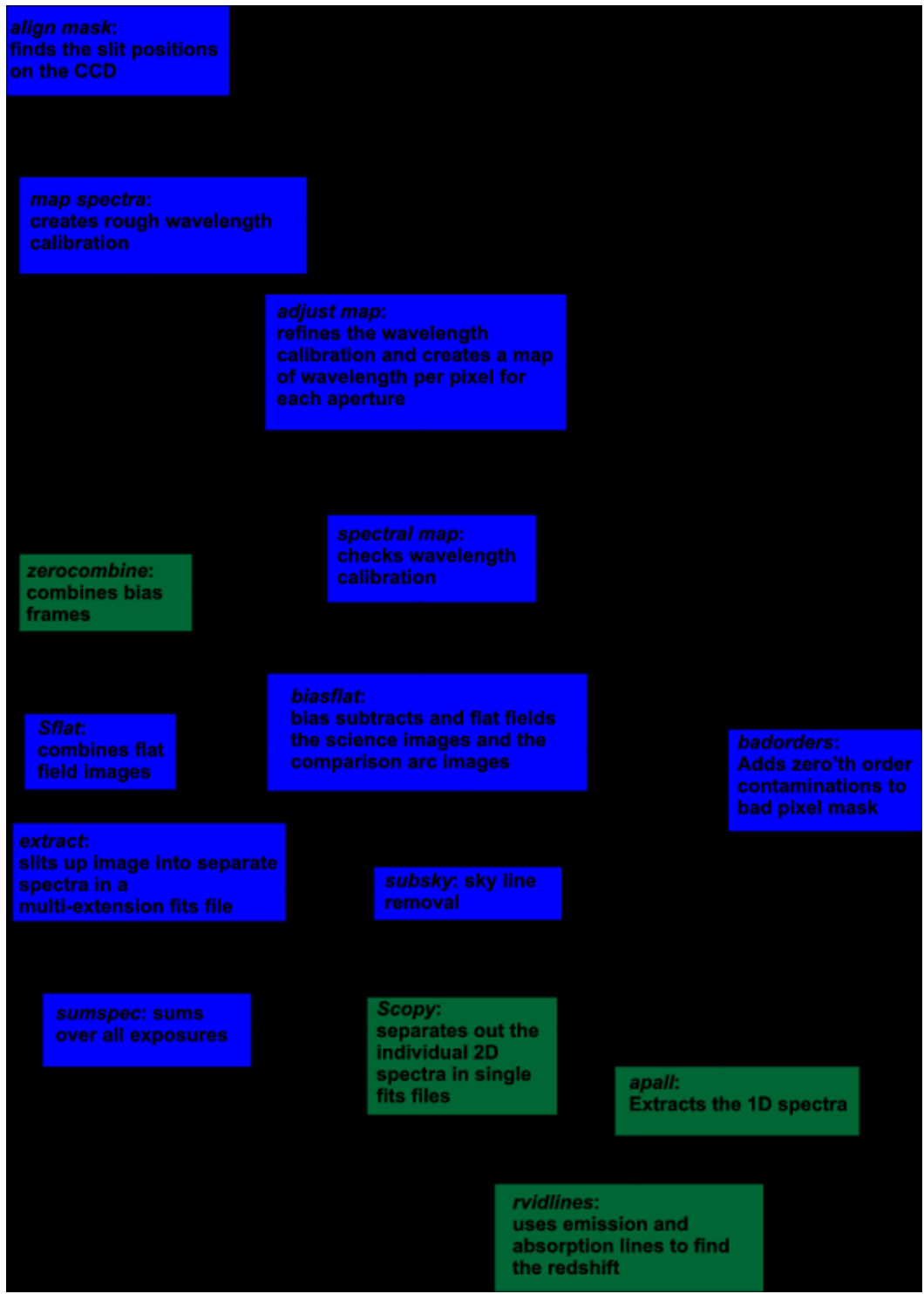}
\caption[Data reduction flowchart]{\small{Flow chart showing the data reduction processes. Boxes in yellow show Cosmos software processes. Boxes in peach show IRAF processes. The plain ovals show input files needed and the plain box shows a necessary output file.}}
\label{fig:datereduction}
\end{figure}

A flow chart of the whole data reduction process can be seen in Figure \ref{fig:datereduction}. The yellow boxes show processes which were run in the Cosmos software, specifically designed to reduce IMACS data. The peach boxes show processes which were run using IRAF. The plain ovals are the necessary input files and the plain rectangular box shows the output file which is needed for the rest of the data reduction.

\clearpage

\clearpage

\section{Flux and Magnitude Calibration of IMACS Spectra}

\subsection{Flux Calibration}
Standard stars are used to calibrate the flux density of spectra. The flux density of the standard star is known and spectra of the standard stars taken on the night of the observations is compared to give a conversion factor for the flux for all spectra. However, during the observations, there were some problems with the observations. Because of this, two of the three standard stars taken did not lie on the slits. The third standard star was reduced and analysed. However, when compared to the published values for this star, the spectrum looked considerably different and because of the pointing problems, it was not certain if this was indeed the standard star. Because of the lack of spectra from a standard star, the calibration was completed by an indirect method. Though this is not ideal, as this method will not take into account any effect on the night of the observations (such as atmospheric turbulence and cloud), the calibration is necessary. The spectra needed to be corrected for atmospheric extinction, detector response, count-to-energy conversion, mirror area, and exposure time.

\begin{figure}[!ht]
\centering
\includegraphics[scale=0.6]{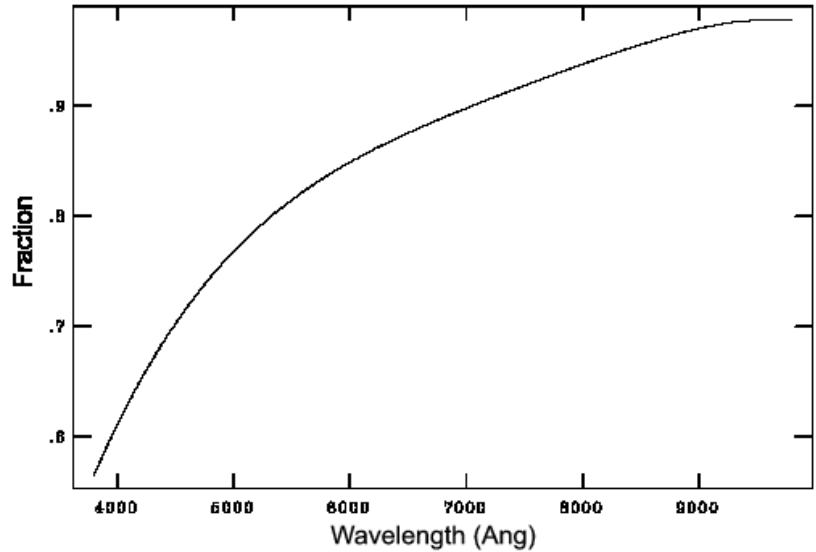}
\caption[Atmospheric Extinction]{\small{The atmospheric extinction as a function of wavelength.}}
\label{fig:extinct}
\end{figure}

The extinction graph (Figure \ref{fig:extinct}) corrects for atmospheric extinction. The values for the extinction were taken from \textit{ctioextinct} in the IRAF database. Cerro Tololo Inter-American Observatory (CTIO) is on a site near the Magellan site where the observations of the spectra were taken, allowing extinction values from this site have been used. These values for the extinction are given in magnitudes per airmass. To correct for the airmass, the equations for the atmospheric extinction is considered and rearranged to get Equation \ref{I/I0}.
\begin{equation}
\frac{I}{I_0} = 10^{-\frac{\Delta m}{2.5}} = e^{-\tau}
\label{I/I0}
\end{equation}

where $\Delta m$ is the difference in magnitudes, given in the CTIO file in magnitudes per airmass, I is the observed intensity after extinction, $I_0$ is the incident intensity before extinction, and $\tau$ is the optical depth. This is for airmass equal to 1. To convert the extinction values due to the airmass during observations,  $e^{-\tau}$ is changed to $[e^{-\tau}]^{sec(z)}$, where $z$ is the angle from the zenith. The value for the airmass used was an average of the airmass values from all of the masks taken from the image header files, which were similar for all of the observations. Figure \ref{fig:extinct} shows the atmospheric extinction curve used, taken from CTIO.

\begin{figure}[!ht]
\centering
\includegraphics[scale=0.6]{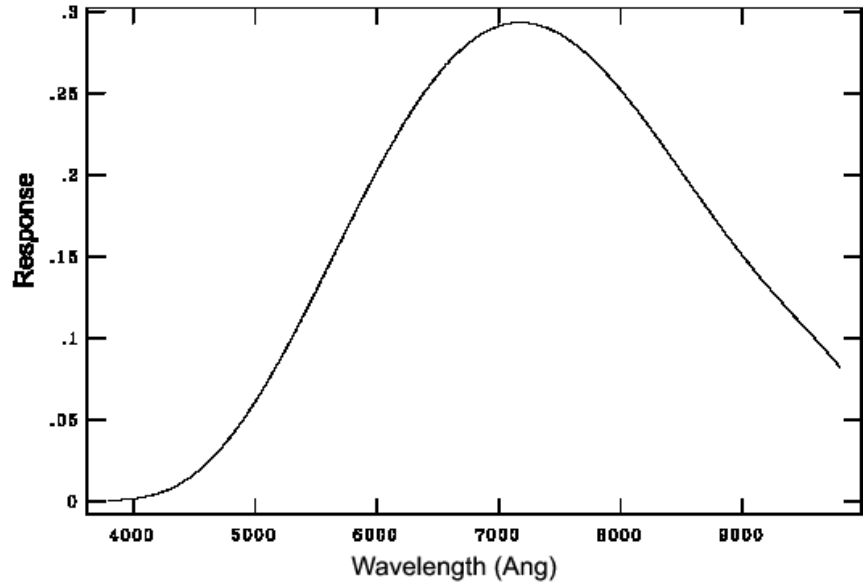}
\caption[CCD response curve]{\small{The response curve for the Magellan Baade telescope.}}
\label{fig:response}
\end{figure}

The spectra also needs to be corrected for the sensitivity and response of the detector and grism, the data for which were taken from the Magellan telescope/Las Campanas Observatory (LCO) website. In IRAF (read in using \textit{onedspec} $\rightarrow$ \textit{sinterp}), a graph of the sensitivity function can be created. As the range covered by the original data is less than that covered by the IMACS spectra, the wavelength range was scaled to cover the IMACS spectral range. Response values at either end of the spectra were linearly extrapolated, so the errors on any measurements at the red and blue edges of the spectra will be larger. Figure \ref{fig:response} shows the response curve of the Magellan Baade telescope.

The spectra on the CCD are in units of analogue to digital units (ADU) and must be converted into energy. On the CCD, one ADU is equal to 5.2 electrons (and one electron corresponds to one photon.) Therefore, the spectra are multiplied by 5.2 to obtain the number of photons received.  The energy of one photon is given using Equation \ref{energyconvert} (where $h$ is Planck's constant, $6.626\times10^{-34}$Js$^{-1}$, $c$ is the speed of light, $3\times10^8$ms$^{-1}$ and $\lambda$ is the wavelength).
\begin{equation}
\mbox{Energy} =  \frac{hc}{\lambda} 
\label{energyconvert}
\end{equation}

\begin{figure}[!ht]
\centering
\includegraphics[scale=0.6]{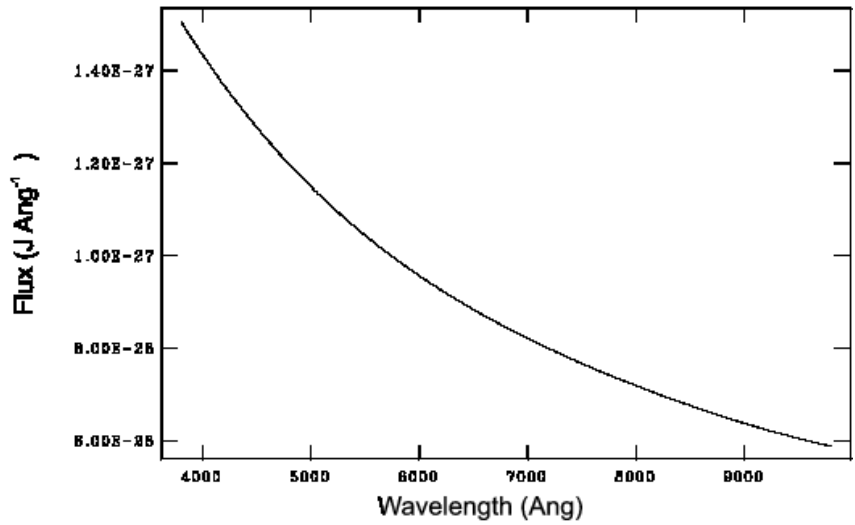}
\caption[Energy calibration]{\small{Converting wavelength to energy.}}
\label{fig:energy}
\end{figure}

Therefore, the value for the energy is multiplied by 5.2 (to convert to the number of photons) and then divided by 2\AA\, which comes from dividing by the wavelength bin size, to convert to the final energy units of J$\AA^{-1}$. A list of the wavelengths covering the spectral range was created and then converted into the equivalent energies. Once the table was created, \textit{sinterp} in IRAF was used to create a graph of this function. Figure \ref{fig:energy} shows the conversion between wavelength and energy using Equation \ref{energyconvert}.

The functions for the extinction, $f_{extinction}$, the response function of the instrument and telescope, $f_{response}$, and energy conversion, $f_{energy}$, are combined to create a single correction function using Equation \ref{eq:functions}.

\begin{equation}
\mbox{final function} = \frac{f_{energy} \times f_{extinction} } {f_{response} } 
\label{eq:functions}
\end{equation}

\begin{figure}[!ht]
\centering
\includegraphics[scale=0.6]{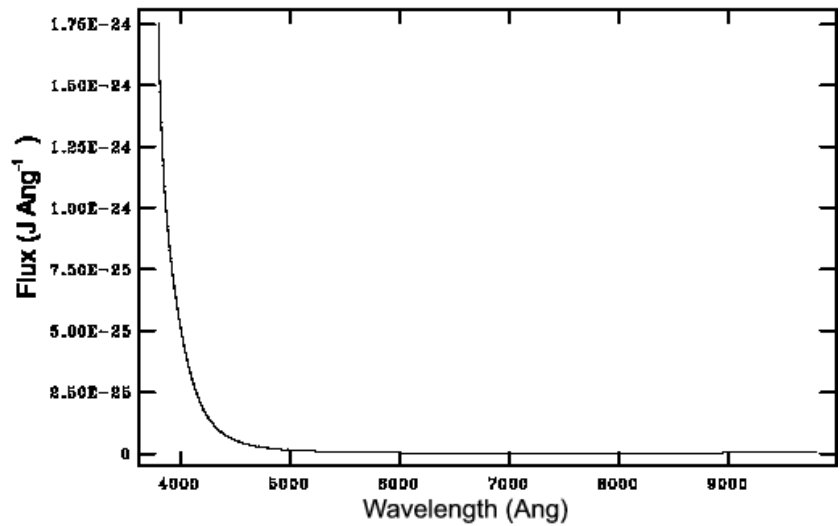}
\caption[Calibration correction]{\small{The calibration correction for the curves described, created using Equation \ref{eq:functions}. }}
\label{fig:finalcalib}
\end{figure}

Figure \ref{fig:finalcalib} shows the final function correction (Equation \ref{eq:functions}) to be applied to the spectra. The spectra were then multiplied by the correction function shown in Figure \ref{fig:finalcalib}. 

The spectra still require further corrections to take into account the mirror size and exposure times. 
\begin{equation}
\mbox{final calibrated spectra} = \frac{\mbox{spectra}}{\mbox{mirror area}\times\mbox{exposure time}}
\end{equation}

The spectra is divided by the exposure time and by the area of the telescope mirror, to give the flux density in astronomical units of Jm$^{-2}$s$^{-1}$\AA$^{-1}$. The exposure time was different for mask 4 as fewer observations were taken. Also the mirror area in m$^2$ was used, with the area for the centre hole removed.

\subsection{Magnitude Calibration}

The spectral lines were measured on the calibrated spectra and used to convert to magnitudes. The main lines measured were $\mbox{[O\textsc{ii}]}\lambda3727$, $\mbox{[O\textsc{iii}]}\lambda5007$, H$\beta\lambda$4861, H$\alpha\lambda6563$, $\mbox{[S\textsc{ii}]}\lambda\lambda6716,6731$ and $\mbox{[N\textsc{ii}]}\lambda6583$, which were used in the classification and to calculate the star formation rates. The equivalent width and the flux values of each of the emission lines were recorded and integrated over wavelength, giving units of Jm$^{-2}$s$^{-1}$.
These values were then converted to intrinsic fluxes, according to the standard $\Lambda$CDM cosmology. The function for the luminosity distance (Equation \ref{eq:lumindist}) in Topcat\footnote{http://www.star.bris.ac.uk/$\sim$mbt/topcat/} was used with the cosmological parameters from \citet{Hinshaw2009} ($H_0=72 $ kms$^{-1}$Mpc$^{-1}$, $\Omega_M=0.27$, $\Omega_{\Lambda}=0.73$). 
Finally the values were converted into ergs$^{-1}$ and corrected with the AB magnitude correction, so star formation rates could be estimated. AB magnitudes are a monochromatic magnitude system which defines a magnitude for a single frequency ($\nu$) or wavelength ($\lambda$) (\citealt{Smith2002}). The AB magnitude system is defined by \citet{Oke1974} as:
\begin{equation}
\mbox{AB}_{\nu} = -2.5 \log F_{\nu} - 48.6
\label{AB_nu_erg}
\end{equation}

where $AB_{\nu}$ is the AB magnitude at a given frequency, $\nu$, and the flux, $F_{\nu}$ is expressed in units of ergs$^{-1}$cm$^{-2}$Hz$^{-1}$. Or if the flux is in units of Js$^{-1}$m$^{-2}$Hz$^{-1}$, 
\begin{equation}
\mbox{AB}_{\nu} = -2.5 \log F_{\nu} - 56.1
\label{AB_nu_J}
\end{equation}

Most of the objects were also imaged with the CFHT using MegaCam in the $r$ and $z$ passbands. An AB magnitude calibration can be estimated from the $r$ magnitudes. 
The passband of the MegaCam $r$ filter has a central wavelength of 6250\AA\ and a FWHM of 1382\AA, giving a wavelength range of 5559--6941\AA. The continuum of the spectra was measured over this range, taking the average of points at each end of the passband and at the central wavelength of the passband, avoiding any emission and absorption features. A conversion factor between the known $r$ magnitudes and the estimated $r$ magnitudes from the spectra was created using Equation \ref{AB_nu_erg} or \ref{AB_nu_J}.

As our AB magnitude values are in units of wavelength, Equation \ref{eq:Fnu} can be used to convert from wavelength to frequency units.
\begin{equation}
F_{\nu} = \frac {\lambda^2}{c} F_{\lambda} 
\label{eq:Fnu}
\end{equation}
where the speed of light, $c$, is expressed in terms of Angstroms, as $3 \times 10^{18}$ \AA s$^{-1}$. 
This gives:

\begin{eqnarray}
\mbox{AB}_{\nu}   & = & -2.5 \log\left( {\lambda^2 F_{\lambda}}\over{c}\right) - 56.1 \nonumber \\
           & = & -2.5 \log\left(\lambda^2 F_{\lambda}\right) + 2.5\log(c) - 56.1 \nonumber \\
           & = & -2.5 \log (\lambda^2 F_{\lambda}) - 9.91  \label{eq:AB_lambda}
\end{eqnarray}

In our calculations, AB$_{\nu}$ was taken as the $r$ magnitude value from the MegaCam data, and $\lambda$ is taken as the central wavelength for the passband of the data (i.e., 6250\AA). Rearranging Equation \ref{eq:AB_lambda}, a value for $F_{\lambda}$ is calculated at the central wavelength of the passband. A conversion factor, $C$, is needed to convert between the known magnitudes and those taken from the estimated magnitudes from the spectra. The value for $C$ was found using Equation \ref{corr_factor} where $g_{\lambda}$ is the continuum level and $F_{\lambda}$ is the flux found using the known AB magnitudes from Equation \ref{eq:AB_lambda}. 

\begin{equation}
F_{\lambda} = Cg_{\lambda}
\label{corr_factor}
\end{equation}

To determine the correction factor, the values for $F_{\lambda}$ against the continuum values were plotted, with the gradient giving the correction factor using linear regression.

\begin{figure}[!ht]
\centering
\includegraphics[scale=0.5,angle=-90]{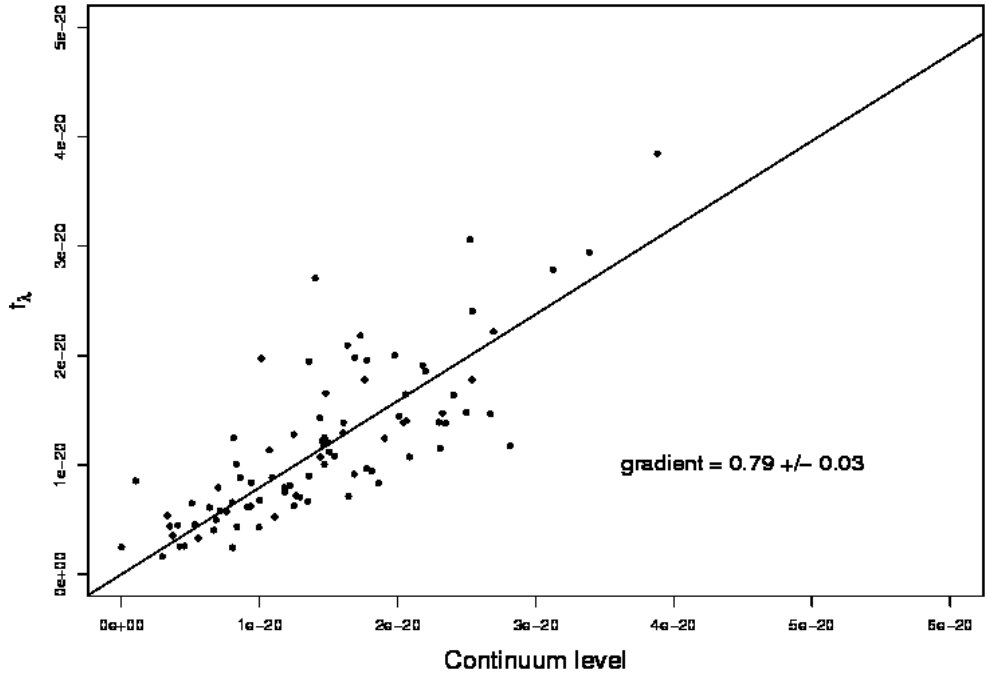}
\caption[Conversion factor to calibrate the flux magnitude.]{\small{Continuum level versus $F_{\lambda}$ from known magnitude values from CFHT. The fitted lines gives the conversion factor, C.}}
\label{fig:conversplot}
\end{figure}

As can be seen from Figure \ref{fig:conversplot}, which shows the continuum level against the flux, $F_{\lambda}$, the correction value $C$ was calculated as 0.79 $\pm$ 0.03, which, given that this value is close to 1, indicates the calibration process was relatively accurate. The final correction is to multiply the estimated flux values by 0.79. This was completed for a random selection of objects. Figure \ref{fig:conversplot} shows the data points from 87 spectra out of a possible 110 measured. 

Three points were excluded from the fit due to abnormal continuum values. These objects were located on the images from the Faint Sky Variability Survey (FSVS) \citep{Groot2000} as this catalogue gives good deep images for these objects. The spectra with an abnormal continuum value were found to be contamination from nearby bright stars or another object lying close by, so flux will have ``bled over" into the spectra, giving an inaccurate result. Therefore, these points were excluded to remove this source of error.

\section{Galaxy Classification}\label{sect:classif}

Galaxies can be classified using their emission lines as different excitation mechanisms in different classes of galaxies will produce different ratios of emission lines. The emission from H\textsc{ii}-like regions (which includes planetary nebulae and star forming regions) is due to ionisation caused by hot young OB stars. The gas surrounding the stars in the interstellar medium (ISM) is ionised by UV photons from the OB stars. AGN show a wider range of ionisations than would be possible with only OB stars as the ionisation source. To produce the emission lines seen, the source must have a harder spectrum, which must extend further into the UV \citep{Osterbrock2006} and, for AGN, this is most likely to come from the accretion disk of the central black hole. The continuum for AGN is featureless, extends over a broad wavelength range, and can be fit with a power law (see Section \ref{sect:simulations} for an example of the power laws used). This group includes Seyfert galaxies and narrow-line radio galaxies.

The different ratios give information about various conditions within the galaxy. For example, 
\begin{itemize}
\item $\mbox{[O\textsc{iii}]}\lambda5007 \over H\beta\lambda4861$ gives a measure of the mean level of ionisation and temperature, 
\item $\mbox{[O\textsc{i}]}\lambda6300 \over \mbox{H}\alpha\lambda6563$ and $\mbox{[S\textsc{ii}]}\lambda\lambda6716,6731 \over \mbox{H}\alpha\lambda6563$ both give the importance of the partially ionised zone which is produced by high-energy photo-ionisation \citep{Osterbrock2006},
\item $\mbox{[N\textsc{ii}]}\lambda6583$ line is stronger in AGN so this line is also used as a classification line.
\end{itemize}
Therefore, by plotting $\mbox{[O\textsc{iii}]}\lambda5007 \over \mbox{H}\beta\lambda4861$ against $\mbox{[S\textsc{ii}]}\lambda\lambda6716,6731 \over \mbox{H}\alpha\lambda6563$ or $\mbox{[N\textsc{ii}]}\lambda6583 \over \mbox{H}\alpha\lambda6563$, different galaxy types lie in different regions on the plot. These diagrams were originally proposed by \citet{Baldwin1981}, and called BPT plots, and then later refined, in particular, by \citet{Veilleux1987}. The emission lines do not need to be reddening corrected, because they are insensitive to reddening  due to the small wavelength separation between the wavelength within the line ratios. 

\begin{figure}[!ht]
\centering
\includegraphics[scale=0.55,angle=-90]{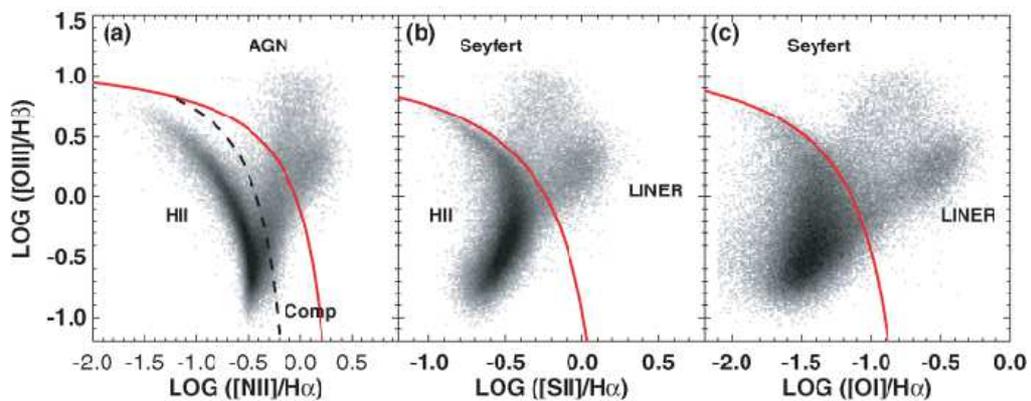}
\caption[Example BPT plots.]{\small{Examples of BPT plots used to classify galaxies and AGN using different emission lines. (\citealt{Kewley2006}) }}
\label{fig:BPTeg}
\end{figure}

An example of a BPT plot for different emission ratios can be seen in Figure \ref{fig:BPTeg}, and is taken from \citet{Kewley2006}. The plots show the different populations of galaxies based on their characteristic emission line ratios. 

It was not possible to classify all of the spectra due to missing lines, either because they were out of the wavelength range of the spectra, or in some cases, have been covered/removed from the spectra due to the reduction process. For those with the lines available, both the equivalent widths and the flux for the  $\mbox{[O\textsc{ii}]}\lambda3727$, $\mbox{[O\textsc{iii}]}\lambda5007$, H$\beta\lambda4861$, H$\alpha\lambda6563$, $\mbox{[S\textsc{ii}]}\lambda\lambda6716,6731$ and $\mbox{[N\textsc{ii}]}\lambda6583$ lines were measured. When visible, the $\mbox{[O\textsc{i}]}\lambda6300$ line was also measured. However, this line was only visible on a few spectra and was usually faint, so this line has not been used as a classification line. The classification criteria were taken from \citet{Veilleux1987}, and from \cite{Yan2006}. The classification using the $\mbox{[N\textsc{ii}]}\lambda6583$ emission line is preferred over that using the [S\textsc{ii}] lines. The [S\textsc{ii}] line is a doublet so the sum of the two lines was used to classify the object. However, both lines are not always present or the lines may be blended causing an inaccurate line flux measurement. 

The IMACS spectra have been classified as either star forming galaxies or AGN, with no distinction is made between LINERs or Seyfert galaxies. AGN are classed using Equations \ref{eq:AGN_NIIclass} and \ref{eq:AGN_SIIclass} (\citealt{Kewley2006}). Galaxies were only classed using Equation \ref{eq:AGN_SIIclass} if the [N\textsc{ii}] emission line was not available, which occurred for 32 galaxies. 129 galaxies were classified using the [N\textsc{ii}] emission line.

\begin{equation}
\log([O\textsc{ii}]/H\beta) > \frac{0.61}{log\left(\frac{[N\textsc{ii}]}{H\alpha}\right) - 0.05} + 1.3
\label{eq:AGN_NIIclass}
\end{equation}

\begin{equation}
\log([O\textsc{ii}]/H\beta) > \frac{0.72}{log\left(\frac{[S\textsc{ii}]}{H\alpha}\right) -0.32} + 1.3
\label{eq:AGN_SIIclass}
\end{equation}

\begin{figure}[!ht]
\centering
\includegraphics[scale=0.5,angle=-90]{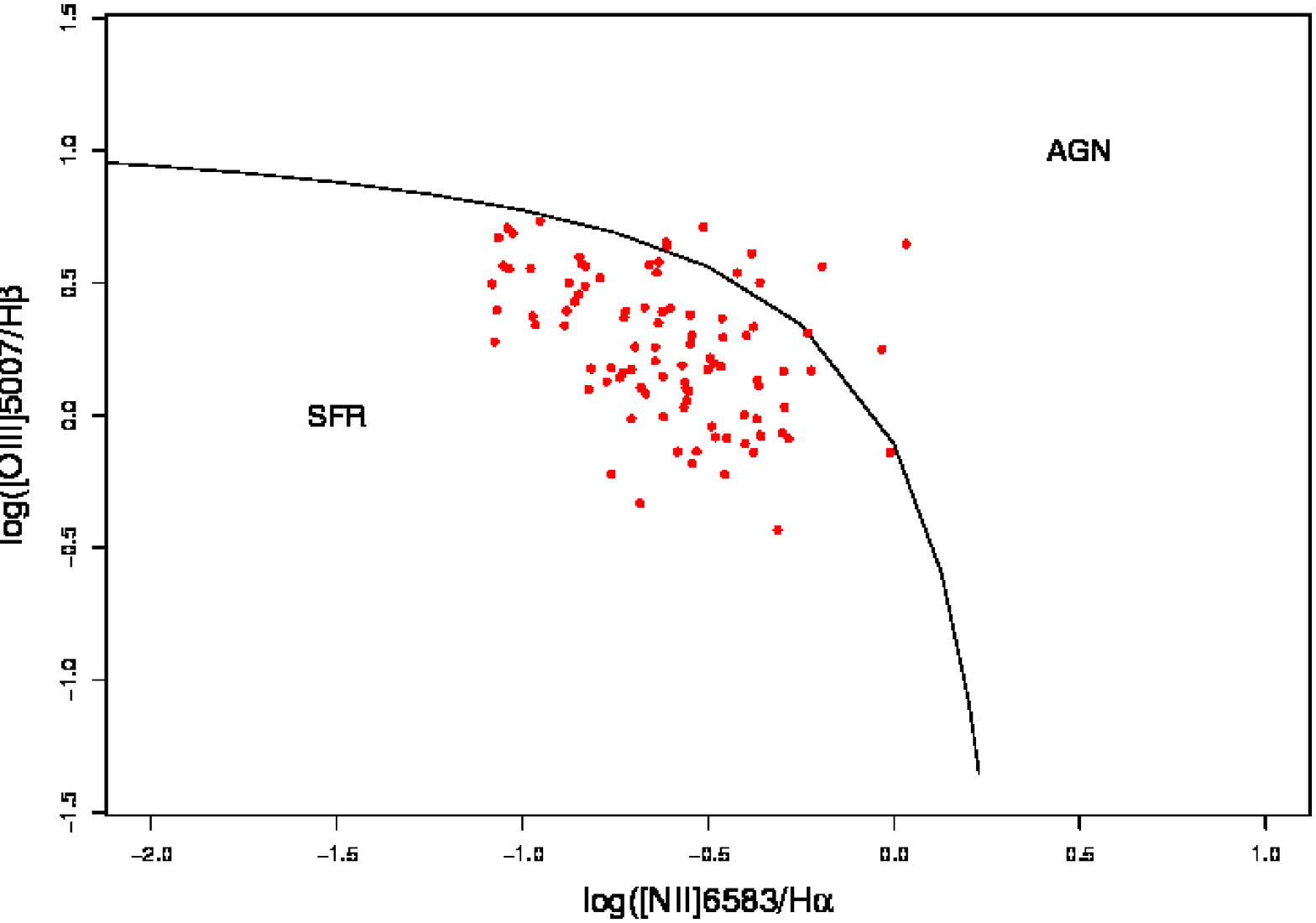}
\caption[Classification using N\sc{ii}.]{\small{Classification using the [N\textsc{ii}]$\lambda6583$ line for the IMACS spectra. The line shows the theoretical separation between the star forming galaxies and the AGN. }}
\label{fig:NII_class}
\end{figure}

\begin{figure}[!ht]
\centering
\includegraphics[scale=0.5,angle=-90]{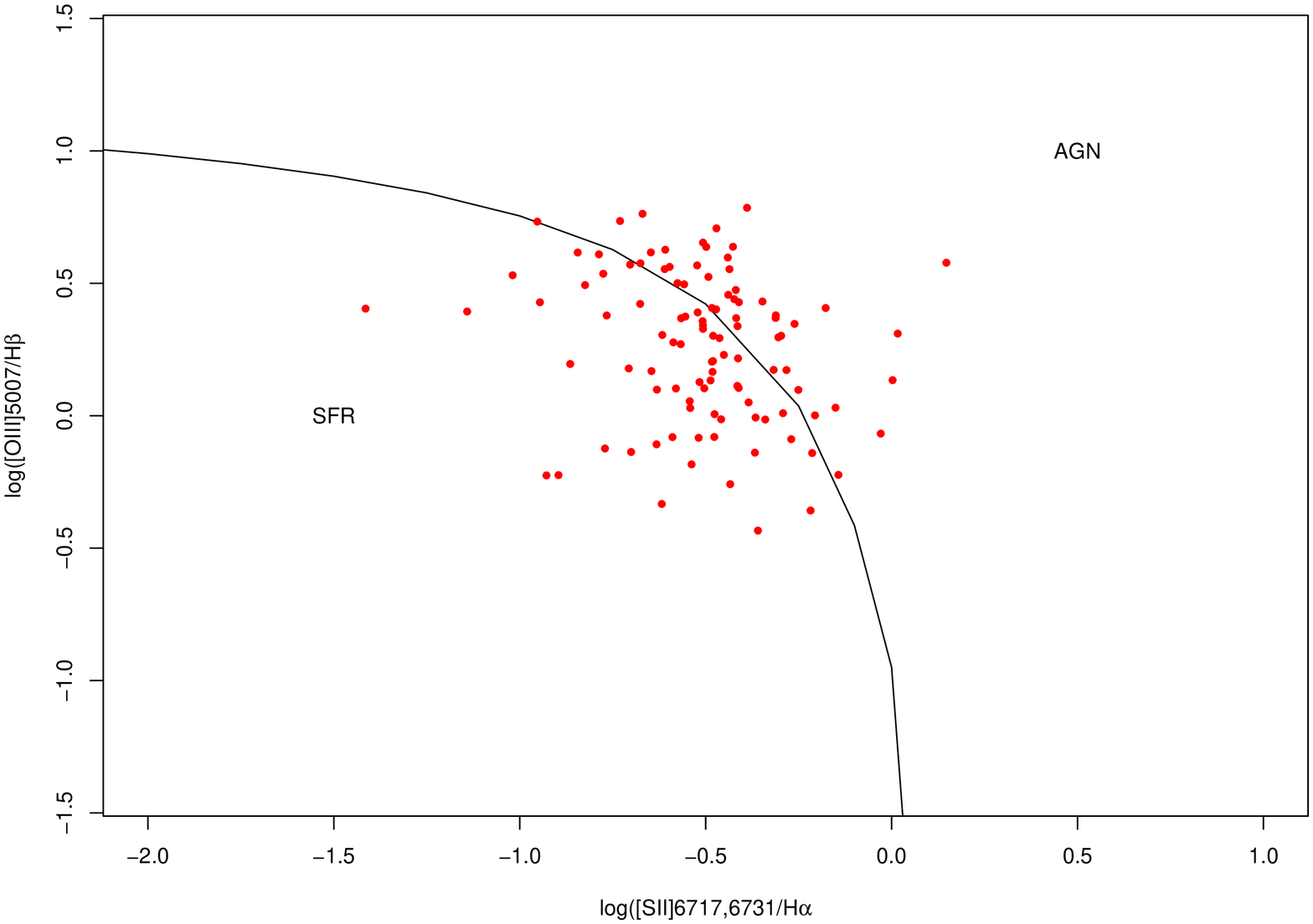}
\caption[Classification using S\sc{ii}.]{\small{Classification using the [S\textsc{ii}]$\lambda\lambda6717,6731$ doublet for the IMACS spectra. The line shows the theoretical separation between the star forming galaxies and the AGN. }} 
\label{fig:SII_class}
\end{figure}

Figure \ref{fig:NII_class} shows the BPT plot using [N\textsc{ii}] to classify the star forming galaxies and the AGN from the IMACS spectra, while Figure \ref{fig:SII_class} uses the [S\textsc{ii}] emission lines. The lines show the separations between star forming galaxies and AGN, taken from \citet{Kewley2006}, who use the line ratios from \citet{Veilleux1987}. BPT plots can be broken down further into other galaxies types such as low-ionization nuclear emission-line regions (LINERs). This, however, is not necessary for our purposes. From Figure \ref{fig:NII_class}, it appears most of the IMACS objects are star-forming galaxies. This is less clear for Figure \ref{fig:SII_class} which uses the [S\textsc{ii}] emission lines, which is expected as this diagram is more prone to errors for the reasons mentioned previously. However, if the [N\textsc{ii}] line is not present (for example, due to a zero'th contamination over the [N\textsc{ii}] line), the [S\textsc{ii}] emission lines can be used to classify the galaxies.

Using the above classification, 33 galaxies were classed as AGN and 70 were classed as star forming galaxies. The galaxies without the lines required to classify the galaxy using the BPT plot will be classed as star forming galaxies. None of the galaxies show evidence of the broad emission lines seen in AGN spectra, so this will be a good assumption. 

\section{Extinction Correction}

The flux of the spectral lines has not been corrected for reddening due to interstellar dust in both the observed galaxy and in intergalactic space. This was estimated using Equation \ref{reddening_eq} (\citealp{Osterbrock2006}):

\begin{equation}
\frac{F(\mbox{H}\alpha)_o}{F(\mbox{H}\beta)_o}=\frac{F(\mbox{H}\alpha)_e}{F(\mbox{H}\beta)_e} 10^{-c[f(\mbox{H}\alpha)-f(\mbox{H}\beta)]}
\label{reddening_eq}
\end{equation}

where $F(\lambda)_0$ is the unreddened intrinsic flux (i.e., the observed flux), $F(\lambda)_e$ is the emitted flux, and $f(\lambda)$ is the reddening curve taken from \citet{Osterbrock2006}, and requires the H$\alpha$ line to be visible. For H\textsc{ii}-like regions, the known ratio for ${F}(\mbox{H}\alpha)_e \over \mbox{F}(\mbox{H}\beta)_e$ is 2.86 (assuming case B recombination and a temperature of 10000K; \citealt{Osterbrock2006, Tresse2002}).

The only unknown is $c$, which is a measure of the extinction and is different for each galaxy, as it depends on the dust within the galaxy and between the viewer and the galaxy. It is found by rearranging Equation \ref{reddening_eq} to give Equation \ref{extinction} and calculating for each individual source.

\begin{equation}
\label{extinction}
c = \frac{ \log\frac{F(\mbox{H}\alpha)_o}{F(\mbox{H}\beta)_o} - \log (2.86)}{-(f(\alpha) - f(\beta))}
\end{equation}

where $F(\lambda)_O$ is the observed flux and $f(\lambda)$ is the taken from the reddening curve. 
\clearpage
Table \ref{reddening} uses interstellar extinction lines to estimate values for the reddening curve, using the standard curve, and R=3.1 \citep{Osterbrock2006}. The reddening values for [N\textsc{ii}] and [S\textsc{ii}] were taken as the same as for H$\alpha$ due to the close proximity of the lines. 

\begin{table}
\centering
\caption[Extinction factors for emission lines.]{\small{The extinction factors for emission lines, with $R$=3.1 from \citealt{Osterbrock2006}, using the standard curve for typical diffuse interstellar medium.}}
\begin{tabular}{c|c c }
Emission line                        & $f$(H$\lambda$)-$f$(H$\beta)$ & $f(\lambda)$ \\ \hline
H$\alpha$                            & -0.346                        & 0.818        \\
H$\beta$                             &  0.00                         & 1.164        \\
$\mbox{[O\textsc{iii}]}\lambda$ 5007 &  0.304                        & 1.122        \\
$\mbox{[O\textsc{ii}]}\lambda$ 3727  &  0.724                        & 1.542        
\label{reddening}
\end{tabular}
\end{table}

$R$ is ratio of the visual extinction at $V$ (i.e $A(V)$) and the colour excess between $B$ (taken at 4350\AA) and $V$ (5550\AA) values, $E(B-V)$ (Equation \ref{eq:reddeningR}). A standard value for $R$ for our galaxy is $R=3.1$.

\begin{equation}
R = \frac{A(V)}{(E(B-V))} 
\label{eq:reddeningR}
\end{equation}
To correct for the interstellar extinction, Equation \ref{eq:I_o} is used.

\begin{equation}
F(\lambda)_e = F(\lambda)_o \mbox{exp}[C f(\lambda)]
\label{eq:I_o}
\end{equation}

where $F(\lambda)_e$ is the emitted flux (i.e, the flux which would be received if there were no extinction), $F(\lambda)_o$ is the observed flux and $C$ is a measure of the extinction and related to $c$ by $c=0.434C$.

Therefore, Equation \ref{extinction} is used to calculate $c$ which is used to calculate the intrinsic flux, F$(\lambda)_e$.

The main problem with the reddening correction method described above is the dependence on H$\alpha$, needed to find the measure of the extinction, $c$. For objects with $z>0.5$, the H$\alpha$ emission line moves out of the optical window. So for higher redshift objects, another method which uses [O\textsc{ii}]$\lambda3727$ is used. To obtain accurate star formation rates, the flux for [O\textsc{ii}]$\lambda3727$ needs to be reddening corrected. To test which method for reddening correction is the most reliable, the [O\textsc{ii}] lines were reddening corrected using different methods, for objects where the H$\alpha$ line was visible.

\subsection{Average Extinction Method to Correct [O\sc{ii}]}
If an object has both H$\alpha$ and [O\textsc{ii}] in the spectra, the extinction correction can be estimated using the H$\alpha$ flux. However, for objects with $z>0.5$, this is not possible. For this reason, the measure of the extinction correction, $c$, has been calculated for all objects containing both H$\alpha$ and [O\textsc{ii}] in the spectra. A mean of these extinction correction values, $c$, is then applied to objects with redshift of $z>0.5$.

\begin{figure}[!ht]
\centering
\includegraphics[scale=0.65]{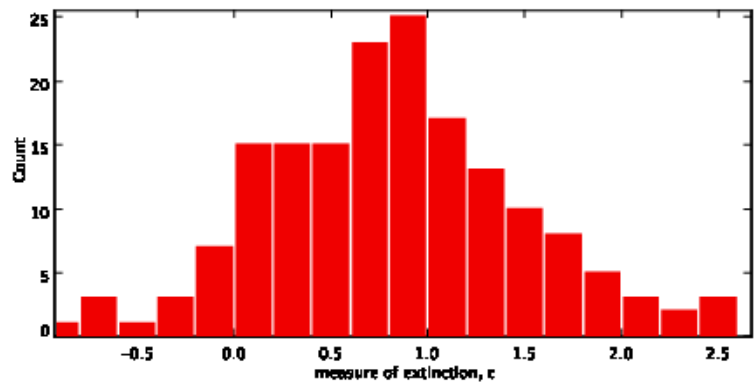}
\caption[Histogram of the extinction values.]{\small{The histogram shows the spread of values for the extinction, $c$, taken from spectra with $z<0.5$. }}
\label{fig:c_histogram}
\end{figure}

An average of the values for $c$ from the objects with H$\alpha$ was found to be 0.82 with a standard deviation of 0.66. This method will not be perfect due to the extinction factor being dependent on an individual galaxy's properties and the range of values seen is large. Figure \ref{fig:c_histogram} show a histogram of the values and shows a profile with a main peak at this value.

\begin{figure}[!ht]
\centering
\includegraphics[scale=0.5,angle=-90]{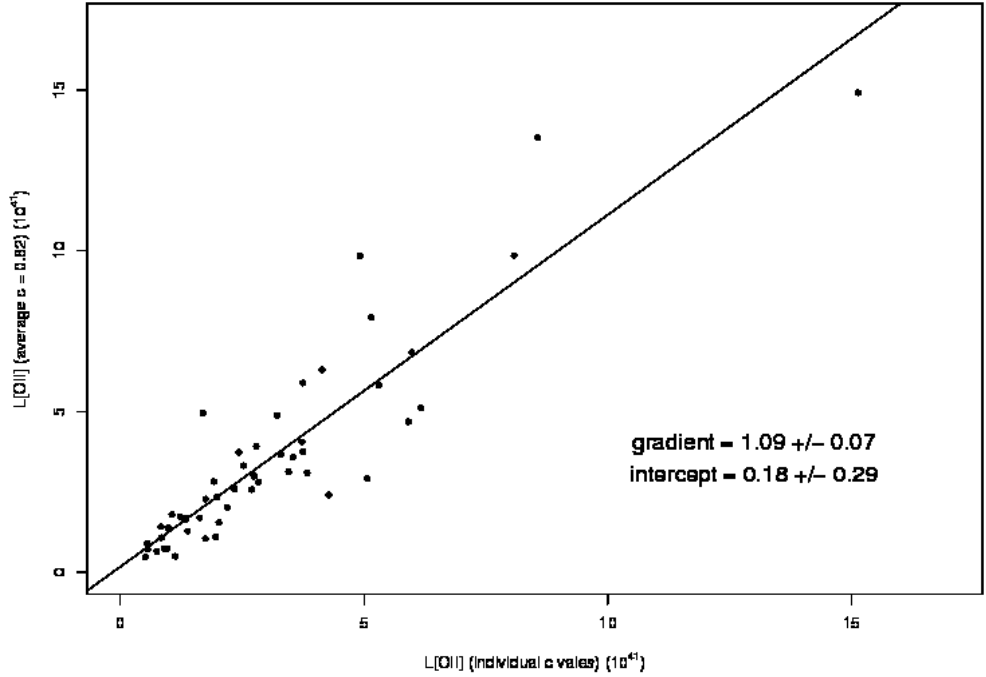}
\caption[Compare estimated to calculated reddening values. ]{\small{Comparing L([O\textsc{ii}]) with an average value of $c$ for the reddening to the L([O\textsc{ii}]) with individual values for $c$, for objects with $z<0.5$ containing the [O\textsc{ii}] and H$\alpha$ emission line. }}
\label{fig:av_red}
\end{figure}

Figure \ref{fig:av_red} shows the corrected [O\textsc{ii}] fluxes using the individual values for $c$ against the corrected [O\textsc{ii}] fluxes using the average value of $c$, $c$=0.82. There is not a significant difference between the corrected fluxes using individual and an average value for c for the extinction.

If an object has an individual value for the extinction correct, $c$, this individual value will be used. Otherwise the [O\textsc{ii}] flux will be corrected using the average value for the extinction.

\subsection{Reddening Curve Extinction Method for [O\sc{ii}]}
It has been suggested by \citet{Kewley2004} that using Equation \ref{OII_red} and the reddening curve from Cardelli, Clayton \& Mathis (1989) will give realistic reddening corrected values. Equation \ref{OII_red} shows the relationship between the observed and redden corrected values for the [O\textsc{ii}] luminosities.

\begin{equation}
\mbox{L([O\textsc{ii}])}_i(ergs^{-1})=\mbox{L([O\textsc{ii}])}_o(\mbox{ergs}^{-1})\times10^{0.4k_{[OII]}E(B-V)}
\label{OII_red}
\end{equation}

where L([O\textsc{ii}])$_i$ is the intrinsic [O\textsc{ii}] luminosity, L([O\textsc{ii}])$_o$ is the observed [O\textsc{ii}] luminosity, and $k_{[O\textsc{ii}]}$=4.771, a constant from the reddening curve from \citet{Cardelli1989}. \citet{Kewley2004} calculated the best fit for $E(B-V)$ for their data sample as:

\begin{equation}
E(B-V)=(0.174\pm0.035)\log(L(\mbox{[O\textsc{ii}]})_i)-(6.84\pm1.44)
\label{E(B-V)}
\end{equation}

When this is substituted into Equation \ref{OII_red}, the value for the intrinsic [O\textsc{ii}] luminosity is given by:

\begin{equation}
L(\mbox{[O\textsc{ii}]})_i=3.11\times10^{-20}L(\mbox{[O\textsc{ii}]})_o^{1.495}
\label{OII_corr}
\end{equation}

\subsection{Comparison of [O{\sc{ii}}] Reddening Correction methods}
Both the extinction correction using the average extinction value from \citet{Osterbrock2006} and using an average $E(B-V)$ value from \citet{Kewley2004} were used to calculate the intrinsic luminosity and the results compared using the objects which contained both [O\textsc{ii}] and H$\alpha$ within the spectra. 

\begin{figure}[!ht]
\centering
\includegraphics[scale=0.5,angle=-90]{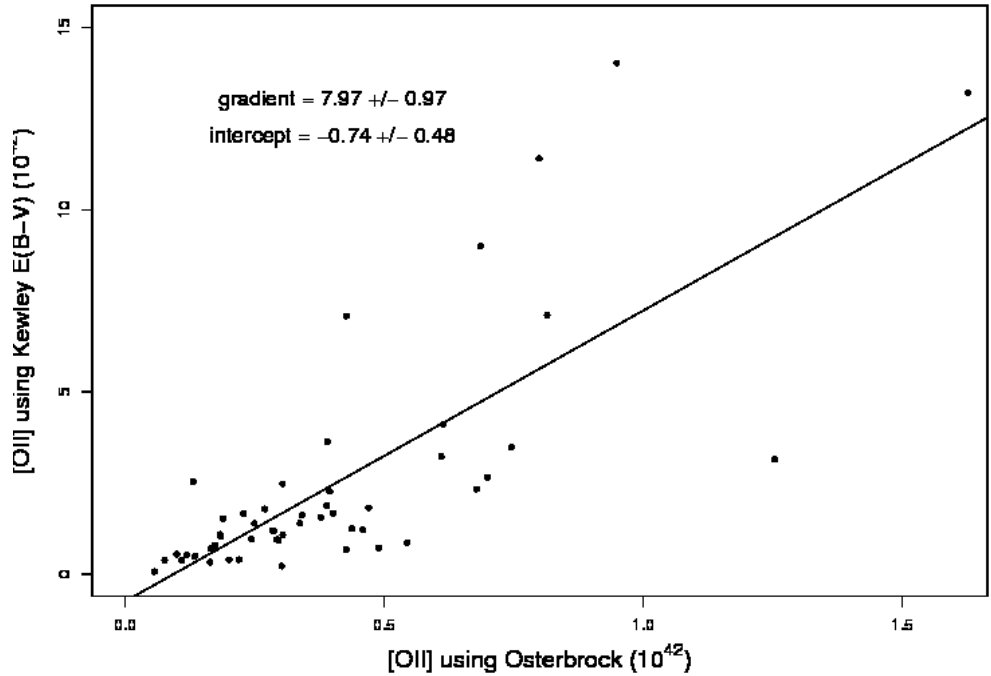}
\caption[Comparing different reddening correction methods.]{\small{Comparing of the [O\textsc{ii}] flux when corrected using the reddening correction method from \citet{Kewley2004} to the [O\textsc{ii}] flux when corrected using an average of all the estimated $c$ values using the method from \citet{Osterbrock2006} for objects with $z<0.5$. }}
\label{fig:kewley_red}
\end{figure}

Figure \ref{fig:kewley_red} shows the reddening corrected [O\textsc{ii}] fluxes using the best fit $E(B-V)$ value from \citet{Kewley2004} on the ordinate axis. The values for the reddening corrected [O\textsc{ii}] fluxes using a mean of the estimated values $c$ values from the \citet{Osterbrock2006} method are on the abscissa axis. This Figure shows a large difference between the [O\textsc{ii}] fluxes using the methods from \citet{Kewley2004} and \citet{Osterbrock2006}. The line is the best fit line. 

As can be seen in Figure \ref{fig:av_red} and \ref{fig:kewley_red}, using an average value for the extinction coefficient, $c$, appears to give a more reliable and predictable comparison to the extinction values corrected using the H$\alpha$ line. Even if the outliers were to be removed from Figure \ref{fig:kewley_red} (these points have some of the largest fluxes for H$\beta$; further investigation is needed to determine the reason for this), the [O\textsc{ii}] reddening corrected values would still be over estimated by a factor of $\sim7$, which can be seen from the gradient of the fitted line in Figure \ref{fig:kewley_red}. 

It should be noted that the calculation for $E(B-V)$ is empirical and, therefore, should not be applied blindly to another sample of galaxies as they may have different properties, which would give variations on the extinction value obtained. \citet {Kewley2004} take galaxies from the Nearby Field Galaxy Survey (NFGS) which contains 198 nearby galaxies. The galaxies in the IMACS data have larger redshifts so are further away and likely to be affected by dust extinction due to an increase in the volume of intergalactic space, therefore, an increase in intergalactic dust. The galaxies within the IMACS data may have different properties to the NFGS sample, producing a different value of $E(B-V)$ if it could be calculated. 

An averaged value of the extinction coefficient, $c$, will therefore be used to complete the reddening correction on [O\textsc{ii}] for the objects  with $z>0.5$, which do not have the H$\alpha$ line within their spectra.

\section{Star Formation Rates}
The star formation rate (SFR) can be calculated using several different emission lines. The Balmer emission lines, especially the H$\alpha$ line, are a direct star formation probe as they scale directly with the total ionising flux. Therefore, the Balmer lines are the most popular and accurate method used to calculate the SFR. Lines such as the [O\textsc{ii}]$\lambda3727$ forbidden line can be used, though this line is affected by metallicity and reddening. Also the Ultra-Violet (UV) flux can give a measure of the SFR. Each method has different strengths and weaknesses as they measure slightly different aspects of SFRs. 

Different star formation indicators give different information about the stellar populations in the galaxies. For example, H$\alpha$ measures the SFR on time-scales of $\sim$10Myrs as opposed to the UV SFR which measures time scales of $\sim$100Myrs, so therefore, measures an older stellar population. The emission lines and the UV continuum are also sensitive to different stellar masses with emission lines being more sensitive to lower mass stars (\citealt{Gilbank2010}).

\subsection{Aperture Correction}
The IMACS observations were taken using rectangular slits so some light will be lost as the slit will only collect part of the light from the source. This loss is an inherent problem with slits. The slit has a width of 2 arcsecs. Therefore any object with a Petrosian radius $>$1 arcsec (which was true for all the objects in the sample) will be affected by slit loss and the estimated SFR  will be an underestimate of the true value. 
To correction for the aperture size, the SFR per arcsec$^2$ was calculated for each object using the slit size and the Petrosian radius for each object. The Petrosian radius is defined as radius at which the Petrosian ratio, $R_p$, is equal to 0.2. The Petrosian ratio is defined as the ratio of the local surface brightness with radius $r$ from the centre of the galaxy and the mean surface brightness with radius $r$, \footnote{http://cas.sdss.org/dr4/en/help/docs/algorithm.asp?key=mag\_petro}.
Then assuming the galaxies are circular with radius equal to the Petrosian radius, the total SFR rate was produced. 

\begin{figure}[!ht]
\centering
\includegraphics[scale=0.5]{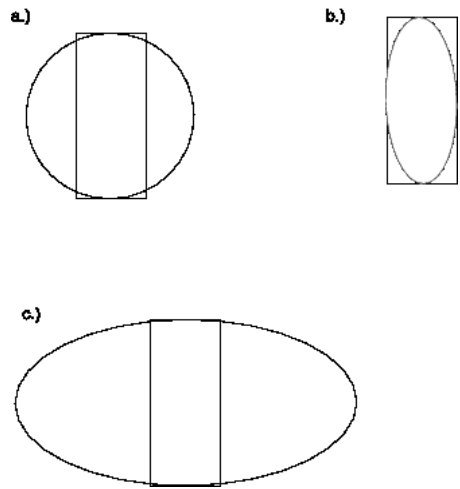}
\caption[Potential errors from galaxy shape]{\small{Examples of potential errors from using assumptions about galaxy shape. }}
\label{fig:apertures}
\end{figure}

There were two main assumptions used in this method. It was assumed that the galaxies are circular. This assumption will mean that, depending on the orientation of the galaxy with respect to the slit, some of these values will be an under or over estimate if the galaxy is elliptical. An example of this can be seen in Figure \ref{fig:apertures} which shows the assumptions of (a) a circular cluster. However, if (b) the cluster is elliptical and aligned along the slit, the SFR will be over-estimated as the light received is under-estimated. If the slit is aligned at 90$^{\circ}$ to the cluster's major axis, the SFR could be under-estimated (c).

It was also assumed that the slit was centred on the centre of the galaxy. This means the slit will be placed across the central bulge, which is brighter than the outskirts of the galaxy. Assuming this is representative of the whole galaxy will give an overestimate of the SFR as the stellar population varies across the radius of the galaxy and the rate is likely to be higher in the bulge than the outskirts. However, correcting for the aperture size in this way will give an upper limit on the SFR while the uncorrected SFR will give a lower limit.

\subsection{H$\alpha$ Star Formation Rate}

To calculate the SFR, the H$\alpha$ line is used, as the H$\alpha$ luminosity is directly proportional to the number of photons from the Lyman continuum in young stars. The H$\alpha$ line gives an indication of the number of ionising stars, which, when an initial mass function (IMF) is assumed (for \citet{Kennicutt1998}, a Salpeter IMF was used) and a solar abundance, gives the star formation rate. Other Balmer lines, such as H$\beta$ and H$\gamma$, could be used though these lines are weaker and more sensitive to stellar absorption. The H$\alpha$ line is still subject to metallicity dependencies and uncertainties from dust corrections though not as heavily as other indicators. The method assumes a solar abundance within the system and that the ionised gas traces all the massive star formation. Equation \ref{Kennicutt_H} is the equation most commonly used for calculating the SFR using H$\alpha$.

\begin{equation}
SFR(M_{\odot} \mbox{yr}^{-1}) = 7.9 \times 10^{-42} L(\mbox{H}\alpha) (\mbox{ergs}^{-1})   \qquad \textrm{(Kennicutt 1998)}
\label{Kennicutt_H}
\end{equation}

Due to metal-rich stars producing less H$\alpha$, the H$\alpha$ will overestimate the SFR for high-metallicity galaxies (\citealt{Gilbank2010}). 

\subsection{[O{\sc{ii}}] Star Formation Rate}
For objects with $z>0.5$, the H$\alpha$ emission line is redshifted out of the range of the optical spectrum so other emission lines are used as SFR indicators. Another SFR indicator is the [O\textsc{ii}]$\lambda3727$ forbidden line \citep{Kennicutt1998, Kewley2004}. Forbidden lines are not directly linked to the luminosity but are more sensitive to metallicities and abundances so are more prone to errors and uncertainties. Unfortunately these methods have yet to agree (e.g., \citealt{Gilbank2010}), and produce different SFR estimates as the ratio of [O\textsc{ii}]/H$\alpha$ varies and is individual to a galaxy due to variations in the gas fractions and excitation differences. Different formulae have been created using separate samples (so therefore producing distinct values). \citet{Kewley2004} used high resolution spectra and a different reddening correction method than \citet{Kennicutt1998} and took the metallicity abundance into account (Equation \ref{Kewley_abun}). The abundances were taken from galaxies in the Nearby Field Galaxies Survey (NFGS). The methods in \citet{Kennicutt1998} are quoted to agree within a factor of two with the methods in \citet{Kewley2004} (\citealt{Kewley2004}).

\begin{eqnarray}
SFR(M_{\odot} yr^{-1}) & = & (1.4 \pm 0.4)\times 10^{-41} L_{[O\textsc{ii}]} (\mbox{erg s}^{-1})   \qquad \textrm{(Kennicutt 1998)} \label{Kennicutt_O} \\
SFR(M_{\odot} yr^{-1}) & = & (6.58 \pm 1.65)\times 10^{-42} L_{[O\textsc{ii}]} (\mbox{erg s}^{-1}) \qquad \textrm{(Kewley 2004)}  \label{Kewley}  \\
SFR(M_{\odot} yr^{-1}) & = & (9.53 \pm 0.91)\times 10^{-42} L_{[O\textsc{ii}]} (\mbox{erg s}^{-1}) \qquad \textrm{(Kewley2004,Z)}\label{Kewley_abun}
\end{eqnarray}

\subsection{Ultra-Violet Star Formation Rate}\label{sect:UVSFR}
The UV SFR measures the continuum of young stars. The UV needs to be corrected for dust extinction but is insensitive to metallicity (e.g. \citealt{Glazebrook1999,Kewley2004}). However, more evolved stars also produce UV emission which contributes to the UV luminosity observed and needs to be corrected for. This last correction is more prominent at lower redshifts as the stellar populations become more evolved \citep{Gilbank2010}.

UV data was taken in the Near-UV (1344-1786\AA) and Far-UV (1771-2831\AA) by GALEX.
The values given for the NUV and FUV fluxes were given in units of counts per second. To convert to units of $\mu$Jy, the FUV was multiplied by 108 and NUV by 36. This is to convert to the units of flux of $\mbox{ergs}^{-1}\mbox{m}^{-2}\mbox{Hz}^{-1}$ needed to calculate the SFR.

\begin{eqnarray}
\mbox{Jy}    & = & 10^{-26}\mbox{Wm}^{-2}\mbox{Hz}^{-1}  \nonumber\\
             & = & 10^{-26}\mbox{Js}^{-1}\mbox{m}^{-2}\mbox{Hz}^{-1} \nonumber\\
\mu\mbox{Jy} & = & 10^{-32}\mbox{Js}^{-1}\mbox{m}^{-2}\mbox{Hz}^{-1} \nonumber \\
             & = & 10^{-25}\mbox{ergs}^{-1}\mbox{m}^{-2}\mbox{Hz}^{-1} \label{eq:Jy}
\end{eqnarray}

The flux is then converted onto a UV luminosity using Equation \ref{eq:lumin}.
\begin{equation}
L = 4\pi D_L^2 F
\label{eq:lumin}
\end{equation}
 
where $L$ is the luminosity, $F$ is the flux and $D_L$ is the luminosity distance. This gives the luminosity in units of ergs$^{-1}$Hz$^{-1}$.

The UV luminosities need to be corrected for dust extinction. Most methods to correct for dust extinction in the UV use infra-red data. However infra-red observations are not available for this data. \citet{Salim2007} measured the SFR for a large sample of optically selected galaxies and fit SEDs to obtain estimates for the dust attenuation, which can then be used on the FUV, when it is not possible to fit an SED (for example when the data is not available as in this case). From this they obtained:

\begin{eqnarray}
A_{FUV}    & = & 3.32 \times (FUV-NUV)^0 + 0.22 \qquad \textrm{$(FUV-NUV)^0 < 0.95$} \\
           & = & 3.37                           \qquad \textrm{$(FUV-NUV)^0 \geq 0.95$} \label{eq:Afuv}
\end{eqnarray}

where $A_{FUV}$ is the dust attenuation in units of magnitudes and the $(FUV - NUV)^0$ are in the rest frame colours. This equation does not assume the objects have classifications. This can then be applied to the GALEX data and the UV SFRs obtained using Equation \ref{eq:UV_SFR} from \citet{Salim2007}.

\begin{equation}
SFR_{FUV}(M_{\odot} yr^{-1}) =  1.08 \times 10^{-28} \times L_{FUV}^0 \times 10^{0.4A_{FUV}}
\label{eq:UV_SFR}
\end{equation}

where $L_{FUV}^0$ is the rest frame FUV luminosity in units of ergs$^{-1}$Hz$^{-1}$ and $A_{FUV}$ is the dust attenuation given by Equation \ref{eq:Afuv}.

\subsection{Star Formation Rate Comparison}
\begin{figure}[!h]
\centering
\includegraphics[scale=0.5,angle=-90]{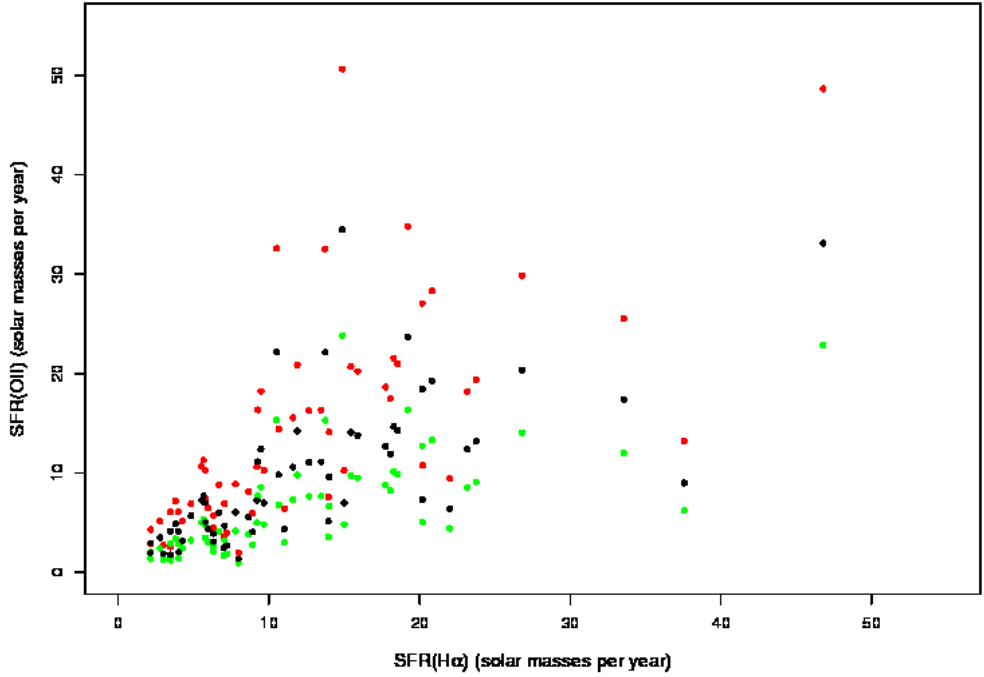}
\caption[Comparing range of SFR using O\sc{ii}]{\small{Comparing the SFR using [O\textsc{ii}]$\lambda3727$ with those found using H$\alpha\lambda6563$. The red points use Equation \ref{Kennicutt_O}. The green points use Equation \ref{Kewley} and the black points use Equation \ref{Kewley_abun}. }}
\label{fig:SFRs}
\end{figure}

The Kennicutt method using H$\alpha$ (Equation \ref{Kennicutt_H}) and the three methods for estimating SFR with [O\textsc{ii}] (Equations \ref{Kennicutt_O}- \ref{Kewley_abun}) were all used to calculate the SFR so a comparison could be made. Figure \ref{fig:SFRs} uses a sub-sample containing only objects with both [O\textsc{ii}] and H$\alpha$ in the spectra and shows the measurements for the SFR from the [O\textsc{ii}] line for the three methods (using Equations \ref{Kennicutt_O} - \ref{Kewley_abun}) against the measurements for the SFR from the H$\alpha$ line. The red points are the SFRs calculated using Equation \ref{Kennicutt_O} \citep[e.g.][]{Kennicutt1998}, the green points are using Equation \ref{Kewley}(\citealt{Kewley2004} without abundance) and the black points use Equation \ref{Kewley_abun} (\citealt{Kewley2004} with abundance corrections). As can be seen in Figure \ref{fig:SFRs}, the different methods give a range of possible star formation rates. 

\begin{figure}[!h]
\centering
\includegraphics[scale=0.5,angle=-90]{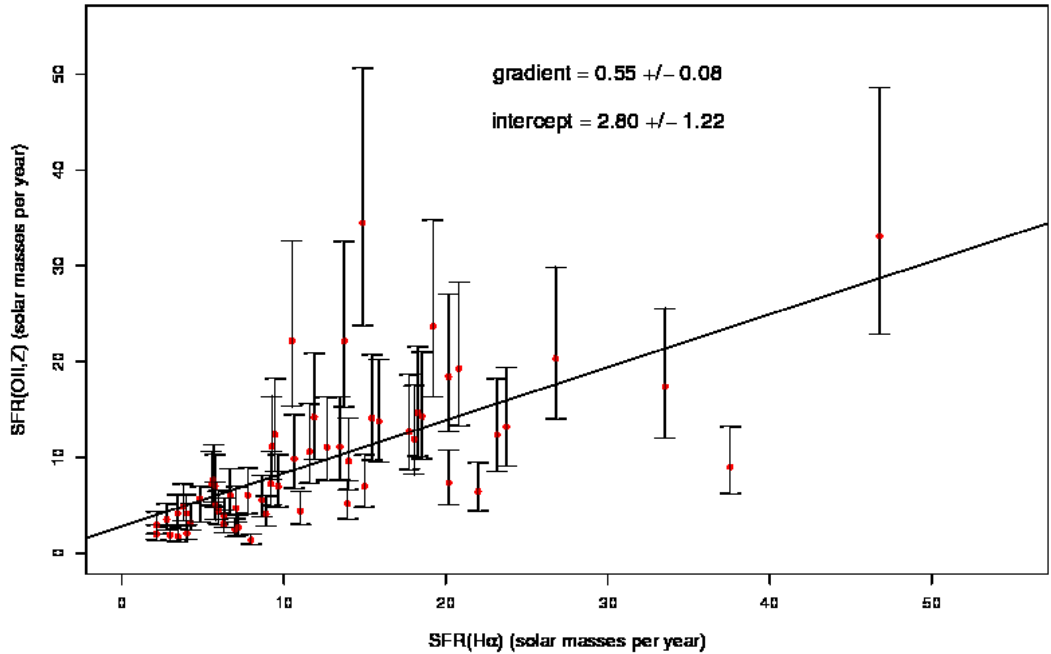}
\caption[SFR(O\sc{ii}) range and SFR(H$\alpha$) comparison]{\small{Comparing the SFR([O\textsc{ii}]) calculated using Equation \ref{Kewley_abun} with SFR(H$\alpha$).}}
\label{fig:SFR_comp2}
\end{figure}

For later comparisons, the [O\textsc{ii}] from the Equation \ref{Kewley_abun} (i.e, corrected for abundance) will be used and Equations \ref{Kennicutt_O} and \ref{Kewley} will be used to give the upper and lower limits for the SFR[O\textsc{ii}]. Figure \ref{fig:SFR_comp2} shows the comparison between the SFR using H$\alpha\lambda6563$ to the SFR values obtained using [O\textsc{ii}] from the Equation \ref{Kewley_abun}. The fitted line shows $[\mbox{O\textsc{ii}}] = m*\mbox{H}\alpha + c$, where $m$ is the gradient and $c$ is the intercept. The gradient is 0.55, which suggests that the SFR for the [O\textsc{ii}] is found to be twice that of the SFR found using the H$\alpha$. The intercept is not zero therefore there are some systematic errors. When comparing SFR at $z>0.5$ (i.e, the H$\alpha$ has been redshifted out of the wavelength range of the spectra), the SFR[O\textsc{ii}] calculated by Equation \ref{Kewley_abun} can be adjusted using the systematic error to give comparable values to the H$\alpha$ SFRs. The error bars on Figure \ref{fig:SFR_comp2} show the upper and lower limits for the SFR([O\textsc{ii}]) using the equations from Kennicutt (\citealt{Kennicutt1998}) and Kewley without abundance (\citealt{Kewley2004}), respectively. 

The star formation rates can be compared to known star formation rates of other objects. For example, the Milky Way has a current star formation rate of 4 M$_{\odot}$yr$^{-1}$ (\citealt{Dekel2009}) while dusty sub-millimetre galaxies (which have the most extreme formation rates) have SFRs of up to $\sim $1000 M$_{\odot}$yr$^{-1}$ (\citealt{Dekel2009}).\newline

To check that the difference between SFRs determined from [O\textsc{ii}] and H$\alpha$ is not a side effect of the calibration, the SFRs were split into redshift slices. The IMACS instrument is less sensitive at the blue end of the spectra, and could affect the measurements for the [O\textsc{ii}] lines, especially for low redshifts. Therefore, the galaxies with higher redshifts should show more consistency between values for H$\alpha$ and [O\textsc{ii}] star formation rates, as the [O\textsc{ii}] line will move out of the effected region at higher redshifts. 

\begin{figure}[!h]
\centering
\includegraphics[scale=0.5,angle=-90]{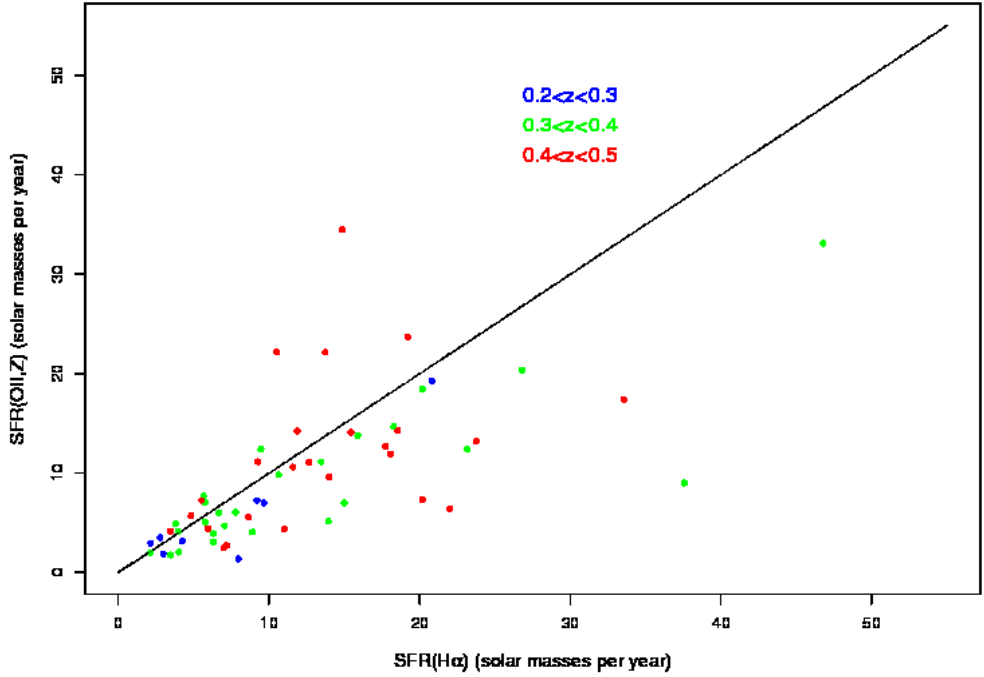}
\caption[SFR(O\sc{ii}) and SFR(H$\alpha$ comparison with redshift]{\small{Comparing the SFR from [O\textsc{ii}] from Equation \ref{Kewley_abun} with SFR from H$\alpha$ for a range of redshifts. The line is SFR(H$\alpha$)=SFR([O\textsc{ii}]).}}
\label{fig:SFRz}
\end{figure}

Figure \ref{fig:SFRz} shows the SFR[O\textsc{ii}] (using Equation \ref{Kewley_abun}) over different redshifts ranges. The plotted line is SFR(H$\alpha$)=SFR([O\textsc{ii}]). The spread in SFRs and the discrepancy between the H$\alpha$ and [O\textsc{ii}] SFR values can still be seen. However the discrepancies increase for the highest redshift range $0.4<z<0.5$. The H$\alpha$ line will be approaching the edge of the spectral wavelength range at this redshift, so the H$\alpha$ estimate is likely to be affected by a decrease in sensitivity at the edge of the wavelength range. This would also be an affect at the lowest redshift range. However, more galaxies in the redshift range would be needed to test this.

Another factor affecting the SFR and the differences in the methods is the range of masses for the galaxies. \citet{Gilbank2010} showed that the mass has a different effect on the SFR measured using the H$\alpha$ and [O\textsc{ii}]. Figure 3 in the Gilbank paper shows how, for a range of masses, the SFR rates from H$\alpha$ and [O\textsc{ii}] are more equal at higher mass galaxies (e.g.\ $log(M_{*})=10.2$) than at lower masses (e.g.\ $\log(M_*)=9.6$), indicating the sample of galaxies may contain a large number of low mass galaxies.

\begin{figure}[!h]
\centering
\includegraphics[scale=0.5,angle=-90]{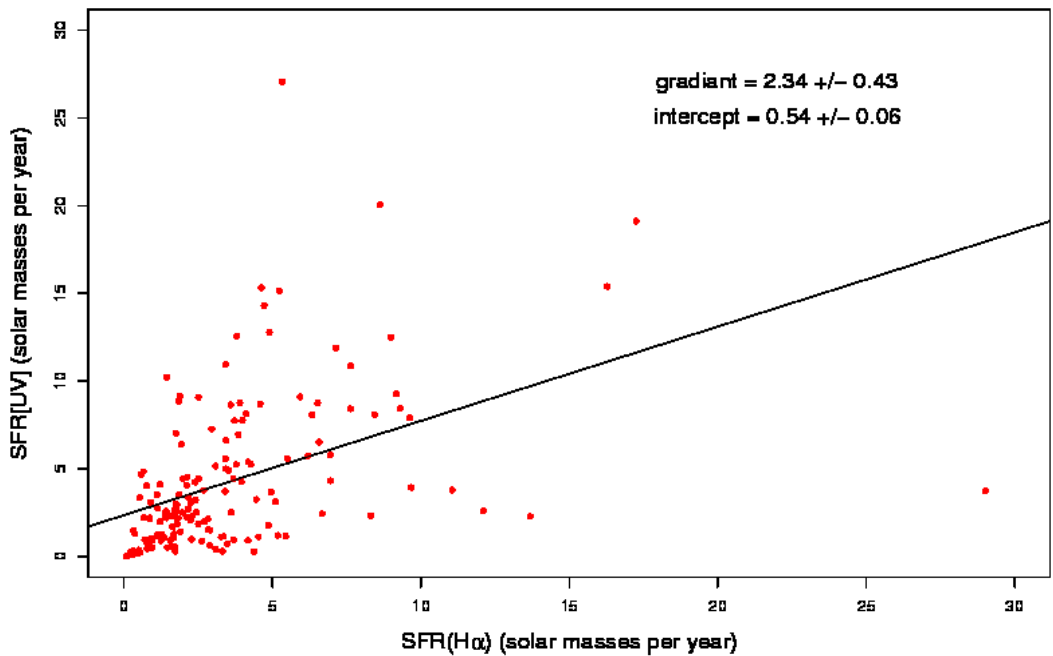}
\caption[SFR(UV) and SFR(H$\alpha$ comparison]{\small{Comparing the SFR from UV emission to the SFR from H$\alpha$ for a range of redshifts. The line is $SFR(UV) = m*SFR(\alpha) + c$, where $m$ is the gradient and $c$ is the intercept.}}
\label{fig:SFRUV}
\end{figure}

Figure \ref{fig:SFRUV} shows the comparison between the SFR found using the H$\alpha$ emission line and the SFR using the UV emission. The fitted line shows $SFR(UV) = m*SFR(\alpha) + c$, where $m$ is the gradient and $c$ is the intercept. Though the intercept is small, there is a steep gradient. This could be expected as the UV SFR measures an older stellar population on time scales of $\sim$100Myrs while H$\alpha$ measures the SFR on shorter time-scales of $\sim$10Myrs so measures emission from younger stars. A difference in H$\alpha$-UV SFR may be due to the difference in time-scales and stellar populations (e.g., \citealt{Sullivan2000,Sullivan2001}). Galaxies have been found with no evidence of H$\alpha$ emission and evidence of star formation in the UV (\citealt{Salim2007}) suggesting that these SFR indicators give information on different populations.  

Discrepancies may also be due to the model of the dust attenuation used (\citealt{Salim2007}), which will affect the luminosity and therefore, the resultant SFR. Ideally the dust attenuation should be calculated on a galaxy-by-galaxy basis as it is dependent on the individual galaxy. However, this is not always possible and was not possible here due to the lack of infra-red data. 

\section{Validation of the Large Scale Structure at $z\sim0.8$}\label{sect:LSS}
\subsection{Spectroscopic Redshifts}

\begin{figure}[!h]
\centering
\includegraphics[scale=0.9]{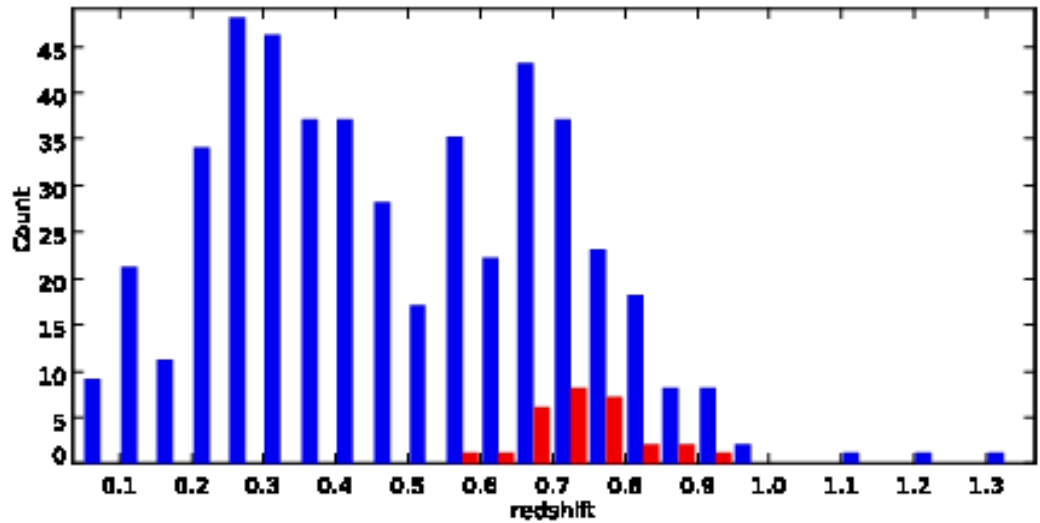}
\caption[Redshift histogram from IMACS]{\small{Histogram of the redshifts from the IMACS data. The blue bars show star forming galaxies selected by \citet{Haberzettl2009} and the red bars show red galaxies.}}
\label{fig:histz}
\end{figure}

Out of a possible 680 spectra, the redshifts for 515 galaxies were calculated using the function \textit{rvidlines} within IRAF. (The other objects were missing either the emission or absorption lines needed to allow a redshift to be determined.) The redshift distribution can be seen in Figure \ref{fig:histz}. The blue bars show the distribution of objects selected to be Lyman Break Galaxies (\citealt{Haberzettl2009}), while the red bars show the redshifts of the red objects. There are obvious peaks in the redshifts at 0.3, 0.4, $\sim$0.5 and 0.7, indicating potential large scale structures. It has been previously thought there may be a structure at $z\sim0.3$ but little work has has been done in this area. 

As can be seen from Figure \ref{fig:histz}, the number of galaxies decrease at higher redshifts. This could mean the sample is incomplete at and above a redshift of 0.7. This may also be due to the limiting magnitude of $i<21.3$ in the SDSS data, which cut out any fainter sources, leaving only the brightest high redshift objects. For this reason, the lower redshift peaks are more reliable and the peak at redshift 0.7 is more suspicious.

\subsection{SDSS Galaxy Photometric Accuracy}

The SDSS database was used to obtain both spectroscopic and photometric redshifts for the galaxies within the field. These galaxy redshifts can be used to estimate the redshifts of galaxy clusters within the area. Photometric redshifts are not as accurate as spectroscopic redshifts as they rely on the colour of the galaxy which in turn depends on the properties of the galaxy. The more filters or wavebands used in a photometric redshift estimate, the more accurate it is. 

To test the accuracy of the SDSS photometric redshifts, they were matched by position to the IMACS spectroscopic redshifts. To ensure the correct galaxies were selected, the difference in sky position given by SDSS and IMACS was purposely selected at $0.001^{\circ}$. The two redshifts were plotted against each other to show the correlation and the spread of redshifts calculated. 

\begin{figure}[!h]
\centering
\includegraphics[scale=0.55,angle=-90]{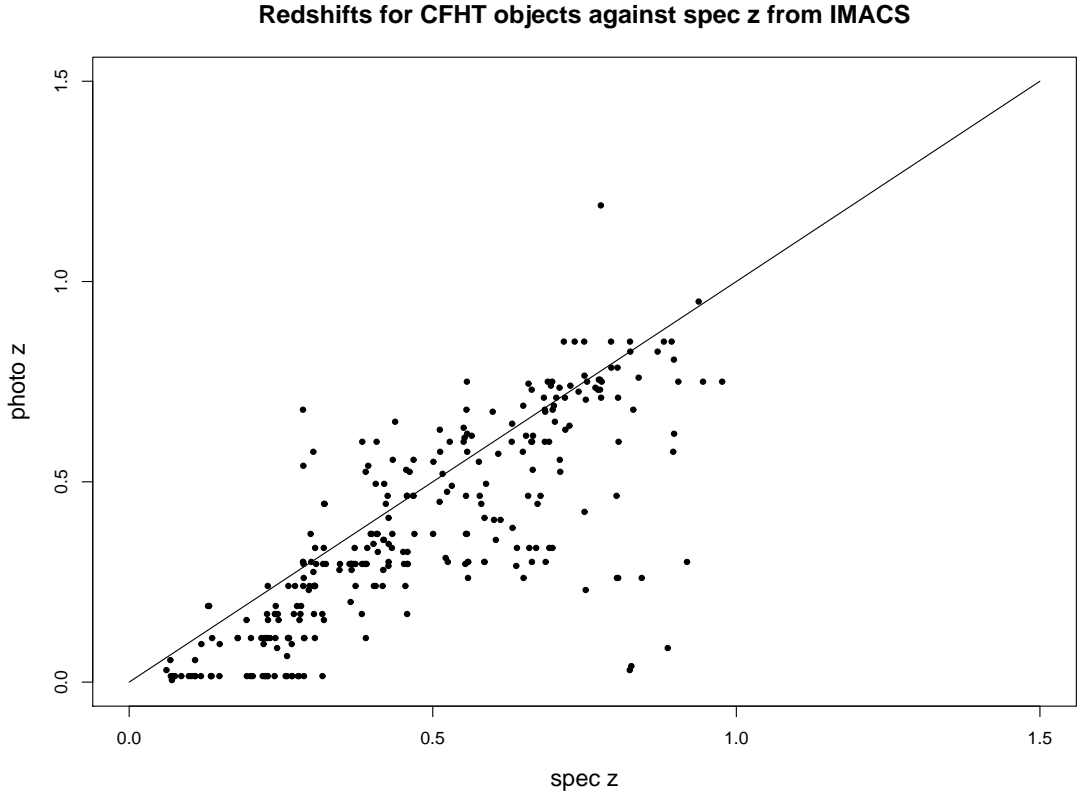}
\caption[Comparing spectroscopic and photometric redshifts using IMACS redshifts]{\small{Comparing the photometric redshifts of SDSS galaxies to their spectroscopic redshifts from the IMACS spectroscopic data. The line fitted is the $z_{spec}=z_{phot}$ line.}}
\label{fig:CHFTcomp}
\end{figure}

Figure \ref{fig:CHFTcomp} shows the photometric and spectroscopic redshifts of galaxies plotted against each other. The line of $z_{spec}=z_{phot}$ is shown in solid black and intercepts both axes at zero. Though there is scatter, the photometric redshifts best estimate the spectroscopic redshift between $0.4<z<0.8$. Below this redshift, the photometric redshifts are an underestimate. It is difficult to determine a trend above $z\sim0.8$ due to the lack of data points. The line of points in the bottom left corner is due to a lack of photometric redshifts. There appears to be systematic differences between the redshifts, with the photometric redshifts being lower than the spectroscopic counter-parts. 

\begin{figure}[!h]
\centering
\includegraphics[scale=0.55,angle=-90]{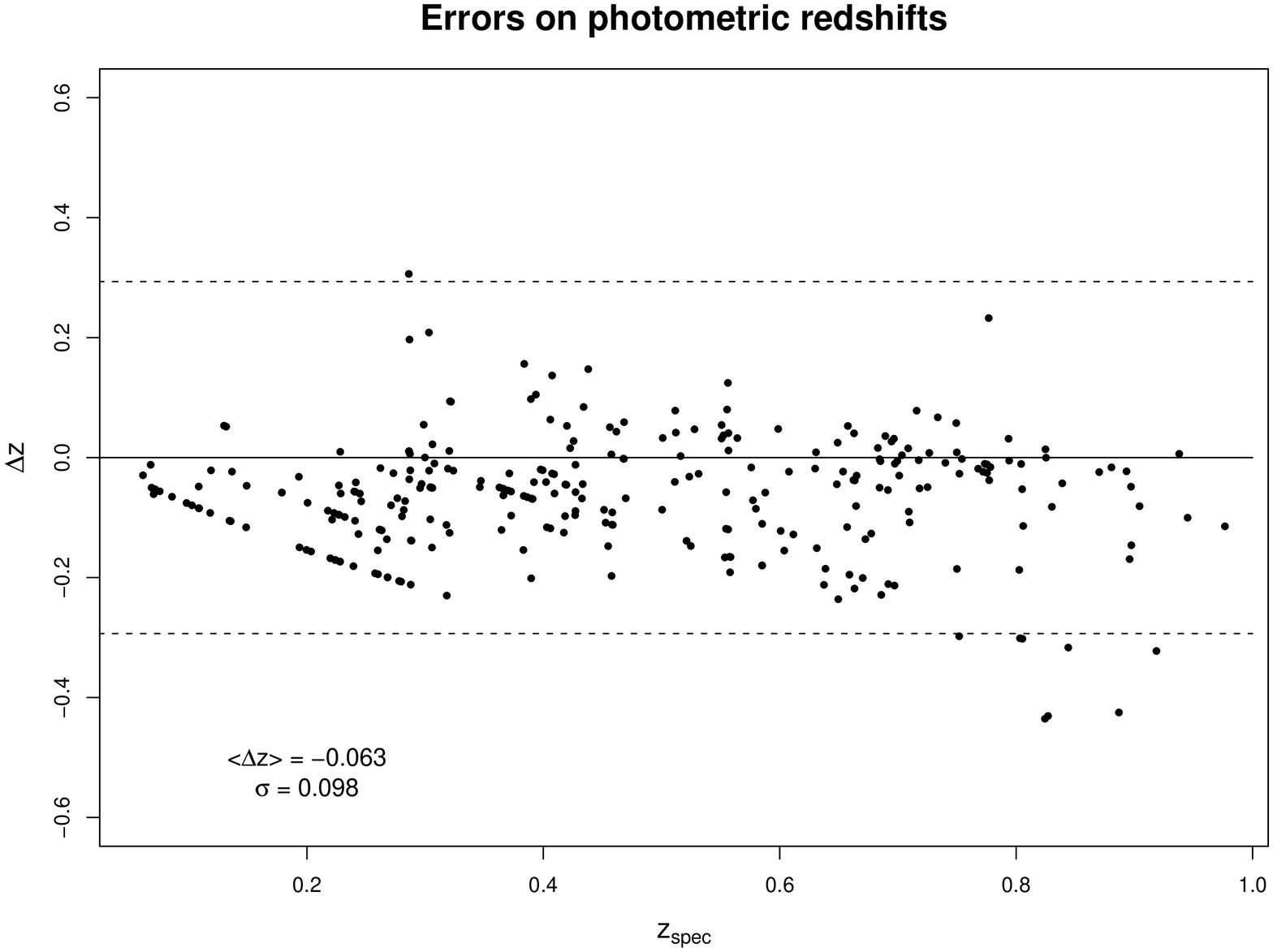}
\caption[Errors on photometric redshifts]{\small{The errors between the photometric redshifts and spectroscopic redshifts. The dashed lines show the $3\sigma$ boundaries.}}
\label{fig:CFHTerr}
\end{figure}

The error on the photometric redshifts, $\Delta z$, using Equation 2.1, is plotted against the spectroscopic redshift with lines showing the 3$\sigma$ levels (Figure \ref{fig:CFHTerr}). This shows the photometric redshifts increase above $\sim0.8$.  

\subsection{CFHT Cluster Red Sequence Fitting}\label{sect:redsequence}

Imaging data using $r$ and $z$ band filters on MegaCam on the CFHT were taken over two fields, one of which is partly covered by the IMACS spectra.

From this, clusters were selected by Ilona S\"ochting using the same method as was used to select the COSMOS clusters (\citealt{Sochting2011}), but instead of photometric redshifts, overlapping colour slices in the $R-z$ vs $z$ space were used to resolve clusters in the radial direction. 
Only 18 cluster members had photometric or spectroscopic redshifts. To estimate redshifts for all the clusters, the red sequence was used to calibrate the colour to the redshift of the cluster. The red sequence comes from the fact that cluster contain populations of galaxies which have a very tight relationship in colour-magnitude space. 
An example of this can be seen in Figure \ref{fig:redseqfit}

So to obtain redshift estimates for all of the clusters within the CFHT catalogue, the red sequence was fit to each cluster.

\subsubsection{Finding the Red Sequence}

The cluster red sequence method uses the colour of red elliptical galaxies to find the best linear fit to the galaxy distribution in a colour-magnitude diagram. This colour changes as the redshift increases, giving a natural redshift estimate. 
To obtain redshifts for these clusters, the colour ($R-z$) at a reference $z$ magnitude of $z=16$ is used as a proxy. The $R-z$ vs $z$ colour magnitude diagram was the best to use in this case as it best samples the intermediate redshifts (\citealt{Gladders2000}). However, since this linear fit has a slope and a zero point, a reference point is needed on the $z$ magnitude axis, which is read as a zero point. $z=16$ is a convenient point (S\"ochting, private communication). 
For each cluster, a line is fitted to the red sequence. This is done by eye and was double checked by an independent observer to agree whether this is the best fit of the sequence. Once the best fit line is found, the value of the intercept at $z = 16$ is found. 

\subsubsection{Redshift Accuracy}
Once the $R-z$ values at the zero point, z = 16 mag, have been found, they need to be compared to known redshifts (either spectroscopic or reliable photometric estimates) to give the calibration between $R-z$ and redshift. The IMACS data covers roughly a third of one field of the CFHT data and provides spectroscopic redshifts for clusters. A SDSS search was also performed, finding objects with both spectroscopic and photometric redshifts. 

The photometric redshifts are less reliable than the spectroscopic redshifts and for a single cluster, there was usually a large spread in redshift values. Therefore, it was not possible to select a reliable single value for a cluster with a large spread. However, some clusters did have multiple members within a small redshift range, suggesting reliable photometric redshifts. For a reliable photometric redshift, five or more members were needed, showing a small spread in redshift, the average giving the redshift, and the spread giving the error. Where possible, a spectroscopic and photometric redshift were obtained to show the accuracy of the photometric determination from the red sequence method (Table \ref{tab:photspec_matches}). The average difference between the photometric redshifts from SDSS and spectroscopic redshifts  from IMACS is 0.0985 with a spread of 0.110. However three of the points (Clusters 5, 12 and 32) have large errors on the photometric redshifts which is given by the spread of the individual member redshifts. If the three points with large spread are removed, the average difference is reduced to 0.0219 with a spread of 0.0190. For this, the errors on the  photometric redshifts used are $\sim$0.1.

\begin{table}[!ht]
\caption[Photometric and spectroscopic galaxies for cluster redshifts]{\small{The photometric and spectroscopic redshifts for 7 of the CFHT clusters and the differences between the two values. }}
\centering
\begin{tabular}{ c c c c c c }
Field & Cluster & Photo z & Phot z err & Spec z  & Difference \\ \hline
LQG1  &   5   	& 0.365   &  0.046    & 0.108   & 0.257      \\
LQG1  &   6  	& 0.3175  &  0.0184   & 0.264   & 0.0535     \\
LQG1  &   7     & 0.199   &  0.009    & 0.196   & 0.003      \\
LQG1  &   12  	& 0.588	  &  0.0308   & 0.305   & 0.283	     \\
LQG1  &   32	& 0.753   &  0.0823   & 0.691   & 0.062	     \\
LQG2  &   7     & 0.266   &  0.0023   & 0.283   & 0.017	     \\
LQG2  &   9     & 0.274   &  0.008    & 0.288   & 0.014      \\
\end{tabular}
\label{tab:photspec_matches}
\end{table}

Using both spectroscopic and photometric cluster redshifts, 18 clusters were used to plot the red sequence in order to derive the redshift for the other clusters.

\subsubsection{Calibrating the Colour and Redshifts}

\begin{figure}[h!]
\centering
\includegraphics[scale=0.45]{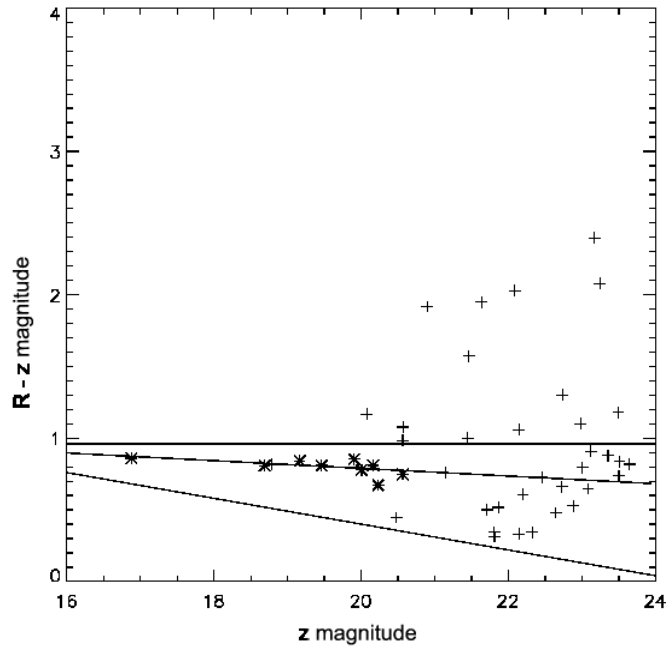}
\caption[Red sequence for a cluster]{\small{The red sequence fit to a galaxy cluster. The external lines show the colour cuts used to select the cluster and the central line shows the fit to the red sequence. The asterisks indicate the core cluster members and the crosses mark all galaxies residing with the cluster boundary.}}
\label{fig:redseqfit}
\end{figure}

Figure \ref{fig:redseqfit} shows the red sequence fit to a galaxy cluster. The external lines show the colour cuts used to select the cluster and the central line shows the fit to the red sequence. The asterisks indicate the core cluster members and the crosses mark all galaxies residing with the cluster boundary. Note that all galaxies which reside both within the cluster boundary and within the colour filter are considered final cluster members even if they haven't been included in the original core selection.

The value for the zero point of the red sequence line is plotted against the redshift. It is this correlation between the colour and redshift which will give the relationship used to estimate the redshifts for other clusters, which do not have a redshift, from their colours.

\begin{figure}[h!]
\centering
\includegraphics[scale=0.55]{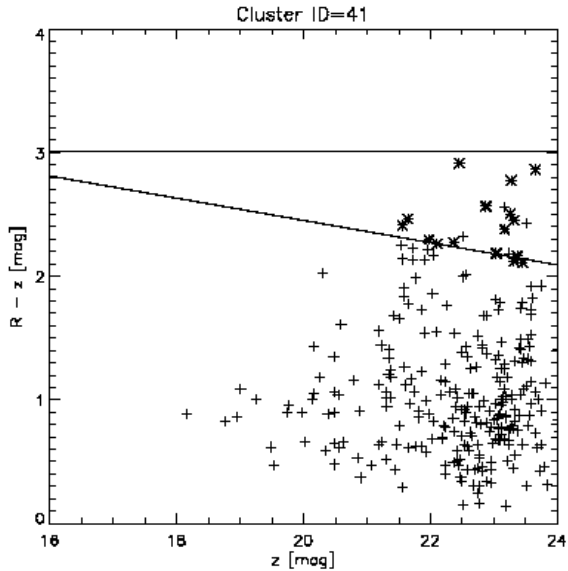}
\caption[Colour cuts for cluster selection]{\small{The colour cuts used to select cluster members. The crosses indicate possible cluster members from the first colour cut and the stars show the final members selected from the final colour cut.}}
\label{fig:blue}
\end{figure}

This method is less accurate at higher redshifts as the fraction of blue galaxies increases and the number of galaxies lying within the filters decreases. This is due to the evolution time needed to create the red sequence. This means the distribution of the galaxies on the colour cut plot becomes wider and the fitting is more prone to error. (See Figure \ref{fig:blue}.)

To find the redshift-colour relation, initially a straight line fit was used, but this proved to be an inadequate fit to the data. Five different models were used (linear, squared, and cubic, along with log and inverse) to fit to the clusters. (See Appendix \ref{append1} for the different models tested and the statistical tests used). To test the fitting of the different models, various statistical tests were performed as different tests have different strengths. These tests included the residual standard error which gives the standard deviation of the differences between the positions of the points and the predicted position from the fit. An F-test and an Akaike's Information Criterion (AIC) test were performed. The F-test uses the null hypothesis that the two samples (here the known redshift values and the predicted redshifts from a model) are not significantly different. If the calculated value for F from the test is greater than critical value for the F-test, the null hypothesis is rejected. The p-value  from the F-test gives the probability that if the null hypothesis is rejected, it is a mistake. So a small p-value means it is correct to reject the null hypothesis. The AIC test is a penalised log-likelihood and is a way of comparing models, rather than a hypothesis test. This test gives a measure of the goodness of the fit and ranks the models in order of their fit to the data, with the model with the lowest AIC value giving the best fit. 

\begin{figure}[!h]
\centering
\includegraphics[scale=0.45,angle=-90]{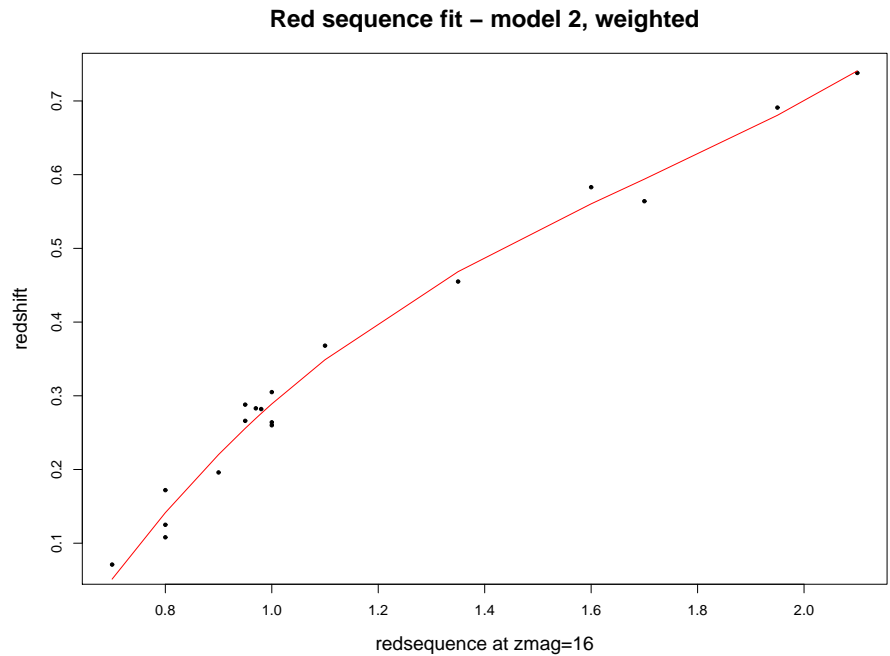}
\caption[Best fit model for the red sequence]{\small{The best model fit for the redshift to the $z$ magnitude.}}
\label{fig:redshiftfit}
\end{figure}

The best fit model (using a variety of tests) was a cubic model with the form $y=ax + bx^{2} + cx^{3} + d$. From the fit, the values are given as $a = 0.21 \pm 0.22$, $b = -1.07 \pm 0.41$, $c = 2.15 \pm 0.10$, and $d = -1.00 \pm 0.22$. This model was also weighted, giving a weight of 2 to the spectroscopic redshifts and 1 to the points with photometric redshifts. Figure \ref{fig:redshiftfit} shows the redshift plotted against the zero point of the red sequence. The points shown in Figure \ref{fig:redshiftfit} are a combination of spectroscopic points from SDSS and IMACS and some photometric points from SDSS. The line shows the model fitted to this data.

As the highest redshift value measured by either the IMACS data or the SDSS search was only 0.7, this fit should only be trusted up to a redshift level of around $0.74-0.75$.

Looking at the difference between the spectroscopic redshifts and the estimated redshifts from the red sequence, the average difference between the value is 0.035 with a spread of 0.028, which suggests that the red sequence will give a reliable fit. 

\section{Summary}

Spectra of a selection of 680 star forming galaxies, red galaxies, and AGN were taken using the IMACS instrument on the Magellan Baade telescope. From the 680 spectra taken, the redshifts of 515 galaxies were calculated. The objects in the spectra were classified and the star formation rates calculated, where possible.

Different methods of reddening correction have been tested. An average value for the extinction for [O\textsc{ii}] in spectra where the H$\alpha$ emission line is within the spectra wavelength was applied to spectra without a H$\alpha$ emission line to correct for the dust extinction.

The star formation rates were calculated using the H$\alpha$ and [O\textsc{ii}] emission lines, and the UV emission from GALEX, and compared. There is a systematic difference between the H$\alpha$ SFRs and the SFRs found using [O\textsc{ii}]. This may be due to the extinction correction. However [O\textsc{ii}] is more affected by metal abundances within the galaxy, which would also affect the SFR value found. 

There is a large difference between the H$\alpha$ SFRs and the SFRs using UV. This may be due to the dust attenuation model used, as an averaged value of the dust extinction was used due to the lack of infra-red data. However, there are also differences between the stellar populations and the time-scales measured by the H$\alpha$ and UV emission, which could also account for these differences.

\chapter{The Environments of AGN and Star Forming Galaxies}\label{sect:chap6}

This chapter will study the environments of 33 objects, which were classified as AGN using the IMACS spectra. The environments will be studied with respect to 649 nearby star forming galaxies found in the IMACS data and the WiggleZ Dark Energy Survey. The WiggleZ Dark Energy Survey had provided an additional 112 star forming galaxies to the sample of star forming galaxies from the IMACS spectra, though the actual SFRs are not available.

The IMACS and WiggleZ data sets are not exhaustive surveys of this area, so there may be other star forming galaxies and AGN not accounted for here. Conclusions will be drawn from the data available. 

Results have been compared against a control sample of AGN created from a set of AGN with randomised positions (the same method used for the quasar control sample to compare to galaxy clusters; see Section \ref{sect:simulations}).

\section{AGN}

There is only one quasar from the DR7QSO catalogue in the field of the IMACS spectra, with $z<1.0$. This quasar has a redshift of $z=0.56$ and lies near the bottom left edge of the field. There are two galaxies from the IMACS spectra nearby. One unfortunately does not have a spectroscopic redshift due to missing lines. The other galaxy has a redshift of 0.42 and has been classified as a star forming galaxy. There are also no quasars available in this area of the sky from the 2dF survey.

There are 33 AGN classified using the BPT plot (Section \ref{sect:classif}). Table \ref{tab:IMACS_AGN} shows the mask the object is found in, the positions, redshifts, SFR(H$\alpha$), and absolute $r$ magnitudes of the AGN from the IMACS spectra.

\begin{longtable}{ c c c c c c c}
\caption[AGN from IMACS]{\small{Galaxies classed as AGN from the IMACS spectra.}}\\
Name      & Mask & RA(J2000)    & DEC(J2000)  & redshift & SFR(H$\alpha$)& Absolute $r$       \\   
          &      &              &             &          & $M_{\odot}$yr$^{-1}$ & magnitude   \\ \hline               
  obj4865 & 1    &  10:46:12.06 & 00:21:54.67 &  0.424  & 2.172  &  -19.059       \\
  obj4388 & 1    &  10:46:20.81 & 00:21:41.51 &  0.267  & 1.817  &  -17.233      \\
  obj5091 & 1    &  10:46:24.54 & 00:22:00.95 &  0.343  & 1.136  &  -18.464        \\
  obj2138 & 1    &  10:46:55.81 & 00:20:46.58 &  0.272  & 5.304  &  -18.690   \\
  obj5572 & 3    &  10:47:04.04 & 00:22:13.31 &  0.468  & 7.444  &  -15.078        \\
  obj5222 & 1    &  10:47:05.69 & 00:22:04.34 &  0.287  & 2.265  &  -21.391        \\   
  obj3202 & 4    &  10:47:05.70 & 00:21:13.24 &  0.345  & 1.020  &  -19.946     \\ 
  obj2262 & 1    &  10:47:07.16 & 00:20:49.06 &  0.347  & 3.402  &  -18.740    \\
  obj2250 & 1    &  10:47:14.08 & 00:20:47.76 &  0.201  & 3.147  &  -17.221   \\
  obj4565 & 4    &  10:47:14.48 & 00:21:46.73 &  0.407  & 1.185  &  -19.696       \\
  obj7093 & 3    &  10:47:15.49 & 00:22:57.67 &  0.399  & 6.309  &  -18.532      \\
  obj4339 & 4    &  10:47:16.94 & 00:21:40.47 &  0.304  & 1.406  &  -17.799       \\
  obj3909 & 4    &  10:47:21.74 & 00:21:30.39 &  0.344  & 1.050  &  -18.053       \\
  obj4890 & 1    &  10:47:28.45 & 00:21:55.25 &  0.288  & 2.198  &  -17.922        \\
  obj3271 & 4    &  10:47:32.23 & 00:21:14.93 &  0.179  & 0.407  &  -16.792      \\ 
  obj4715 & 3    &  10:47:33.46 & 00:21:49.85 &  0.318  & 1.392  &  -17.990       \\
  obj4571 & 1    &  10:47:35.84 & 00:21:46.83 &  0.417  & 4.314  &  -19.772     \\ 
  obj6093 & 3    &  10:47:36.39 & 00:22:27.89 &  0.385  & 0.926  &  -19.041         \\
  obj5104 & 3    &  10:47:39.94 & 00:22:01.40 &  0.320  & 5.618  &  -17.845          \\
  obj4385 & 4    &  10:47:40.83 & 00:21:42.96 &  0.383  & 1.005  &  -22.633     \\
  obj4769 & 3    &  10:47:44.81 & 00:21:51.98 &  0.306  & 2.609  &  -18.179       \\
  obj4704 & 1    &  10:47:47.75 & 00:21:50.43 &  0.406  & 1.336  &  -18.509       \\ 
  obj5294 & 3    &  10:47:50.15 & 00:22:05.92 &  0.287  & 3.230  &  -19.606        \\
  obj4648 & 3    &  10:48:05.34 & 00:21:48.97 &  0.307  & 0.170  &  -19.149      \\
  obj6918 & 3    &  10:48:11.93 & 00:22:52.36 &  0.262  & 17.900 &  -24.341        \\
  obj3246 & 4    &  10:48:12.03 & 00:21:13.63 &  0.264  & 0.572  &  -15.048     \\
  obj5866 & 3    &  10:48:19.44 & 00:22:22.88 &  0.402  & 3.897  &  -19.050        \\ 
  obj4487 & 4    &  10:48:24.46 & 00:21:44.70 &  0.118  & 0.192  &  -18.643      \\
  obj5667 & 4    &  10:48:27.34 & 00:22:16.84 &  0.398  & 1.729  &  -21.067       \\
  obj5046 & 4    &  10:48:27.92 & 00:21:59.11 &  0.227  & 0.837  &  -17.134      \\
  obj7062 & 3    &  10:48:31.62 & 00:22:55.71 &  0.390  & 2.028  &  -22.266       \\ 
  obj2403 & 4    &  10:48:33.04 & 00:20:53.03 &  0.423  & 1.273  &  -18.871    \\
  obj4436 & 4    &  10:48:35.07 & 00:21:42.48 &  0.371  & 1.488  &  -16.847     
\label{tab:IMACS_AGN}
\end{longtable}

There are no AGN classified at redshifts $>0.5$, as the H$\alpha$ line is needed for the classification and is redshifted out of the wavelength range of the spectra at $z>0.5$. Therefore, the environments of AGN with $z<0.5$ will be studied. 

\section{Star Forming Galaxies}
Star formation in normal galaxies such as the Milky Way is in the range $\sim$0-20 M$_{\odot}$yr$^{-1}$, while starburst galaxies have SFRs up to $\sim$100 M$_{\odot}$yr$^{-1}$. Ultra-luminous infra-red galaxies (ULIRGs) have much higher rates, forming up to $\sim$1000 M$_{\odot}$ per year (\citealt{Grebel2011}). Therefore, in this Chapter, we will define :
\begin{itemize}
\item galaxies with a SFR of 0-20 M$_{\odot}$yr$^{-1}$ as being normal star forming galaxies,
\item galaxies with a SFR of 20-100 M$_{\odot}$yr$^{-1}$ as being high star forming galaxies (starbursts), and
\item galaxies with a SFR of $>$100 M$_{\odot}$yr$^{-1}$ as being ultra-high star forming galaxies (ULIRGs).
\end{itemize}

\subsection{WiggleZ Dark Energy Survey}

Spectroscopic redshifts for star forming galaxies are available from the WiggleZ Dark Energy Survey (\citealt{Glazebrook2007, Drinkwater2010}) using the AAOmega spectrograph on the 3.9-m Anglo-Australian Telescope (AAT). The AAOmega spectrograph uses optical fibres to take up to 400 spectra simultaneously over a 2 deg diameter field of view, covering a wavelength range of 3700-8750\AA\ (\citealt{Glazebrook2007}). The aim of the WiggleZ Dark Energy Survey is to study intermediate redshift ($0.2<z<1.0$) UV-selected star forming galaxies in order to determine large-scale structure and is used to test the predictions of the ‘cosmological constant’ model of dark energy (\citealt{Drinkwater2010}). The survey covers 1000 deg$^2$ along the equatorial plane over 221 nights. 

The survey contains the redshifts of $\sim$240,000 emission line galaxies with $\sim$90\% of the galaxies within the redshift range $0.2<z<1.0$. The median redshift is 0.6  and an error of $\sigma_{z} = 0.0042$ for the lowest quality spectra and $\sigma_{z} = 0.0022$ for the highest quality spectra (\citealt{Drinkwater2010}). Only spectra with a quality value of 3 or above have been used. Spectra with quality values of 1 or 2 have either no redshift or the redshift is uncertain. Spectra with a quality value of 3 (which indicates a reasonably confident redshift) have a redshift error of $\sigma_{z} = 0.0042$. Spectra with a quality of 4 (redshift with multiple emission lines in agreement) has an error of $\sigma_{z} = 0.0031$, and a quality of 5 (excellent redshift with high signal-to-noise) has an error of $\sigma_{z} = 0.0022$ (\citealt{Drinkwater2010}).

The galaxies from the WiggleZ survey have been used to augment the sample of galaxies within the area of sky covered by the IMACS data. These galaxies have redshifts but the SFRs have not been calculated and the spectra are not publicly available.

\section{Large Scale Structure}\label{sect:peaks}

\begin{figure}[!ht]
\centering
\includegraphics[scale=0.85]{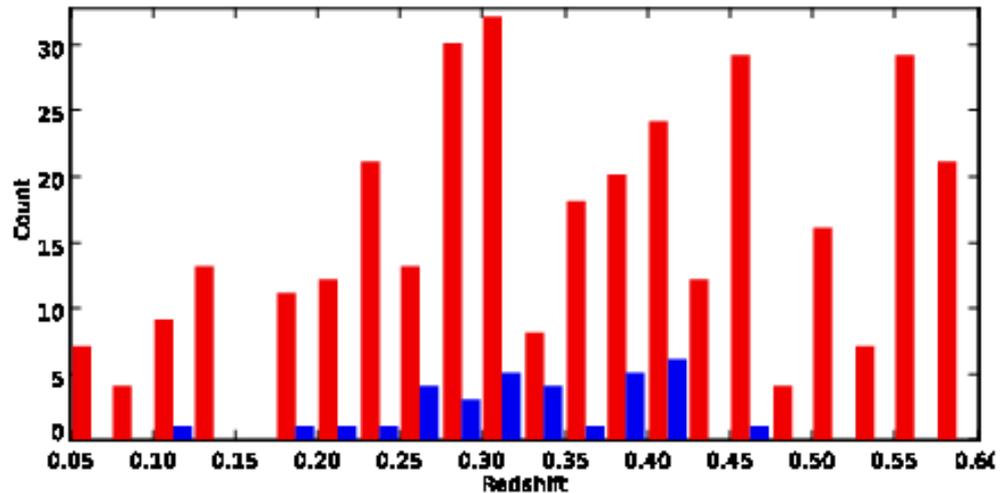}
\caption[Distributions of redshifts of AGN and star forming galaxies]{\small{The distribution of redshifts for star forming galaxies (red) and AGN (blue) within the area of sky covered by the IMACS spectra.}}
\label{fig:zdist}
\end{figure}

Figure \ref{fig:zdist} shows the redshift distribution for star forming galaxies (red), including galaxies taken from the WiggleZ survey, and AGN (blue) in the redshift range $0.05<z<0.6$. There is a peak in the number of star forming galaxies and an increase in the number of AGN at $z\sim 0.3$, suggesting large scale structure. There are also smaller peaks in the number of galaxies and AGN at $z\sim0.25$ and $z\sim0.4$. These peaks in the number of galaxies and AGN will be studied later in this Chapter.

\section{Quasar proximity to Star Forming Galaxies} 
In Section \ref{sect:separations}, the positions and redshifts of the AGN and galaxy clusters were used to calculate the separations between quasars and the centre of galaxy cluster. The same method has been used here to calculate the 2D projected separations between AGN and star forming galaxies, at the epoch of the AGN. The galaxies within a redshift range of the AGN, set by the redshift errors on the galaxies, are selected and the 2D projected distances between the AGN and the galaxy calculated. For star forming galaxies, the  errors on the redshift of the galaxies were used to select the redshift range. The errors on the redshifts for the WiggleZ galaxies are between 0.002 and 0.004, depending on the quality of the spectra. These errors are larger than the redshift errors estimated for the IMACS spectra, which are of the order $10^{-4}$. Therefore, a redshift range of $\pm$0.004 was used to take into account the redshift errors on the WiggleZ star forming galaxies.

The significance of any relation in the section is tested in Section \ref{sect:AGN_KStest} using the two dimensional KS test, which is described in Chapter \ref{2dks}.

Table \ref{tab:ANG_SFR_sep} shows the 2D projected separations between the AGN and the closest galaxy, the name of the closest galaxy, the SFR of the closest galaxy, and the number of galaxies with a 2D projected separation of $<$1 Mpc of the AGN.

\begin{longtable}{ c c c c c c}
\caption[2D separations for IMACS AGN]{\small{2D projected separations between AGN and the closest star forming galaxy.}}\\
Name     & Closest      & 2D projected         & SFR(H$\alpha$)           &  $<$1Mpc     \\ 
         & Galaxy       & separation (Mpc)     & (M$_{\odot}$yr$^{-1}$)                   \\ \hline
obj2138  & obj2422      & 0.518                & 15.001                               & 2       \\
obj2250  & obj2298      & 0.398                &                                      & 2     \\
obj2262  & obj2090      & 0.500                &                                      & 2      \\
obj2403  & obj2764      & 0.838                &  4.825                               & 3    \\
obj3202  & obj3054      & 0.277                &                                      & 8    \\
obj3246  & obj3060      & 0.547                &  0.534                               & 4   \\
obj3271  & obj2838      & 0.681                &  9.647                               & 3    \\
obj3909  & obj3875h     & 0.298                &                                      & 4    \\
obj4339  & obj4453      & 0.554                &  6.678                               & 2     \\
obj4385  & obj4159      & 0.840                &  3.967                               & 1    \\
obj4388  & obj4096      & 1.067                &  5.057                               & 0     \\
obj4436  & obj4687      & 0.579                & 20.173                               & 1     \\
obj4487  & obj5153      & 0.592                &  0.780                               & 8     \\ 
obj4565  & obj4914      & 0.772                & 13.999                               & 1     \\
obj4571  & obj4736      & 0.641                &                                      & 3     \\ 
obj4648  & obj4796      & 0.387                &  4.065                               & 2     \\
obj4704  & obj4847      & 0.817                & 12.219                               & 1     \\ 
obj4715  & obj4572      & 0.236                &                                      & 5     \\
obj4769  & obj4832      & 0.260                & 13.809                               & 6     \\ 
obj4865  & obj4913      & 0.119                &                                      & 2     \\
obj4890  & obj4960      & 0.253                &                                      & 5     \\
obj5046  & obj5108      & 0.150                &  1.420                               & 6     \\
obj5091  & obj5078      & 0.349                & 26.225                               & 2     \\
obj5104  & obj4960      & 0.644                &                                      & 2     \\
obj5222  & obj5295      & 0.184                & 16.439                               & 3     \\
obj5294  & obj5048      & 0.632                &  5.652                               & 2     \\
obj5572  & obj5685      & 0.497                & 23.744                               & 4     \\
obj5667  & obj6117      & 1.027                & 11.734                               & 0     \\
obj5866  & obj6117      & 0.532                & 11.734                               & 1     \\
obj6093  & obj5902      & 0.675                & 32.471                               & 1     \\
obj6918  & obj6985      & 0.096                &  8.831                               & 4     \\
obj7062  & obj7498      & 1.742                & 13.934                               & 0     \\
obj7093  & obj6589      & 1.341                &                                      & 0
\label{tab:ANG_SFR_sep}
\end{longtable}

\begin{figure}[!ht]
\centering
\includegraphics[scale=0.5,angle=-90]{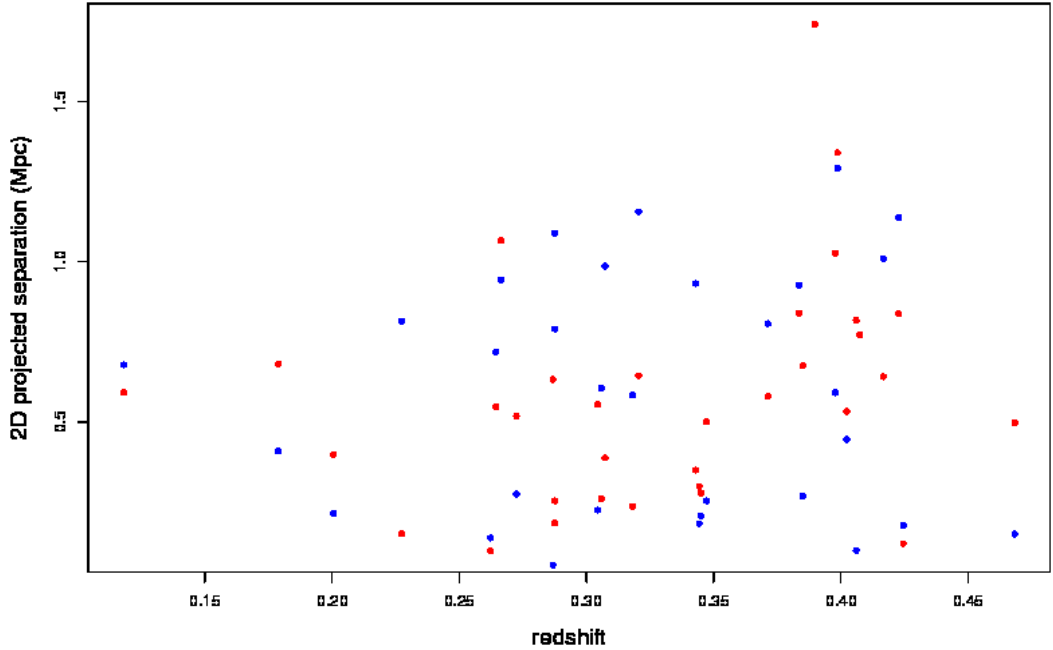}
\caption[2D separations for IMACS AGN]{\small{The 2D projected separations between an AGN and the closest star forming galaxy as a function of redshift for the observed data (red) and control data (blue).}}
\label{fig:z_2dsep}
\end{figure}

Figure \ref{fig:z_2dsep} shows the 2D projected separations (for the epoch of the AGN) between an AGN and the closest star forming galaxy as a function of redshift for the observed data (red) and control data (blue). A control data set of AGN has been created by randomising the RAs and DECs of the AGN. This method of randomising the positions of the AGN is the same method as used for the quasars in the previous chapters and is described in more detail in Section \ref{sect:simulations}. 
There appears to be no difference between the control and observed samples. 

The median 2D projected separation is 0.532 Mpc for the observed sample and 0.551 Mpc for the control sample. The significance is tested in Section \ref{sect:AGN_KStest}. Only three AGN lie near star forming galaxies which can be classed as starburst galaxies (i.e, star formation rates of 20-100 M$_{\odot}$yr$^{-1}$). These are obj4436, obj6093, and obj5091. Table \ref{tab:starbursts} shows the details for these three AGN and the closest starburst galaxy. The brightest of these AGN lies nearest to the starburst galaxy with the highest star formation rate. However, a larger sample is needed to determine whether this is a real effect.

\begin{table}[!ht]
\caption[AGN and starburst galaxy 2D separation information]{\small{AGN information lying close to starburst galaxies.}}
\centering
\begin{tabular}{c | c c c c}
AGN     & redshift & Absolute $r'$ & 2D projected & SFR(H$\alpha$)    \\
        &          & magnitude     & separation (Mpc) & M$_{\odot}$yr$^{-1}$  \\ \hline
obj4436 & 0.371    & -16.847       & 0.579        & 20.173            \\
obj5091 & 0.343    & -18.464       & 0.346        & 26.225             \\   
obj6093 & 0.285    & -19.041       & 0.675        & 32.471
\end{tabular}
\label{tab:starbursts}
\end{table}

\begin{figure}[!ht]
\centering
\includegraphics[scale=0.5,angle=-90]{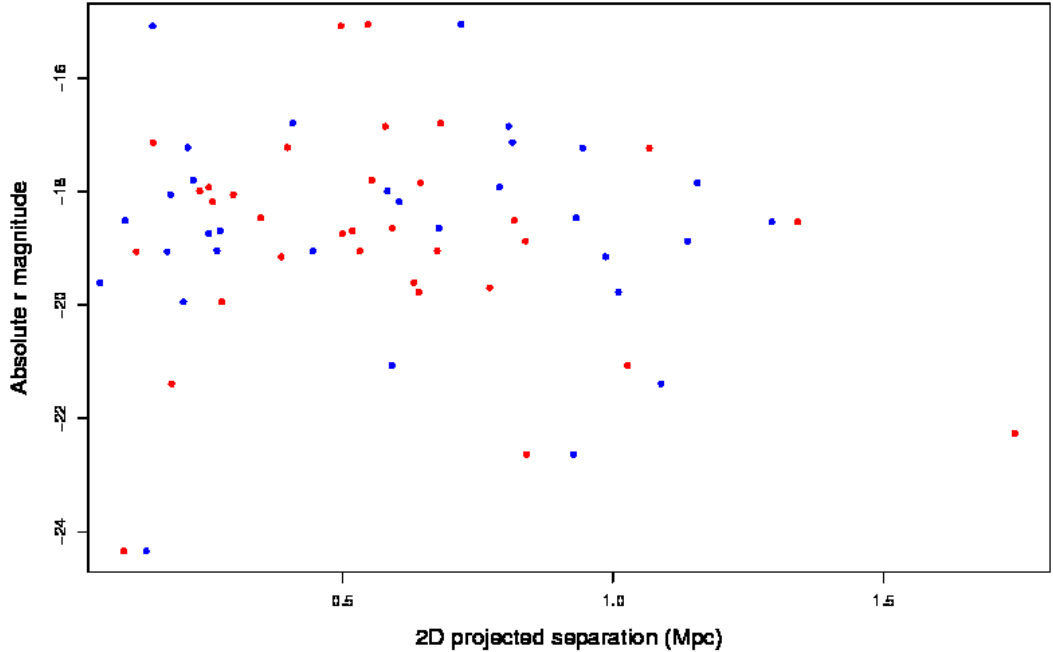}
\caption[2D separation as a function of magnitude for IMACS AGN]{\small{The 2D projected separations between an AGN and the closest star forming galaxy as a function of the absolute $r'$ magnitude of the AGN for the observed data (red) and control data (blue).}}
\label{fig:mag_2dsep}
\end{figure}

Figure \ref{fig:mag_2dsep} shows the 2D projected separations between AGN and the closest star forming galaxy as a function of the absolute $r'$ magnitude of the AGN for the observed data (red) and control data (blue). The brightest AGN classified from the IMACS spectra has the smallest 2D projected separation between the AGN and closest star forming galaxy. However, the next two brightest AGN lie at larger separations than most of the fainter AGN. All of these pairs are from the IMACS data.

\subsection{3D Separations}

Spectroscopic redshifts of the star forming galaxies within the field enable the calculation of the 3D separations between the AGN and the nearest star forming galaxy. The same method as that described in Section \ref{3Dsep} to find the 3D separations between the quasars and the closest galaxy cluster is used here. The errors on the redshifts of the star forming galaxies from the IMACS spectra were calculated in \textit{rvidlines}. The errors on the redshift of the WiggleZ galaxies are dependent on the quality of the spectra. Spectra with a quality value of 3 (which indicates a reasonably confident redshift) has a redshift error of $\sigma_{z} = 0.0042$. Spectra with a quality of 4 (redshift with multiple emission lines in agreement) has an error of $\sigma_{z} = 0.0031$, and a quality of 5 (excellent redshift with high signal-to-noise) has an error of $\sigma_{z} = 0.0022$ (\citealt{Drinkwater2010}).

\begin{figure}[!ht]
\centering
\includegraphics[scale=0.5,angle=-90]{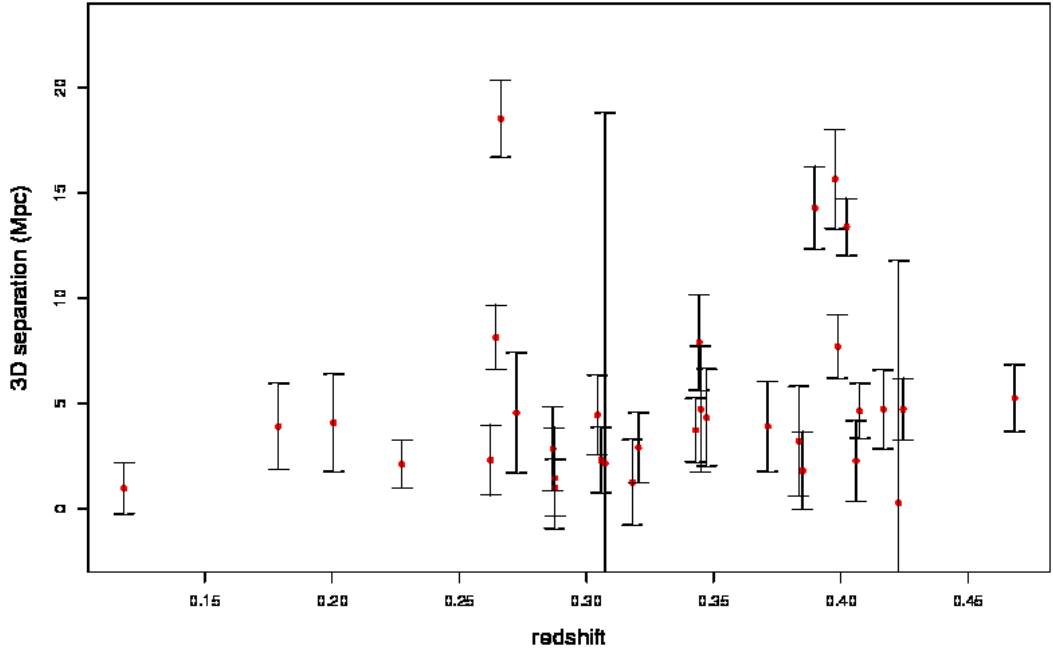}
\caption[3D separations for IMACS AGN]{\small{The 3D separations between an AGN and the closest star forming galaxy with errors as a function of redshift.}}
\label{fig:z_3dsep}
\end{figure}

Figure \ref{fig:z_3dsep} shows the 3D separations and the errors on the 3D separations as a function of redshift. Unlike with the galaxy clusters, the error on the galaxy redshifts are significantly lower, giving small errors on the 3D separations so the 3D separations can be used. 

The median 3D separation for the observed sample is 3.91 Mpc and for the control is 2.38 Mpc. The significance is tested in Section \ref{sect:AGN_KStest}.

\begin{figure}[!ht]
\centering
\includegraphics[scale=0.5,angle=-90]{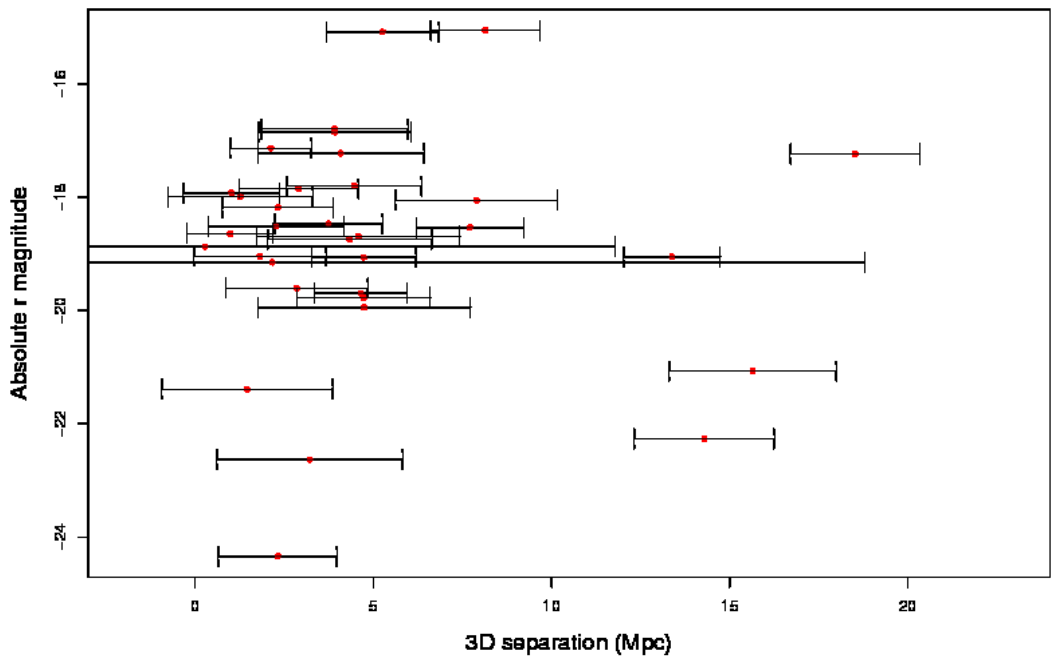}
\caption[3D separations as a function of magnitude for IMACS AGN]{\small{The 3D separations between an AGN and the closest star forming galaxy with errors as a function of the absolute $r'$ magnitude of the AGN.}}
\label{fig:mag_3dsep}
\end{figure}

Figure \ref{fig:mag_3dsep} shows the 3D separations between an AGN and the closest star forming galaxy with errors as a function of the absolute $r'$ magnitude of the AGN. Most of the AGN lie within a 3D separation of $<$10 Mpc, regardless of the absolute magnitude of the AGN. The brightest AGN in the IMACS sample has the smallest 3D separation between it and the closest star forming galaxy, with a star formation rate of 8.831 M$_{\odot}$yr$^{-1}$. However, the other bright AGN lie over a wide range of 3D separations. A few of the points on Figure \ref{fig:mag_3dsep} have large error bars. The closest star forming galaxy to these AGN is from the WiggleZ catalogue, which has larger errors on the redshifts than the errors on the redshift for the galaxies from the IMACS spectra. 

\clearpage
\subsection{AGN and SFR of Closest Galaxy}

\begin{figure}[!ht]
\centering
\includegraphics[scale=0.5,angle=-90]{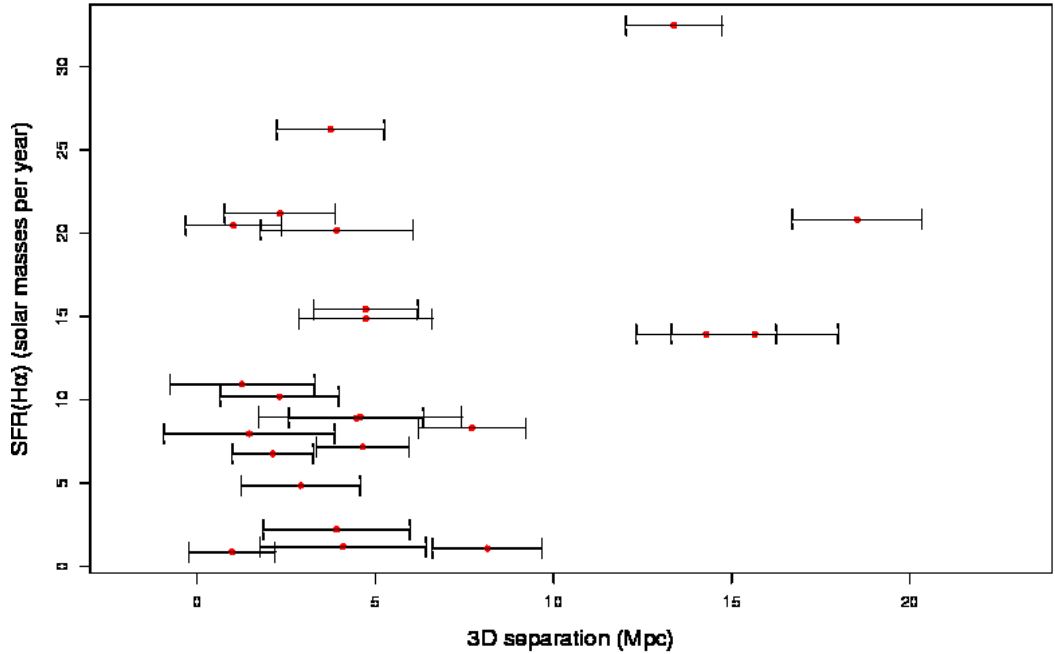}
\caption[3D separations as a function of closest galaxy SFR]{\small{The 3D separation between an AGN and the closest star forming galaxy with errors as a function of the SFR(H$\alpha$) of the closest galaxy.}}
\label{fig:sfr_3dsep}
\end{figure}

Figure \ref{fig:sfr_3dsep} shows the 3D separations between an AGN and the closest star forming galaxy (with errors) as a function of the SFR determined from the H$\alpha$ emission of the closest galaxy. There does not appear to be any relation between the distance between the AGN and the star forming galaxy, and the strength of the SFR of the galaxy.

\begin{figure}[!ht]
\centering
\includegraphics[scale=0.5,angle=-90]{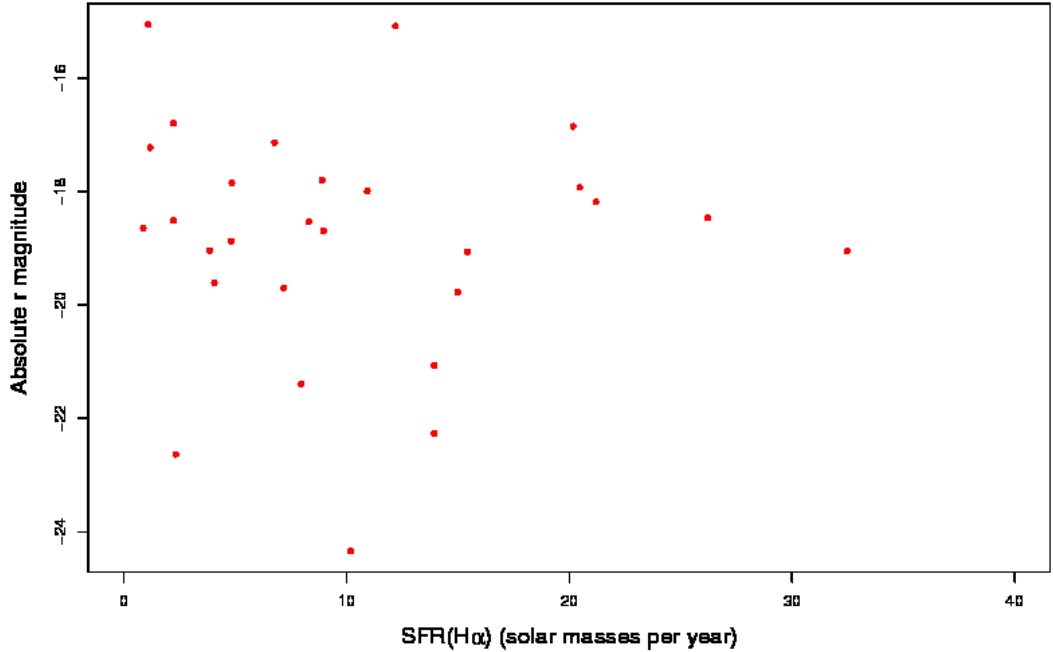}
\caption[Closest galaxy SFR as a function of quasar magnitude]{\small{The SFR(H$\alpha$) of the closest galaxy as a function of the absolute $r$ magnitude of the closest AGN.}}
\label{fig:sfr_mag}
\end{figure}

Figure \ref{fig:sfr_mag} shows the SFR calculated using the H$\alpha$ emission line of the closest galaxy as a function of the absolute $r'$ magnitude of the AGN. There appears to be no relation between the AGN magnitude and the SFR of the closest galaxy. 

The AGN which lie close to star burst galaxies (obj4436, obj6093, and obj5091) have 3D separation between the AGN and starburst galaxies of between 1.8 and 4 Mpc. Table \ref{tab:3Dstarburst} shows the 3D separations and the SFR of the starburst galaxies. 
\begin{table}[!ht]
\caption[AGN and starburst galaxy 3D separation information]{\small{AGN lying close to starburst galaxies. }}
\centering
\begin{tabular}{c c c}
AGN      & 3D separation (Mpc) & SFR(H$\alpha$) (M$_{\odot}$yr$^{-1}$) \\ \hline
obj4436  & 3.92 $\pm$ 1.12     & 20.17  \\
obj6093  & 3.74 $\pm$ 1.51     & 26.23   \\
obj5091  & 1.82 $\pm$ 1.84     & 32.47  
\end{tabular}
\label{tab:3Dstarburst}
\end{table}

\section{Clusters}

There are 14 clusters from the CFHT data (described in Section \ref{sect:redsequence}), 5 of which have redshifts of $z<0.5$, and 3 of which all lie at the same redshift, $z=0.289$. Table \ref{tab:data_cluster} shows the positions, number of members, cluster richness and orientation of the major axis (using the inertia tensor; Section \ref{sect:clustMorph}) for the three clusters at redshift $z=0.289$. The errors on the orientation angle (from the inertia tensor) were found using the method described in Section \ref{sect:angerr}. 

\begin{table}[!ht]
\caption[Details on the clusters at $z=0.289$]{\small{Details on the clusters lie at $z=0.289$ from CFHT data. }}
\centering
\begin{tabular}{c c c c c c }
Cluster  & RA (J2000)  & DEC (J2000)  & Members & Richness & Inertia angle  \\ \hline
LQG1\_10  & 10:48:41.29 & +05:15:51.62 & 15      & 6        & -65.34 $\pm$ 19.7   \\
LQG1\_12  & 10:46:44.12 & +05:38:38.82 & 18      & 9        & -45.34 $\pm$ 9.06   \\
LQG1\_13  & 10:47:36.12 & +05:44:23.75 & 14      & 6        & -62.99 $\pm$ 18.42  
\end{tabular}
\label{tab:data_cluster}
\end{table}

\begin{figure}[!ht]
\centering
\includegraphics[scale=0.5,angle=-90]{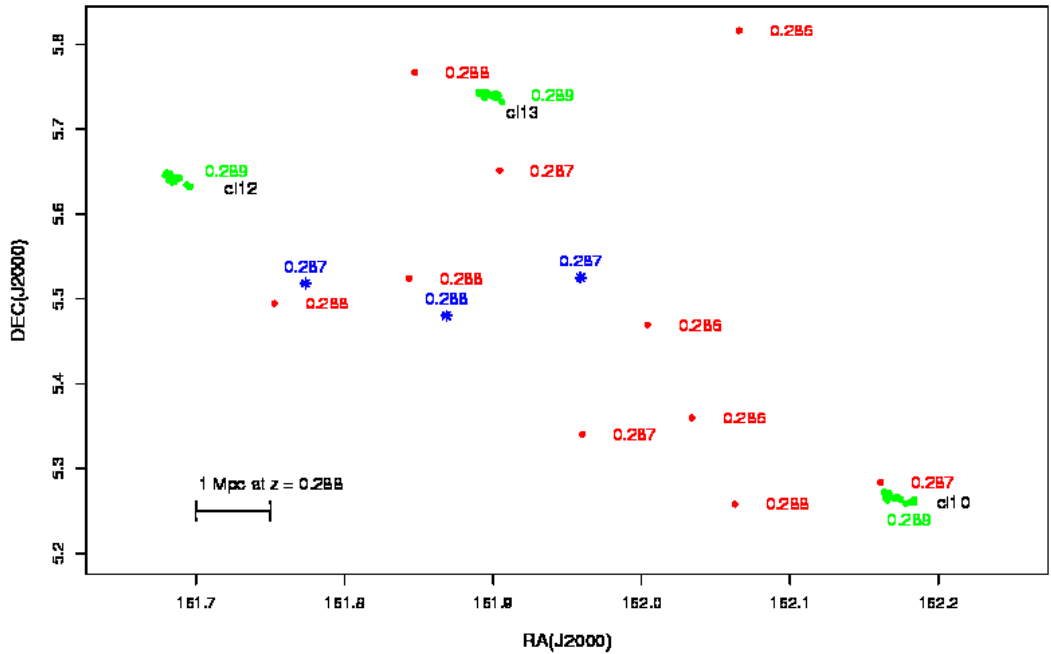}
\caption[Structure at $z\sim$0.288]{\small{Positions of AGN (blue asterisks), star forming galaxies (red circles) and galaxy clusters from CFHT data (green circles) at $z\sim0.288$.}}
\label{fig:imacs_structure}
\end{figure}

Figure \ref{fig:imacs_structure} shows three clusters with $z\sim0.288$, found in the CFHT data by Ilona S\"ochting. The redshifts were estimated using the red sequence method described in Section \ref{sect:redsequence}. The red circles show star forming galaxies from the IMACS spectra and the blue asterisks show the positions of the AGN at $z\sim0.28$. The AGN have absolute $r'$ magnitudes of -19.61, -17.92, and -21.39, and 7 of the star forming galaxies have SFRs available, with an average of SFR(H$\alpha$) = 7.04 M$_{\odot}$yr$^{-1}$. The AGN appear to lie along the major axes of the clusters and a line connecting the clusters, potentially within a filament, which would be defined by the clusters and by the star forming galaxies in this region at the same redshift as the clusters and AGN. 

\begin{figure}[!ht]
\centering
\includegraphics[scale=0.5,angle=-90]{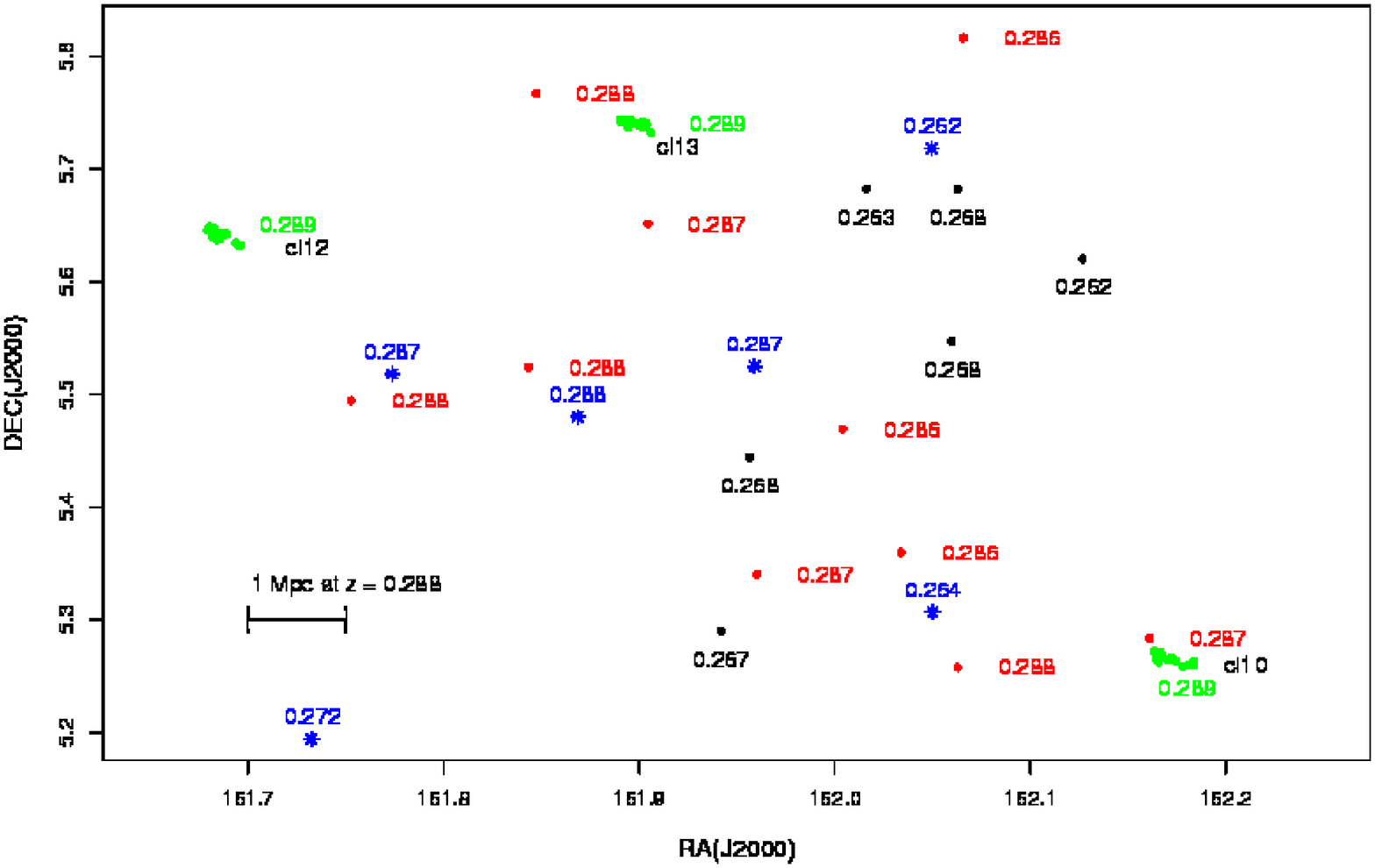}
\caption[Extended structure at $z\sim$0.288]{\small{Positions of AGN (blue asterisks), star forming galaxies (red and black circles) and galaxy clusters from CFHT data (green circles) at $z\sim0.288$.}}
\label{fig:imacs_all}
\end{figure}

Figure \ref{fig:imacs_all} includes the positions and redshifts of other AGN (blue asterisks) and star forming galaxies (black circles) at a slightly lower redshift but also in this region of the sky. This Figure includes the quasar with a redshift of $z=0.262$, which has been classed as a quasar due to an absolute $r'$ magnitude of -24.34 mag. This AGN lies near 2 star forming galaxies at the same redshift, but has a slightly lower redshift than the nearby clusters. 

Table \ref{tab:AGN_seps} shows the positions, redshifts and absolute $r'$ magnitudes of the six quasars lying between $0.26<z<0.29$, along with the 2D projected distances and the 3D distances to the three galaxy clusters with redshift $z=0.289$. This table shows small 2D and 3D separations between the top three AGN in the table and the clusters, suggesting they could be part of the same structure.
However, the redshifts of the clusters have been estimated using the red sequence. To determine how close the redshifts of the quasars and star forming galaxies are, the cluster redshifts need to be spectroscopically confirmed. 

\begin{table}
\caption[Details on objects within $z\sim0.288$ structure]{\small{AGN information for AGN in the redshift range $0.26<z<0.29$, at the epoch of the AGN.}}
\centering
\begin{tabular}{c | c c | c c c | c c c}
AGN     & redshift & Absolute $r'$ & \multicolumn{3}{c}{2D separation (Mpc)} & \multicolumn{3}{c}{3D separation (Mpc)} \\
        &          & magnitude     & cl10  & cl12  & cl13                        & cl10  & cl12  & cl13     \\ \hline
obj4890 & 0.288    & -17.922       & 7.50  & 4.96  & 5.29                        & 8.36  & 6.17  & 6.44     \\
obj5222 & 0.287    & -21.391       & 7.37  & 2.40  & 3.99                        & 12.06 & 8.04  & 9.03    \\   
obj5294 & 0.287    & -19.606       & 5.25  & 4.66  & 3.49                        & 10.05 & 9.55  & 8.68     \\ \hline
obj2138 & 0.272    & -18.690       & 6.67  & 6.79  & 8.59                        & 64.45 & 64.47 & 64.84   \\
obj3246 & 0.264    & -15.048       & 2.40  & 9.85  & 8.53                        & 94.17 & 94.62 & 94.56    \\
obj4388 & 0.267    & -17.233       & 11.33 & 4.50  & 8.35                        & 87.35 & 86.68 & 86.99    \\
obj6918 & 0.262    & -24.341       & 6.88  & 5.44  & 2.77                        & 102.13 & 101.98 & 101.77 
\end{tabular}
\label{tab:AGN_seps}
\end{table}

There is a cluster lying at $z\sim0.39$, with an AGN at $z=0.34$ at a 2D projected distance of 0.65 Mpc but a 3D distance of 154 Mpc. However, the errors on the cluster redshift are likely to be large as the redshifts have been estimated using the red sequence and spectroscopically confirmed. 
There is another cluster at $z\sim0.17$. However, this cluster lies near the edge of the IMACS field, and there are no quasars nearby.

\section{Contour plots}

\begin{figure}[!ht]
\centering
\includegraphics[scale=0.5,angle=-90]{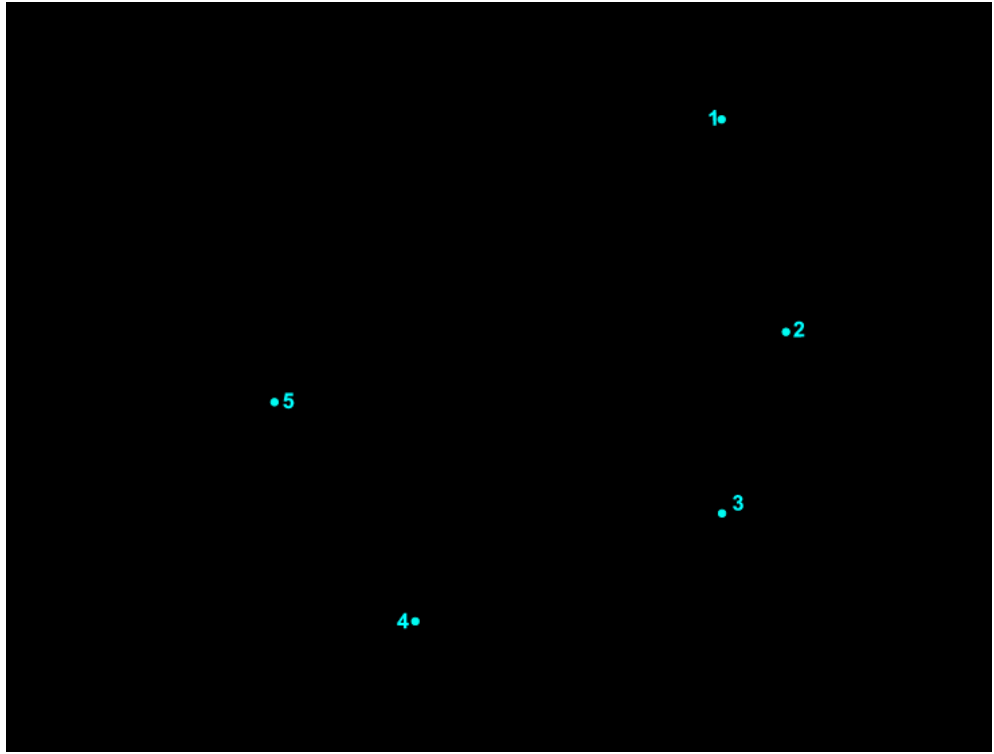}
\caption[Contour plot for $0.225<z<0.275$ galaxies]{\small{Contour plot showing the positions of the AGN with respect to the density of the local star forming galaxies for $0.225<z<0.275$. The contours start at the mean density and each contour corresponds to doubling the density above the mean density. There are 35 star forming galaxies used in the contours in this plot.}}
\label{fig:lowpeak}
\end{figure}

\begin{figure}[!ht]
\centering
\includegraphics[scale=0.5,angle=-90]{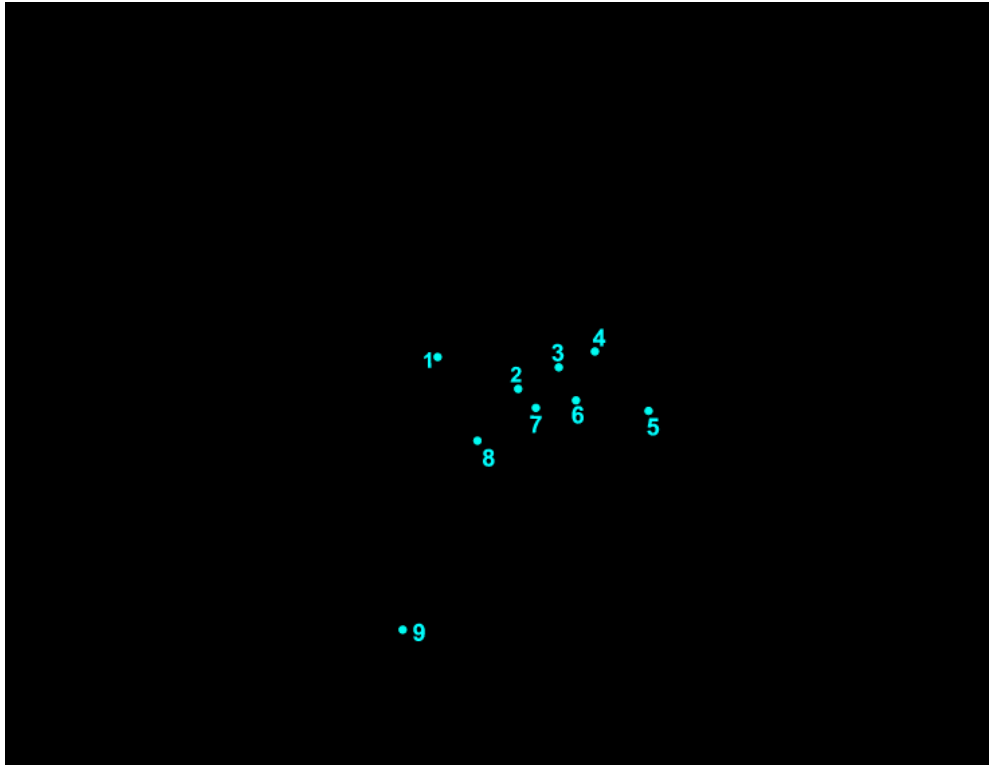}
\caption[Contour plot for $0.275<z<0.325$ galaxies]{\small{Contour plot showing the positions of the AGN with respect to the density of the local star forming galaxies for $0.275<z<0.325$. The contours start at the mean density and each contour corresponds to doubling the density above the local mean density. There are 63 star forming galaxies used in the contours in this plot. }}
\label{fig:midpeak}
\end{figure}

\begin{figure}[!h]
\centering
\includegraphics[scale=0.5,angle=-90]{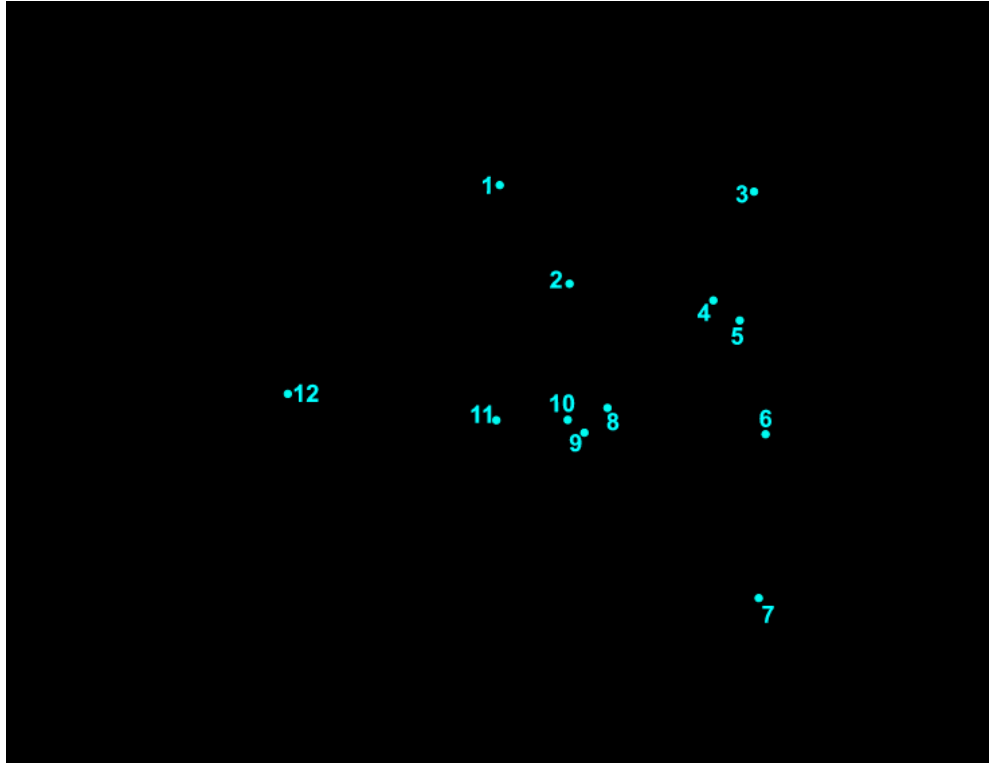}
\caption[Contour plot for $0.375<z<0.425$ galaxies]{\small{Contour plot showing the positions of the AGN with respect to the density of the local star forming galaxies for $0.375<z<0.425$. The contours start at the mean density and each contour corresponds to doubling the density above the local mean density. There are 46 star forming galaxies used in the contours in this plot. }}
\label{fig:highpeak}
\end{figure}

In Section \ref{sect:peaks}, three peaks in the number of galaxies and AGN in small redshift ranges were identified at $z\sim0.25$, $z\sim0.3$, and $z\sim0.4$. The contours on Figures \ref{fig:lowpeak} - \ref{fig:highpeak} are created using a Gaussian kernel sigma of 0.02 deg, and each contour line shows a doubling of the density above the local mean density. The contours use the star forming galaxies from both the IMACS data and the WiggleZ data. 

Each Figure shows an empty area in the top left corner. This void is due to the shape of the IMACS spectra which is ``L'' shaped and has no spectra is the region $RA < 161.74$ and $DEC > 5.58$.

Figure \ref{fig:lowpeak} shows a contour plot of the local density of star forming galaxies in the region of sky covered by the IMACS data at $z\sim0.25$. The star forming galaxies have been used to create the contours and the red points show the positions of the AGN. Table \ref{tab:lowpeak} shows the magnitudes, 2D projected separations and 3D separations of the AGN numbered on Figure \ref{fig:lowpeak}. AGN 5 (obj4388) appears to be isolated. However, this AGN is near the edge of the field so it is not possible to determine whether this AGN is indeed isolated. AGN 3 and 4 (obj3246 and obj2138) lie near galaxy contours but not on the contours. AGN number 1 and 2 on Figure \ref{fig:lowpeak} (obj6918 and obj5946) lie within a contoured area. AGN number 1 (obj6918), which lies near the centre of the contours, has one of the brightest magnitudes ($M_r = -24.34$) in the AGN sample from IMACS. 

\begin{table}[!h]
\caption[Details on AGN on $0.225<z<0.275$ contour plot]{\small{AGN information from Figure \ref{fig:lowpeak} at the epoch of the AGN.}}
\centering
\begin{tabular}{ c c c c c }
Number        & Name    & Absolute $r$  & 2D projected     & 3D separation \\
              &         & Magnitude   & separation (Mpc) & (Mpc)            \\ \hline
1             & obj6918 & -24.34      & 0.096            & 2.32 $\pm$ 1.21 \\
2             & obj5046 & -17.13      & 0.150            & 2.12 $\pm$ 0.8  \\
3             & obj3246 & -15.05      & 0.547            & 8.14 $\pm$ 1.08 \\
4             & obj2138 & -18.69      & 0.518            & 4.58 $\pm$ 1.50 \\
5             & obj4388 & -17.23      & 1.067            & 18.52 $\pm$ 1.29
\end{tabular}
\label{tab:lowpeak}
\end{table}

Figure \ref{fig:midpeak} shows a contour plot of the local density of star forming galaxies at $z\sim0.3$. The red points show the positions of the AGN. Table \ref{tab:midpeak} shows the magnitudes, 2D projected separations, and 3D separations of the quasars numbered on Figure \ref{fig:midpeak}. AGN number 9 (obj2138) lies on the edge of contour lines. The other AGN in Figure \ref{fig:midpeak} all lie around a central area of structure. AGN numbers 1 and 7 (obj5222 and obj4715) appear to lie at the centre of the contours, while AGN numbers 2 and 6 (obj4890 and obj4769) lie between the peaks, in regions of rapidly changing density. AGN numbers 3, 4, 5 and 8 (obj5104, obj5294, obj4648 and 4339 respectively) lie on the outskirts of the contours, suggesting they lie on the edges of the changes in density.  

\begin{table}[!h]
\caption[Details on AGN on $0.275<z<0.325$ contour plot]{\small{AGN information from Figure \ref{fig:midpeak} at the epoch of the AGN.}}
\centering
\begin{tabular}{ c c c c c}
Number        & Name     & Absolute $r$           & 2D projected    & 3D separation \\
              &          & Magnitude            & separation (Mpc) & (Mpc)  \\ \hline
1             & obj5222  & -21.39               & 0.184            & 1.46 $\pm$ 1.72\\
2             & obj4890  & -17.92               & 0.253            & 1.02 $\pm$ 0.95 \\
3             & obj5104  & -17.85               & 0.644            & 2.91 $\pm$ 1.20 \\
4             & obj5294  & -18.61               & 0.632            & 2.02 $\pm$ 1.28 \\
5             & obj4648  & -19.15               & 0.387            & 4.20 $\pm$ 1.53 \\            
6             & obj4769  & -18.18               & 0.260            & 2.32 $\pm$ 1.10 \\
7             & obj4715  & -17.99               & 0.236            & 1.27 $\pm$ 1.45 \\
8             & obj4339  & -17.80               & 0.554            & 4.46 $\pm$ 1.35 \\
9             & obj2138  & -18.69               & 0.518            & 4.57 $\pm$ 1.50
\end{tabular}
\label{tab:midpeak}
\end{table}

Figure \ref{fig:highpeak} shows a contour plot of the local density of star forming galaxies for $z\sim0.4$. Table \ref{tab:highpeak} shows the magnitudes, 2D projected separation and 3D separations of the quasars numbered on Figure \ref{fig:highpeak}. AGN numbers 1, 3, 6 and 12 (obj7093, obj7062, obj4436 and obj4865 respectively) appear to lie in isolated regions. However, these areas are on the edges of the field so it is not possible to determine if there is any other structure outside the field of view and whether they are actually isolated. AGN number 11 (obj4565) does appear to be isolated, lying between areas of higher density. 

\begin{table}[!h]
\caption[Details on AGN on $0.375<z<0.425$ contour plot]{\small{AGN information from Figure \ref{fig:highpeak} at the epoch of the AGN.}}
\centering
\begin{tabular}{ c c c c c }
Number        & Name    & Absolute $r$  & 2D projected    & 3D separation \\
              &         & Magnitude   & separation (Mpc) & (Mpc)  \\ \hline
1             & obj7093 & -18.53               & 1.431 & 7.71 $\pm$ 1.07 \\
2             & obj6093 & -19.04               & 0.675 & 1.81 $\pm$ 1.31 \\
3             & obj7062 & -22.27               & 1.742 & 14.29 $\pm$ 1.40\\
4             & obj5866 & -19.05               & 0.532 & 13.84 $\pm$ 0.96 \\
5             & obj5667 & -21.07               & 1.027 & 15.64 $\pm$ 1.66 \\            
6             & obj4436 & -16.85               & 0.579 & 3.92 $\pm$ 1.52 \\
7             & obj2403 & -18.87               & 0.838 & 15.69 $\pm$ 1.53 \\
8             & obj4704 & -18.51               & 0.817 & 2.27 $\pm$ 1.39 \\
9             & obj4385 & -22.63               & 0.840 & 3.22 $\pm$ 1.84 \\
10            & obj4571 & -19.77               & 0.641 & 4.73 $\pm$ 1.33 \\
11            & obj4565 & -19.70               & 0.772 & 4.65 $\pm$ 0.94 \\
12            & obj4865 & -19.06               & 0.199 & 4.73 $\pm$ 1.05
\end{tabular}
\label{tab:highpeak}
\end{table}  

\section{Statistics}\label{sect:AGN_KStest}
The two dimensional KS test has been used on the AGN-galaxy pairs in the IMACS field to test the significance of any relationships between separations, the magnitude, redshift and SFRs. Table \ref{tab:SFR_stats} shows the D and p-values from this test. The top half of the table uses the 2D projected separations between the galaxy and the AGN. The bottom half of the table uses the 3D separations. 

\begin{table}[!h]
\caption[2D KS test for AGN-galaxy pairs]{\small{D and p-values from two dimensional KS test for distributions of different characteristics of AGN-galaxy pairs for the IMACS field.}}
\centering
\begin{tabular}{c | c c }
Distribution                                          & D                    & p-value          \\ \hline
2D quasar-galaxy separation - redshift                & 0.182                & 0.720            \\
2D quasar-galaxy separation - Absolute $r$ magnitude  & 0.197                & 0.625            \\
2D quasar galaxy separation - SFR(H$\alpha$)(2D)      & 0.285                & 0.351            \\       
Absolute $r$ magnitude      - SFR(H$\alpha$)(2D)      & 0.283                & 0.362            \\ \hline
3D quasar-galaxy separation - redshift                & 0.320                & 0.311            \\
3D quasar-galaxy separation - Absolute $r$ magnitude  & 0.379                & 0.031            \\
3D quasar galaxy separation - SFR(H$\alpha$)(3D)      & 0.497                & 0.027           \\
Absolute $r$ magnitude      - SFR(H$\alpha$)(3D)      & 0.465                & 0.040            
\label{tab:SFR_stats}
\end{tabular}
\end{table}

Using a significance level of 0.01, the distributions for the observed and control samples are not significantly different for any of the distributions with the 2D projected separations or the 3D separations.
\clearpage
However, as mentioned previously, the catalogues of star forming galaxies used here are not complete samples. The star forming galaxies in the IMACS field were not selected to cover the entire field at the redshift range $0<z<0.5$, which covers the range of these AGN.

\section{Summary}

None of the relations for the 2D projected separations or the 3D separations between an AGN and the closest star forming galaxy are statistically significant. 

Three of the AGN lie at the same redshifts as three of the clusters from the CFHT data. These clusters were found by Ilona S\"ochting and the redshifts were estimated using the red sequence. Three AGN and 10 star forming galaxies lie at the same redshift, and are potentially part of the same structure. The spectroscopic redshifts of the clusters will support this result.   

Taken individually, the brightest AGN, which would be classed as a quasar, has the smallest 3D separation between the quasar and a star forming galaxy and lies in the middle of contours in Figure \ref{fig:lowpeak} so in a high density region. This quasar has $M_r = -24.34$ and is the only AGN from the IMACS sample bright enough to be classed as a quasar. This quasar can be seen in Figure \ref{fig:imacs_all} lying near star forming galaxies and potentially near a filament structure. 

The contour plots show AGN lying in various different areas. In the high redshift slice ($0.375<z<0.425$), all AGN appear to avoid the high density regions, preferring to lie on the edges of contours, therefore on the edges of the mass distributions. In the redshift slice for $0.275<z<0.325$, most AGN also lie on the edges on the contours. However, there are a few AGN which lie in the centres of the contours, i.e., in the middle of the mass distributions. There are less AGN in the lowest redshift slice ($0.225<z<0.275$) so it is difficult to determine the preferred environment. Two of the five AGN lie in the middle of the contours, while two lie on the edges. This change in preferred positions may indicate a change in preferred small-scale environment with redshift over the range $0.225<z<0.425$. However, the data used for the contours comes from the IMACS spectra, which is not a complete sample of star forming galaxies in this area. A more complete sample of the star forming galaxies in this region of sky would be needed properly studying this possible evolution with redshift.

\chapter{Ultra-Strong Fe{\sc{ii}} Emitters}

In this Chapter, we present 15 quasars (12 from Hectospec spectra and 3 from SDSS) which show evidence of strong to ultra-strong UV Fe\textsc{ii} emission. The quasars are all within an area of two deg$^{2}$ covering a portion of the CCLQG ($\overline{z}=1.28$) and two other LQGs (at $\overline{z}=1.11$ and $\overline{z}=$1.54). This area has a very high density of quasars in a small area, which is the only other known feature in this area at this redshift. The quasars span a redshift range of $1.11<z<1.67$, four of which are confirmed members of LQGs (qso48, qso425, qso26 and qso22).

The data used in this section was taken by Lutz Haberzettl and was initially intended to be used to provide more quasars to be used in the studies described in Chapters 3 and 4. Unfortunately, this is not work. However, further study showed a high number of quasars with an excess of ultra-strong Fe\textsc{ii} emission and it was decided to investigate this further.  

The cosmology used is $H_0=70$kms$^{-1}$Mpc$^{-1}$, $\Omega_m=0.27$ and $\Omega_{\Lambda}=0.73$.

\section{Fe{\sc{ii}} Emission}

\begin{figure}[!ht]
\centering
\includegraphics[scale= 0.6]{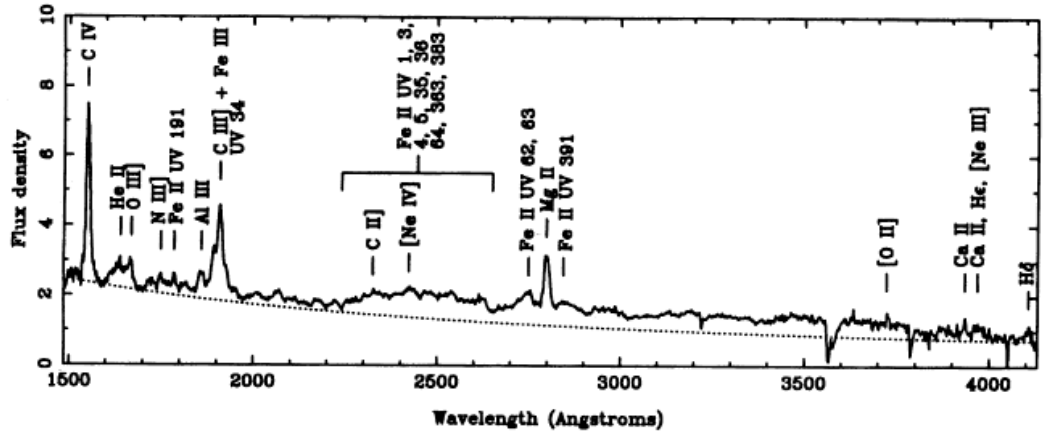}
\caption[Example of Ultra-strong UV Fe{\sc{ii}} emission ]{\small{Example of ultra-strong Fe{\sc{ii}} emission in quasar 2226-3905 from \citet{Graham1996}.}}
\label{fig:grahamspec}
\end{figure}

Iron emission can be seen in Active Galactic Nuclei (AGN) and quasars in the optical and ultra-violet at varying levels. Few quasars have been found to be strong or ultra-strong UV Fe\textsc{ii} emitters, suggesting this strength of emission is rare. The most notable ultra-strong UV Fe\textsc{ii} emitters are:
\begin{itemize}
\item IZw1, a Seyfert galaxy \citep{Bruhweiler2008,Vestergaard2001},
\item 2226-3905 \citep{Graham1996},
\item 0335-336 from \citet{Weymann1991}, and
\item Mrk 376 and Mrk 486 (Seyfert galaxies). 
\end{itemize}
All of these show ultra-strong Fe\textsc{ii} emission in the rest-frame region between 2255{\AA} and 2650{\AA}. An example of this can be seen in Figure \ref{fig:grahamspec}, taken from \citep{Graham1996}. 

The presence of ultra-strong Fe\textsc{ii} has been shown to be a general characteristic of low-ionization Broad Absorption Line (BAL) quasars \citep{Weymann1991,Duc2002}. BAL quasars make up $\sim$10\% of the quasar population. Iron low-ionisation BAL (FeLoBAL) quasars are even more rare, making up 1.5-2.1\% of the entire quasar population (\citealt{Dai2010}). Although both 2226-3905 and IZw1 have no BAL features, they do show evidence of emission features seen in low-ionisation BAL quasars \citep{Graham1996}. 

Several mechanisms have been suggested for producing UV Fe\textsc{ii} emission in AGN and quasars; the relative importance of each is still unknown. 
It is assumed that the Fe\textsc{ii} is produced within the Broad Line Region (BLR) and by the environment within this region \citep{Shields2010}. However, the emission strength can not be explained by standard photoionization cloud models (e.g. \citealt{Collin2000}). The most commonly accepted physics of Fe\textsc{ii} emission come from \citet{Elitzur1985}. The observed UV Fe\textsc{ii} emission is most likely caused by the interplay of different mechanisms \citep{Elitzur1985, Sigut2004, Osterbrock2006}. Collisional excitation can excite the Fe\textsc{ii} to a few eV above ground level, whereas resonance fluorescence (from both the continuum and Ly$\alpha$) can excite Fe\textsc{ii} to 5-10 eV above ground level. 

Simulations of Fe\textsc{ii} emitting regions have suggested that the Fe\textsc{ii} abundance alone may not be the main factor influencing the strength of the UV Fe\textsc{ii} emission seen, though still important \citep{Sigut2003}. There are several non-abundance factors such as the gas density in the BLR and the strength of the radiation field \citep{Sameshima2009}. Large Fe abundances and high densities alone can not reproduce the high level of Fe emission seen (\citealt{Baldwin2004}).

Ly$\alpha$ excitation (or fluorescence) is thought to be fundamental in enhancing UV Fe\textsc{ii} emission \citep{Sigut2003}, accounting for most of the identified Fe\textsc{ii} emission between 2000{\AA} and 3000{\AA}. Sigut and Pradhan (1998) found Ly$\alpha$ fluorescent excitation can more than double the UV flux.
The Fe\textsc{ii} flux strength can be strongly modified by increasing the microturbulence within the BLR in the AGN \citep{Vestergaard2001,Sigut2003, Sigut2004, Verner2003, Verner2004, Bruhweiler2008}. \citet{Bruhweiler2008} explain this by considering the spread of the line absorption coefficient over a larger wavelength, which in turn increases the flux absorption by the Fe\textsc{ii}. This increases the effectiveness of radiative pumping, producing enhanced Fe\textsc{ii} emission. They also found an increase in the photoionizing flux increased the predicted Fe\textsc{ii} flux (as was expected), and while increasing the hydrogen density did have an effect, it was smaller than the influence from the other two factors. A number of mechanisms may be involved in increasing the Fe\textsc{ii} emission. For example, an increase in the microturbulence increases the fluorescence in the BLR \citep{Wang2008}, which both increase the strength of the Fe\textsc{ii} emission. Although, it has been proposed that the weak emission between 2800 and 2900{\AA} indicates that collisional excitation is more likely as high gas temperatures would give rise to strong Fe\textsc{ii} lines \citep{Wang2008}.

Models have been compared to the spectrum of IZw1 and can predict the overall spectrum but there are discrepancies with the observed spectra at other wavelengths such as 1800-2000{\AA}, and especially for Fe\textsc{iii} \citep{Sigut2003,Sigut2004}.

Large Quasar Groups (LQGs) are some of the largest structures seen in the Universe and can span 50-200 $h^{-1}$ Mpc. These clusters of quasars exist at high redshifts, presumably trace the mass distribution, and are potentially the precursors of the large structures seen at the present epoch, such as super-clusters \citep{Komberg1996}. There are $\sim$40 published examples of LQGs. The observations for this Chapter were taken in the direction of the Clowes-Campusano LQG (CCLQG) \citep[][1994]{Clowes1991} which lies at a redshift of $z\sim1.3$, spans $\sim$100-200$h^{-1}$ Mpc. 

Three different LQGs have been found in this area. The CCLQG lies at $\overline{z}=1.28$, contains 34 members, and is statistically significant. There is another layer of quasars at $\overline{z}=1.54$, which is not statistically significantly (with 21 members), and there is a new layer at $\overline{z}=1.11$, which has recently been found and contains 38 members (\citealt{Clowes2011}; private communication). 
The latest discussion of these LQGs can be found in \citet{Clowes2011}.

\section{Data}
The spectra were taken with the Hectospec instrument, a multiobject optical spectrograph, fed by 300 optical fibres, mounted at the 6.5m Multiple Mirror Telescope (MMT) on Mount Hopkins, Arizona. The aim of this data was to study the impact of the LQG on galaxies, using blue Lyman-Break galaxies (LBGs) and red-selected galaxies at redshifts $z\sim0.8$ and $z\sim1.3$. These galaxies will help reduce the redshift uncertainties on the structures, and provide insight into the star formation activity of the brightest galaxies and their relation to quasars. 

The Hectospec observations, taken over nine nights, include 30 quasars taken from the Sloan Digital Sky Survey (SDSS) photometric catalogue created by \citet{Richards2009}, and objects selected from a set of previous observations on the Anglo-Australian Telescope (AAT) where the data had insufficient signal-to-noise. The quasars were selected to have magnitudes brighter than $r\sim$20.1. Objects with photometric redshifts between 0.6 and 1.8 were selected, giving priority to quasars which are likely to be within the CCLQG. Further information on the objects (such as magnitudes) was taken from the SDSS database. Table \ref{tab:feIIobslog} shows the dates, fields, and exposures times for the Hectospec observations.

\begin{table}[!ht] 
\caption[Hectospec observing log]{\small{Observing log for the Hectospec data.}}
\centering
\begin{tabular}{ c c c c  }
Date        & RA (J2000) & DEC (J2000) & Exposure (s)      \\ \hline
17.02.2010  & 10:50:16.9 & +04:37:12 & 5400        \\
18.02.2010  & 10.50:16.9 & +04:37:12 & 5400  \\
19.02.2010  & 10:50:06.9 & +04:29:16 & 5094  \\
06.04.2010  & 10:50:06.9 & +04:29:16 & 5400  \\
07.04.2010  & 10:48:31.8 & +05:23:29 & 7200  \\
09.04.2010  & 10:48:31.8 & +05:23:29 & 5400  \\
10.04.2010  & 10:48:38.9 & +05:25:57 & 5400  \\
11.04.2010  & 10:48:38.9 & +05:25:57 & 5400  \\
11.04.2010  & 10:49:57.0 & +04:30:01 & 5400  \\
12.04.2010  & 10:49:57.0 & +04:30:01 & 1800
\end{tabular}
\label{tab:feIIobslog}
\end{table}

The spectra from Hectospec cover 3900{\AA} to 9100{\AA} and have a dispersion of 1.2{\AA} per pixel.
These spectra were reduced using the Hectospec pipeline reduction by Sophia Mitchell, a student at the University of Cincinnati.
The redshifts and errors were found using \textit{rvidlines} within IRAF and can be seen in Table \ref{tab:Fedata}. The columns in Table \ref{tab:Fedata} show the names, RA, DEC, and redshift data for the quasars. Visual inspection of the spectra showed evidence of an unusual frequency of quasars with strong UV Fe\textsc{ii} emission.

\section{Spectra}\label{spectra}

Two methods have been used previously to measure the strength of the UV Fe\textsc{ii} emission.

\begin{figure}[!ht]
\centering
\includegraphics[scale=0.6]{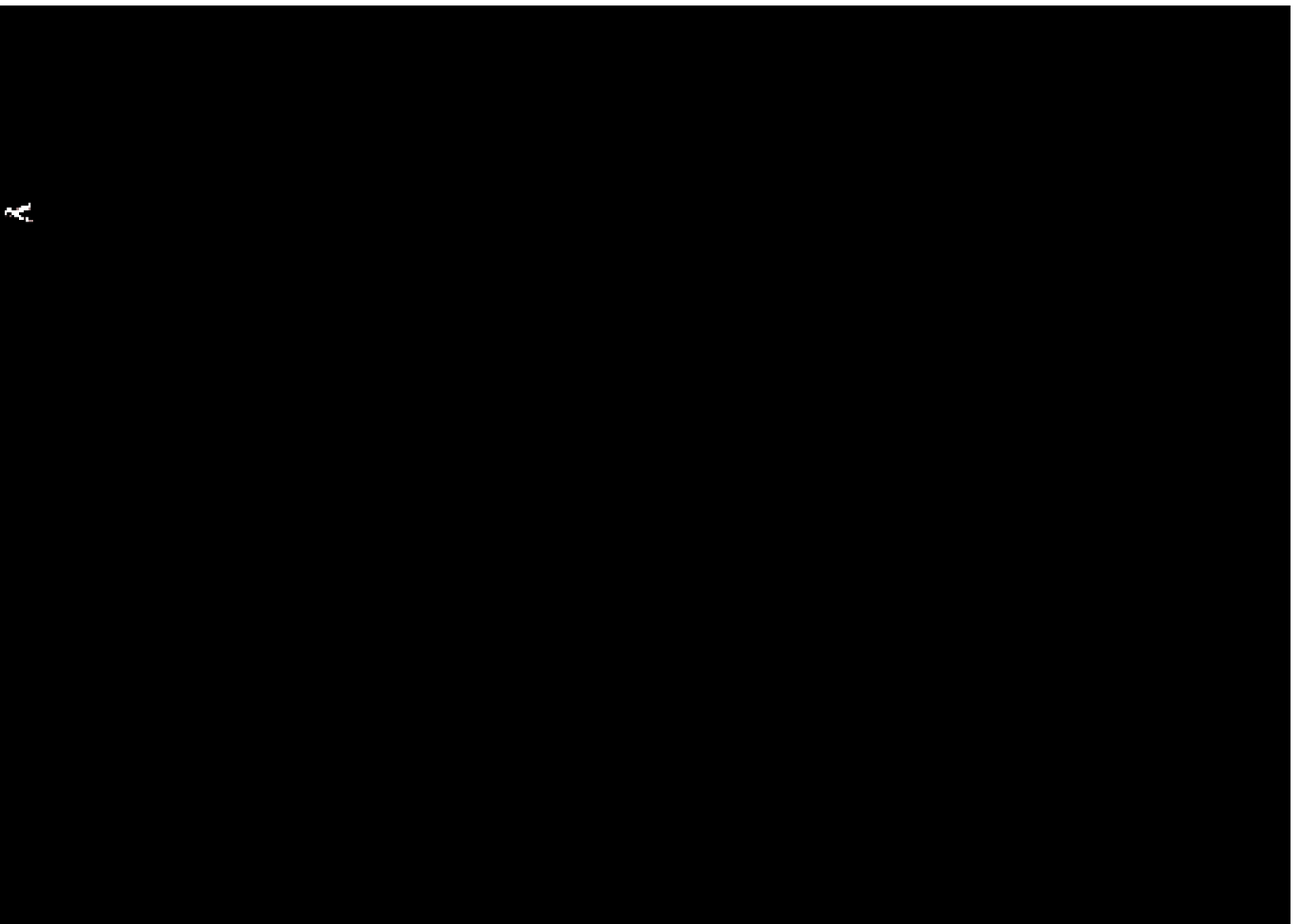}
\caption[Example of Weymann et al. measuring method]{\small{Demonstration of \citet{Weymann1991} method.}}
\label{fig:weymann}
\end{figure}

\cite{Weymann1991} calculate the equivalent width (EW) between 2255 and 2650{\AA} (W2400) with respect to an effective continuum level. The continuum level is found in two wavelength ranges, 2240-2255{\AA} and 2665-2695{\AA}. A straight line is then drawn between the centres of these two wavelength ranges to create the effective continuum. This effective continuum is used to find the EW between 2255 and 2650{\AA} (W2400). The same method is used to find the EW between 2040 and 2130{\AA} (W2070). The W2070 measurement uses the wavelength ranges 1975-2000{\AA} and 2140-2155{\AA} to measure the effective continuum. Figure \ref{fig:weymann} shows the measurements taken for the Weymann method to calculate W2400. ``a'' indicates the two small wavelength ranges used to calculate the effective continuum, ``b''. Then the EW, ``c'', is found. The same method is used to calculate W2070 using different wavelengths. 

\begin{figure}[!ht]
\centering
\includegraphics[scale=0.6]{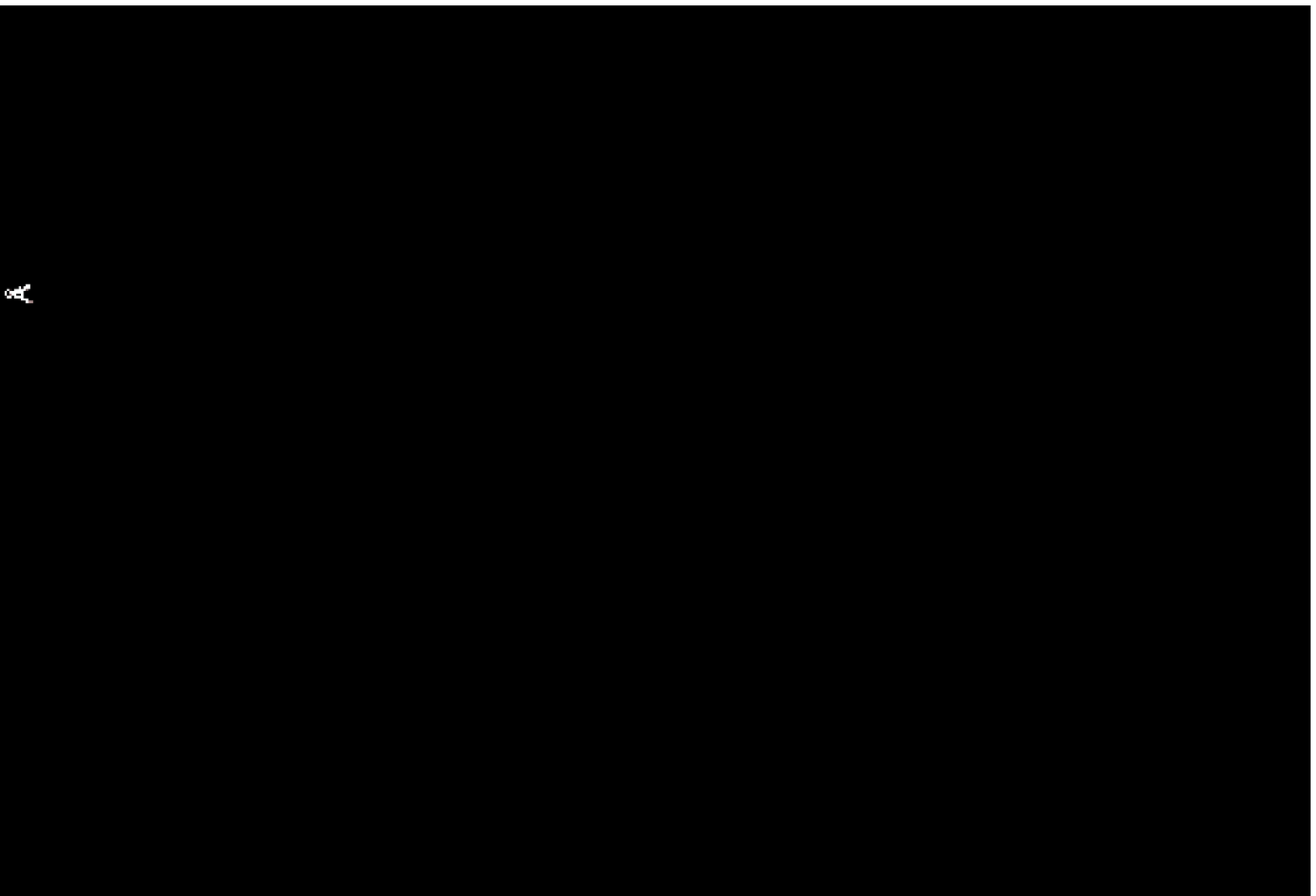}
\caption[Example of Hartig \& Baldwin measuring method]{\small{Demonstration of \citet{Hartig1986} method.}}
\label{fig:HB}
\end{figure}

\citet{Hartig1986} estimated the continuum at three different wavelengths, 2010, 2200, and 2650 {\AA} in the rest frame. By linearly interpolating the continuum over the ranges 2010 - 2200{\AA} and 2200 - 2650{\AA}, the continuum at 2080{\AA} and 2450{\AA}  can be estimated. The monochromatic flux at these wavelengths are also measured (HB2080 and HB2450 respectively). The flux above the continuum is calculated (by subtracting the continuum level from the measured flux) and then divided by the continuum level to give a measure of the Fe\textsc{ii} emission. Figure \ref{fig:HB} shows the measurements taken to calculate the Fe\textsc{ii} strength with the Hartig \& Baldwin method. ``a'' denotes the continuum levels used and ``b'' shows the flux measurements. 

Both of these methods were used to compare the quasars from Hectospec and SDSS to values for previously published ultra-strong UV Fe\textsc{ii} quasars and the composite spectra from \citet{VandenBerk2001}. The SDSS composite excludes BAL quasars. Given the redshifts of the Hectospec quasars, the BAL region of the spectra lies outside the spectral window so it is not possible to determine whether they are BALs. Therefore, we also used a high luminosity BAL composite and a low luminosity composite, both from \citet{Brother2001} using the First Bright Quasar Survey (FBQS) and the Large Bright Quasar Survey (LBQS), to compare to BAL samples. Table \ref{tab:Fedata} shows the Fe\textsc{ii} measurements using both methods for the quasars within the redshift range $1.1<z<1.7$, from the Hectospec data, labelled qsoxxx, and quasars in the DR7QSO catalogue (\citealt{Schneider2010}) in the area of sky covered by the Hectospec observations, labelled with a 17 digit number. The values for the strength of the Fe\textsc{ii} for the composite spectra are also shown for comparison. 

Using the values for the median equivalent width of the Fe\textsc{ii} bumps in Table \ref{tab:Weymann}, the representative average value for the Fe\textsc{ii} equivalent width was taken as 30, with any value greater than 45 being classed as ultra-strong Fe\textsc{ii} and anything between 30 and 45 classed as strong. Using this system, eight quasars were classed as ultra-strong (seven from Hectospec data and one from SDSS DR7QSO), with seven classed as strong Fe\textsc{ii} emitters (five Hectospec and two DR7QSO) from a sample size of 34 quasars within a two deg$^{2}$ area, all within the redshift interval of $1.1<z<1.7$.

The quasars in \citet{Weymann1991} are split into groups of BAL and non-BAL. BAL quasars tend to have stronger Fe\textsc{ii} emission. The mean values for the equivalent with of the Fe\textsc{ii} emission and the RMS errors ({\AA}) for their BAL and non-BAL samples are given (see Table \ref{tab:Weymann}). Typical broad absorption lines are at shorter wavelengths than C\textsc{iv} so are below the wavelength range of our spectra. We are unable to classify our quasars as either non-BAL or BAL so have compared our results to both samples.

\begin{table}[!ht] 
\caption[Median EWs and RMS of Fe{\sc{ii}} from literature]{\small{The median EWs and RMS errors ({\AA}) on the EW on the samples from \citet{Weymann1991}. }}
\centering
\begin{tabular}{ c | c c c c }
Sample & 2400median ({\AA}) & 2400RMS ({\AA}) & 2070median ({\AA}) & 2070RMS ({\AA}) \\ \hline
All     & 29.86       &  15.71   &  4.86       &  2.60   \\
non-BAL & 29.51       &  11.45   &  4.33       &  2.09   \\
BAL     & 33.10       &  18.30   &  4.90       &  2.85   \\
\end{tabular}
\label{tab:Weymann}
\end{table}

\section{Fe{\sc{ii}} Results}

The details of the quasars studied in the area as the CCLQG can be seen in Table \ref{tab:Fedata}. The table contains all of the quasars within the area $161.5<RA<163.5$ and $4.0<DEC<6.0$, which is the sky area covered by the Hectospec data. The columns in Table \ref{tab:Fedata} show the the names, RA, DEC, and redshift data for the quasars as well as the measurements of the Fe\textsc{ii} strength using both the Weymann, and Hartig \& Baldwin methods. The quasars have been split into 3 groups. These groups are based on the mean W2400 equivalent width from Weymann, giving the average Fe\textsc{ii} value of 30 (supported by the SDSS composite spectra value of 27, within errors) and class values above 45 (i.e., the average plus one RMS) as ultra-strong quasars. Any objects with a W2400 value between 30 and 45 is classed as strong and anything below 30 classed as weak. The last section at the bottom of Table \ref{tab:Fedata} shows other quasars from the Hectospec spectra which are not at the redshift of the LQGs but for which it was possible to measure the UV Fe\textsc{ii} emission.

\begin{landscape}
\begin{longtable}{c c | c c c c c c c}
\caption[Properties of Hectospec quasars]{\small{Properties of the Hectospec quasars along with the properties for any other quasars within the field from the SDSS DR7QSO catalogue (\citealt{Schneider2010}). The columns show the the names, RA, DEC, redshift, and Fe{\sc{ii}} measurements using both the Weymann (\citealt{Weymann1991}), and Hartig \& Baldwin (\citealt{Hartig1986}) methods.}} \label{tab:Fedata} \\
Group          & Quasar             & Redshift & RA (J2000)  & DEC (J2000)  & W2400 & W2070 & HB2080 & HB2450 \\ \hline
Ultra-strong   & qso412             & 1.1593   & 10:49:47.35 & +04:17:46.35 & 56.08 & 0.81  & 0.001  & 0.187  \\
               & qso425             & 1.2303   & 10:48:00.41 & +05:22:09.90 & 56.01 & 4.86  & 0.078  & 0.28   \\
               & 587728879806972003 & 1.607    & 10:49:14.33 & +04:14:28.65 & 54.34 & 7.26  & 0.135  & 0.293  \\
               & qso27              & 1.3145   & 10:49:30.46 & +05:40:46.20 & 53.75 & 4.25  & 0.128  & 0.142  \\
               & qso421             & 1.665    & 10:48:15.94 & +05:50:07.80 & 53.32 & 6.3   & 0.105  & 0.189  \\
               & qso417             & 1.6533   & 10:49:26.83 & +04:23:34.80 & 49.03 & 5.12  & 0.105  & 0.254  \\
               & qso29              & 1.4166   & 10:49:21.07 & +05:09:48.30 & 47.78 & 6.72  & 0.115  & 0.223  \\
               & qso41              & 1.434    & 10:51:31.94 & +04:51:24.90 & 47.53 & 5.02  & 0.066  & 0.147  \\ \hline
Strong         & 587732701256548431 & 1.593    & 10:52:51.72 & +05:57:33.90 & 43.33 & 4.52  & 0.113  & 0.215  \\
               & 587732701256089663 & 1.503    & 10:48:40.35 & +05:59:12.98 & 37.94 & 4.51  & 0.08   & 0.130  \\
               & qso217             & 1.6222   & 10:49:58.92 & +04:27:23.40 & 37.64 & 1.44  & 0.07   & 0.198  \\
               & qso48              & 1.2166   & 10:50:10.06 & +04:32:49.20 & 35.33 & 4.22  & 0.097  & 0.139  \\
               & qso219             & 1.3491   & 10:49:34.71 & +05:48:36.00 & 35.15 & 4.38  & 0.009  & 0.061  \\
               & qso221             & 1.665    & 10:48:15.94 & +05:50:07.80 & 33.92 & 5.51  & 0.102  & 0.036  \\
               & qso410             & 1.4184   & 10:50:00.36 & +04:51:57.90 & 31.39 & 1.18  & 0.067  & 0.118  \\ \hline
Weak           & 587732576700596362 & 1.689    & 10:52:55.65 & +05:51:12.93 & 32.8  & 0.87  & 0.086  & 0.131  \\
               & 588010358541910265 & 1.552    & 10:51:54.14 & +04:10:59.55 & 29.94 & 2.09  & 0.008  & 0.015  \\
               & 588010359615651877 & 1.608    & 10:51:41.91 & +04:58:27.90 & 29.42 & 5.12  & 0.099  & 0.13   \\
  	       & qso49              & 1.1314   & 10:50:07.90 & +04:36:59.70 & 28.46 & 3.27  & 0.085  & 0.137  \\
               & qso45              & 1.3171   & 10:50:36.10 & +04:56:11.40 & 27.81 & 3.53  & 0.078  & 0.097  \\
               & 588010359615914006 & 1.132    & 10:53:52.73 & +05:00:43.92 & 26.89 & 6.49  & 0.091  & 0.054  \\      
               & qso22              & 1.2164   & 10:50:30.77 & +04:30:55.05 & 26.5  & 3.41  & 0.074  & 0.085  \\
               & 588010360688869424 & 1.228    & 10:46:56.71 & +05:41:50.25 & 24.57 & 3.02  & 0.045  & 0.102  \\
               & 587728880343711798 & 1.696    & 10:47:51.89 & +04:37:09.90 & 24.49 & 0.66  & 0.065  & 0.169  \\
               & qso420             & 1.2381   & 10:48:40.85 & +04:09:38.55 & 24.42 & 10.86 & 0.068  & 0.013  \\
               & 587728879270494315 & 1.193    & 10:52:49.68 & +04:00:46.50 & 24.12 & 4.41  & 0.056  & 0.145  \\
               & qso26              & 1.111    & 10:49:32.23 & +05:05:31.50 & 23.16 & 1.77  & 0.052  & 0.106  \\
               & 588010360152064102 & 1.334    & 10:47:33.17 & +05:24:55.05 & 20.42 & 3.72  & 0.09   & 0.097  \\
               & qso413             & 1.2948   & 10:49:43.30 & +04:49:48.75 & 19.37 & 2.27  & 0.012  & 0.091  \\
               & 587728881417650297 & 1.517    & 10:49:38.35 & +05:29:31.95 & 18.83 & 2.43  & -0.029 & 0.138  \\
               & 587728881417715881 & 1.307    & 10:50:18.12 & +05:28:26.40 & 18.46 & 2.3   & 0.053  & 0.015  \\
               & 588010360152653920 & 1.528    & 10:52:43.99 & +05:26:22.95 & 15.74 & 3.87  & 0.108  & 0.048  \\
               & 587728881417846834 & 1.488    & 10:51:18.60 & +05:33:31.65 & 13.24 & 3.46  & 0.087  & 0.098  \\
               & 588010358541451372 & 1.606    & 10:47:47.64 & +04:05:27.90 & 7.58  & 1.55  & -0.003 & 0.054  \\ \hline
Composites     & SDSScomposite      & NA       & -           & -            & 27.0  & 2.02  & 0.059  & 0.103  \\
               & HighBAL            & NA       & -           & -            & 21.98 & 2.89  & 0.037  & 0.096  \\
   	       & lowBAL             & NA       & -           & -            & 18.15 & 5.45  & 0.075  & 0.064  \\ \hline
Outside        & qso223             & 2.042    & 10:49:16.38 & +05:48:26.03 & 27.81 & 6.16  & 0.068  & 0.186  \\
$1.1<z<1.7$    & qso218             & 2.227    & 10:49:58.91 & +04:27:23.34 & 19.56 & 2.98  & 0.065  & 0.196  \\
               & qso210             & 1.738    & 10:49:14.94 & +05:14:52.64 & 19.13 & 2.89  & 0.064  & 0.083  \\
               & qso225             & 1.730    & 10:48:05.38 & +05:39:37.26 & 17.67 & 0.35  & 0.020  & 0.090  \\
               & qso222             & 1.949    & 10:49:18.41 & +04:59:59.06 & 10.07 & 5.28  & 0.077  & 0.013  \\
               & qso215             & 1.93     & 10:51:47.95 & +04:43:11.96 & 6.05  & 6.72  & 0.119  & 0.015  
\end{longtable}
\end{landscape}          

\begin{figure}[!h]
\centering
\subfigure[Ultra-strong Fe{\sc{II}} emission]{
\includegraphics[scale=0.4]{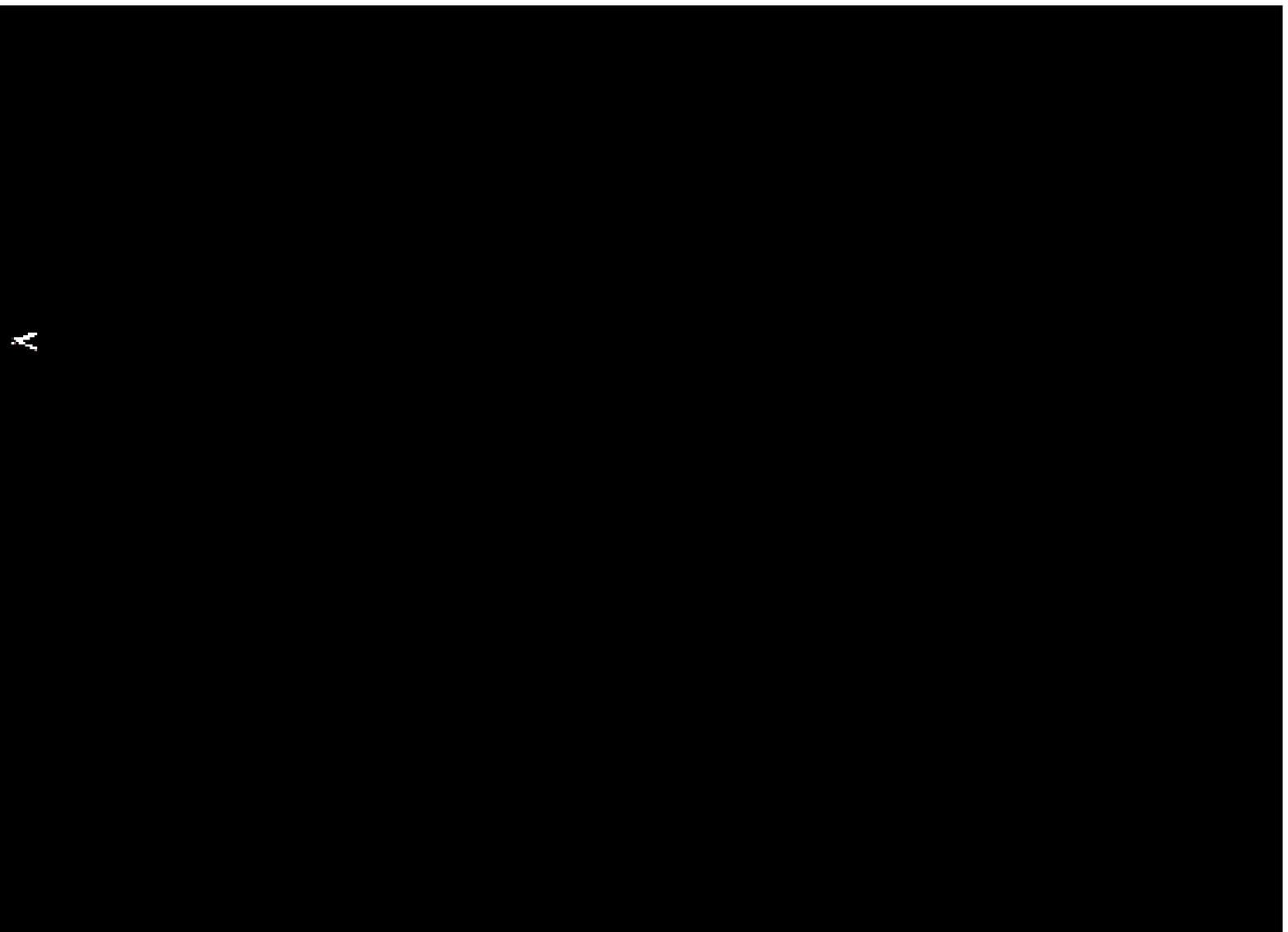}
\label{fig:USfe}
}
\subfigure[Strong Fe{\sc{II}} emission]{
\includegraphics[scale=0.4]{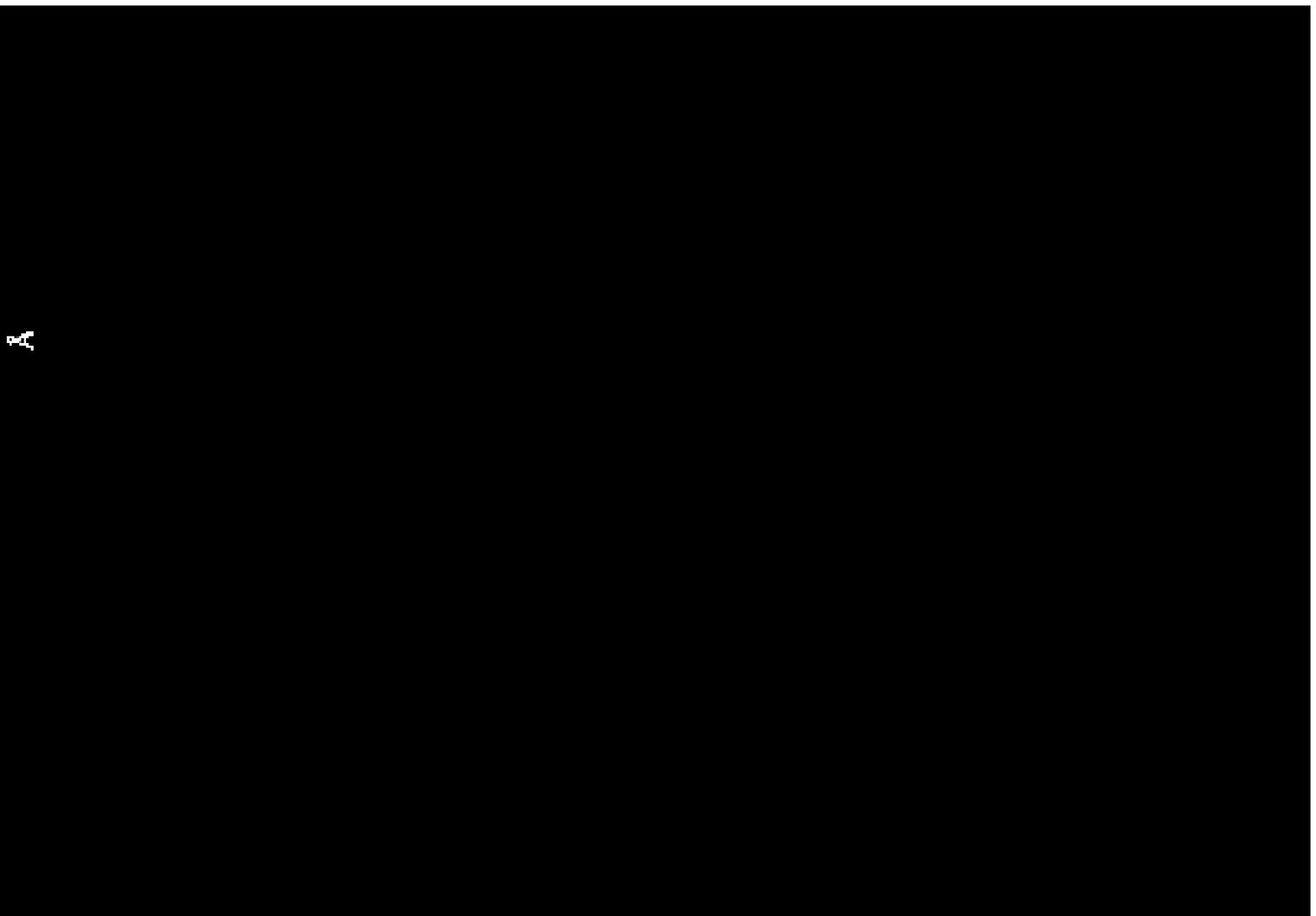}
\label{fig:Sfe}
}
\subfigure[Weak Fe{\sc{II}} emission]{
\includegraphics[scale=0.4]{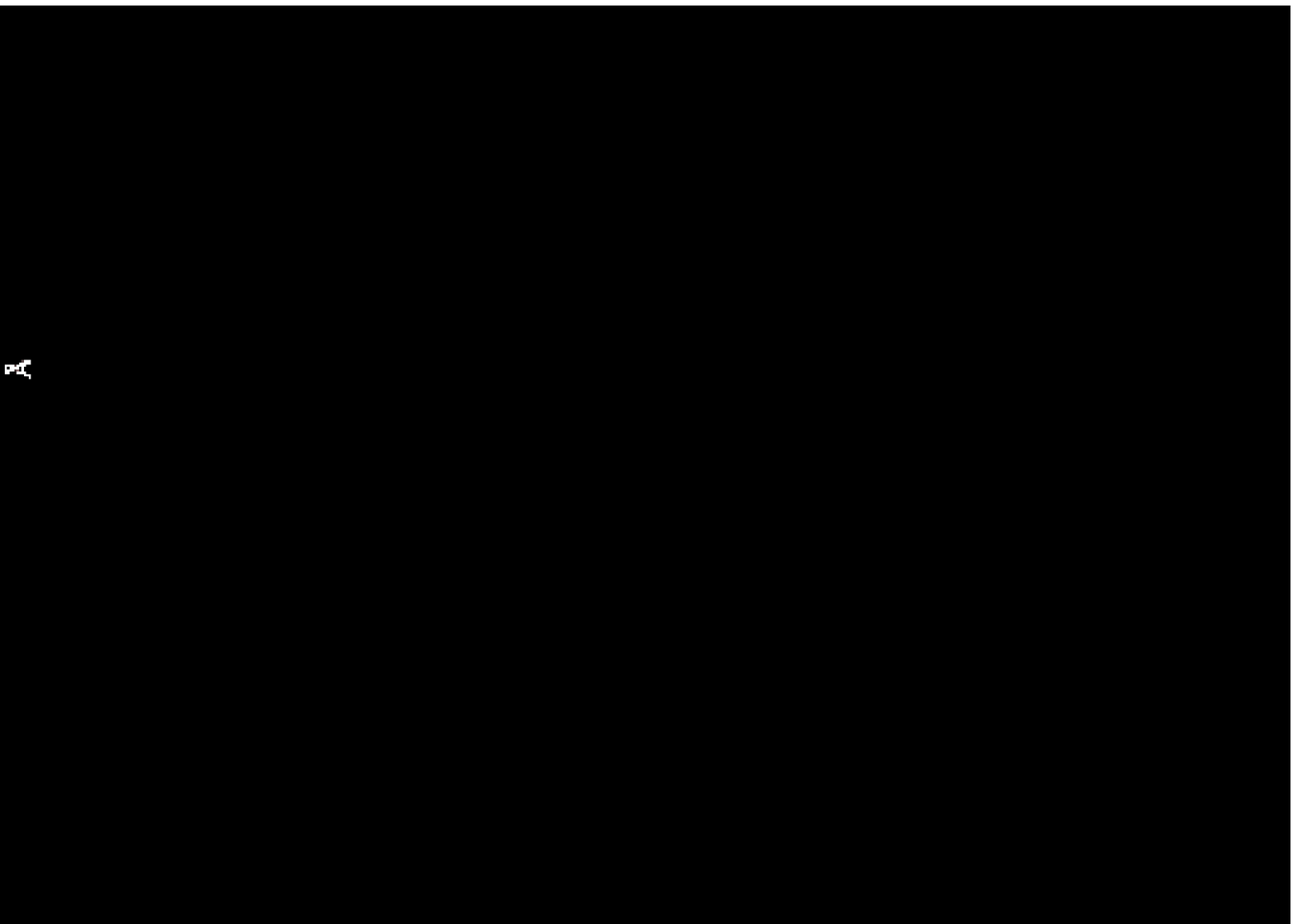}
\label{fig:Wfe}
}
\caption[Examples of levels of Fe{\sc{ii}} emission]{\small{Example of \subref{fig:USfe} Ultra-strong, \subref{fig:Sfe} strong, and \subref{fig:Wfe} weak UV Fe\textsc{ii} emission.}}
\label{fig:festrength}
\end{figure}

Figure \ref{fig:festrength} shows some examples of ultra-strong \subref{fig:USfe}, strong \subref{fig:Sfe}, and weak \subref{fig:Wfe} UV Fe\textsc{ii} emission.

\begin{figure}[!h]
\centering
\subfigure[$u-g$ against $g-r$]{
\includegraphics[scale=0.24,angle=-90]{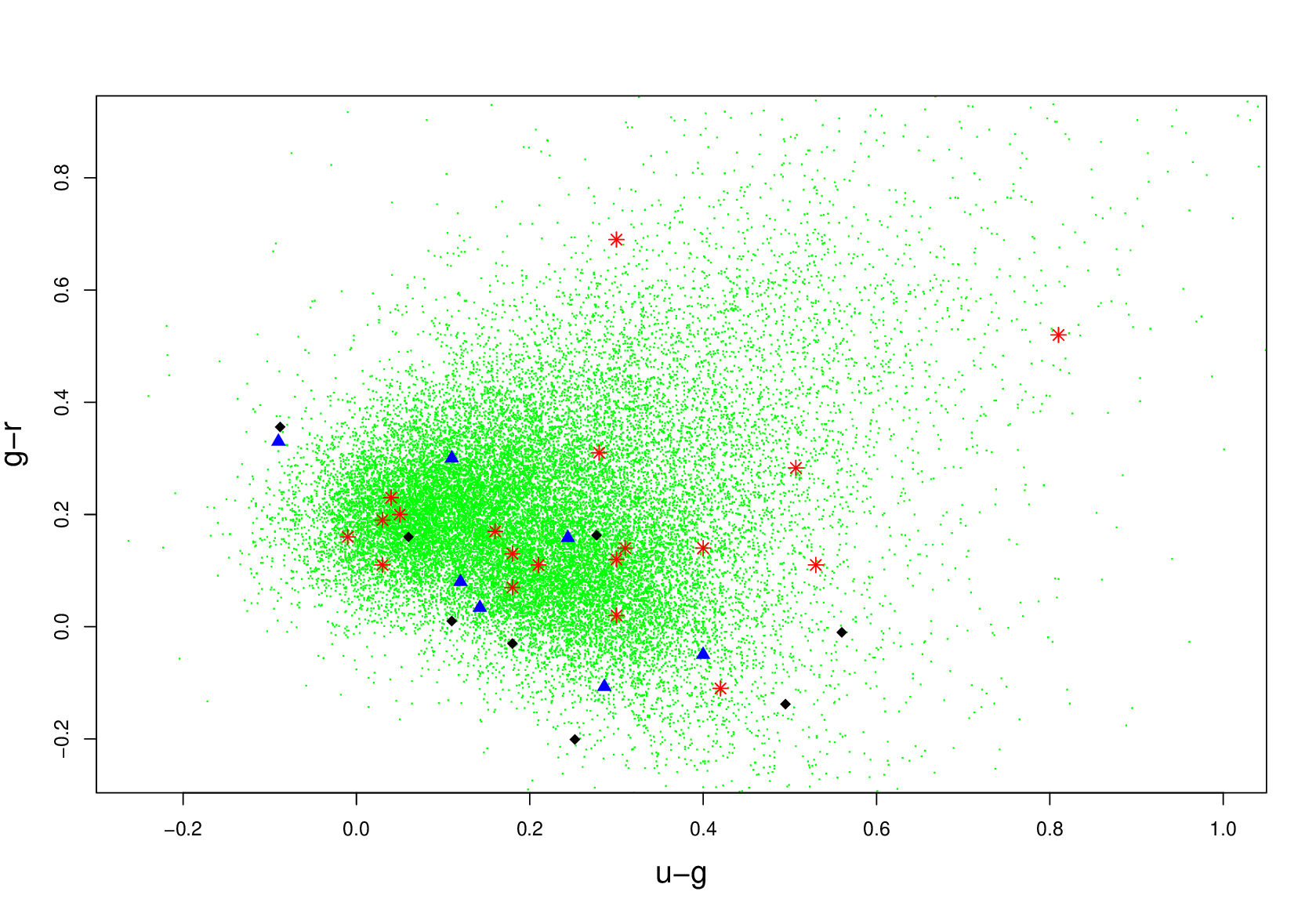}
\label{fig:ug-gr}
}
\subfigure[$g-r$ against $r-i$]{
\includegraphics[scale=0.24,angle=-90]{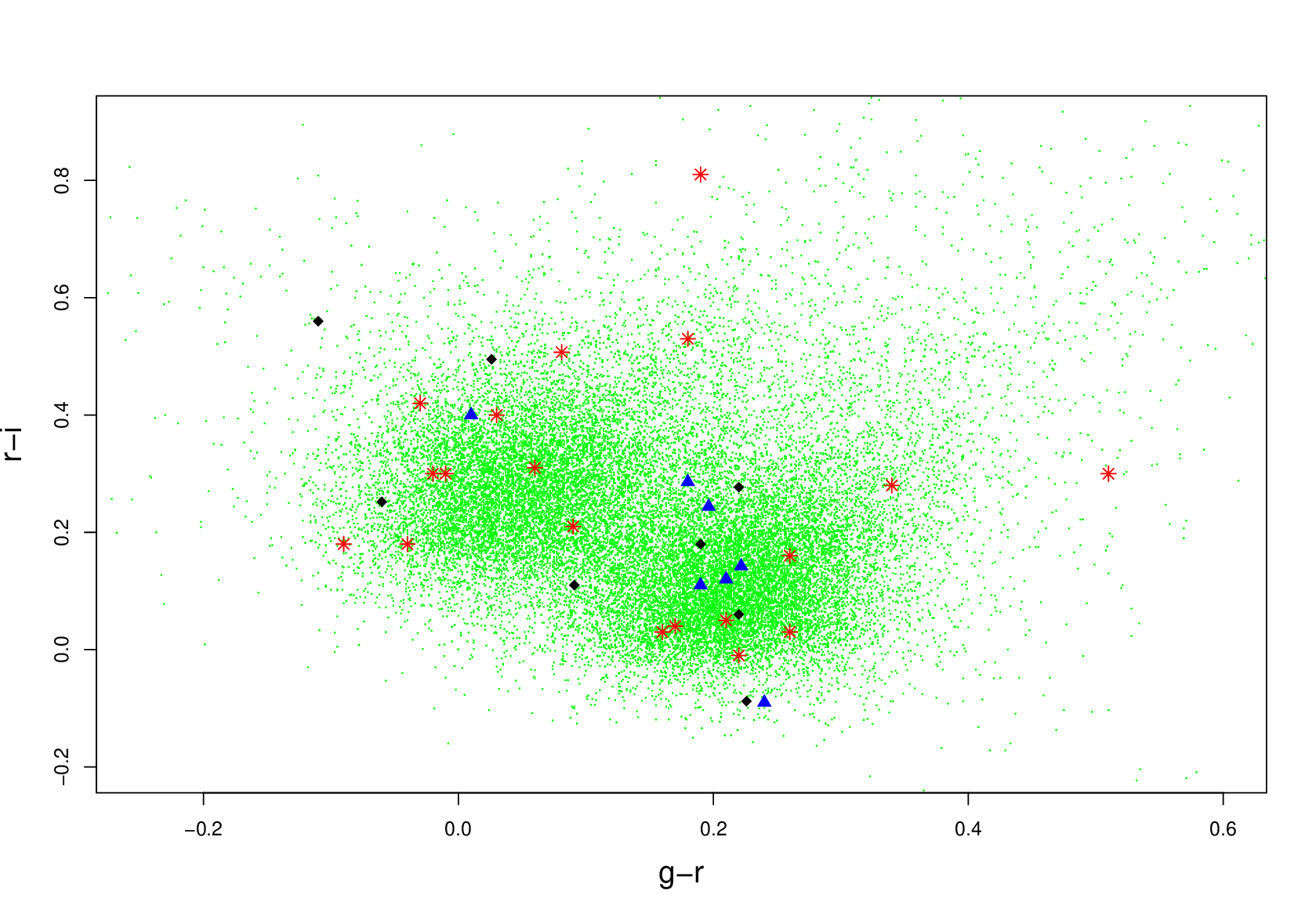}
\label{fig:ri-gr}
}
\subfigure[$r-i$ against $i-z$]{
\includegraphics[scale=0.24,angle=-90]{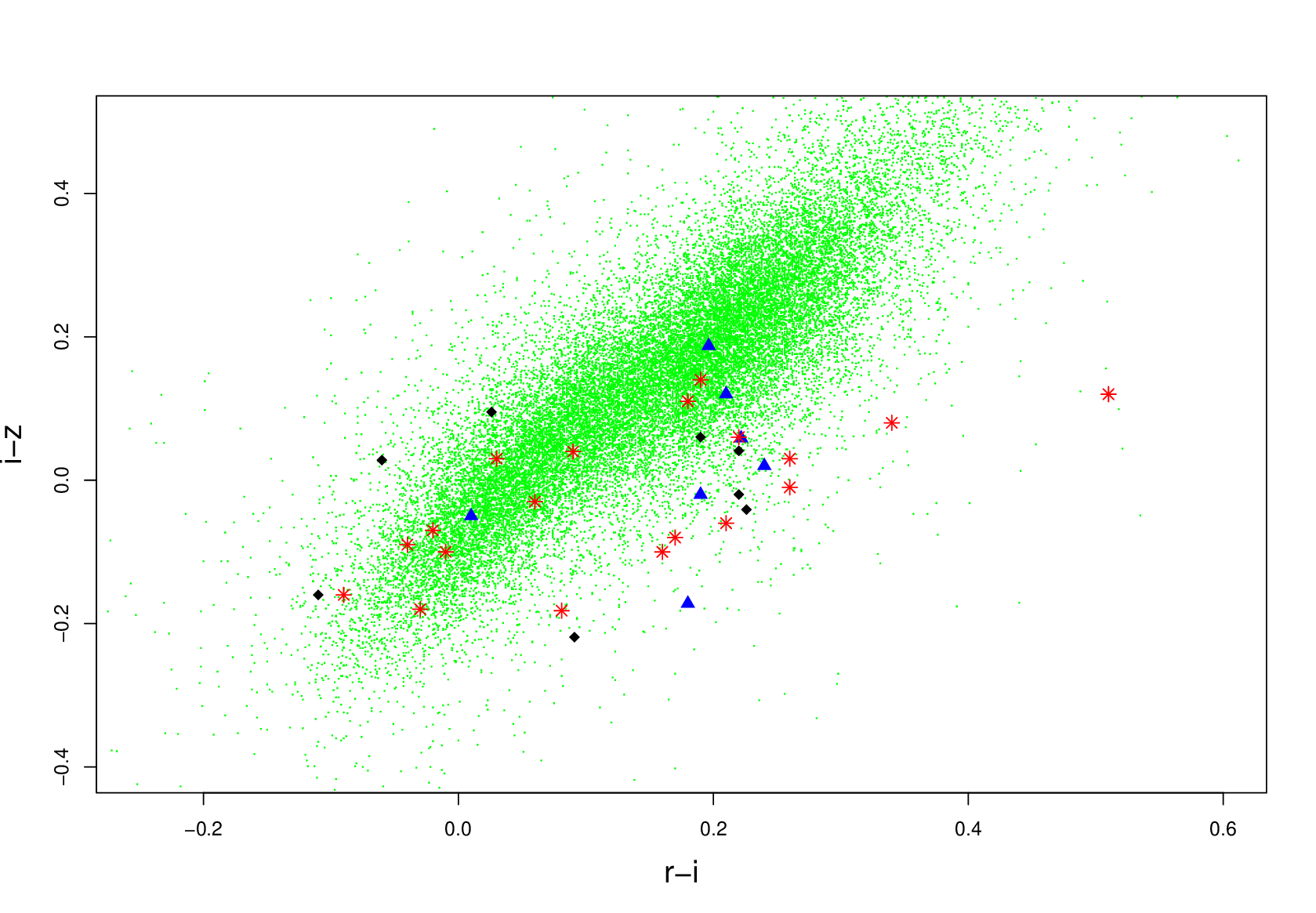}
\label{fig:iz-ri}
}
\subfigure[$u-g$ against $r-i$]{
\includegraphics[scale=0.24,angle=-90]{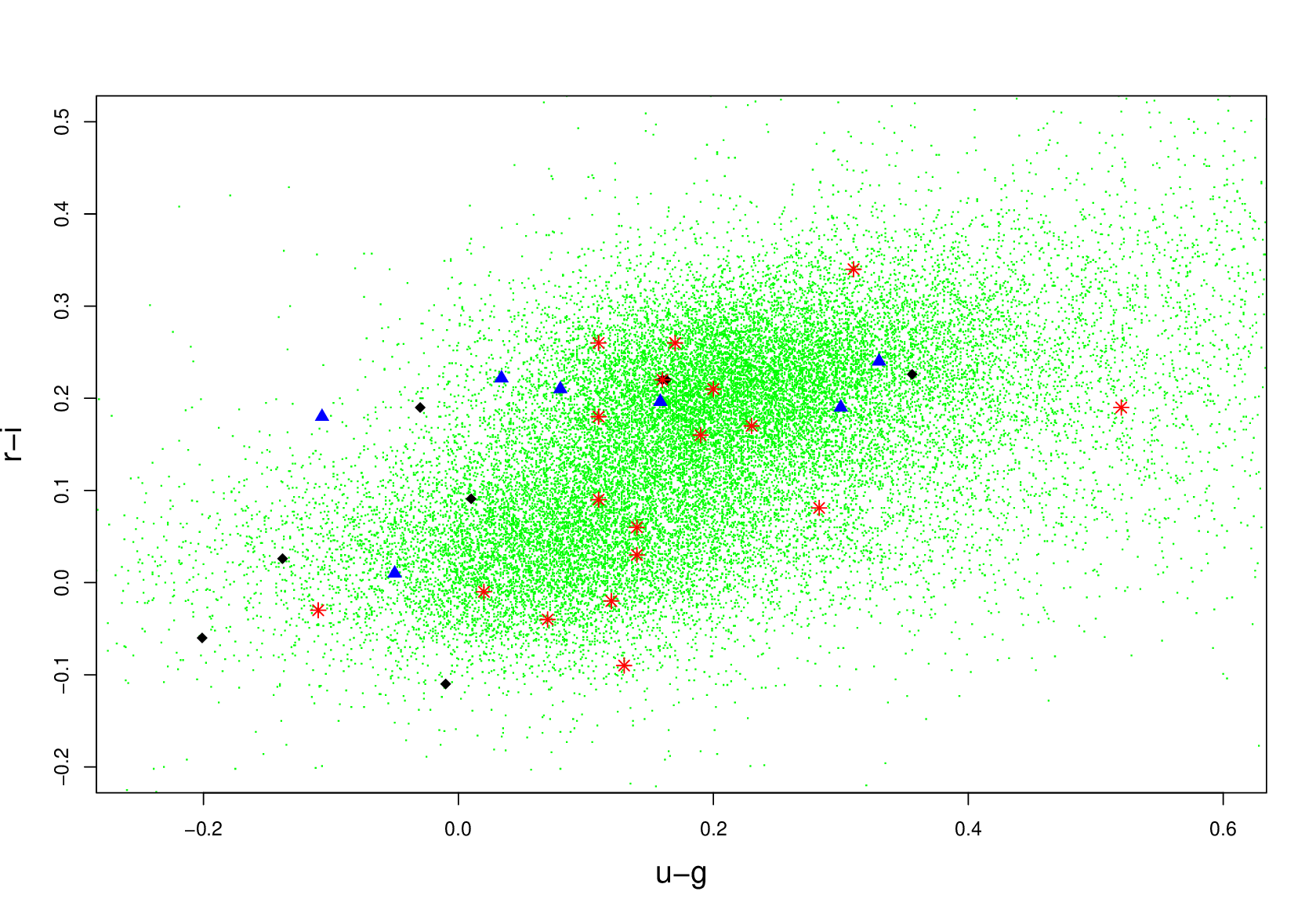}
\label{fig:ug-ri}
}
\subfigure[$u-g$ against $i-z$]{
\includegraphics[scale=0.24,angle=-90]{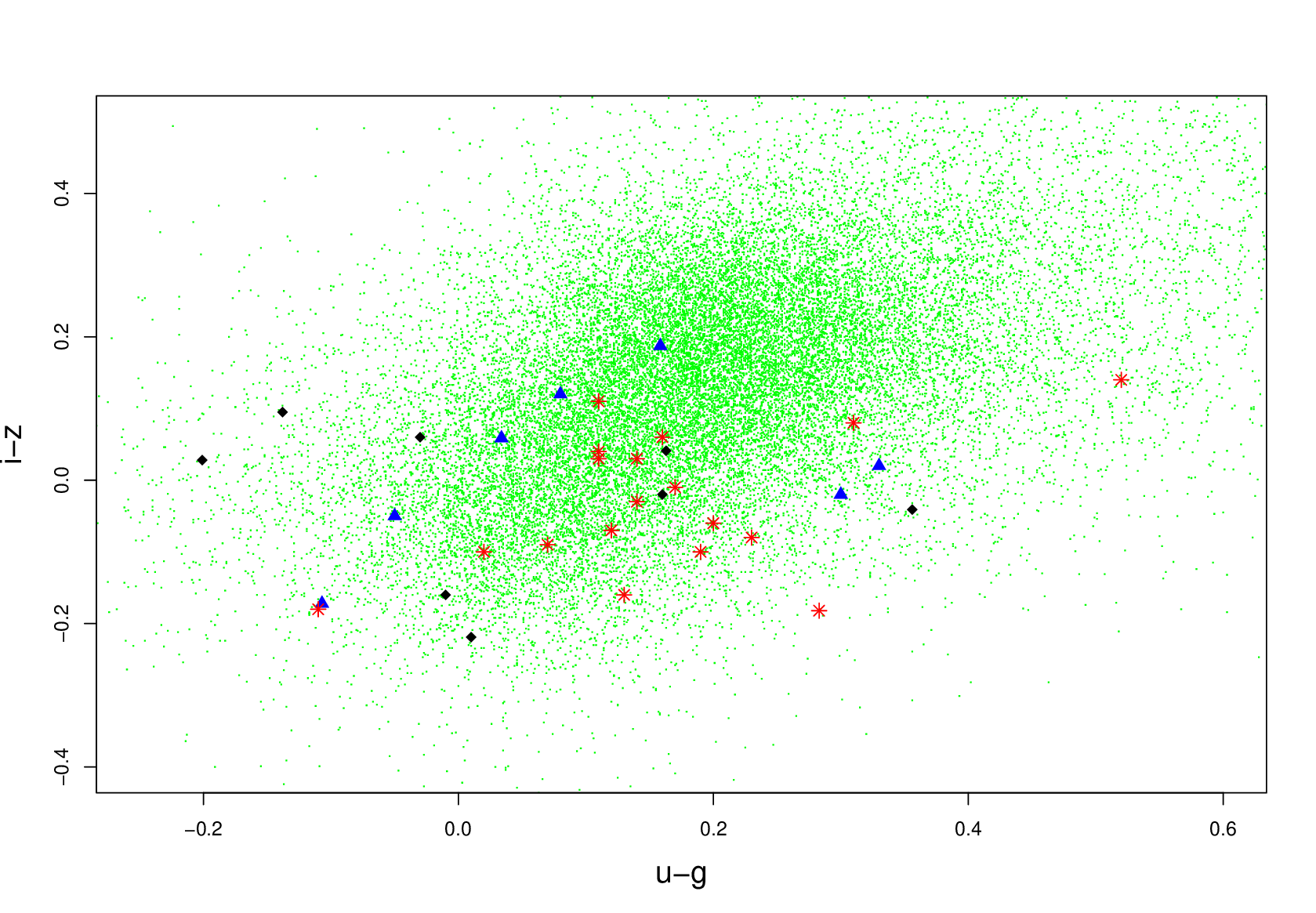}
\label{fig:ug-iz}
}
\subfigure[$g-r$ against $i-z$]{
\includegraphics[scale=0.24,angle=-90]{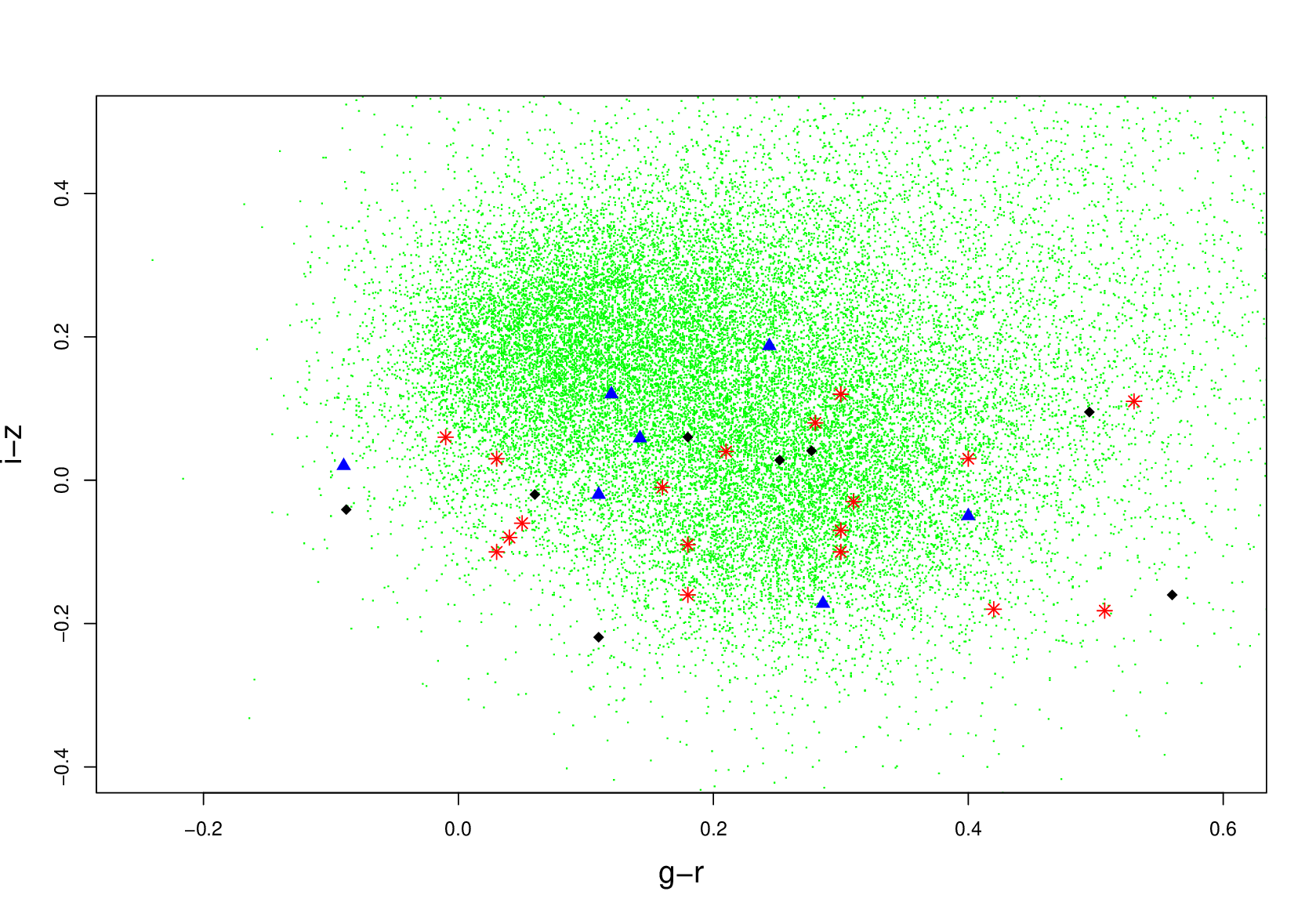}
\label{fig:gr-iz}
}
\caption[Positions of AGN on two-colour diagrams]{\small{Two-colour diagrams for Hectospec quasars. The background green points are taken from the SDSS DR7QSO catalogue. The black diamonds show the ultra-strong quasars (W2400$>$45 ), the blue triangles show the positions of strong Fe\textsc{ii} emitters (30$<$W2400$<$45) while the red stars show the positions of the weaker emitters (W2400$<$30).}}
\label{fig:colours}
\end{figure}

\clearpage
The $u, g, r, i, z$ magnitudes were taken from the SDSS database and the colours can be seen in Figure \ref{fig:colours}. The green points were taken from the DR7QSO catalogue. The points overlaid are the Hectospec quasar sample within the redshift range of the LQGs with the red stars showing quasars with lower than average Fe\textsc{ii} strengths (W2400$<$30), blue triangles showing the strong Fe\textsc{ii} emitters (30$<$W2400$<$45), and the black diamonds showing the ultra-strong emitters (W2400$>$45). There is a possible trend for the strong and ultra-strong Fe\textsc{ii} emitters to lie outside the main area of the colour plots indicated by the green points from the DR7QSO catalogue. However, this trend is weak so it will not be possible to select strong Fe\textsc{ii} emitters solely on colour cuts

\clearpage

\subsection{Quasar Location}

\begin{figure}[!h]
\centering
\includegraphics[scale= 0.65]{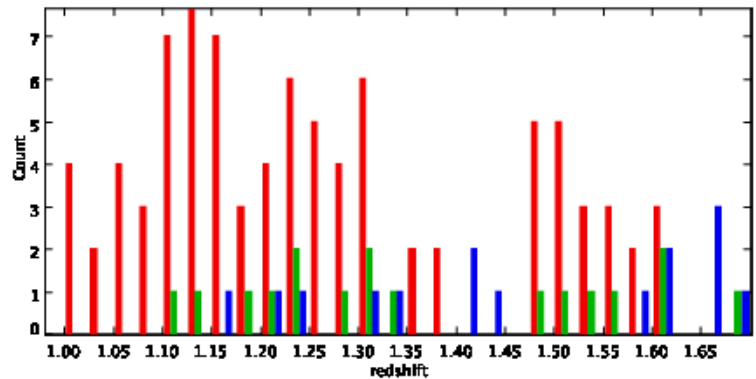}
\caption[Redshift positions Fe\sc{ii} emitters with respect to LQGs]{\small{Redshift location for the LQGs (red) (\citealt{Clowes2011}; private communication) and the positions in redshift of the strong and ultra-strong Fe{\sc{ii}} emitters (W2400$>$30; blue) and weak Fe{\sc{ii}} emitters (W2400$<$30; green) from Table \ref{tab:Fedata}.}}
\label{fig:zlocation}
\end{figure}

The strong and ultra-strong UV Fe\textsc{ii} emitting quasars are in the area of the 3 LQGs. This area consists of potentially 3 layers of quasars at different redshifts; $\overline{z}=1.11, \overline{z}=1.28$ and $\overline{z}=1.54$. Figure \ref{fig:zlocation} shows the redshifts of the strong and ultra-strong Fe{\sc{ii}} emitters (W2400$>$30; blue) and weak Fe{\sc{ii}} emitters (W2400$<$30; green) with respect to the redshifts of the quasars in these LQGs (red). 

Figure \ref{fig:zlocation} shows that the strong and ultra-strong Fe\textsc{ii} emitters appears to prefer to lie between the LQGs rather than in the groups. The weak emitters do not appear to lie in any preferential redshift positions.

\section{Comparing to Control Data}
\subsection{Weymann Quasars}
The SDSS survey has spectra from three of the quasars in the Weymann paper (one non-BAL and two BAL quasars). Measurements of these three quasars were taken in order to compare values quoted in \citet{Weymann1991} to the equivalent widths measured using current spectra. With only three matching spectra (seen in Table \ref{Weymann_matches}), it is only possible to say the measurements match within less than 35\%. It is also possible that the strength of the Fe\textsc{ii} has varied over the years between when the spectra were taken.

\begin{table}
\caption[Measurements of previously published Fe{\sc{ii}} emitters ]{\small{Measurements from \citet{Weymann1991} with matching SDSS measurements and the percentage difference as a fraction of the largest measurement.}}
\centering
\begin{tabular}{ c | c c c }
Quasar     & Re-measured W2400 & Weymann W2400 & \% Difference \\ \hline
0043+0048  & 35.21        & 44.66         & 21.2       \\
1216+1103  & 25.41        & 19.37         & 23.7       \\
1227+1215  & 27.90        & 42.14         & 33.8       
\end{tabular}
\label{Weymann_matches}
\end{table}

\subsection{Control Quasar Samples from SDSS}
The spectra described above have been taken in the direction of the CCLQG. To assess whether the environment of the LQG is likely to be a factor in the number of ultra-strong and strong Fe\textsc{ii} emitting quasars, control samples were taken from the quasars in Stripe 82, in areas which do not contain any previously known LQGs. The field control samples were taken from SDSS stripe 82 which has a similar limiting magnitude as our control samples. Four two deg$^{2}$ samples were taken, containing in total 128 quasars within the redshift range 1.1$<z<1.7$. 

The Fe\textsc{ii} 2400 bumps on the spectra were measured with the Weymann, and Hartig and Baldwin methods as used on our spectra. In these control samples, three quasars on average were found to be classed as ultra-strong per 2 deg$^2$ field, and five per 2 deg$^2$ field as strong. 

Using a minimal spanning tree based method for finding LQGs (as discussed in \citealt{Clowes2011}), only seven of the quasars from the Stripe 82 samples were found to be part of candidate groups, and none of these had a W2400 measurement greater than 31. However, given the limited width (therefore area covered) of Stripe 82, we can only say for certain which quasars are definitely part of a quasar group. Other quasars within our control samples may also be part of groups, containing members which lie beyond the edges of Stripe 82.

\begin{figure}[!h]
\centering
\includegraphics[angle=-90,scale=0.5]{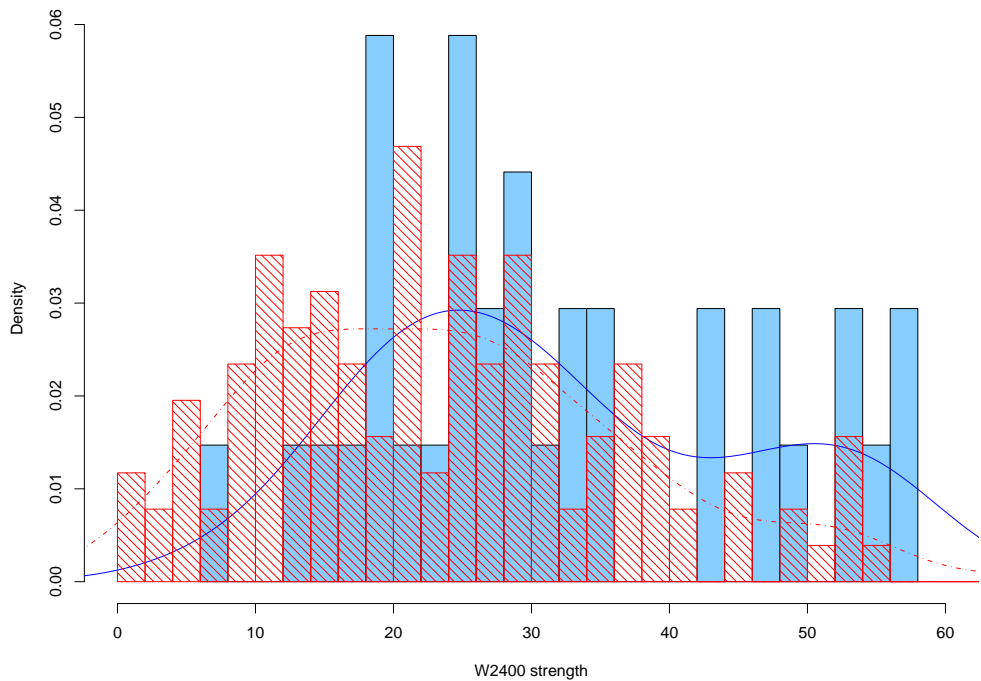}
\caption[Comparing Hectospec Fe\sc{ii} emitters to control sample]{\small{Frequency of W2400 strength in the control sample and our observed quasars. The figure shows frequency density of W2400 strengths from the Hectospec quasars and SDSS quasars (solid blue) and from the Stripe 82 control sample (red hatches). The fitted lines show the probability distribution of the sample. The dashed red line shows the distribution for the control sample while the solid blue line shows the distribution for the Hectospec and SDSS sample.}}
\label{fig:fehist}
\end{figure}

Figure \ref{fig:fehist} shows the distribution of Fe\textsc{ii} strengths from the equivalent widths measured using the Weymann method in solid blue and the blue line shows the density function of the distribution for the observed quasars in the CCLQG field. The red hatched columns show the distribution of the control samples and the red dotted line shows the density function of the distribution for the control fields from Stripe 82. The solid blue line shows two concentrations, one at an equivalent width of $\sim25$ and another at a higher Fe\textsc{ii} strength of $\sim50$. The red dotted line from the control samples only shows one peak at a slightly lower value than the first blue peak. The red hatched bars show evidence of more quasars with lower Fe\textsc{ii} emission in the control sample than is seen in the sample in the direction of the CCLQG. There is also an excess of quasars with higher Fe\textsc{ii} emission in the CCLQG than is seen in the control samples from Stripe 82.

To test the significance, a Mann-Whitney test has been used. The Mann Whitney test assesses whether there is a significant difference between medians of two independent samples. It uses the null hypothesis that the distribution in both of the samples is the same. The alternative hypothesis can be either that the samples are simply different, or that one sample is shifted in one direction. The two samples are combined and ranked from lowest to highest. The samples are then sorted back into their original groups and the ranks for each sample are summed. If the distributions were the same (i.e., the null hypothesis is correct), the sum of the ranking should be the same for each sample. 

The Mann-Whitney test gives level of confidence of 0.9973 (p-value of 0.0027) that the distribution in our control sample is shifted to higher strengths than the control sample.
Given that the number of ultra-strong Fe\textsc{ii} emitters is significant, the overdensity of ultra-strong Fe\textsc{ii} emitters is $\sim3\times$ that of the control samples from Stripe 82.

\section{Discussion}

\subsection{Location of Fe{\sc{ii}} Emission}
It was previously thought that UV and optical Fe\textsc{ii} are emitted from the BLR around the quasar. Some studies have found that the line widths of H$\beta$ (also emitted from the BLR) are larger than the widths for Fe\textsc{ii} (e.g., \citealt{Marziani2003,Vestergaard2005}). This suggests that these two emission lines do not originate from the same area and the Fe\textsc{ii} originates further out in the BLR in an intermediate line region which lies at the outer edge of the BLR. The kinematics of this area are dominated by infall and may be a transition area between the torus, and the BLR and accretion disk (\citealt{Hu2008a,Hu2008b}). The infall of material could cause an increase in the microturbulence.

\subsection{Microturbulence}
Microturbulence is the non-thermal random motions within a cloud's line emitting region (\citealt{Bottorff2000a,Bottorff2000b}) and affects the line widths in the quasar spectra. The presence of microturbulence can explain the smooth line profiles and is needed in some cases to reproduce observed line ratios (\citealt{Bottorff2000b}). A typical quasar BLR could support a number of individual microturbulent motions with a velocity range between 0 kms$^{-1}$ (thermal broadening only) to $\sim10^4$ kms$^{-1}$. The microturbulence in clouds can be very high if the cloud is magnetically confined.  

There is an inverse relation between the velocity shift in Fe (which causes the broad lines seen) due to the motions in the clouds and the Eddington ratio, $L_{bol}/L_E$, where $L_{bol}$ is the bolometric luminosity and $L_E$ is the Eddington luminosity. This suggests that the Eddington ratio is a potential driver for microturbulence and therefore, the increase in Fe\textsc{ii} emission (\citealt{Hu2008b}).  

\citet{Bruhweiler2008} looked at the effects of abundance, ionising flux, and microturbulence and showed that microturbulence may have the greatest effect on Fe\textsc{ii} emission. Microturbulence increases the line widths which allows line photons to escape more easily, increasing the importance of continuum pumping \citep{Baldwin2004} and fluorescence efficiency \citep{Netzer1983}.

\subsection{Iron Abundance}
Iron is primarily produced in Type Ia supernovae from long lived intermediate-mass white dwarf binaries, which means that, due to the time scales of white dwarf lifetime evolution, it will only be found in substantial amounts at cosmic times greater than 1 Gyr \citep{Bruhweiler2008}. If the iron were produced in the host galaxy of the quasar, this host galaxy would have to have had a large starburst around 1 Gyr previous to produce excess amounts of iron. The iron, from nearby galaxies, can also be distributed to the intergalactic medium (IGM) via winds and falls onto the quasars. The iron abundance would be dominant if there were a number of star forming galaxies within the close vicinity of the quasar and the winds were strong enough to distribute the metals widely.

Iron emission has also been associated with radio galaxies, and in particular the jets and plumes (\citealt{Hlavacek2011}). In 4C+55.26, a large plume-like feature from one of the X-ray cavities shows an unusual amount of Fe emission. If associated with SN, it would require 2.5 SN1a per century to create the metal enrichment seen. It is likely both SN1a ($\sim70$\%) and SNII ($\sim30$\%) contribute to the galaxy enrichment (\citealt{Sanders2006}). It is suggested that this material could be created through SNe and moved out of the galaxy by the central AGN. However, it is still unclear why a large amount of iron would have accumulated in some AGN.

\citet{Barai2011} show that AGN winds may be more effective than supernovae (SNe) in moving metals into the IGM, and will be a significant contributor to IGM metal enrichment through cosmic time. \citet{Cen1999} found that metallicity is strongly dependent on the local density with higher density regions having higher metallicities (over area of 50-100$h^{-1}$ Mpc). However, when including anisotropy of the AGN outflows into the model, the outflows from the AGN will enrich the lower-density regions first as the outflow follows the path of least resistance \citep{Barai2011}. Baria found that the AGN outflows could be responsible for 10-20\% of the observed enrichment values of the IGM. They did not include other enrichment methods such as supernovae.

At high redshifts, the outflow will encounter a more dense IGM environment than at late times. However, though the expansion of the metals due to the pressure from the outflow will stop quickly when it hits the IGM, the material will be carried along with the Hubble expansion, allowing it to cover large distances. \citet{Kirkpatrick2011} studied metal enriched outflows from AGN and found that the outflows would have difficulty moving metal enriched materials out further than 1 Mpc from the AGN, unless the black hole mass was greater than $10^9M_{\odot}$.

\subsection{Ly$\alpha$ pumping}
While collisional excitation can produce the optical Fe\textsc{ii} lines, the temperature is unlikely to be high enough to produce the UV Fe\textsc{ii} emission produced by the 8-10 eV levels of Fe\textsc{ii}. In producing the UV emission, continuum and Lyman-$\alpha$ pumping are likely to be significant. 
The Ly$\alpha$ from star-forming galaxies and from the other quasars causes Ly$\alpha$ pumping, which increases the Fe\textsc{ii} emission in the quasar. This is seen in symbiotic stars and may be more prevalent in quasars due to the stronger UV radiation from the central source. Fluorescent excitation by Ly$\alpha$ can more than double the Fe\textsc{ii} flux in both  the UV and in the optical \citep{Sigut2003}. This would again require the quasars to be in close proximity to star forming galaxies.

Using the DR7QSO catalogue, the flux contribution from neighbouring quasars to each quasar can be calculated simply using the 3D distances between the Hectospec quasars with strong Fe\textsc{ii} emission and the SDSS quasars within the redshift range and area. This uses the luminosity (from the SDSS $r$-band magnitudes) of the objects to calculate the emission received. Without Ly$\alpha$ emission line measurements (not found within our spectra), it is not possible to calculate the Ly$\alpha$ contribution alone.

To take into account the lifetime of a quasar and the light travel time, only the flux from objects within a distance of $c\times$3 quasar lifetimes was considered, assuming a lifetime of 10$^{8}$ years \citep{Hopkins2008}. This uses the assumption that a LQG had always been a LQG so there would naturally be an excess of Ly$\alpha$ within the system. The flux for every quasar within a distance of $3\times c$ is summed at the positions of each of the strong and ultra-strong quasars. 
The flux from the additional quasars is of the order of $10^{-5}$J s$^{-1}$ m$^{-2}$, which is negligible in comparison to the flux from the central black hole ($\sim10^{38}$J s$^{-1}$ m$^{-2}$). 

However, these calculations do not take into account Ly$\alpha$ or flux from others sources such as star-forming galaxies so will be an underestimate of the actual flux received at the quasar. \citet{Haines2001} found a quasar within the LQG residing between two merging clusters and lying within a band of star-forming galaxies. If this is the case with the quasars studied here, the flux from the star-forming galaxies will play a much larger role than the Ly$\alpha$ flux from other nearby quasars.

\section{Summary}

The area of the CCLQG has a significantly higher number of strong and ultra-strong UV Fe\textsc{ii} emitters than the control fields tested. The only unusual feature in this region is the existence of 3 LQGs. Previously, there has been no indication that the environment within a LQG is unique. However, the high level of Fe emission may indicate there is an environmental difference which would cause this metal increase. 

If quasars in LQGs trace the LSS, it is likely there will be an increase in the number of clusters in this area (as clusters are also believed to trace the LSS). This would mean there may be more major mergers in this region as there is a higher density. A large number of major mergers may lead to a high number of starburst galaxies. This would later lead to a larger number of supernova and an increase in the iron abundance. 
An increase in the number of star forming galaxies would also raise the Ly$\alpha$ flux density and would increase the level of Ly$\alpha$ pumping.

There is currently no data on the other objects in these regions such as star forming galaxies and galaxy clusters. Without more data, it is not possible to comment on exactly what effect the environment has on these quasars.

\chapter{Conclusions and Future Work}

\section{Summary of Conclusions}

\subsection{Quasar-Cluster Proximity as a Function of Redshift}

Using the 2D KS test on both fields together, the distribution for the observed data for the 2D projected separation between the quasar and the closest cluster centre as a function of redshift proved to be significantly different from the control data sample. In Figure \ref{fig:sep_each}, there appears to be a deficit of quasars lying close to cluster centres for $0.4<z<0.8$, which is not seen as prominently in the control sample (Figure \ref{fig:sep_closestsim}). 
This would indicate quasars at $0.4<z<0.8$ prefer to lie in less dense environments. 
When the fields are taken separately, this difference in the distributions is still significant for each field, indicating this relation is not dependent on the field. This would agree with previous work that AGN and quasars avoid high density areas (e.g., \citealt{Popesso2006,Silverman2009,Gavazzi2011}).

The observed distribution of the 2D projected separation between the quasar and the BCG in the closest cluster (Figure \ref{fig:sepbcg_obs}) is also significantly different from the distribution in the control sample (Figure \ref{fig:sepbcg_sim}). The observed quasars prefer to lie further away from the BCG than in the control sample. When the fields are taken separately, this significant difference is still found in the Stripe 82 but not in the COSMOS field. 

The observed distribution of the 2D separation between the quasars and the closest cluster centre and the quasar orientation angle with respect to the cluster major axis is significantly different from the control sample for the COSMOS field but not for the Stripe 82 field.
This may be a product of the cluster detection mechanism. The angle of the cluster major axis is affected by the distribution of the cluster members, which is affected by the cluster member selection process. The same detection method needs to be run on both fields to test if the differences in the orientation distributions for the different fields are due to the cluster detection method.

For the 2D separations between the quasar and the cluster richness, the difference between observed and control samples is significant for the COSMOS field, but not for the Stripe 82 field. This difference may be due to the difference in cluster selection criteria. The COSMOS catalogue uses 8 as the minimum number of members as opposed to Stripe 82 which uses 5 minimum members to define a galaxy cluster. 

\citet{Lietzen2009} found that groups of galaxies with a quasar lying at less than 2Mpc tended to be poorer than on average. In the COSMOS and Stripe 82 samples, the average cluster richness for clusters with a quasar lying at $<$2 Mpc is 11.6 compared to a richness of 8.9 for clusters lying at $>$2Mpc from a quasar. This suggests, for the richest clusters, quasars lie closer to the cluster centre. However, the quasar sample used by \citet{Lietzen2009} used only quasars in the redshift range $0.078<z<0.172$, which is below the redshift range for most of this work. When samples of $0<z<0.2$ and $0.2<z<0.4$ were compared, there was a change in the trend for quasars to lie closer to a poorer cluster at $z\sim$0.2. However, richer clusters are found at higher redshifts due to the limited field size, so this change in the trend may be due to selection effects. Larger fields would be needed, so richer clusters at lower redshifts could be found, to test this result.

\citet{Lietzen2009} also found that the richest clusters lay between 5 and 15$h^{-1}$ Mpc from a quasar. In contrast, in the COSMOS and Stripe 82 samples, 13 out of 14 of the richest clusters (richness $>$ 40) lie at 2D projected separations of $<5.5$ Mpc from a quasar; the 14th lies at 8.5 Mpc from a quasar. Given the small redshift range covered by \citet{Lietzen2009}, this change in the trend may be due to selection effects. A larger number of quasar-cluster pairs in the redshift range $0<z<0.4$ would be needed to support this result.

Using the separation ratio (which is the ratio of the 2D projected separation between the quasar and the closest cluster centre and the mean radius of the cluster), the quasar is estimated to lie inside a cluster for 34 out of 677 quasar-cluster pairs (i.e., 5\%). This is likely to be an overestimate, as the separation ratio relies on the 2D project distance between the quasar and the cluster centre. Given the large errors on the 3D separations, it is not possible to determine whether the quasar does lie in the cluster or whether the separation ratio is due to a projection effect. More accurate redshifts for the galaxy clusters are needed to fully determine exactly where these quasars lie. 
However, as more quasars are likely to lie outside cluster than this result suggests, this result likely to be an overestimate and and still supports the idea that quasars avoid the highest density areas (e.g. \citealt{Sochting2002,Strand2008,Lietzen2009,Lietzen2011}). 

\citet{Haines2001} and \citet{Sochting2002} found quasars located between close, potentially merging clusters of galaxies. If the galaxy clusters align along filaments, and the merging is due to infall along those filaments, a quasar would be expected to lie at an orientation angle of $\sim0^{\circ}$ with respect to the cluster major axis. This would indicate the quasars lay within the filament and would likely lie along a line connecting the major axes of the two clusters. However, there is no evidence that quasars have any preferred orientation angle with respect to the closest cluster major axis and this does not evolve with redshift. So while some quasars may be triggered by the tidal forces created by merging galaxy clusters (and potentially major mergers when the galaxy clusters do collide), this is not a dominate mechanism (as suggested by \citealt{Shirasaki2009}). And this mechanism is also not the preferred mechanism at any redshift. 

There is no obvious relation between the orientation angle between a quasar and the major axis of the closest galaxy cluster and the richness of the cluster.

The Landy-Szalay estimator shows that at small separations between the quasar and the closest cluster centre, the clustering is greater in the COSMOS field than in the Stripe 82 field. There is also an increase in the clustering at $r=8-9$ Mpc in the Stripe 82 field at low redshifts ($0<z<0.4$), which is not shown in the COSMOS field. This may be a result of the smaller field size for COSMOS.

\subsection{Quasar-Cluster Proximity as a Function of Luminosity}

Using the t-test and comparing the means for faint ($M_r > -23.0$ mag) and bright ($M_r < -23.0$ mag) quasars, there is no difference between the two magnitude samples for the 2D separations or the cluster richness. 

For the bright quasar sample, there is no difference between the observed and control samples from either the t-test or the 2D KS test. This suggests that quasars with $M_r < -23$ mag do not lie in any preferred position with respect to clusters or with respect to the cluster richness. This may be due to the limited number of quasars with $M_r < -23$ mag in the COSMOS and Stripe 82 quasar samples. 

However, there is a difference between the distributions for the 2D separations for faint quasars in the observed and control samples, shown in the results from both the t-test and the 2D KS test. This suggests that faint quasars lie in preferred positions with respect to galaxy clusters. The 2D KS test shows that there is a difference between the distribution of the 2D projected distance between the quasar and the closest cluster centre as a function of the quasar absolute $r'$ magnitude for faint observed and control quasars. Figure \ref{fig:sep_abs_faint} shows that observed quasars with $M_r>-21$ mag prefer to lie closer to the closest cluster centre than quasars from the control sample. 

However, this result is dependent on the sample size used in the t-test. Using a reduced sample for the faint quasars to match the size of the bright quasar sample, the significance between the faint observed and control quasars for the 2D projected separations between the quasar and the closest cluster centre, and the quasar and the closest cluster member, disappeared. The normalised histograms in Figures \ref{fig:hist_faint} - \ref{fig:hist_bright} do suggest a possible difference between the observed and control quasar samples for both the faint and to a lesser extent brighter quasars. A larger sample of both faint and bright quasars would clarify this difference and significance. A larger number of control samples would also help to improve the statistical significance.

Using the separation ratio, the quasars within a cluster are brighter at higher redshifts ($z>0.8$) than lower redshift ($z<0.6$). However, as mentioned previously, the separation ratio is based on the 2D separation between the quasar and the closest cluster centre, as the errors on the 3D separations are too large to determine if the quasar does lie inside the cluster.  

\citet{Strand2008} found that brighter quasars ($M_i < -23.25$ mag) lay in denser environments than dimmer quasars ($M_i > -23.25$ mag). For the Stripe 82 and COSMOS samples, using the same magnitude boundary, no difference in the environment is found. The brighter quasars lie on average, 6.2$\pm$4.0 Mpc from the closest cluster centre, while the dimmer quasars lie 5.8$\pm$3.8 Mpc from the closest cluster centre. This does not alter if the redshift range is reduced to $z<0.6$, to match that used by \citet{Strand2008}. 

These is no change with redshift (over the range $0<z<1.2$) for the positions of the quasars with respect to the cluster or the cluster richness as a function of absolute quasar magnitude. There is also no preferred orientation between the quasar and the cluster major axis for either bright or faint quasars. 

\subsection{Treatment of Statistical Significance}

To improve the statistical significance of the results in Chapters 3 and 4, a larger number of control samples should be used. Due to time limitations, only one control sample was used. With only one control sample, it is possible for trends to appear in the control sample which would cause the results for the statistical significance to be inaccurate. For this, the random sampling of the RAs and DECs would be repeated  multiple times (ideally $\sim1000$, using the number of control samples used in previously literature as a reference). Increasing the number of control samples would give a better indication of the significance of any trends observed and improve the statistics. 

Using a larger number of control samples, it would be possible to improve the conclusions drawn from Figures 3.10-3.13 and Figures 4.6-4.11. These figures show the distribution of angles between the quasar and the closest cluster major axis.  In these Figures, one control sample is not enough to accurately show if any trends seen are real. A larger number of control samples would create a band of possible angles of the histogram. Any peaks in the histograms over or below this band is likely to be significant.

The two-point correlation function works best with a large number of random samples (often 1000+). In Chapter 3, the Landy-Szalay estimator was used because, of the three estimators studied, the Landy-Szalay estimator works best with a limited number of random samples (\citealt{Kerscher2000}). However, the results from the two-point correlation function would also be improved by a larger number of control samples.

\subsection{Star Forming Galaxies}
Spectra of a selection of 680 star forming galaxies, red galaxies, and AGN were taken using the IMACS instrument on the Magellan Baade telescope. From the 680 spectra taken, the redshifts of 515 galaxies were calculated. The objects in the spectra were classified and the star formation rates calculated, where possible. 33 AGN were classified using the BPT plot. Only one of the AGN is brighter enough to be classed as a quasar. 70 objects from the IMACS spectra were classified as star forming galaxies. The process used to select the objects for the IMACS spectra selected for star forming galaxies. Therefore, if not classed as an AGN, and the spectra does not show any evidence of broad emission lines (which are a signature of AGN), the galaxies were also classed as star forming galaxies.

The star formation rates were calculated using the H$\alpha$ and [O\textsc{ii}] emission lines, and the UV emission from GALEX, and compared. There is a systematic difference between the H$\alpha$ SFRs and the SFRs found using [O\textsc{ii}], a well known problem (e.g., \citealt{Gilbank2010}). This may be due to the extinction correction. However, [O\textsc{ii}] is more affected by metal abundances within the galaxy, which would also affect the SFR value found (\citealt{Kewley2004}). 

There is a large difference between the H$\alpha$ SFRs and the SFRs using UV. This may be due to the dust attenuation model used, as an averaged value of the dust extinction was used due to the lack of infra-red data (\citealt{Salim2007}). However, there are also differences between the stellar populations and the time-scales measured by the H$\alpha$ and UV emission, which could also account for these differences (\citealt{Gilbank2010}). 

\subsection{The Environment of AGN and Star Forming Galaxies}

None of the relations for the 2D projected separations or the 3D separations between an AGN and the closest star forming galaxy are statistically significant. 

Three of the AGN lie at the same redshifts as three of the clusters from the CFHT data. These clusters were found by Ilona S\"ochting and the redshifts were estimated using the red sequence. Three AGN and 10 star forming galaxies lie at the same redshift, and are potentially part of the same structure at $z\sim0.289$. The redshifts of the clusters need to be spectroscopically confirmed to support this result. The three galaxy clusters have the same orientation angle (within errors). 
The AGN appear to lie within the filament and are relatively faint (with magnitudes of -17, -19, and -21). 

Taken individually, the brightest AGN, which would be classed as a quasar, has the smallest 3D separation between the quasar and a star forming galaxy and lies in the middle of contours in Figure \ref{fig:lowpeak}. This quasar has $M_r = -24.34$ mag and is the only AGN from the IMACS sample bright enough to be classed as a quasar. 
This quasar does not lie close to the clusters, but does lie close to star forming galaxies at the same redshift. As quasars are expected to have different fuelling mechanisms than low-luminosity AGN (\citealt{HopkinsHernquist2009}), quasars are likely to reside in different environments to AGN. Quasars are believed to be triggered by major mergers, which will occur in galaxy groups, which are not as density as galaxy clusters but contain galaxies with a supply of gas and have a lower velocity dispersion. This quasar appears to lie in a less dense region, surrounded by a few star forming galaxy and could be part of a galaxy group.  

Contour plots of this area of sky show a possible evolution of preferred environment over the redshift range $0.225<z<0.425$.
The contour plots show AGN lying in various different areas. In the high redshift slice ($0.375<z<0.425$), all AGN appear to avoid the high density regions, preferring to lie on the edges of contours, therefore on the edges of the mass distributions (as seen in \citealt{Sochting2004}). In the redshift slice $0.275<z<0.325$, most AGN also lie on the edges of the contours. However, there are a few AGN which lie in the centres of the contours, i.e., in the middle of the mass distributions. There are less AGN in the lowest redshift slice ($0.225<z<0.275$) so it is difficult to determine the preferred environment. Two of the five AGN lie in the middle of the contours, while two lie on the edges, and one is too close to the edge of the field to determine its environment. 

This change in preferred positions may indicate a change in preferred small-scale environment with redshift over the range $0.225<z<0.425$. 
However, the data used for the contours comes from the IMACS spectra, which is not a complete sample of star forming galaxies. A more complete sample of the star forming galaxies in this region of sky would be needed properly studying this possible evolution with redshift.

\subsection{Ultra-Strong Fe{\sc{ii}} Emitters}

The area of the CCLQG has a significantly higher number of strong and ultra-strong UV Fe\textsc{ii} emitters than the control fields tested. The only unusual feature in this region is the existence of three LQGs (two of which are statistically significant), including the CCLQG (\citealt{Clowes2011}). Previously, there has been no indication that the environment within a LQG is unique. However, the high level of Fe emission may indicate there is an environmental difference which would cause this metal increase. 

It appears that the strong and ultra-strong UV Fe\textsc{ii} emitters may lie between LQGs, rather than within the groups.

If quasars in LQGs trace the LSS, it is likely there will be an increase in the number of clusters in this area (as clusters trace the LSS). This would mean there may be more major mergers in this region as there is a higher density. A large number of major mergers may lead to a high number of starburst galaxies which would then later lead to a larger number of supernova and an increase in the iron abundance. 
An increase in the number of star forming galaxies would also create an increase in the Ly$\alpha$ in the region, which would increase Ly$\alpha$ pumping \citep{Sigut2003}. Using other quasars in the area, the level of emission received from outside sources is negligible compared to the emission from the quasar's central source. However, this does not include emission from any star forming galaxies near the quasar, which would increase the Ly$\alpha$ emission.

Another mechanism to increase the Fe\textsc{ii} emission is microturblence (e.g., \citealt{Vestergaard2001,Verner2004,Bruhweiler2008}). This is likely to occur in the intermediate line region, close to the central source. More work on the close environments around these quasars is needed to assess how the environment in a LQG could increase the microturbulence in the centre of the host galaxy. 

There is currently no data on the other objects in these regions such as star forming galaxies and galaxy clusters. Without more data, it is not possible to comment on exactly what effect the environment has on these quasars.

\subsection{Methods for Environments}
Two methods have been used to study the positions of quasars with respect to galaxy clusters, the two point correlation function and a simpler method, which does not involve binning the data. The two point correlation function has been used to compare the clustering from the COSMOS and Stripe 82 fields to other published data. This allows a comparison of the fields and the effect of binning the data. 

As the two point correlation function involves binning the data, there are problems when there are small numbers in a bin. This is also a problem with viewing the data and using statistical tests, such as the t-test and the KS test, and can only be resolved with a larger data set.

Using simple plots and statistics gives a good idea of the significant of the relationships between different characteristics of quasar-cluster pairs (such as richness and absolute magnitude), something that can not be done using only the two point correlation function without choosing bins. When binning the data, the results are dependent on the binning selected, which could smear out any results or create false results. 

Angle method is a good indication of the orientation, which is not given in the correlation function. Finding the orientation angle of the quasar with respect to the closest cluster major axis will give a better indication of the position of a quasar around clusters. Clusters trace the large scale structure and lie along filaments. Angular information gives more details about where quasars lie in relation to filaments and large scale structures. The two point correlation function does not provide angular information.

\section{Future Work}
\subsection{Unusual Spectra}
During the IMACS observations, there were a few spectra which have not been fully explained.

\begin{figure}[!ht] 
\centering
\includegraphics[scale=0.7]{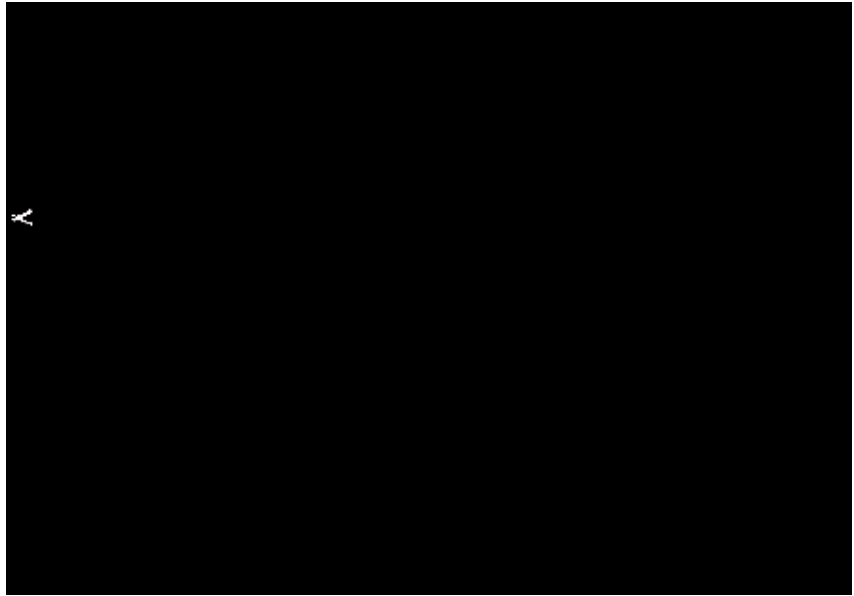}
\caption[High redshift spectra with usual emission line]{\small{Spectra for QSOcand\_02, a high redshift quasar, from the IMACS spectra}}
\label{fig:QSOcand02} 
\end{figure}

Figure \ref{fig:QSOcand02} shows the spectrum of a quasar at $z=2.157\pm0.007$. The redshift for this quasar was calculated from the lines shown in Figure \ref{fig:QSOcand02}. However, the emission line at 9540\AA\ has not been identified. At $z=2.157$, this emission line would be at 3021\AA. The spectra has been checked to ensure this emission line is real and not an artefact of the data reduction process. This line was not seen in the other spectra.

It was possible to estimate the UV Fe\textsc{ii} emission for the 2255-2650\AA\ bump and was estimated to have an equivalent width of 35.1 using \citet{Weymann1991} and a monochromatic flux value of 0.28 using \citet{Hartig1986}. However, there is a gap in the spectra due to a zero'th order contamination in the middle of this wavelength range, so these values are only estimates. It is not possible to measure the weak Fe\textsc{ii} bump at 2040-2130\AA, as the whole wavelength range from 2052\AA\ to 2164\AA\ is missing. If these values are accurate, these Fe\textsc{ii} emission values would mean this quasar had an average Fe\textsc{ii} emission. However, better spectra with no missing areas would be needed to accurately assess the Fe\textsc{ii} emission strength for this quasar.

\begin{figure}[!ht]
\centering
\subfigure[Object 3444]{
\includegraphics[scale=0.5]{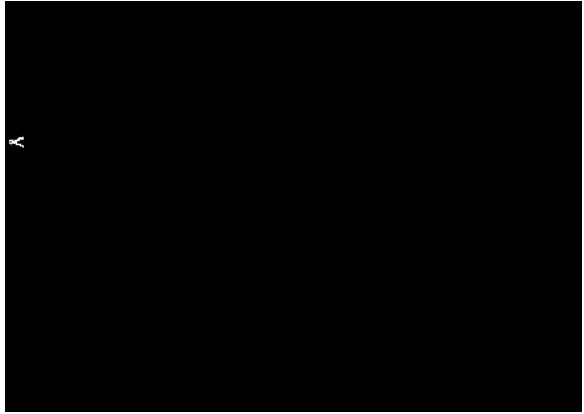}
\label{fig:obj3444}
}
\subfigure[Object 3867]{
\includegraphics[scale=0.5]{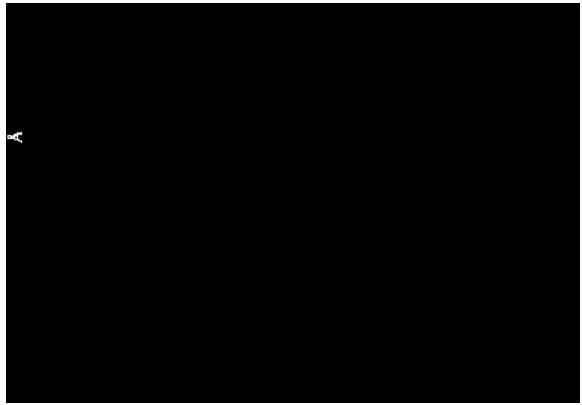}
\label{fig:obj3867}
}
\subfigure[Object 6455]{
\includegraphics[scale=0.5]{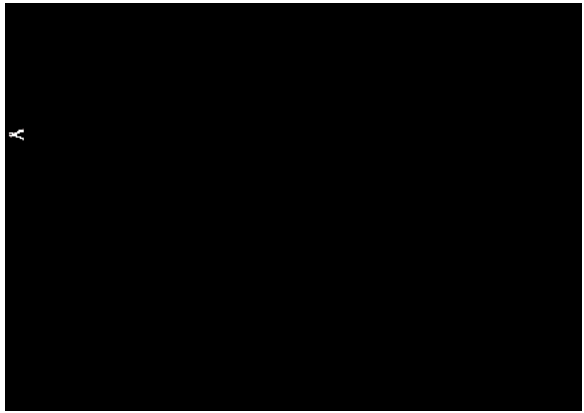}
\label{fig:obj6455}
}
\caption[Unidentified spectra]{\small{Examples of objects where the spectra show strong flux but no emission lines from the IMACS spectra.}}
\label{fig:no_em_lines}
\end{figure}

There were also 17 spectra with strong fluxes but no emission lines so it was not possible to calculate the redshifts. Figure \ref{fig:no_em_lines} shows a sample of the spectra from these objects. These objects do sometimes show evidence of weak absorption lines but no emission lines. More work will be needed to identify these objects.   
 
\subsection{Quasar Environments}

A minimal spanning tree could be used to connect the clusters and determine any filamentary structure in the COSMOS and Stripe 82 fields. The method of calculating the angular orientation of the quasar can then be used to assess the position of the quasar with respect to these filaments. The angular information could be extended to the orientation of the quasar with respect to more than the closest cluster to show the wider environment.  

There are quasars within the COSMOS and Stripe 82 fields, which did not lie within 15 Mpc of a galaxy cluster. It would be interesting to study these, and if possible determine, if the small scale environment around these quasars is different to that of quasars lying within 15 Mpc of a cluster. This would help determine whether the small scale or large scale environment has more of an effect on a quasar, and whether this changes with redshift and luminosity.

For the filament at $z=0.289$ in the IMACS spectra, spectroscopic redshifts of the clusters would confirm this filament. Also, with a more complete sample of star forming galaxies, it would be possible to determine if there were any other galaxies in this area, and determine other strucutures like this. 
  
Radio data on the quasars within this filament would help to study the environments of radio AGN. Whether radio loud quasars lie in different environments is still under debate. However, it is thought that radio quiet quasars avoid peaks in density (\citealt{Sochting2004}) so if these AGN are lying within a filament, they should be radio loud.

\clearpage
\subsection{Fe{\sc{ii}} Quasars}

SDSS spectra cover the wavelength range 3850-9000\AA. As the UV Fe\textsc{ii} bump is at 2255-2650\AA , the Fe\textsc{ii} feature will be visible within the redshift range $0.8<z<2.3$. The SDSS DR7QSO catalogue contains 70,980 quasar spectra within this redshift range. The UV Fe\textsc{ii} emission has been measured using an automated program written by Srinivasan Raghunathan at the Universidad de Chile. These spectra need to be checked as some have extremely strong UV Fe\textsc{ii}, which is unlikely to be real emission. However, within this catalogue, there will be quasars with strong and ultra-strong Fe\textsc{ii} emission. The environments and positions of these quasars with respect to the LSS need to be studied. 

Information about the properties of galaxies and clusters of galaxies in the environment of a LQG is needed to study how these quasars could have acquired an excess of iron. If the galaxies in the area of the LQG show evidence for an increased level of star formation or past starbursts, this excess of star forming galaxies could produce an iron excess. Ideally, the redshifts and SFR of galaxies within clusters at $1.1<z<1.7$ could be used to study the exact environment in LQGs. However, cluster detection at $z>1.1$ is difficult, with most clusters being detected using X-ray or \textit{Spitzer} mid-IR (\citealt{Foley2011} and references therein). 

The Webster LQG (\citealt{Webster1982}) contains 4 quasars and lies at a redshift of $z=0.37$. At this redshift, it will be possible to obtain redshifts and other characteristics (such as star formation rates) for galaxies in and around a LQG.  This could then be applied to the higher redshift LQG, under the assumption that they have similar characteristics at higher redshifts. If possible, the Fe\textsc{ii} emission will be measured but this will depend on the sensitivity in the UV of the telescope used. Spectra from this LQG will allow more information to be gathered about the exact environments in LQGs.

\chapter{Statistics tests and models}\label{append1}

\section{The Models}
\begin{figure}[!h]
\centering
\includegraphics[scale=0.5,angle=-90]{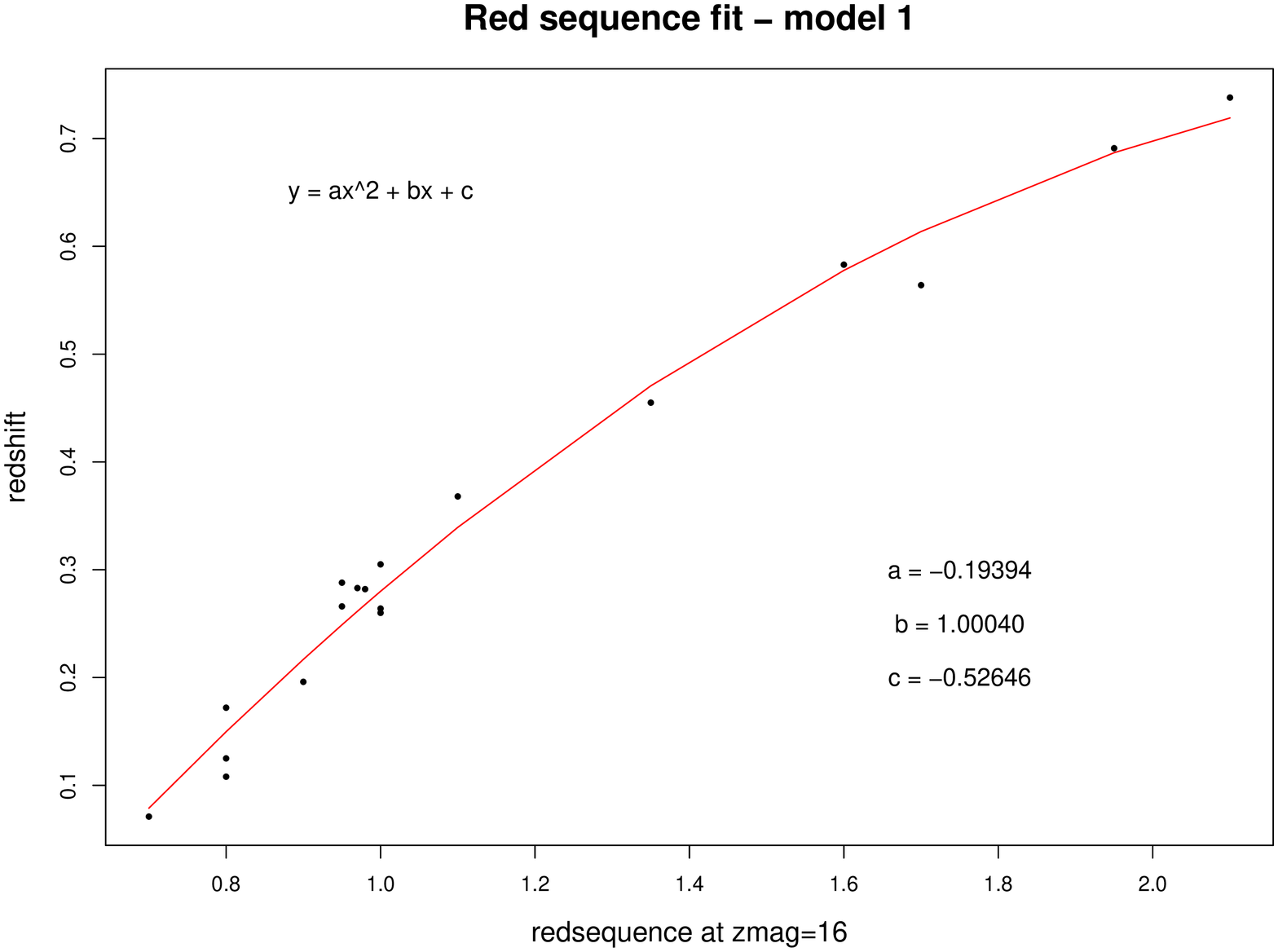}
\caption[Fitting model 1 to red sequence]{\small{Model 1}}
\label{model1}
\end{figure}

\begin{figure}[!h]
\centering
\includegraphics[scale=0.5,angle=-90]{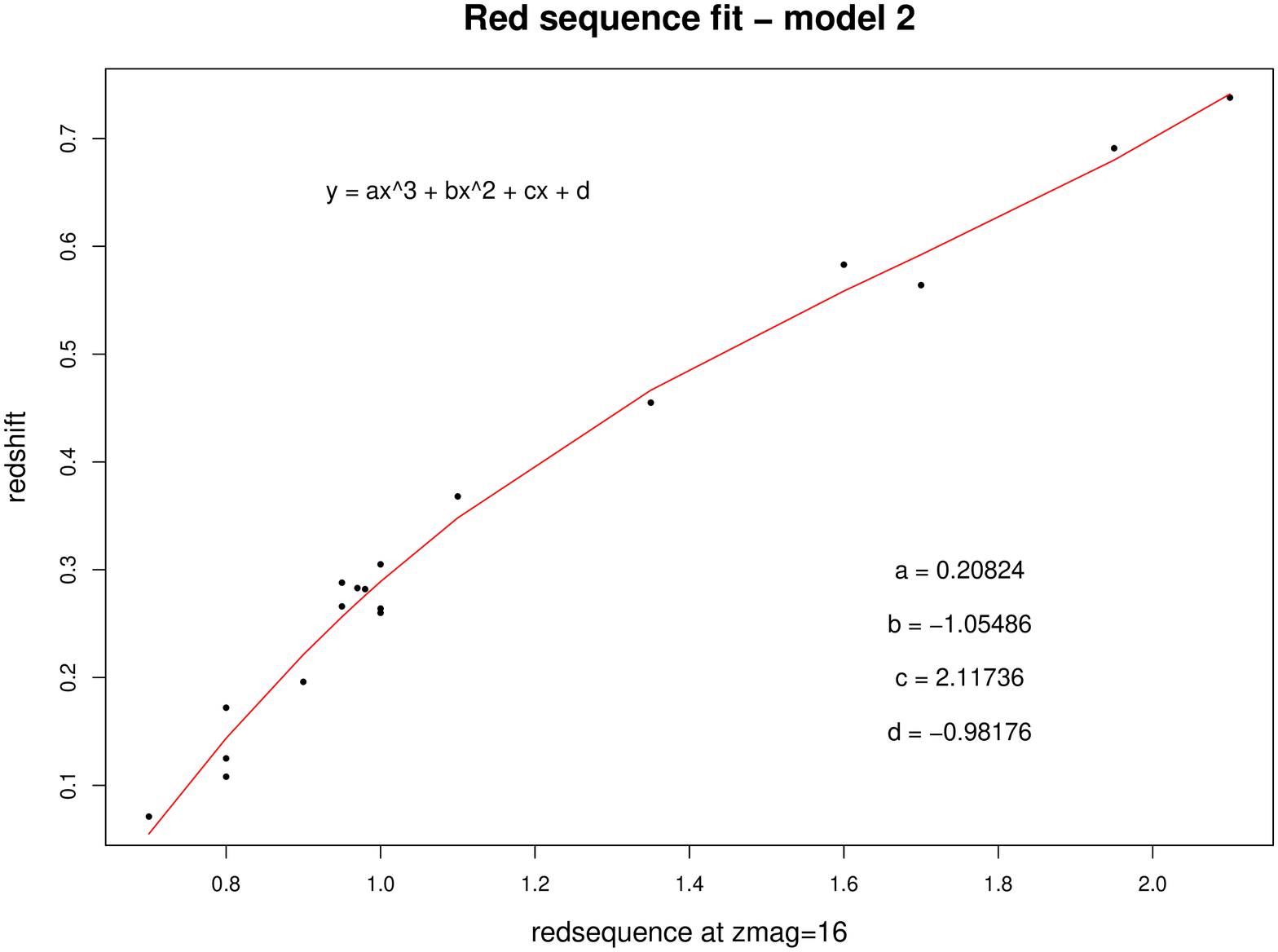}
\caption[Fitting model 2 to red sequence]{\small{Model 2}}
\label{model2}
\end{figure}

\begin{figure}[!h]
\centering
\includegraphics[scale=0.5,angle=-90]{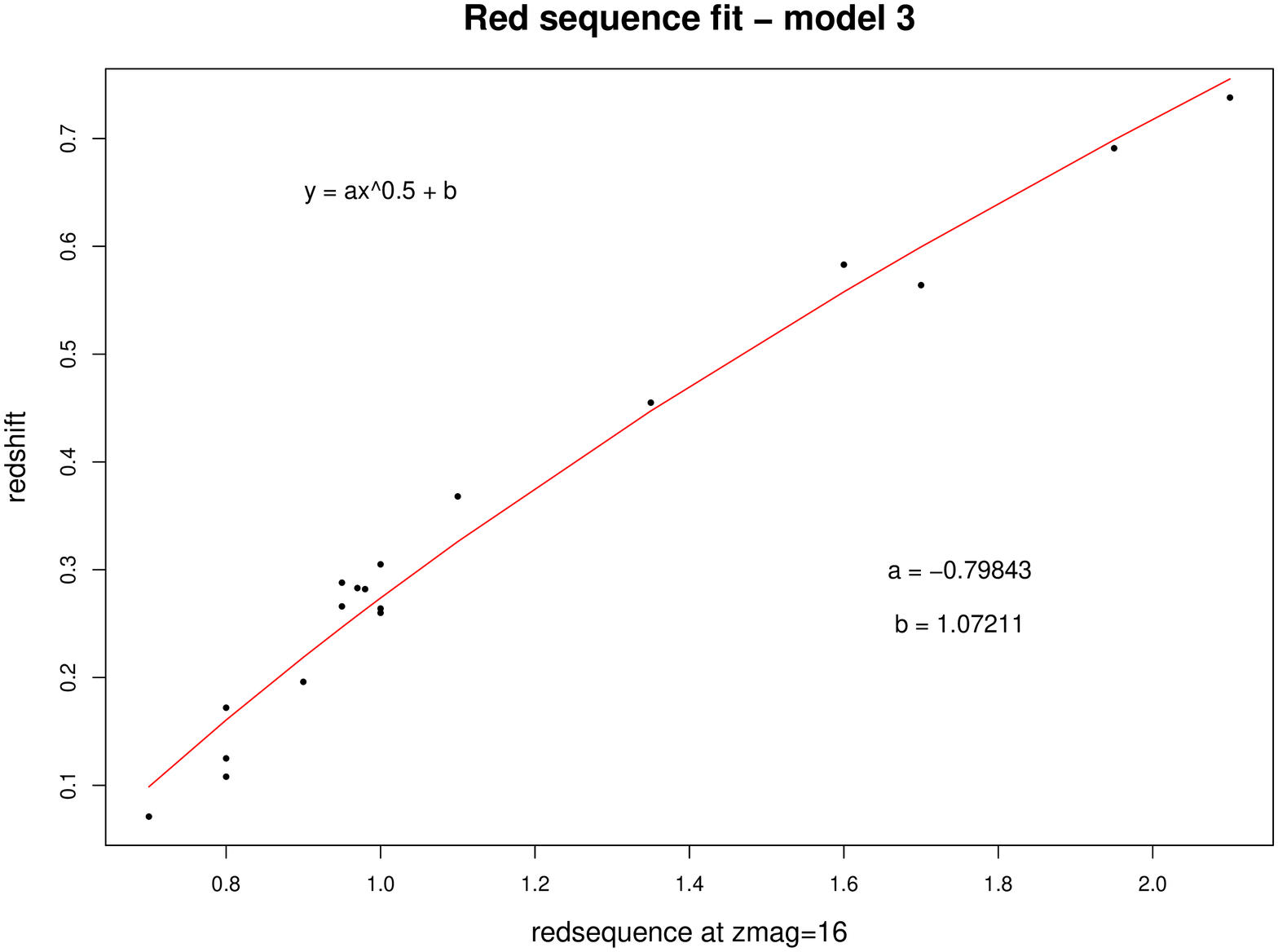}
\caption[Fitting model 3 to red sequence]{\small{Model 3}}
\label{model3}
\end{figure}

In order to fit the red sequence, a number of different models were tested. Four final models were selected to test. The models are given by:
\begin{equation}
\mbox{Model 1} : y = ax^2 + bx + c  \qquad \textrm{(Figure \ref{model1})}
\end{equation}

\begin{equation}
\mbox{Model 2} : y = ax^3 + bx^2 + cx + d \qquad \textrm{(Figure \ref{model2})}
\end{equation}

\begin{equation}
\mbox{Model 3} : y = ax^{1/2} + b  \qquad \textrm{(Figure \ref{model3})}
\end{equation}

Figures \ref{model1} to \ref{model3} show the value of the red sequence of a cluster at $zmag = 16$ against the redshift of the galaxy cluster. The fitted line shows the model fit. 

\begin{figure}[!h]
\centering
\includegraphics[scale=0.5,angle=-90]{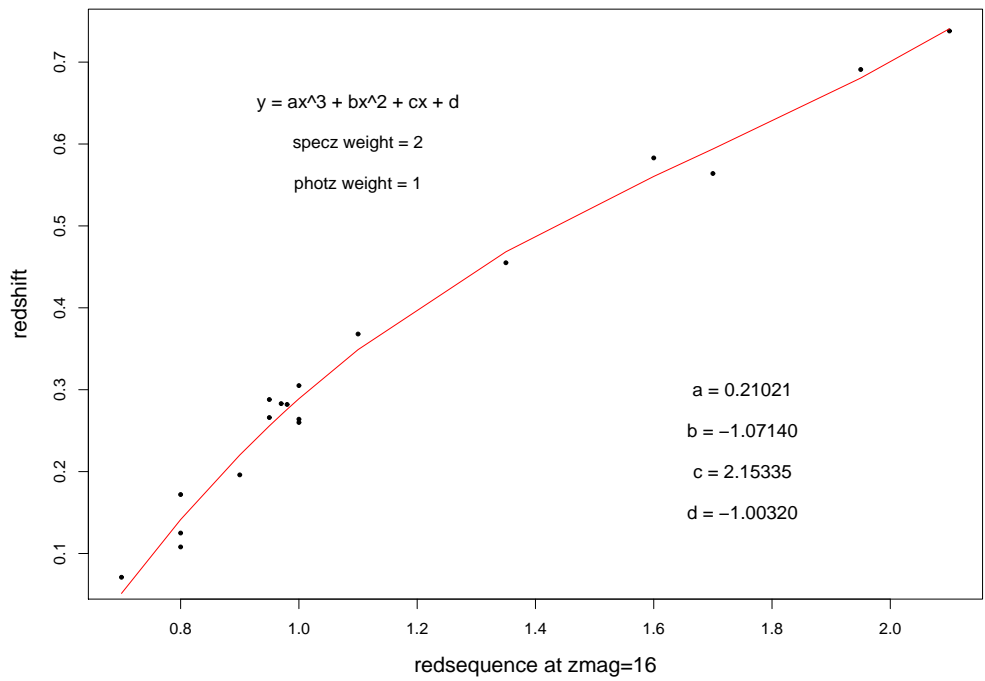}
\caption[Fitting a weighted model 2  to red sequence]{\small{Model 2 with weights}}
\label{model2w}
\end{figure}

A weighted version of model 2 was also tested. As the spectroscopic redshifts are more likely to be accurate than the photometric redshifts, a weight of two was given to points with spectroscopic redshifts and a weight of one to points with photometric redshifts. Figure \ref{model2w} shows the fit of the model to the data for the weighted version of model 2.

\clearpage
\section{Statistical Tests}
The residual standard error is the square root of the error variance, and measures the deviation from the data sample and the fitted values from the model.

Multiple R$^2$ is the fraction of the total variance explained by the model, and should be close to one. R$^2$ is defined as the ratio of the sum of squares explained by a regression model and the "total" sum of squares around the mean\footnote{http://www.people.vcu.edu/~nhenry/Rsq.htm}.

The Fisher's F-test compares the difference between 2 variances, and divides the larger one by the smaller. If the ratio is one, then the variances are the same. The F-statistic should, therefore, be close to one for the best fit model. A larger ratio (F$>>$1) indicates the models are significantly different (as the top of the ratio of the variances will be a lot larger than the bottom). When comparing the models here, the model with the smallest value from the F-test will be the one which fits best to the observed data.


The Akaike's Information Criterion (AIC) test is a penalised log-likelihood and is a means of comparing models, rather than a hypothesis test. This test gives a measure of the goodness of the fit and ranks the models in order of their fit to the data, with the model with the lowest AIC value giving the best fit. 
A value for the AIC test is found using:
 
\begin{equation}
\label{AIC}
\mbox{AIC} = -2 \times \log(-\mbox{likelihood}) + 2(p+1)
\end{equation}

where $p$ is the number of parameters in the model. This can therefore be used to compare models.

\section{Selecting the Model}

Each model was compared to the data to judge how well it fitted the data using the Residual standard error, multiple R$^2$, and F-test. Table \ref{tab:tests} shows the results from the statistics tests the fit of each of the models to the data.

\begin{table}[!ht]
\caption[Results from statistics test for red sequence fit]{\small{The statistical tests results for each model for fit to data.} }
\centering
\begin{tabular}{ c | c c c }
Model                & Residual standard error & Multiple R$^2$ & F-test  \\ \hline 
Model 1              & 0.02715                 & 0.9831         & 435.9   \\
Model 2              & 0.02451                 & 0.9871         & 358.1    \\
Model 3              & 0.02944                 & 0.9788         & 738.2    \\
Model 2 with weights & 0.03116                 & 0.9897         & 446.6    
\end{tabular}
\label{tab:tests}
\end{table}

Comparing the values for Table \ref{tab:tests}, model 2 appears to be the best fit to the data points. Model 2 has the lowest value for the F-test, suggesting it has the least significant difference between the model and the data. It also has the lowest residual standard errors and the lowest p-value. 

When viewing the fits of the models (Figures \ref{model1} - \ref{model2w}), the fit near the high redshift end needs to be considered. Ideally, this fit should allow us to extrapolate to high redshift and obtain higher redshift clusters. Model 1 fits well at the high redshift end but the function appears to flatten at higher values for the red sequence fit. This would mean the redshifts would not increase pass a certain value of $zmag$. Model 2 fits well but the function appears to start increasing at around $zmag$ = 1.95, making it difficult to predict what will happen at a higher value of $zmag$. Model 3 appears more predictable at higher values of $zmag$ and therefore, at higher redshifts, but does not appear to fit as well as the other models at lower redshifts.

As model 2 appears to fit the best at lower redshifts, weights were added to the model to adjust the fit to give more weight to the spectroscopic points, which are more likely to be accurate than the photometric points. This does adjust the fit at higher values of $zmag$, though not by a large amount.

The models were also compared to one another. Table \ref{tab:comptests} shows the comparison of the fit of each model to the fits from the other models, using anova in R.

\begin{table}[!ht]
\caption[Comparison of models]{\small{Comparison of models using different statistical tests compared between models. }}
\centering
\begin{tabular}{ c  c c c c }
Model                & RSS      & F-test  &  Pr($>$F)  &   AIC     \\ \hline
Model 1              & 0.01106  &         &            &  -74.026  \\
Model 2              & 0.00841  & 4.4090  & 0.05436    &  -76.954  \\
Model 3              & 0.01387  & 4.5451  & 0.03012    &  -71.947  \\
Model 2 with weights & 0.01359  & 0.2352  & 0.79345    &  -77.141  
\end{tabular}
\label{tab:comptests}
\end{table}
 
Pr($>$F) is the p-value for the F-test. A large p-value indicates the model is not significantly different to the data, which is required in this case. 
The residual sum of squares (RSS) is the a measure of the discrepancy between the fitted model and the data, with a small RRS meaning the model is a good fit to the data. 

Model 2 with weights has the largest p-value and the lowest value for the AIC, suggesting that this is the best of the four models to use. Therefore, model 2 with weighted values for the redshift will be used to correlate the red sequence.

\chapter{Appendix 2}

The disk included with this thesis contains the data used in the Chapters.

\section{Quasar-cluster pairs}\label{databaseinfo}
 
The file \emph{quasar-cluster\_pairs.txt} contains the quasar-cluster pairs in used in Chapters 2-4. 

The columns are as follows:
\begin{itemize}
\item \emph{Field}: the quasar is from the COSMOS or Stripe 82 field
\item \emph{idq}:   quasar ID
\item \emph{zq}:    quasar redshift
\item \emph{ra\_q}:  RA(J2000) for the quasar
\item \emph{dec\_q}: DEC(J2000) for the quasar
\item \emph{idc}:   ID for the closest cluster 
\item \emph{zc}:    cluster redshift
\item \emph{zcerr}:  error on the cluster redshift
\item \emph{ra\_c}:  RA(J2000) for the cluster centre
\item \emph{dec\_c}: DEC(J2000) for the cluster centre
\item \emph{sep\_qc\_zq}: separation between quasar and closest cluster centre using the quasar redshift at the quasar epoch
\item \emph{degree}: angle between the line running through constant RA of the quasar and the cluster centre
\item \emph{UMAG}: u band magnitude
\item \emph{GMAG}: g band magnitude
\item \emph{RMAG}: r band magnitude
\item \emph{IMAG}: i band magnitude
\item \emph{ZMAG}: z band magnitude
\item \emph{absMr}: absolute $r$ band magnitude using $\alpha$=0.5 and Equation \ref{eq:absmag}
\item \emph{members}: Number of galaxies within the cluster
\item \emph{richness}: Number of galaxies with magnitude between the magnitude of the BCG, M$_{BCG}$, and M$_{BCG}$+3
\item \emph{inertia\_ang}: angle of the orientation of the cluster major axis using the inertial tensor method between 0 and 180$^{\circ}$
\item \emph{average}: average of the angles for the cluster orientation from calculating the inertial tensor errors 
\item \emph{stan\_dev}: standard deviation of the orientation of cluster angles from calculating the errors
\item \emph{maj\_axis\_Mpc}: length of the cluster major axis in Mpc
\item \emph{min\_axis\_Mpc}: length of the cluster minor axis in Mpc
\item \emph{meanRadius}: mean of the length of the major and minor axes in Mpc
\item \emph{ra\_bcg}: RA(J2000) of the brightest cluster galaxy
\item \emph{dec\_bcg}: DEC(J2000) of the brightest cluster galaxy
\item \emph{sep\_bcg}: separation between the quasar and the BCG using the redshift of the quasar at the quasar epoch
\item \emph{dist\_redge}: distance between the quasar and the right edge of the field
\item \emph{dist\_ledge}: distance between the quasar and the left edge of the field
\item \emph{dist\_tedge}: distance between the quasar and the top edge of the field
\item \emph{dist\_bedge}: distance between the quasar and the bottom edge of the field
\item \emph{sepRatio}: sep\_qc\_zq divided by the mean of the major and minor axes
\item \emph{closest\_gal}: separation between the quasar and the closest cluster member using the redshift of the quasar at the quasar epoch
\item \emph{degree\_clqso}: angle between the cluster major axis and the quasar
\item \emph{idc\_3D}: the ID of the cluster with the shortest 3D separation
\item \emph{ra\_3D}: the RA(J2000) of the cluster with the shortest 3D separation
\item \emph{dec\_3D}: the DEC(J2000) of the cluster with the shortest 3D separation
\item \emph{zc\_3D}: the redshift of the cluster with the shortest 3D separation
\item \emph{zcerr\_3D}: the error on the redshift of the cluster with the shortest 3D separation
\item \emph{sep\_3D}: 3D separation between quasar and closest cluster centre
\item \emph{sep\_3D\_err}: error on the 3D separations using zerr\_3D
\end{itemize}

\section{Quasar-cluster control pairs}\label{databaseinfo_control}
 
The file \emph{quasar-cluster\_pairs\_control.txt} contains the quasar-cluster pairs from the control fields in used in Chapters 2-4. 

The columns are as follows:
\begin{itemize}
\item \emph{Field}: the quasar is from the COSMOS or Stripe 82 field
\item \emph{zq}:    quasar redshift
\item \emph{ra\_q}:  RA(J2000) for the quasar
\item \emph{dec\_q}: DEC(J2000) for the quasar
\item \emph{idc}:   ID for the closest cluster 
\item \emph{zc}:    cluster redshift
\item \emph{zcerr}:  error on the cluster redshift
\item \emph{ra\_c}:  RA(J2000) for the cluster centre
\item \emph{dec\_c}: DEC(J2000) for the cluster centre
\item \emph{sep\_qc\_zq}: separation between quasar and closest cluster centre using the quasar redshift at the quasar epoch
\item \emph{degree}: angle between the line running through constant RA of the quasar and the cluster centre
\item \emph{absMr}: absolute $r$ band magnitude using $\alpha$=0.5 and Equation \ref{eq:absmag}
\item \emph{members}: Number of galaxies within the cluster
\item \emph{richness}: Number of galaxies with magnitude between the magnitude of the BCG, M$_{BCG}$, and M$_{BCG}$+3
\item \emph{inertia\_ang}: angle of the orientation of the cluster major axis using the inertial tensor method between 0 and 180$^{\circ}$
\item \emph{average}: average of the angles for the cluster orientation from calculating the inertial tensor errors 
\item \emph{stan\_dev}: standard deviation of the orientation of cluster angles from calculating the errors
\item \emph{maj\_axis\_Mpc}: length of the cluster major axis in Mpc
\item \emph{min\_axis\_Mpc}: length of the cluster minor axis in Mpc
\item \emph{meanRadius}: mean of the length of the major and minor axes in Mpc
\item \emph{ra\_bcg}: RA(J2000) of the brightest cluster galaxy
\item \emph{dec\_bcg}: DEC(J2000) of the brightest cluster galaxy
\item \emph{sep\_bcg}: separation between the quasar and the BCG using the redshift of the quasar at the quasar epoch
\item \emph{dist\_redge}: distance between the quasar and the right edge of the field
\item \emph{dist\_ledge}: distance between the quasar and the left edge of the field
\item \emph{dist\_tedge}: distance between the quasar and the top edge of the field
\item \emph{dist\_bedge}: distance between the quasar and the bottom edge of the field
\item \emph{sepRatio}: sep\_qc\_zq divided by the mean of the major and minor axes
\item \emph{closest\_gal}: separation between the quasar and the closest cluster member using the redshift of the quasar at the quasar epoch
\item \emph{degree\_clqso}: angle between the cluster major axis and the quasar
\item \emph{idc\_3D}: the ID of the cluster with the shortest 3D separation
\item \emph{ra\_3D}: the RA(J2000) of the cluster with the shortest 3D separation
\item \emph{dec\_3D}: the DEC(J2000) of the cluster with the shortest 3D separation
\item \emph{zc\_3D}: the redshift of the cluster with the shortest 3D separation
\item \emph{zcerr\_3D}: the error on the redshift of the cluster with the shortest 3D separation
\item \emph{sep\_3D}: 3D separation between quasar and closest cluster centre
\item \emph{sep\_3D\_err}: error on the 3D separations using zerr\_3D
\end{itemize}

\section{IMACS Spectra}

The file \emph{IMACS\_spectra.txt} contains the input parameters used, and all of the parameters derived from the methods described in Chapter 5. 

The columns are as follows:
\begin{itemize}
\item \emph{Name}: Object name
\item \emph{Mask}: Number of mask used in observations
\item \emph{RA}: Right ascension (J2000)
\item \emph{DEC}: Declination (J2000)
\item \emph{Redshift}: Estimated galaxy redshift
\item \emph{zerror}: Estimated redshift error
\item \emph{Lines}: Number of lines used to estimate the redshift
\item \emph{Quality}: Quality of the spectra
\item \emph{Class}: Object classification
\item \emph{SFR(Ha)}: SFR using the H$\alpha$ emission line in M$_{\odot}$yr$^{-1}$
\item \emph{SFR([OII])\_Ken}: SFR using the [O\textsc{ii}] emission line and Equation from \citet{Kennicutt1998} in M$_{\odot}$yr$^{-1}$
\item \emph{SFR([OII])\_Kew)}: SFR using the [O\textsc{ii}] emission line and Equation from \citet{Kewley2004} in M$_{\odot}$yr$^{-1}$
\item \emph{SFR([OII],Z)}: SFR using the [O\textsc{ii}] emission  and abundances values with Equation from \citet{Kewley2004} in M$_{\odot}$yr$^{-1}$
\item \emph{SFR(FUV)}: SFR using far ultra-violet emission from GALEX in M$_{\odot}$yr$^{-1}$
\end{itemize}

Objects only have a classification if they have been classified using a BPT plot. The letter h, m, l which appear after some of the object names indicate there was more than one spectrum within the split and stand for high, middle and low (the position within the slit). As it was not known which of spectra were the intended target, all of the spectra were reduced and analysed.



\addcontentsline{toc}{chapter}{References}
\bibliographystyle{apj}
\bibliography{refs}

\end{document}